# Path-Controlled Secure Network Coding

Masahide Sasaki,  Te Sun Han, *Life Fellow, IEEE*,  Mikio Fujiwara,  Kai Li,  Oliver Hambrey, and  Atsushi Esumi

*Abstract*—Multicast for securely sharing confidential data among many users is becoming increasingly important. Currently, it relies on duplicate-and-forward routing and cryptographic methods based on computational security. However, these approaches neither attain multicast capacity of the network, nor ensure long-term security against advances in computing (information-theoretic security: ITS). Existing ITS solutions—quantum key distribution (QKD), physical layer security (PLS), and secure network coding (SNC)—still fail to enable scalable networks, as their underlying assumptions, such as trusted nodes and wiretap thresholds, gradually become invalid as the network grows. Here, we develop an efficient multi-tree multicast path-finding method to address this issue, integrating it with universal strongly ramp SNC. This system, path-controlled universal strongly ramp SNC (PUSNEC), can be overlaid onto QKD/PLS networks, enabling multicast capacity, ITS, and scalability. We derive the maximum leakage information to an eavesdropper under the probabilistic wiretap network assumption and demonstrate secure multicast in multi-hop networks through numerical simulations. Our quantitative analysis of the secrecy-reliability tradeoff highlights a practical approach to achieving secure, reliable multicast on a global scale.

## I. INTRODUCTION

**T**HE demand for secure multicast of confidential data has increased in both terrestrial and space networks. While established unicast techniques can be adapted for reliable and secure multicast in principle, such an extension is inefficient. IP multicast, utilizing duplicate-and-forward routing on distribution trees, offers more efficiency than repeated unicast use but still cannot achieve multicast capacity defined by the maximum flow of the network [1]. IPsec protocols can provide security protection but fall short of ensuring information-theoretic security (ITS), leaving them vulnerable to "store now, decrypt later" attacks with advanced computing.

To achieve ITS, quantum key distribution (QKD) [2]–[4] and physical layer security (PLS) [5]–[8] offer viable solutions. QKD enables the generation of keys that, when used with a one-time pad (OTP), guarantee perfect secrecy, making eavesdropping impossible regardless of any hardware technology and computational power. PLS, based on channel models that limit eavesdropper's access, is effective for directional wireless channels and often achieves a higher secrecy rate than QKD [9]–[16].

Since QKD and PLS themselves are basically point-to-point link technology, networking is made by concatenating QKD/PLS links via *trusted nodes* and by relaying OTP-encrypted packets in a multi-hop fashion. QKD networks have expanded from intracity optical fiber [17]–[19] to continental [20], and space-based implementations [21]–[23]. PLS networks are expected to be deployed in next-generation wireless networks [24].

However, applying a QKD/PLS-based network (QKD/PLSN) to a multicast distribution tree presents significant challenges. If the trusted node condition is breached (*node compromise*) even in a single relay node, the multicast fails entirely. As the network expands, the risk of node compromises—whether due to natural disasters or security breaches—grows, making complete trust in all nodes progressively impractical. Additional methods that relax the trusted node requirement, such as distributed relay with path redundancy [25]–[27], should be considered. However, achieving efficient multicast remains an unsolved challenge.

A promising approach is to introduce secure network coding (SNC) to the QKD/PLSN. SNC realizes multicast capacity and ITS under the wiretap network assumption, where an eavesdropper (Eve) can tap a limited number of links up to a certain threshold $\mu$. Basic theoretical frameworks of SNC were established about a decade ago [28]–[33].

A source node (Alice) encodes a message into multiple packets for secrecy protection and error-control, which are then distributed to terminal nodes (Bobs) via intermediate nodes (Charlies). Charlies perform network coding by linearly combining incoming packets and forwarding them to other nodes [34]–[36], enabling multicast capacity [1], [34].

When a maximum rank distance (MRD) code [37], [38] is employed at Alice, secrecy and error-control coding can be designed independently of network coding at Charlies (the *universality* property), enabling efficient error-control capability for multicast [32].

This scheme was extended to a ramp scheme in [39], which improves the coding rate while allowing Eve to obtain partial information about the message, yet prevents her from knowing the message itself. This property can be rigorously characterized by *strong ramp secrecy* in [40] (see Supplementary Note 1 in [41] on related works). We call the scheme *universal strongly ramp* SNC (usr-SNC).

While the usr-SNC enhances multicast security and efficiency, its application in multi-hop networks presents additional challenges. In such networks, where the total number of links far exceeds Alice's outgoing links, the risk of wiretapping

This work was supported in part by "Research and Development for Construction of a Global Quantum Cryptography Network (JPJ008957)" in "R&D of ICT Priority Technology (JPMI00316)" of Ministry of Internal Affairs and Communication (MIC), Japan. *(Corresponding author: Masahide Sasaki.)*

M. Sasaki, T. S. Han and M. Fujiwara are with National Institute of Information and Communications Technology (NICT), Nukui-kitamachi 4-2-1, Koganei, Tokyo,184-8795, Japan (email: psasaki@nict.go.jp).

K. Li and A. Esumi are with Siglead Inc., Nakagawachuo 1-38-10, Tsuzuki, Yokohama, 224-0003, Japan (email: atsushi.esumi@siglead.com).

O. Hambrey is with Siglead Europe Ltd., The TechnoCentre, Puma Way, Coventry, CV1 2TT, United Kingdom.

beyond the threshold $\mu$ increases as the number of hops grows. Indeed, if a node is compromised, all its incoming links are eavesdropped, allowing Eve, who can launch node-based attacks, to easily jeopardize the wiretap network assumption. Increasing the number of Alice's outgoing links (equivalently the MRD code length) does not necessarily enhance security; on the contrary, it may even elevate wiretap risks because the number of Charlies also increases, thereby amplifying the probability of node compromise.

This dilemma highlights the need for careful network topology design. However, most existing research on the usr-SNC has primarily focused on identifying effective codes for general network topologies, while investigations into wiretap risks and their relationship with network structure have been largely lacking. In fact, the scalability of usr-SNC in large-scale networks has yet to be clearly demonstrated, posing a significant barrier to its practical application.

We address this issue by developing an efficient multicast path-finding method. In usr-SNC-based multicast, Alice must establish multiple link-disjoint paths to each Bob. To mitigate the risk of node compromise, it is essential to control the number of input links (indegree) to each Charlie below a certain threshold. Consequently, constrained multi-tree multicast path finding is required. However, such path finding problems are generally NP-hard [42], [43]. Even heuristic algorithms have not been reported so far.

To overcome this challenge, we develop a simple yet versatile path-finding method tailored to emerging network infrastructure, such as reconfigurable optical fiber networks based on wavelength-division multiplexing (WDM) switching and mega-constellation satellite networks in low-earth orbits (LEOs), which support rich and flexible connectivity. Specifically, we introduce the concept of a cross-hop (X-hop) grid with an indegree of less than three and develop an algorithm to minimize the total number of links called rank-preserving upstream path search (RAPUS). Importantly, network coding dramatically simplifies this multicast path finding.

By integrating RAPUS-based path control with the usr-SNC, referred to as path-controlled usr-SNC (PUSNEC), and overlaying the PUSNEC on QKD/PLSNs, scalable multicast networks can be realized that achieve ITS and exhibit high resilience against node compromise (Supplementary Note 2-Section II in [41]).

We develop methods for quantifying information leakage to Eve, based on a probabilistic node-compromise model that goes beyond the conventional threshold assumption. This approach allows us to select appropriate coding specifications and multicast paths, based on the actual state of the network. We conduct extensive numerical simulations in a global multicast network model and demonstrate the feasibility of reliable multicast with ITS.

## II. OVERVIEW OF PUSNEC

Our objective is to achieve secure and reliable 1-to-$N$ multicast transmission from Alice to Bob 1 through Bob $N$, who are scattered across diverse directions and distances. This is accomplished by constructing distributed paths that traverse Charlies and reach each Bob, and then applying appropriately designed usr-SNC atop these paths (see Supplementary Note 2 in [41] for the background and a basic introduction to the PUSNEC).

We make the following assumptions throughout the system:

i. Alice's and Bobs' nodes work as the trusted nodes.

ii. The message in each link is encrypted by a key generated by QKD or PLS in the OTP manner. For simplicity, each link has the same unit capacity, and the necessary number of keys are available for each link.

iii. Charlies are designed to be the trusted nodes. However, each of them may be compromised with a small but nonzero probability. Once a node is compromised, Eve gets the OTP keys and reads all packets sent to that node.

iv. A transmitter and a receiver in each link are authenticated by appropriate methods, ensuring no spoofing occurs at any node [1].

### A. Construction of multicast graph

The first step is to establish a set of *indegree-constrained* multicast paths, in which Alice is connected to each Bob via $n_0$ *link-disjoint* paths. The parameter $n_0$ is referred to as the degree of distribution. Note that paths arriving at *different* Bobs may be link-joint, node-joint, or both. The algorithm is intentionally designed to allow such sharing, thereby minimizing the total number of links in the network. The resulting $N$ sets of paths collectively form a multi-tree multicast graph $\mathcal{G} = (\mathcal{V}, \mathcal{E})$, where $\mathcal{V}$ is the set of nodes and $\mathcal{E}$ is the set of links (edges). The details of RAPUS-based path finding are explained later in Section IV.

Note that the number of outgoing links from Alice (the outdegree denoted by $n_1$) is not necessarily equal to the distribution degree $n_0$; rather, $n_1 \geq n_0$, and tends to increase as the spatial spread of Bobs widens.

This behavior stands in contrast to the conventional universal SNC [32] and its ramp extension [39], which do not explicitly take network topology into account, where the outdegree from Alice was assumed to be fixed at $n_0$.

### B. Encoding at Alice

Let $\boldsymbol{u}$ denote the message that Alice wishes to multicast to Bob 1 through Bob $N$. This message may represent a codeword obtained by source encoding a given source message. To facilitate distributed transmission, Alice represents the message as a row vector and divides it into $k_0$ blocks $\boldsymbol{u} = \begin{bmatrix} u_0 & u_1 & \ldots & u_{k_0-1} \end{bmatrix}$, where each block $u_i$ is an element of the extension field $\mathbb{F}_{q^m}$ of degree $m$.

Since every extension field $\mathbb{F}_{q^m}$ can be regarded as a vector space over the ground field $\mathbb{F}_q$, each block $u_i$ can be viewed as an $m$-dimensional column vector over $\mathbb{F}_q$ and is referred to as a *packet* (of length $m$). In addition, Alice prepares a random sequence $\boldsymbol{r} = \begin{bmatrix} r_0 & r_1 & \ldots & r_{\mu_0-1} \end{bmatrix}$ called the *masking* key to deceive Eve, where each $r_i \in \mathbb{F}_{q^m}$.

---

[1]One such method is the Wegman-Carter message authentication scheme based on a trusted third-party verifier in [44], which provides ITS and network scalability

Both $u_i$'s and $r_i$'s are assumed to be independent and uniformly distributed over $\mathbb{F}_{q^m}$. The length of the sequences $\begin{bmatrix} \boldsymbol{u} & \boldsymbol{r} \end{bmatrix}$ is denoted as $k \equiv k_0 + \mu_0$, referred to as the message dimension. These sequences form a set of row vectors of length $k$ over $\mathbb{F}_{q^m}$, which we denote as $\mathbb{F}_{q^m}^k$.

Alice first performs the secrecy and error-control coding based on MRD codes over $\mathbb{F}_{q^m}$ as an *outer* code. As for MRD codes, a well-studied subclass, Gabidulin code, is adopted.

Let $\text{Gab}[n, k]$ denote a Gabidulin code of length $n$ ($\leq m$) and message dimension $k$ ($\leq n$), constructed using a *linearized* polynomial $f(x) = f_0 x^{[0]} + f_1 x^{[1]} + \cdots + f_{k-1} x^{[k-1]}$ [45] and a set of evaluation points $g_0, g_1, ..., g_{n-1} \in \mathbb{F}_{q^m}$ that are linearly independent over the ground field $\mathbb{F}_q$. Here, the notation $[i] \equiv q^i$ indicates the $i$-th Frobenius $q$-power. The generator matrix is given by

$$G = \begin{bmatrix} g_0^{[0]} & g_1^{[0]} & \cdots & g_{n-1}^{[0]} \\ g_0^{[1]} & g_1^{[1]} & \cdots & g_{n-1}^{[1]} \\ \vdots & \vdots & \ddots & \vdots \\ g_0^{[k-1]} & g_1^{[k-1]} & \cdots & g_{n-1}^{[k-1]} \end{bmatrix} \in \mathbb{F}_{q^m}^{k \times n}. \quad (1)$$

The code length is set to $n = k_1 + n_0$, where $k_1 \geq k_0$. The value of $k_1$ is chosen to minimize decoding complexity for $\text{Gab}[n, k]$, by ensuring that the integer pair $(q, n)$ supports the use of *optimal self-dual normal bases*, which greatly simplify arithmetic operations. Ideally, if such bases are available when $k_1 = k_0$, strong ramp secrecy can be achieved with minimal code length. However, these bases exist only for specific combinations of $(q, n)$; thus, given $k_0$ and $n_0$, we select the shortest possible $n$ that yields Gabidulin codes satisfying the required conditions, referred to as *good* Gabidulin codes (see Supplementary Note 3 in [41]).

In contrast, the original Silva-Kschischang scheme [32] sets the code length to $n = n_0$ and defines the coefficient vector as $\boldsymbol{f} = \begin{bmatrix} \boldsymbol{u} & \boldsymbol{r} \end{bmatrix}$. This design, however, does not always guarantee strong ramp secrecy by the same reason in [46]. Kurihara et al. proposed to use a $[k_0 + n_0, k]$ MRD code to solve this issue [39], following a similar approach to the Reed–Solomon code construction by Nishiara and Takizawa [47].

We extend this concept to good Gabidulin codes. The outer encoding proceeds in two steps.

*1) Pre-encoding:* Alice's message $\boldsymbol{u}$ and the masking key $\boldsymbol{r}$ are converted to the coefficient vector as

$$\boldsymbol{f} = \begin{bmatrix} \boldsymbol{u} & \boldsymbol{r} \end{bmatrix} G_1^{-1}, \quad (2)$$

with the $k \times k$ $q$-Vandermonde matrix $G_1$ consisting of the first $k$ columns of $G$. This matrix is nonsingular because $g_0, g_1, ..., g_{k-1}$ are linearly independent [48, Lemma 3.15], and hence Eq. (2) specifies 1-to-1 relation between $\begin{bmatrix} \boldsymbol{u} & \boldsymbol{r} \end{bmatrix}$ and $\boldsymbol{f}$.

*2) Encoding for distributed transmission:* The codeword for $n_0$-degree distribution is generated by $\text{Gab}[n_0, k]$ as

$$\boldsymbol{x} = \boldsymbol{f} G_2 \in \mathbb{F}_{q^m}^{n_0}, \quad (3)$$

where $G_2$ consists of the last $n_0$ columns of $G$.

Alice further applies on $\boldsymbol{x}$, random linear network coding over the ground field $\mathbb{F}_q$ ($\mathbb{F}_q$-RLNC) as an *inner* code generating a codeword $\boldsymbol{w} = \begin{bmatrix} w_0 & w_1 & \ldots & w_{n_1-1} \end{bmatrix}$ with

$$w_j = \sum_{i=0}^{n_0-1} c_i^{(j)} x_i, \quad j = 0, ..., n_1 - 1, \quad c_i^{(j)} \in \mathbb{F}_q, \quad (4)$$

and sends out the $\mathbb{F}_q$-RLNC packets into $n_1$ links. Eq. (4) is denoted as $\boldsymbol{w} = \boldsymbol{x} R_A$ with an $n_0 \times n_1$ random matrix $R_A$. This final process plays a key role in simplifying the RAPUS-based path-finding algorithm (see Section IV).

Note that Eqs. (2) and (3) can be jointly written as

$$\boldsymbol{x}' = \boldsymbol{f} G, \quad (5)$$

where $\boldsymbol{x}' \in \mathbb{F}_{q^m}^n$ denote an output sequence from the $\text{Gab}[n, k]$ encoder specified by $G$. Importantly, the first $k_1$ symbols of $\boldsymbol{x}'$ are deliberately withheld by Alice, and hence treated as erasures by each Bob's $\text{Gab}[n, k]$ decoder (see Supplementary Note 4 in [41]).

### C. Relaying at each Charlie

Charlies ($v$'s) also apply $\mathbb{F}_q$-RLNC on incoming packets, as the *inner* code. Let $in(v)$ and $out(v)$ be the sets of incoming and outgoing links, and $|in(v)|$ and $|out(v)|$ be the numbers of them, called the indegree and outdegree of $v$, respectively. Each Charlie should obey the constraint

$$|in(v)| < k, \quad (6)$$

to maintain tolerance against node-based Eve. Some Charlies execute $\mathbb{F}_q$-RLNC such that $|in(v)| < |out(v)|$, i.e., *link amplification*, in order to support multicast. For every link, the packet transmitted over it can be expressed as a linear combination of $w_0, w_1, \cdots, w_{n_1-1}$, and hence of $x_0, x_1, \cdots, x_{n_0-1}$. A set of coefficients of this linear combination in terms of $\boldsymbol{x}$ is called *global encoding vector* of the link, denoted as $\boldsymbol{c} = \begin{bmatrix} c_0 & c_1 & \ldots & c_{n_0-1} \end{bmatrix}$ where $c_i \in \mathbb{F}_q$. Let input and output links at Charlie $v$ be indexed by $v_1, ..., v_I$ and $v_1', ..., v_O'$, where $|in(v)|$ and $|out(v)|$ are denoted as $I$ and $O$, respectively. A packet in each link is represented as $\boldsymbol{x} \cdot \boldsymbol{c}_{v_i}^T$. The encoding operation can be expressed as the transformation of global encoding vectors $\begin{bmatrix} \boldsymbol{c}_{v_1'} \\ \vdots \\ \boldsymbol{c}_{v_O'} \end{bmatrix} = R_\nu \begin{bmatrix} \boldsymbol{c}_{v_1} \\ \vdots \\ \boldsymbol{c}_{v_I} \end{bmatrix}$, where $R_v$ is an $O \times I$ matrix with random number elements over $\mathbb{F}_q$. The global encoding vector $\boldsymbol{c}$ is attached to the packet, and sent to Bobs.

### D. Decoding at each Bob

Every Bob collects $n_0$ packets, which should be linearly independent over $\mathbb{F}_q$. The codeword that arrives at Bob $t$ from Alice can be generally represented as $\boldsymbol{x} A_t \in \mathbb{F}_{q^m}^{n_0}$, where $A_t$ is an $n_0 \times n_0$ matrix over $\mathbb{F}_q$ and called the transfer matrix. Packet erasures occur in the network, causing rank deficiency which is characterized by $\rho = n_0 - \min_t \{\text{rank } A_t\}$. Packets are corrupted by errors, which cause rank errors in Bobs. We model it in such a way that an erroneous packet is added to the original one on at most $\tau$ links. The received word at Bob $t$ is finally expressed as

$$\boldsymbol{y}_t = \boldsymbol{x}A_t + \boldsymbol{e}_0 D_t \in \mathbb{F}_q^{n_0}, \tag{7}$$

where $D_t \in \mathbb{F}_q^{\tau \times n_0}$ corresponds to the overall linear transformation applied to the erroneous packets $\boldsymbol{e}_0 \in \mathbb{F}_{q^m}^{\tau}$.

A received word $\boldsymbol{y}_t$ of length $n_0$ is first decoded by the $\mathbb{F}_q$-RLNC decoder, which solves $\boldsymbol{y}_t = \tilde{\boldsymbol{y}}_t A_t$ using Gaussian elimination, and is then converted into an outer word $\tilde{\boldsymbol{y}}_t$ of length $n_0$. The $\tilde{\boldsymbol{y}}_t$ is then decoded by the $[n, k]$ Gabidulin decoder, which deals with $\tilde{\boldsymbol{y}}_t$ as a part of the length-$n$ word

$$\boldsymbol{y}_t' = \begin{bmatrix} \boldsymbol{0} & \tilde{\boldsymbol{y}}_t \end{bmatrix} \in \mathbb{F}_{q^m}^n \tag{8}$$

with $k_1$ erasures expressed by $\boldsymbol{0}$, and finally outputs an estimated message $\hat{\boldsymbol{u}}$. The Gab$[n, k]$ decoder can correct $\tau$ rank errors and $\rho$ rank erasures under the condition

$$2\tau + \rho \le n_0 - k, \tag{9}$$

independent of network coding schemes used in Charlies [32]. The details of Gabidulin decoder are explained in Supplementary Note 3 in [41].

### E. Wiretapping by Eve

Eve obtains $\mu$ linearly independent packets over $\mathbb{F}_q$ arriving through the incoming links of compromised Charlies in a noiseless and lossless manner. This idealization considers the cryptographically worst-case scenario, which ensures a strictly guaranteed level of security. Given a fixed $\mu$, the combination patterns of those incoming links are indexed by $\zeta = 1, ..., \nu_\mu$, where $\nu_\mu$ represents the total number of packet configurations.

A wiretapped word can be expressed as

$$\boldsymbol{z}_\zeta^{(\mu)} = \boldsymbol{x} B_\zeta^{(\mu)}, \tag{10}$$

where $B_\zeta^{(\mu)} \in \mathbb{F}_q^{n_0 \times \mu}$ is the wiretap matrix, which is defined in terms of the outer codeword $\boldsymbol{x}$ (not the inner one $\boldsymbol{w}$). Eve can apply any operations on $\boldsymbol{z}_\zeta^{(\mu)}$ over $\mathbb{F}_{q^m}$ with the knowledge of the outer and inner codes which are open to public, while the message $\boldsymbol{u}$ and the masking key $\boldsymbol{r}$ are kept secret from Eve. Eve can tamper with packets, and its effect can be regarded as already accounted for in the erroneous packets in the second term of Eq. (7).

## III. Characterizing Secrecy and Reliability: Probabilistic Wiretap Network Model

Existing studies in SNC often assume a fixed number of wiretapped links $\mu$. However, such threshold-based wiretap models offer limited guidance for practical coding and routing in real-world network operations.

To address this limitation, we introduce a probabilistic model that incorporates node compromise rates $\gamma_v$ for Charlies $v = 1, ..., N_C$, where $N_C$ is the total number of Charlies, as well as erasure rates $\epsilon_i$ and error rates $e_i$ for link $i = 1, ..., |\mathcal{E}|$, where $|\mathcal{E}|$ is the total number of links. We assume that erasure and error events on each link, as well as node compromise events at each Charlie, occur in a statistically independent manner. The node compromise rates $\vec{\gamma} = (\gamma_1, ..., \gamma_{N_C})$ can be quantitatively estimated based on historical network failure records and the level of node protection technologies.

Reliability is measured by the probability that one or more Bobs fail to recover the message, as defined by

$$e_F = \Pr\{\hat{\boldsymbol{u}} \ne \boldsymbol{u} \text{ at one or more Bobs}\}, \tag{11}$$

which we call *frame error rate (FER)*. The $e_F$ depends on $(\epsilon_1, ..., \epsilon_{|\mathcal{E}|})$ and $(e_1, ..., e_{|\mathcal{E}|})$ as well as $n_0$, $k$, and the multicast graph $\mathcal{G}$.

The size of the wiretapped word varies probabilistically within the range $0 \le \mu \le |\mathcal{E}|$, depending on each multicast transmission. Let $Z'$ denote a random variable for all possible Eve's observations [2]. Its realization $\boldsymbol{z}'$ is drawn from a set

$$\boldsymbol{z}' \in \left\{ \boldsymbol{z}_{\zeta,1}^{(\mu)}, ..., \boldsymbol{z}_{\zeta,q^{m\mu}}^{(\mu)} \middle| 0 \le \mu \le |\mathcal{E}|; \zeta = 1, ..., \nu_\mu \right\}, \tag{12}$$

where $\boldsymbol{z}_{\zeta,i}^{(\mu)}$ represents each realization of $\boldsymbol{z}_\zeta^{(\mu)}$.

The probability of occurrence of $B_\zeta^{(\mu)}$ generally depends not only on $\vec{\gamma} = (\gamma_1, ..., \gamma_{N_C})$, but also on the graph $\mathcal{G}$, as well as the coding parameters $n_0$ and $k$. Therefore, it is denoted as $\pi_\zeta^{(\mu)}(\vec{\gamma}, \mathcal{G}, n_0, k)$. We define the $\mu$-wiretap probability as

$$
\begin{aligned}
p(\mu, \vec{\gamma}, \mathcal{G}, n_0, k) &= \Pr\{\mu \text{ packets are wiretapped}\}, \\
&= \sum_{\zeta=1}^{\nu_\mu} \pi_\zeta^{(\mu)}(\vec{\gamma}, \mathcal{G}, n_0, k). \tag{13}
\end{aligned}
$$

Regarding secrecy metrics, several candidates are known, such as the mutual information $I(U; Z')$, the divergence distance $\delta(U; Z')$, and the variational distance $\partial(U; Z')$ between random variables $U$ at Alice and $Z'$ at Eve. Since $\delta(U; Z') \to 0$ implies $I(U; Z') \to 0$ and $\partial(U; Z') \to 0$, the divergence distance is the strongest secrecy metric [49]. Therefore, we adopt the divergence distance. In the case of a noiseless wiretap channel, where $\boldsymbol{u}$ and $\boldsymbol{r}$ are independently and uniformly distributed, this metric coincides with the mutual information $I(U; Z')$ (Supplementary Note 5 in [41]).

Strong ramp secrecy is then characterized by the mutual information between Eve's observation $\boldsymbol{z}'$ and an arbitrarily chosen submessage of $\xi$ symbols $\boldsymbol{u}_{\{\xi\}} = \begin{bmatrix} u_{i_1} & \cdots & u_{i_\xi} \end{bmatrix}$ $(1 \le \xi \le k_0)$

$$I(U_{\{\xi\}}; Z') = \sum_{\mu=0}^{|\mathcal{E}|} I^{(\mu)}(U_{\{\xi\}}; Z'), \tag{14}$$

where

$$
\begin{aligned}
&I^{(\mu)}(U_{\{\xi\}}; Z') \\
&= \frac{1}{q^{m\xi}} \sum_{\boldsymbol{u}_{\{\xi\}}} \sum_{\zeta=1}^{\nu_\mu} \sum_{\boldsymbol{z}_\zeta^{(\mu)}} \pi_\zeta^{(\mu)} P_{Z_\zeta^{(\mu)}|U_{\{\xi\}}}(\boldsymbol{z}_\zeta^{(\mu)}|\boldsymbol{u}_{\{\xi\}}) \\
&\quad \times \log \frac{P_{Z_\zeta^{(\mu)}|U_{\{\xi\}}}(\boldsymbol{z}_\zeta^{(\mu)}|\boldsymbol{u}_{\{\xi\}})}{\frac{1}{q^{m\xi}} \sum_{\boldsymbol{u}_{\{\xi\}}} P_{Z_\zeta^{(\mu)}|U_{\{\xi\}}}(\boldsymbol{z}_\zeta^{(\mu)}|\boldsymbol{u}_{\{\xi\}})}, \tag{15}
\end{aligned}
$$

with the reduced $(\mu, \zeta)$-wiretap channel matrix

$$\hat{P}_{Z_\zeta^{(\mu)}|U_{\{\xi\}}} \equiv \left[ P_{Z_\zeta^{(\mu)}|U_{\{\xi\}}}(\boldsymbol{z}_\zeta^{(\mu)}|\boldsymbol{u}_{\{\xi\}}) \right], \tag{16}$$

---

[2] The notation $Z$ is reserved for Eve's variable in conventional threshold-based wiretap model. See Appendix and Supplementary Note 12 in [41].

which is derived from the original channel matrix by

$$P_{Z_\zeta^{(\mu)}|U_{\{\xi\}}}(\boldsymbol{z}_\zeta^{(\mu)}|\boldsymbol{u}_{\{\xi\}})$$
$$= \frac{1}{q^{m(k-\xi)}} \sum_{\boldsymbol{u}_{\{k_0-\xi\}}} \sum_{\boldsymbol{r}} P_{Z_\zeta^{(\mu)}|UR}(\boldsymbol{z}_\zeta^{(\mu)}|\boldsymbol{u},\boldsymbol{r}). \quad (17)$$

The variables of the function $\pi_\zeta^{(\mu)}$ are suppressed to simplify notation.

**Theorem 1.** *The maximum information on $U_{\{\xi\}}$ available for Eve per single shot use of the probabilistic wiretap channel $P_{Z'|U_{\{\xi\}}}$ is given by*

$$I(U_{\{\xi\}}; Z') = I_L(\vec{\gamma}, \mathcal{G}, n_0, k, \xi) \log_2 q^m, \quad (18)$$

*with the leakage information index (LII)*

$$I_L(\vec{\gamma}, \mathcal{G}, n_0, k, \xi) = I_R(\vec{\gamma}, \mathcal{G}, n_0, k, \xi) + \xi p_L(\vec{\gamma}, \mathcal{G}, n_0, k), \quad (19)$$

*the ramp leakage information index*

$$I_R(\vec{\gamma}, \mathcal{G}, n_0, k, \xi) = \sum_{\mu=k-\xi}^{k-1} (\mu - k + \xi) p(\mu, \vec{\gamma}, \mathcal{G}, n_0, k), \quad (20)$$

*and the perfect leakage probability (PLP)*

$$p_L(\vec{\gamma}, \mathcal{G}, n_0, k) = \sum_{\mu=k}^{|\mathcal{E}|} p(\mu, \vec{\gamma}, \mathcal{G}, n_0, k). \quad (21)$$

*The maximum information $I(U_{\{\xi\}}; Z')$ is attained by the coset-based random guess strategy.*

*Proof.* See Appendix. □

Analytical evaluations of LII and FER in the case of multicast over disjoint multi-hop paths are provided in Supplementary Note 6 in [41].

## IV. CONSTRAINED MULTI-TREE MULTICAST PATH FINDING

The task is to construct a multi-tree multicast graph $\mathcal{G}^{(N)}$ consisting of $N$ sets of $n_0$ link-disjoint paths from Alice to each Bob, while minimizing the total number of links under the indegree constraint at each Charlie. To address this NP-hard problem, we propose a heuristic method that first defines a virtual network, *the X-hop grid*, based on the underlying real network, derives a multi-tree multicast graph $\tilde{\mathcal{G}}^{(N)}$, and finally maps it back onto the real network structure, yielding a solution $\mathcal{G}^{(N)}$.

The graph $\mathcal{G}$ may be derived from a larger base graph $\mathcal{G}_\mathcal{B}$, or even be configured over a given base set of nodes $\mathcal{V}_\mathcal{B}$ by identifying appropriate connecting paths. A typical example of the former arises in reconfigurable optical fiber networks employing WDM switching, while the latter is particularly relevant to mega-constellation satellite networks in LEOs.

We assume that the following properties/capabilities for $\mathcal{G}_\mathcal{B}$ or $\mathcal{V}_\mathcal{B}$, which can be met with the above examples:

1. There are sufficiently many candidate nodes for Charlies who actually participate in the multicast relay.
2. An average distance between nearest candidate nodes is much less than the maximum link distance over which QKD or PLS can be established.

3. Each candidate node is equipped with transmitters and receivers which can support multiple laser communication channels or highly directional microwave channels for QKD/PLS links, typically 2-input-2-output or 1-input-3-output links.

Figure 1 illustrates the four main stages of the proposed method in the terrestrial network model:

a) **Real nodes** (Fig. 1a)
A base set of nodes $\mathcal{V}_\mathcal{B}$ including Alice, Bobs, and many intermediate nodes is given with their geographical position information. Bobs are initially numbered as Bob $1, ..., \bar{N}$ from closest to farthest to Alice.

b) **X-hop grid** (Fig. 1b)
- We construct the X-hop grid consisting of nodes with at most four links, covering the area across Alice and Bobs in $\mathcal{V}_\mathcal{B}$. The grid constant is kept within the QKD/PLS maximum link distance.
- Alice and Bobs in $\mathcal{V}_\mathcal{B}$ are mapped onto their closest nodes in the X-hop grid.
- Sparse areas often exist in $\mathcal{V}_\mathcal{B}$ which paths cannot traverse. So, for each *virtual* intermediate node, we look for a closest *real* node in $\mathcal{V}_\mathcal{B}$ within the grid constant and assign it a candidate of Charlies. The assigned node is excluded from next search onward. If no real node exists, the virtual node and connected links are removed from the X-hop grid.

c) **Multicast path finding** (Fig. 1c)
- The multicast path-finding is performed on the X-hop grid, effectively narrowing the solution space to graphs that satisfy the indegree constraint Eq. (6) with $k = 3$, thus reducing overall search complexity.
- The path-finding process begins by applying the algorithm *RAPUS* to each candidate Bob to find $n_0$ link-disjoint paths to Alice with minimum hops on the X-hop grid. After executing RAPUS for all $N$ Bobs, the one requiring the minimum total number of links is designated as Bob 1, yielding the graph $\tilde{\mathcal{G}}^{(1)}$. Details of the RAPUS are provided in Subsection IV-A.
- For each of the remaining $N - 1$ Bobs, the same procedure is applied. The Bob with the fewest additional links beyond $\tilde{\mathcal{G}}^{(1)}$ is designated as Bob 2, forming $\tilde{\mathcal{G}}^{(2)}$.
- This process is repeated for the remaining Bobs, and the multicast graph grows progressively. Eventually a 1-to-$N$ multicast graph $\tilde{\mathcal{G}}^{(N)}$ is established.

d) **Mapping to real network** (Fig. 1d)
The virtual graph $\tilde{\mathcal{G}}^{(N)}$ on the X-hop grid is mapped back to the graph $\mathcal{G}^{(N)}$ on the real network model, by picking up the closest node in $\mathcal{V}_\mathcal{B}$ for each node in $\tilde{\mathcal{G}}^{(N)}$, thus determining Charlies, and assigning a set of links $\mathcal{E}$ to form the real paths $\mathcal{P}_1^{(t)}, ..., \mathcal{P}_{n_0}^{(t)}$ for $t = 1, ..., N$.
In the derived $\mathcal{G}^{(N)}$, there are often continuous segments of links without any branches or joints, which are unnecessarily bent. Such segments can be smoothed by appropriately reselecting a smaller number of nearby nodes to create straighter ones. Finally, a smoother graph $\tilde{\mathcal{G}}^{(N)}$ with fewer hops can be generated. Such an example is given in the LEO satellite network in Section V.

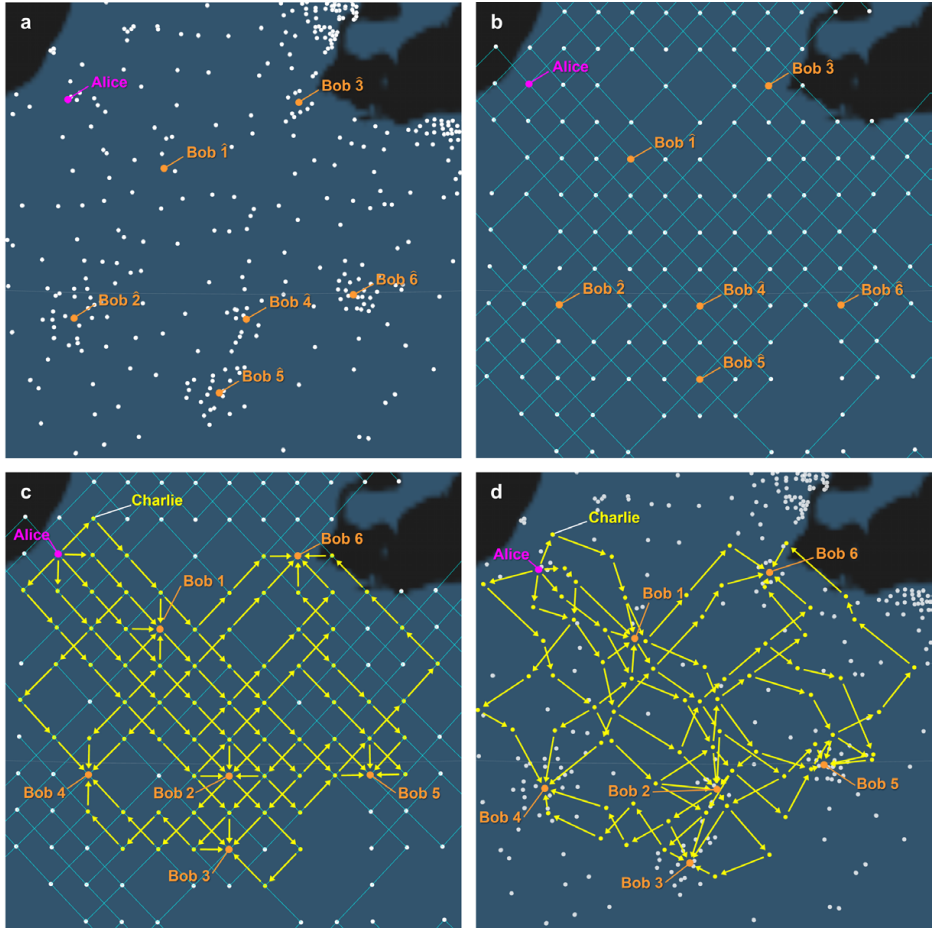

Fig. 1. **Outline of multicast path finding based on the X-hop grid.** Depicted is a case of 1-to-6 directional multicast with $n_0 = 5$. **a** A base set of nodes $\mathcal{V}_\mathcal{B}$ of a real network model in North America. Alice is in South Bend. Bob $\hat{1} \sim \hat{6}$ are in Fort Wayne, Indianapolis, Toledo, Dayton, Cincinnati, Columbus. The white dots are candidate nodes for Charlies. **b** The X-hop grid with some sparse areas. **c** A multicast graph on the X-hop grid obtained by the RAPUS-based path finding. The yellow arrows represent QKD/PLS links constituting the graph. **d** The multicast graph mapped back from the virtual one in the X-hop grid to the real network model.

### A. RAPUS-based multicast path finding algorithm

**Main flow**

A main flow diagram of multicast path finding is shown in Fig. 2a. At the beginning, we add 4 links between Alice and her nearest nodes to the X-hop grid. The outdegree of Alice, $n_1$, is automatically determined in the range of $n_0 \leq n_1 \leq 8$ when the multicast path finding is completed for all Bobs. (Figure 1 corresponds to the case of $n_1 = n_0 = 5$.)

Alice transmits the $\mathbb{F}_q$-RLNC packets in Eq. (4) over $n_1$ links. Unlike directly sending $x_0, x_1, ..., x_{n_0-1}$ over $n_0$ links, the $\mathbb{F}_q$-RLNC approach significantly simplifies path finding, as any link can potentially carry all components $x_0, x_1, ..., x_{n_0-1}$. So, we need not consider individual packet attributes per link, but simply identify $n_0$ link-disjoint paths.

The initial graph $\tilde{\mathcal{G}}^{(0)}$ trivially consists of Alice only. Suppose that the $(t-1)$th multicast graph $\tilde{\mathcal{G}}^{(t-1)}$ from Alice to Bob 1,..., $t-1$ were established and $N-t+1$ Bobs remain unconnected to $\tilde{\mathcal{G}}^{(t-1)}$ yet, those are denoted as Bob

$\hat{t}_1, ..., \hat{t}_{N-t+1}$. Then for each Bob $\hat{t}_i$, we run the RAPUS to produce a graph

$$\tilde{\mathcal{G}}^{(t,\hat{t}_i)} \equiv \tilde{\mathcal{G}}^{(t-1)} \cup \tilde{\mathcal{P}}_1^{(\hat{t}_i)} \cup \cdots \cup \tilde{\mathcal{P}}_{n_0}^{(\hat{t}_i)}. \quad (22)$$

Among $\{\tilde{\mathcal{G}}^{(t,\hat{t}_1)}, ..., \tilde{\mathcal{G}}^{(t,\hat{t}_{N-t+1})}\}$, we select the one whose number of links is minimum. The multicast graph $\tilde{\mathcal{G}}^{(t)}$ from Alice to Bob 1,..., $t$ is thus established.

The above process is performed sequentially from $t = 2, ..., N$, as represented by the nested loops in terms of $t$ and $\hat{t}$ in Fig. 2a, and finally a multicast graph $\tilde{\mathcal{G}}^{(N)}$ is established.

**RAPUS (RAnk-Preserving Upstream path Search)**

A flow diagram of the RAPUS is shown in Fig. 2b. We explain the details of the RAPUS using Fig. 3, which corresponds to the 1-to-6 multicast with $n_0 = 5$ in Fig. 1. Given an input graph $\tilde{\mathcal{G}}^{(t-1)}$ and a Bob $\hat{t}$ unconnected to it, the task is to find a set of $n_0$-link disjoint paths from Alice to Bob $\hat{t}$, $\tilde{\mathcal{P}}_1, ..., \tilde{\mathcal{P}}_{n_0}$, minimizing the number of links unused in $\tilde{\mathcal{G}}^{(t-1)}$. These paths carry $n_0$ linearly independent packets, and Bob must receive enough to *preserve rank* $n_0$ for successful decoding of $\mathbb{F}_q$-RLNC. To ensure this, $n_0$-link disjoint paths are searched *upstream* from Bob, hence the name *rank-preserving upstream path search*.

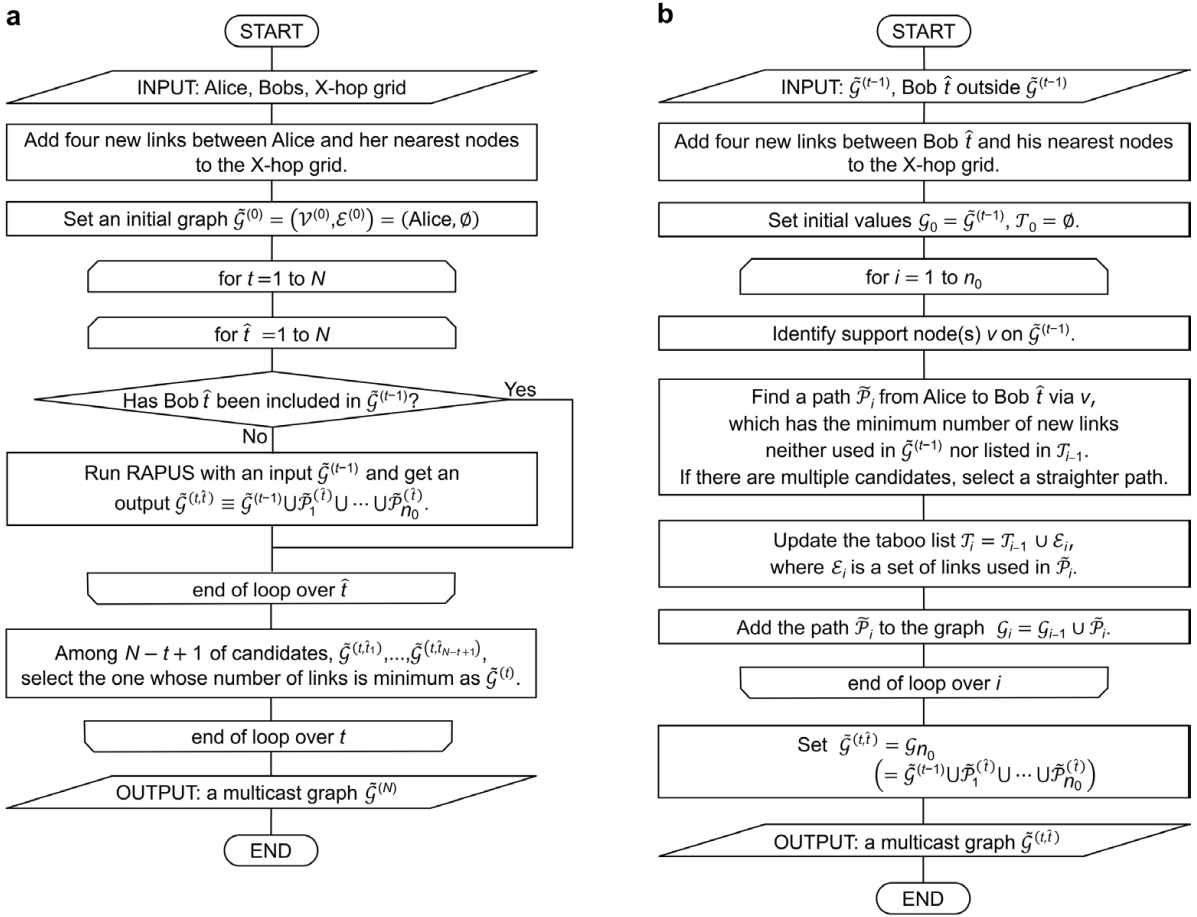

**a**

START

INPUT: Alice, Bobs, X-hop grid

Add four new links between Alice and her nearest nodes to the X-hop grid.

Set an initial graph $\tilde{\mathcal{G}}^{(0)} = (\mathcal{V}^{(0)}, \mathcal{E}^{(0)}) = (\text{Alice}, \emptyset)$

for $t = 1$ to $N$

for $\hat{t} = 1$ to $N$

Has Bob $\hat{t}$ been included in $\tilde{\mathcal{G}}^{(t-1)}$? — Yes

No

Run RAPUS with an input $\tilde{\mathcal{G}}^{(t-1)}$ and get an output $\tilde{\mathcal{G}}^{(t\hat{t})} \equiv \tilde{\mathcal{G}}^{(t-1)} \cup \tilde{\mathcal{P}}_1^{(\hat{t})} \cup \cdots \cup \tilde{\mathcal{P}}_{n_0}^{(\hat{t})}$.

end of loop over $\hat{t}$

Among $N - t + 1$ of candidates, $\tilde{\mathcal{G}}^{(t,1)}, ..., \tilde{\mathcal{G}}^{(t,N-t+1)}$, select the one whose number of links is minimum as $\tilde{\mathcal{G}}^{(t)}$.

end of loop over $t$

OUTPUT: a multicast graph $\tilde{\mathcal{G}}^{(N)}$

END

**b**

START

INPUT: $\tilde{\mathcal{G}}^{(t-1)}$, Bob $\hat{t}$ outside $\tilde{\mathcal{G}}^{(t-1)}$

Add four new links between Bob $\hat{t}$ and his nearest nodes to the X-hop grid.

Set initial values $\mathcal{G}_0 = \tilde{\mathcal{G}}^{(t-1)}$, $\mathcal{T}_0 = \emptyset$.

for $i = 1$ to $n_0$

Identify support node(s) $v$ on $\tilde{\mathcal{G}}^{(t-1)}$.

Find a path $\tilde{\mathcal{P}}_i$ from Alice to Bob $\hat{t}$ via $v$, which has the minimum number of new links neither used in $\tilde{\mathcal{G}}^{(t-1)}$ nor listed in $\mathcal{T}_{i-1}$. If there are multiple candidates, select a straighter path.

Update the taboo list $\mathcal{T}_i = \mathcal{T}_{i-1} \cup \mathcal{E}_i$, where $\mathcal{E}_i$ is a set of links used in $\tilde{\mathcal{P}}_i$.

Add the path $\tilde{\mathcal{P}}_i$ to the graph $\mathcal{G}_i = \mathcal{G}_{i-1} \cup \tilde{\mathcal{P}}_i$.

end of loop over $i$

Set $\tilde{\mathcal{G}}^{(t\hat{t})} = \mathcal{G}_{n_0}$ $\left( = \tilde{\mathcal{G}}^{(t-1)} \cup \tilde{\mathcal{P}}_1^{(\hat{t})} \cup \cdots \cup \tilde{\mathcal{P}}_{n_0}^{(\hat{t})} \right)$

OUTPUT: a multicast graph $\tilde{\mathcal{G}}^{(t\hat{t})}$

END

Fig. 2. **Flow diagrams of multicast path finding on the X-hop grid. a** Main flow diagram. **b** Flow diagram of the RAPUS.

- At the beginning, up to four available links between Bob $\hat{t}$ and the nearest nodes are added to the X-hop grid. Nodes around Bob $\hat{t}$ are grouped into *surrounding node sets* arranged in concentric squares, where the first, second, third surrounding node sets consist of at most 8, 16, 24 nodes, respectively (Fig. 3a). If sparse regions exist near Bob, the corresponding surrounding node sets may contain fewer nodes.

- To find the first path $\tilde{\mathcal{P}}_1$, we run the following processes:

  - We first identify *support* nodes on $\tilde{\mathcal{G}}^{(t-1)}$, denoted by $v_1, ..., v_g$, which can be reached from Bob $\hat{t}$ with the minimum number of hops. This is done by sequentially checking which surrounding node set first connects to $\tilde{\mathcal{G}}^{(t-1)}$. In Fig. 3b, two support nodes $v_1$ and $v_2$ are identified.

  - We next find a path from Alice to Bob $\hat{t}$ via $v_1$ which has the minimum number of new links not used in $\tilde{\mathcal{G}}^{(t-1)}$ yet. We repeat this process for $v_2, ..., v_g$.

  - Among the paths thus found, the one with the minimum number of newly added links is assigned as the first path $\tilde{\mathcal{P}}_1$. If there are multiple candidates (Fig. 3b), the one which is as straight as possible along the X-hop grid is selected (Fig. 3c). If there still remain multiple candidates, one of them is randomly selected.

  - The links used in $\mathcal{P}_1$ are registered to the *taboo list* $\mathcal{T}_1$.

- We search for the second path $\tilde{\mathcal{P}}_2$ in a similar way except that we exclude the links in $\mathcal{T}_1$ from the path finding. Then the support nodes may differ from the previous ones. Once $\tilde{\mathcal{P}}_2$ is determined, the taboo list is updated from $\mathcal{T}_1$ to $\mathcal{T}_2$ by adding the links used in $\tilde{\mathcal{P}}_2$. Using the taboo list is actually the mechanism to find link-disjoint paths (Fig. 3d).

- We repeat the above process until the $n_0$th path $\tilde{\mathcal{P}}_{n_0}$ is determined as in Fig. 3e, f.

- A set of paths thus obtained is output, denoted as $\{\tilde{\mathcal{P}}_1^{(\hat{t})}, ..., \tilde{\mathcal{P}}_{n_0}^{(\hat{t})}\}$.

The RAPUS finally outputs the following union as a multicast graph for Bob $\hat{t}$

$$\tilde{\mathcal{G}}^{(t,\hat{t})} = \tilde{\mathcal{G}}^{(t-1)} \cup \tilde{\mathcal{P}}_1^{(\hat{t})} \cup \cdots \cup \tilde{\mathcal{P}}_{n_0}^{(\hat{t})}. \quad (23)$$

## V. PERFORMANCE OF 5-DEGREE PUSNEC ON A GLOBAL SCALE LEO SATELLITE NETWORK

We present numerical results on the 5-degree PUSNEC over the entire globe using LEO satellite constellations. In this example, Alice is a LEO satellite and Bobs are totally 13 ground stations. Alice prepares 8 $\mathbb{F}_q$-RLNC packets according to Eq. (4) and sends them out into $n_1$ (= 8) links in the omni directions. Each Bob has 5 downlinks from the LEO satellites, and collects 5 packets, but does not relay them to other nodes. It is obviously a waste of resources to perform 5-degree distributions independently in a star-type configuration. Our multicast path finding method provides an efficient multicast graph which has both star-type and ring-type natures by sharing as many links as possible under the indegree constraint (Fig. 4).

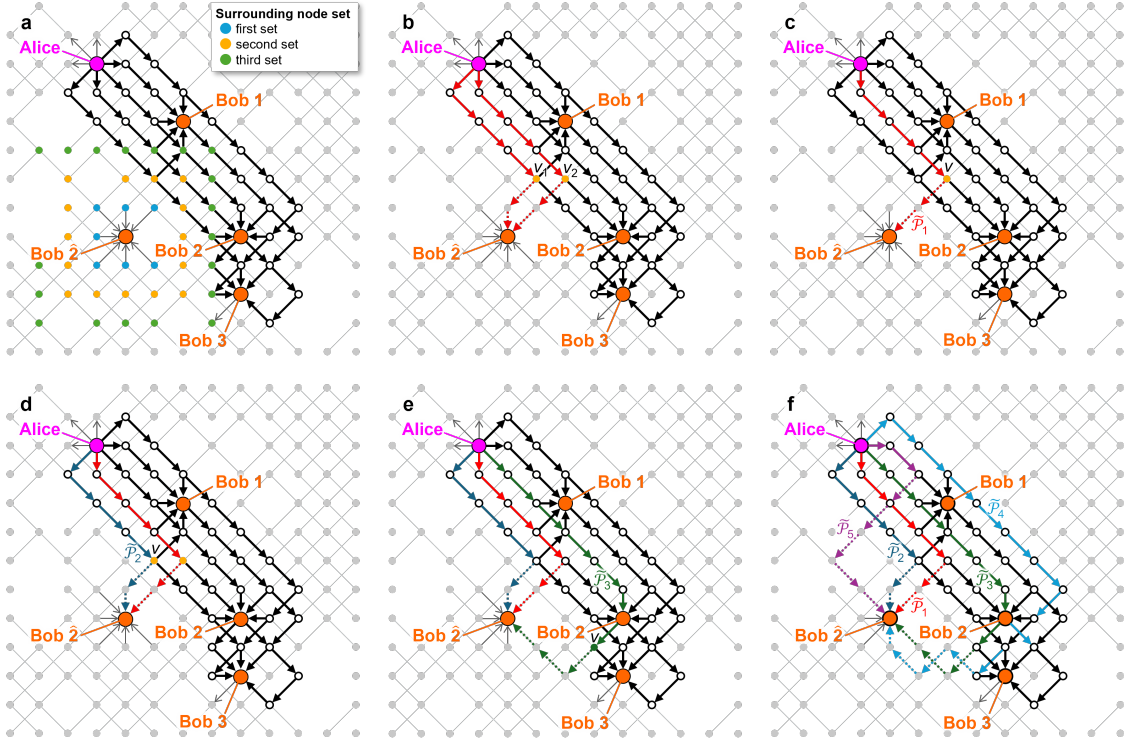

Fig. 3. **RAPUS on the X-hop grid.** Basic concepts and algorithms in the example of Fig. 1. **a** Established multicast graph $\vec{\mathcal{G}}^{(t-1)}$ for Bob 1 $\sim$ Bob $t-1$ ($t=4$), and surrounding node sets around one candidate (Bob $\hat{2}$) of $N-t+1$ remaining Bobs. **b** Two support nodes $v_1$ and $v_2$ which can be reached from Bob $\hat{2}$ by minimum hops are identified. A path from Alice to Bob $\hat{2}$ via $v_1$ with the minimum number of unused links (dotted arrow) in $\vec{\mathcal{G}}^{(t-1)}$ is searched. This is repeated for $v_2$. **c** The straighter path is assigned as the first path $\bar{\mathcal{P}}_1$. **d** The second path $\bar{\mathcal{P}}_2$ is searched in a similar way but excluding the links in the taboo list $\mathcal{T}_1$ from the path finding. **e** The third path $\bar{\mathcal{P}}_3$ is searched in a similar way. **f** The fourth and fifth paths $\bar{\mathcal{P}}_4$ and $\bar{\mathcal{P}}_5$ are searched in a similar way. An output graph $\vec{\mathcal{G}}^{(t,t_i)}$ is determined according to Eq. (23).

Errors and erasures in the satellite network mainly occur in the links with the ground stations, which are exposed to atmospheric turbulence. In contrast, inter-satellite links can be more stable, and hence assumed to be error-free. Therefore, we introduce the link error rate $e_{L1}$ and the link erasure rate $\epsilon$ to the last-hop links to each Bob to model the atmospheric turbulence effects.

We also consider the node erasure rate $\epsilon_N$ to model packet losses or large latency of packet arrival which often occur when the node is overloaded due to congestion. We further introduce the link error rate $e_{L2}$ to all the links to take into account noises caused by jamming the satellites, which is also a likely risk.

We assume that node compromise in each Charlie occurs at a uniform rate $\gamma$, which serves as a reasonable first-step approximation.

To realize the 5-degree PUSNEC, we implemented the Gab$[n, k]$ code suite over $\mathbb{F}_{q^m}$ with $n = 9, 11$, $q = 2^8$ and $m = 27$ or larger if necessary. Incoming sample data is divided into basic units of $mk_0$ bytes, i.e., $\boldsymbol{u} = \begin{bmatrix} u_0 & u_1 & \dots & u_{k_0-1} \end{bmatrix} \in \mathbb{F}_{q^m}^{k_0}$. We took $\mu_0 = 0$ (no masking key) and used the ramp scheme to increase the message rate. For a symbol $u_i$, the remaining symbols function as random data to confuse Eve. In each transmission of $\boldsymbol{u}$, node compromise and erasure/error events were randomly generated, and the coding performances were evaluated with sufficient statistical events (roughly 10 times the reciprocal of PLP or FER).

Figure 5a shows the PLP $p_L(\gamma, k)$ of Eq. (21) as a function of $\gamma$ for $k = 1 \sim 5$. Obviously the meaningful region is $\gamma < 1/N_C$ where $p_L(\gamma, k)$ monotonically decreases as $\gamma$ decreases. For $k = 1$, the quantity $p_L(\gamma, 1)/n_0$ approximately represents the PLP of the conventional QLD/PLSNs without usr-SNC. In this case, it scales linearly with $\gamma$, as $\sim \gamma N_C/n_0$. In contrast, with usr-SNC, the PLP exhibits higher-order dependence on $\gamma$, and the suppression effect becomes more pronounced as $k$ increases.

Figure 5b and c show the FERs, $e_F(\epsilon, k)$ due to the link erasure and $e_F(e_{L1}, k)$ due to the link error for $k = 1 \sim 5$, respectively. For $k = 5$, with no erasure or error correction capability, an error floor as high as $10^{-1}$ persists. In contrast, the FER drops sharply as $k$ decreases to 4 or below. The $e_F(e_{L1}, k)$ curve shifts every two steps of $k$ in Fig. 5c, because error correction requires twice as much resources as erasure correction as expressed by $2\tau + \rho = n_0 - k$.

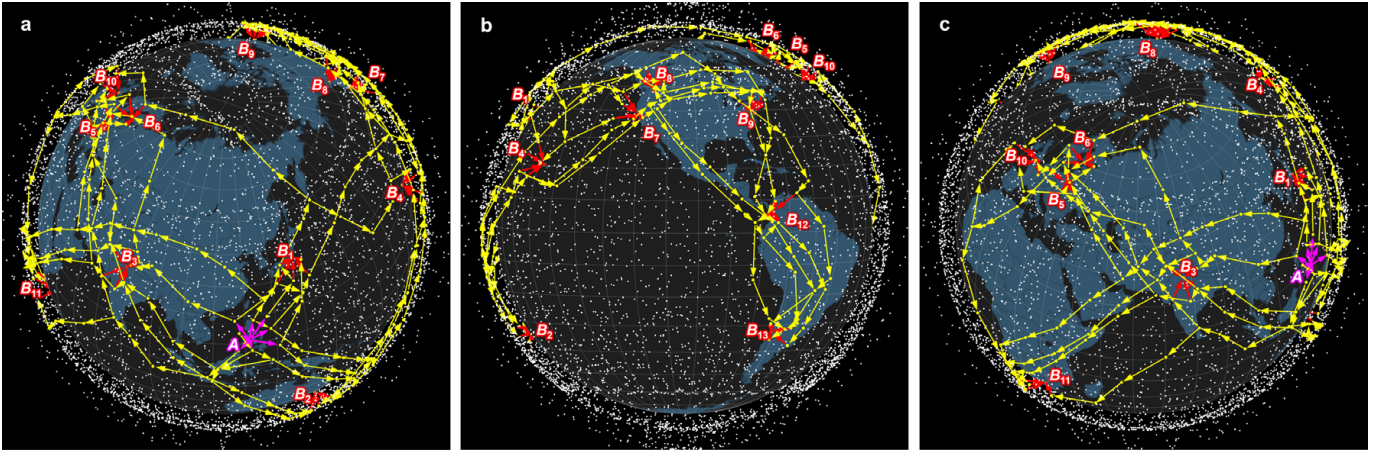

**Fig. 4. The 1-to-13 omni-directional multicast graph for the 5-degree PUSNEC over the entire globe using LEO constellations.** The LEO constellations are modeled based on the satellite distribution data in [50], [51]. The white dots represent LEO satellites. **a**, **b**, and **c** represent the multicast graph viewed from three different directions: above Eastern Asia, the Eastern Pacific, and Western Asia. Alice ($A$) is a LEO satellite above Philippines and sends 8 $\mathbb{F}_q$-RLNC packets into 8 links (purple arrows). Bobs are totally 13 ground stations, which are in Tokyo ($B_1$), Canberra ($B_2$), New Delhi ($B_3$), Honolulu ($B_4$), Warszawa ($B_5$), Helsinki ($B_6$), San Francisco ($B_7$), Vancouver ($B_8$), Washington ($B_9$), London ($B_{10}$), Pretoria ($B_{11}$), Bogota ($B_{12}$), Santiago ($B_{13}$), each of which has 5 downlinks to collects 5 packets from the LEO satellites (red arrows). The yellow arrows represent QKD/PLS links constituting the multicast graph. The details on multicast path finding are given in Supplementary Note 7 in [41]. A 3D interactive view is available in [52].

Thus, the PLP and the FERs are in a tradeoff relation in terms of $k$, which is exemplified in Fig. 5d and e for four sets of $\gamma$ and $\epsilon$. The $k$-dependences of the LII $I_L(\gamma, k, \xi)$ of Eq. (19) for $\xi = 1 \sim 4$ and the FERs are explicitly shown in Fig. 5f and g. Figure 5f shows characteristics similar to those in Supplementary Fig. 4b in [41]. These results provide a design guideline for selecting optimal coding specifications according to the values of $\gamma$, $\epsilon$, and $e_{L1}$.

We further examine reliability performance at a balance point of $k = 3$ by comparing the Gabidulin code with its conventional counterpart, the RS code—a widely adopted standard due to its strong error-correcting capability—under the same coding rate. The code lengths are also selected to be nearly identical in terms of the total number of bits: $\mathrm{Gab}[9, 3]$ over $\mathbb{F}_{256^{27}}$ versus $\mathrm{RS}[243, 81]$ over $\mathbb{F}_{2^8}$. To examine Bob-specific differences, we define the FER for each individual Bob, denoted as FER-$B_t$ ($t = 1 \sim 13$). Figures 6a and b show the FER-$B_t$ as a function of $\epsilon$ and $e_{L1}$, respectively. The FER-$B_t$ for the RS code monotonically decreases and reaches an error floor as $\epsilon$ or $e_{L1}$ decreases. In contrast, the FER-$B_t$ for the Gabidulin code monotonically decreases much faster, and no error floor appears in the plot range. Figures 6c and d show the FER-$B_t$ as functions of $e_N$ and $e_{L2}$, respectively. They exhibit similar characteristics to Figs. 6a and b, but the curves themselves spread more, because we assume that the node erasures and the link errors occur in the entire network.

Thus, the Gabidulin code significantly outperforms the RS code in terms of error and erasure correction capabilities. Furthermore, the decoding time of $\mathrm{Gab}[9, 3]$ was comparable to, or even shorter than, that of $\mathrm{RS}[243, 81]$ (see Supplementary Note 3-Section IV in [41]).

In other words, network-coded multicast inherently contains fatal errors that the RS code—more generally, Hamming-metric codes—cannot handle even when errors or erasures occur solely on the last-hop links to Bobs (Fig. 6a and b).

The error floor of the FER-$B_t$ for the RS code arises due to the *finite-size effect* of the ground field in the $\mathbb{F}_q$-RLNC. Specifically, in the $\mathbb{F}_q$-RLNC operation of Eq. (??), an element of zero may appear with a probability of roughly $1/q$, potentially resulting in rank-1 deficiency in $A_t$ in Eq. (7) after further $\mathbb{F}_q$-RLNC operations. This rank deficiency is often interpreted by the RS decoder as multiple symbol erasures, which may exceed its error correction capability (see illustrative examples in Supplementary Note 8 in [41]). Increasing the code length does not mitigate this issue.

In contrast, the Gabidulin decoder treats such multiple symbol erasures, when linearly dependent, as rank-1 erasures, and correctly recovers the original message, thereby effectively suppressing the error floor.

Therefore, the use of Gabidulin codes is essential to fully exploit the potential of $\mathbb{F}_q$-RLNC for multicast. In particular, existing coding standards recommended for satellite communications must be fundamentally re-evaluated and replaced with MRD codes when aiming to achieve reliable and secure multicast to ground stations via satellite constellations.

The fundamental behaviors shown in Fig. 5—including the flat and monotonically decreasing regions in $p_L(\gamma, k)$ and $e_F(\epsilon, k)$, as well as their tradeoff with respect to $k$—also appear in the 1-to-6 directional multicast over the terrestrial network shown in Fig. 1. By appropriately adjusting $\gamma$ and $\epsilon$, the resulting curves of $p_L(\gamma, k)$, $e_F(\epsilon, k)$, their tradeoffs, and the LII $I_L(\gamma, n_0, k, \xi)$ closely match those obtained for the LEO satellite network (Supplementary Note 9 in [41]). These consistent patterns stem from the uniform indegree constraint on Charlies inherent to the X-hop grid topology, which thus serves as a universal architectural model for achieving reliable

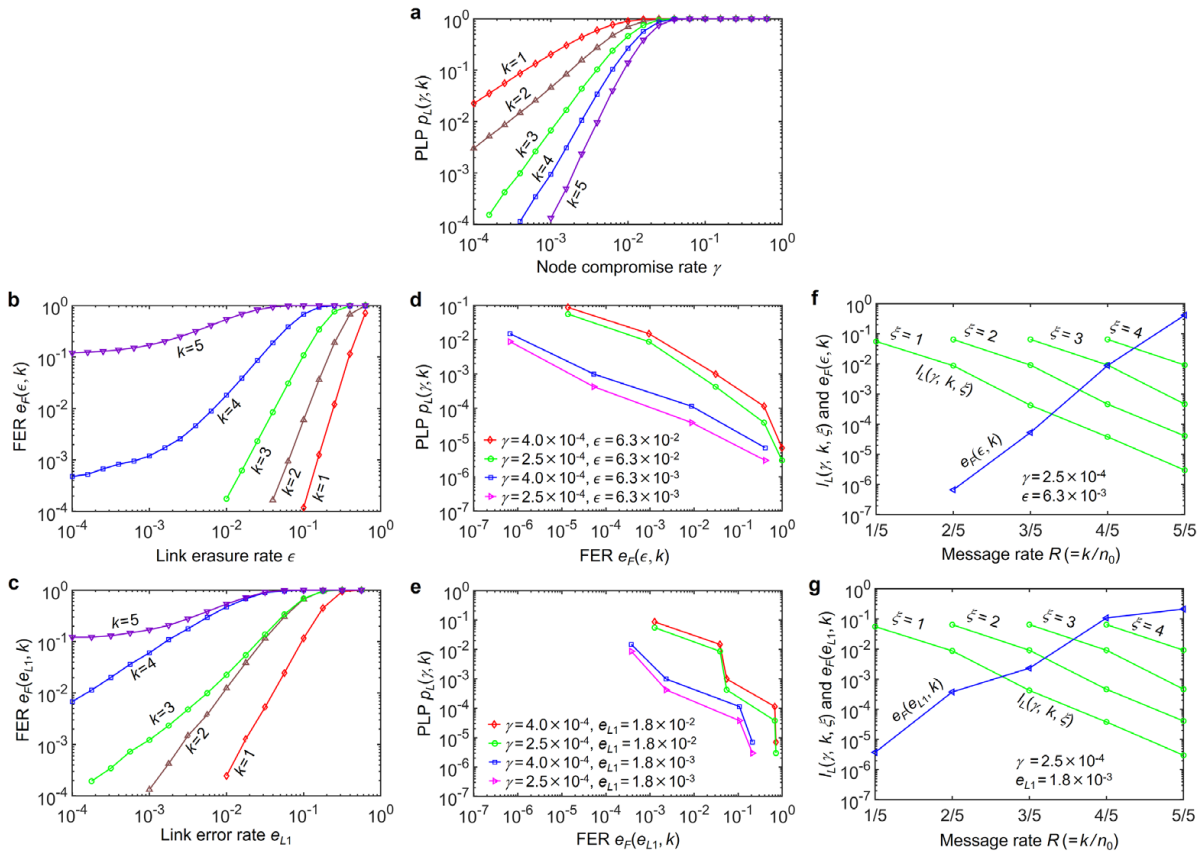

Fig. 5. **Secrecy and reliability performances of 1-to-13 omni-directional multicast with the 5-degree PUSNEC in the LEO constellations.** **a** The PLP $p_L(\gamma, k)$ as a function of the node compromise rate $\gamma$. **b** The FER $e_F(\epsilon, k)$ as a function of the link erasure rate $\epsilon$, setting $e_{L1} = \epsilon_N = e_{L2} = 0$. **c** The FER $e_F(e_{L1}, k)$ as a function of the link error rate $e_{L1}$, setting $\epsilon = \epsilon_N = e_{L2} = 0$. **d** The tradeoff relations between the PLP and the FER due to the link erasure, $e_F(\epsilon, k)$, for four sets of $(\gamma, \epsilon)$ $(e_{L1} = \epsilon_N = e_{L2} = 0)$. **e** The tradeoff relations between the PLP and the FER due to the link error, $e_F(e_{L1}, k)$, for four sets of $(\gamma, e_{L1})$ $(\epsilon = \epsilon_N = e_{L2} = 0)$. **f** The LII $I_L(\gamma, k, \xi)$ and the FER due to the link erasure, $e_F(\epsilon, k)$, as a function of the message rate $R = k/n_0$ for a set of $(\gamma, \epsilon)$ $(e_{L1} = \epsilon_N = e_{L2} = 0)$. **g** The LII $I_L(\gamma, k, \xi)$ and the FER due to the link error, $e_F(e_{L1}, k)$, as a function of the message rate $R = k/n_0$ for a set of $(\gamma, e_{L1})$ $(\epsilon = \epsilon_N = e_{L2} = 0)$

multicast with ITS in both large-scale terrestrial infrastructures and satellite constellations.

We also studied the 6-degree PUSNEC, which enhances secrecy and reliability compared to the 5-degree PUSNEC, but at the expense of network overhead. In practice, it is often difficult to find six link-disjoint paths to certain Bobs, especially when Bobs are located close to each other or when there are sparse areas of Charlies. One possible approach is to deploy several relay satellites in medium Earth orbit (MEO) to configure efficient 6-degree multicast paths. Such a LEO-MEO hybrid satellite network and the performance of the 6-degree PUSNEC over it are provided in Supplementary Note 10 in [41].

## VI. CONCLUSION AND OUTLOOK

In the PUSNEC, the usr-SNC consisting of Gabidulin code and $\mathbb{F}_q$-RLNC provides secure and reliable multicast. The RAPUS-based path finding enhances resilience against node-based Eve, expanding usr-SNC's applicability. $\mathbb{F}_q$-RLNC simplifies the RAPUS-based path finding algorithms. Interestingly, for small-scale networks, it serves to find secure and reliable multicast paths even when conventional IP-based routing fails in finding them. Gabidulin code reduces FER for $\mathbb{F}_q$-RLNC packets. This minimizes resend or re-multicast requests from Bobs, resulting in low latency and simpler multicast architectures. The combination of these three techniques is crucial for achieving IT-secure, reliable, resource-efficient, scalable multicast (Supplementary Note 2-Section II in [41]).

Based on the probabilistic node-compromise model, we derived the maximum leakage information to Eve (in the sense of a single-shot use of the wiretap channel) in terms of the $\mu$-wiretap probability $p(\mu, \tilde{\gamma}, \mathcal{G}, n_0, k)$. Although our numerical simulations assume a uniform node compromise rate $\gamma$, the actual compromise rates $\gamma_v$ vary across nodes. Therefore, future studies must take this variability into account in RAPUS-based path finding, in order to determine optimal paths that minimize the maximum information leakage to Eve, $I(U_{\{\xi\}}; Z')$. Similarly, the capacity and the number of available keys vary across links. These factors should be treated as cost functions and incorporated into the optimization process.

While the PUSNEC itself ensures ITS, it can also be effectively integrated with state-of-the-art post-quantum cryp-tography (PQC). Although the overall security remains within the computational security framework, rank errors and rank erasures—major issues in multicast—can be effectively corrected, significantly enhancing the availability of confidential data. Moreover, the usr-SNC further strengthens security by making message decoding more difficult compared to relying solely on PQC encryption. A benchmark of various cryptographic scheme combinations with the PUSNEC is provided in Supplementary Note 11 in [41].

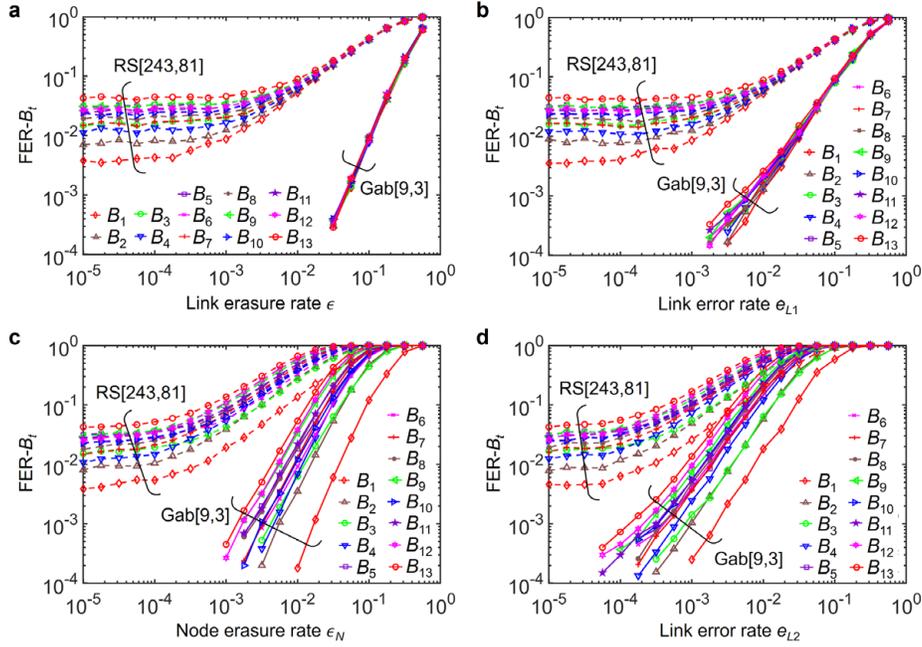

Fig. 6. **Comparison of FER-$B_t$ for Gabidulin code (solid lines) and RS code (dotted lines).** The scheme is 1-to-13 omni-directional multicast with the 5-degree PUSNEC in the LEO constellations. **a** FER-$B_t$ as a function of the link erasure rate $\epsilon$ at each Bob, setting $e_{L1} = e_N = e_{L2} = 0$. **b** FER-$B_t$ as a function of the link error rate $e_{L1}$ at each Bob, setting $\epsilon = e_N = e_{L2} = 0$. **c** FER-$B_t$ as a function of the node erasure rate $\epsilon_N$ at each Bob, setting $\epsilon = e_{L1} = e_{L2} = 0$. **d** FER-$B_t$ as a function of the link error rate $e_{L2}$ at each Bob, setting $\epsilon = e_{L1} = e_N = 0$.

## APPENDIX

From Eqs. (2), (3), and (10), the wiretapped word at Eve is

$$z_\zeta^{(\mu)} = \begin{bmatrix} u & r \end{bmatrix} G_1^{-1} G_2 B_\zeta^{(\mu)}. \quad (24)$$

This relation defines the structure of the wiretap channel $P_{Z_\zeta^{(\mu)}|UR}(z_\zeta^{(\mu)}|u, r)$, and determines Eve's decoding strategy.

We employ an important property of MRD codes with generator matrix $G$: the $k \times k$ matrix $\bar{G} \equiv GT$ is nonsingular for any full-rank matrix $T \in \mathbb{F}_q^{n \times k}$ [38, Theorem 2 and 3]. The $[n, k]$ Gabidulin encoding Eq. (5) is converted to

$$x'_{\{k\}} = f\bar{G}, \quad (25)$$

which specifies 1-to-1 relation between $x'_{\{k\}}$ and $f$. According to Eq. (2), the symbols constituting the word $x'_{\{k\}}$ are independent and uniformly distributed as the elements of $\begin{bmatrix} u & r \end{bmatrix}$ are. The essence of strong ramp secrecy is actually this statistical independence of the symbols constituting $x'_{\{k\}}$ (See the mutual information $I^{(\mu)}(U_{\{\xi\}}; Z)$ in the threshold-based wiretap model in Supplementary Note 12 in [41]).

For $\mu \leq \mu_0$, we take

$$T_\zeta^{(\mu)} = \begin{bmatrix} I_{(k-\mu)\times(k-\mu)} & \mathbf{0}_{k_1 \times \mu} \\ \mathbf{0}_{(n-k+\mu)\times(k-\mu)} & B_\zeta^{(\mu)} \end{bmatrix}, \quad (26)$$

where $I$ is the unit matrix and $\mathbf{0}$'s are the all-zero matrices. We then have $x'_{\{k\}} = \begin{bmatrix} u & r_0 & \dots & r_{\mu_0 - \mu - 1} & z_\zeta^{(\mu)} \end{bmatrix}$, meaning that perfect secrecy of $u$ can be met.

For $\mu_0 < \mu \leq k - 1$, we take

$$T_\zeta^{(\mu)} = \begin{bmatrix} J_{k_0 \times (k-\mu)} & \mathbf{0}_{k_1 \times \mu} \\ \mathbf{0}_{(n-k_0)\times(k-\mu)} & B_\zeta^{(\mu)} \end{bmatrix}, \quad (27)$$

where $J$ is a punctured unit matrix. We then have

$$x'_{\{k\}} = \begin{bmatrix} u_{\{k-\mu\}} & z_\zeta^{(\mu)} \end{bmatrix}, \quad (28)$$

i.e., $u_{\{k-\mu\}}$ and $z_\zeta^{(\mu)}$ are statistically independent. Therefore, in the case of $0 \leq \mu \leq k - \xi$,

$$\sum_{\mu=0}^{k-\xi} I^{(\mu)}(U_{\{\xi\}}; Z') = 0, \quad (29)$$

and Eve concludes that no information can be obtained.

For $k - \xi < \mu$, the mutual information can be expressed as

$$I(U_{\{\xi\}}; Z') = I_2(U_{\{\xi\}}; Z') + I_3(U_{\{\xi\}}; Z'), \quad (30)$$

$$I_2(U_{\{\xi\}}; Z') = \sum_{\mu=k-\xi+1}^{k-1} I^{(\mu)}(U_{\{\xi\}}; Z'), \quad (31)$$

$$I_3(U_{\{\xi\}}; Z') = \sum_{\mu=k}^{|\mathcal{E}|} I^{(\mu)}(U_{\{\xi\}}; Z'). \quad (32)$$

$$
\hat{P}_{Z_\zeta^{(\mu)}|U_{\{\xi\}}} = \frac{1}{q^{m(k-\xi)}}
\begin{array}{c}
\begin{array}{ccccccc}
\boldsymbol{z}_{\zeta,1}^{(\mu)} & \cdots & \boldsymbol{z}_{\zeta,q^*}^{(\mu)} & \cdots & \boldsymbol{z}_{\zeta,q^{m\mu}-q^*+1}^{(\mu)} & \cdots & \boldsymbol{z}_{\zeta,q^{m\mu}}^{(\mu)}
\end{array}\\
\left[
\begin{array}{ccc|c|ccc}
1 & & & & & & \\
\vdots & & & & & & \\
1 & & & & & & \\
& \ddots & & \cdots & & \ddots & \\
& & 1 & & & & 1\\
& & \vdots & & & & \vdots\\
& & 1 & & & & 1
\end{array}
\right]
\end{array}
\begin{array}{l}
\boldsymbol{u}_{\{\xi\}1,1}\\
\vdots\\
\boldsymbol{u}_{\{\xi\}q^{m(k-\mu)},1}\\
\vdots\\
\boldsymbol{u}_{\{\xi\}1,q^*}\\
\vdots\\
\boldsymbol{u}_{\{\xi\}q^{m(k-\mu)},q^*}
\end{array}
. \tag{38}
$$

To describe Eve's strategy for $k - \xi < \mu \le k - 1$, we derive a coset expansion of $\begin{bmatrix} \boldsymbol{u} & \boldsymbol{r} \end{bmatrix}$ in terms of $\boldsymbol{z}_\zeta^{(\mu)}$. Denoting $\bar{G}_\zeta^{(\mu)} \equiv G T_\zeta^{(\mu)}$, Eq. (28) can be rewritten as

$$
\boldsymbol{x}'_{\{k\}} = \begin{bmatrix} \boldsymbol{u} & \boldsymbol{r} \end{bmatrix} G_1^{-1} \bar{G}_\zeta^{(\mu)}. \tag{33}
$$

Denote the first $k - \mu$ rows and the last $\mu$ rows of $\left( G_1^{-1} \bar{G}_\zeta^{(\mu)} \right)^{-1}$ as $H'^{(\mu)}_\zeta \in \mathbb{F}_{q^m}^{(k-\mu)\times k}$ and $H''^{(\mu)}_\zeta \in \mathbb{F}_{q^m}^{\mu \times k}$, respectively. Then Eq. (33) can be rewritten as

$$
\begin{bmatrix} \boldsymbol{u} & \boldsymbol{r} \end{bmatrix} = \boldsymbol{\sigma} + \boldsymbol{\omega}, \tag{34}
$$

$$
\boldsymbol{\sigma} \equiv \boldsymbol{u}_{\{k-\mu\}} H'^{(\mu)}_\zeta, \quad \boldsymbol{\omega} \equiv \boldsymbol{z}_\zeta^{(\mu)} H''^{(\mu)}_\zeta. \tag{35}
$$

All possible sequences of $\begin{bmatrix} \boldsymbol{u} & \boldsymbol{r} \end{bmatrix} \in \mathbb{F}_{q^m}^k$ can be indexed as

$$
\begin{bmatrix} \boldsymbol{u} & \boldsymbol{r} \end{bmatrix}_{ij} = \boldsymbol{\sigma}_i + \boldsymbol{\omega}_j, \tag{36}
$$

where $i = 1, ..., q^{m(k-\mu)}$ and $j = 1, ..., q^{m\mu}$. This can be regarded as a coset expansion of $\begin{bmatrix} \boldsymbol{u} & \boldsymbol{r} \end{bmatrix}$ with coset leaders $\boldsymbol{\omega}_j$. The above relation specifies a structure of the $(\mu, \zeta)$-wiretap channel in a $q^{mk} \times q^{m\mu}$ matrix form

$$
\hat{P}_{Z_\zeta^{(\mu)}|UR}
$$
$$
=
\begin{array}{c}
\begin{array}{cccc}
\boldsymbol{z}_{\zeta,1}^{(\mu)} & \boldsymbol{z}_{\zeta,2}^{(\mu)} & \cdots & \boldsymbol{z}_{\zeta,q^{m\mu}}^{(\mu)}
\end{array}\\
\left[
\begin{array}{cccc}
1 & & & \\
\vdots & & & \\
1 & & & \\
& 1 & & \\
& \vdots & & \\
& 1 & & \\
& & \ddots & \\
& & & 1\\
& & & \vdots\\
& & & 1
\end{array}
\right]
\end{array}
\begin{array}{l}
\begin{bmatrix} \boldsymbol{u} & \boldsymbol{r} \end{bmatrix}_{1,1}\\
\vdots\\
\begin{bmatrix} \boldsymbol{u} & \boldsymbol{r} \end{bmatrix}_{q^{m(k-\mu)},1}\\
\begin{bmatrix} \boldsymbol{u} & \boldsymbol{r} \end{bmatrix}_{1,2}\\
\vdots\\
\begin{bmatrix} \boldsymbol{u} & \boldsymbol{r} \end{bmatrix}_{q^{m(k-\mu)},2}\\
\vdots\\
\begin{bmatrix} \boldsymbol{u} & \boldsymbol{r} \end{bmatrix}_{1,q^{m\mu}}\\
\vdots\\
\begin{bmatrix} \boldsymbol{u} & \boldsymbol{r} \end{bmatrix}_{q^{m(k-\mu)},q^{m\mu}}
\end{array}
, \tag{37}
$$

where the blanks in the matrix means 0. Note that each $\hat{P}_{Z_{\zeta'}^{(\mu)}|UR}$, where $\zeta' \ne \zeta$, is obtained from $\hat{P}_{Z_\zeta^{(\mu)}|UR}$ by some permutation of the rows and columns.

For a submessage $\boldsymbol{u}_{\{\xi\}}$, the reduced channel matrix can be derived by Eq. (17) as in Eq. (38) with $q^* = q^{m(\xi-k+\mu)}$. More specifically, the matrix can be divided into $q^{m(k-\xi)}$ blocks of $q^{m\xi} \times q^*$ matrix, and the coset expansion of $\boldsymbol{u}_{\{\xi\}}$ can be made for *each* of $q^{m(k-\xi)}$ blocks. For the $J$th block, the coset expansion is

$$
\boldsymbol{u}_{\{\xi\}ij} = \boldsymbol{\sigma}_{\{\xi\}i} + \boldsymbol{\omega}_{\{\xi\}j}, \tag{39}
$$

where $i = 1, ..., q^{m(k-\mu)}$ and $j = (J-1)q^* + 1, ..., Jq^*$.

Therefore, given the realization $\boldsymbol{z}_{\zeta,j}^{(\mu)}$—namely, the coset leader $\boldsymbol{\omega}_{\{\xi\}j}$—Eve can identify that the transmitted sequence belongs to the $j$th coset. However, she remains uncertain about which of the $q^{m(k-\mu)}$ possible elements within the coset was sent.

As a result, Eve randomly selects one element from the coset as the message, a strategy referred to as *coset-based random guessing*. The corresponding mutual information can be straightforwardly derived as

$$
I_2(U_{\{\xi\}}; Z') = \sum_{\mu=k-\xi+1}^{k-1} (\mu - k + \xi) p(\mu) \log_2 q^m. \tag{40}
$$

For $\mu \ge k$, $\mu$ symbols constituting $\boldsymbol{z}_\zeta^{(\mu)}$ belong to a set of the symbols used in the $\mathrm{Gab}[n_0, k]$ codeword because the wiretap channel is noiseless. In particular, when $\mu = k$, $\boldsymbol{z}_\zeta^{(\mu)}$ and $\begin{bmatrix} \boldsymbol{u} & \boldsymbol{r} \end{bmatrix}$ have the 1-to-1 relation and the wiretap channel matrix is the unit matrix $\hat{P}_{Z_\zeta^{(\mu)}|UR} = I_{q^{mk} \times q^{mk}}$.

For $\mu > k$, $\mu$ symbols constituting $\boldsymbol{z}_\zeta^{(\mu)}$ are linearly dependent while $k$ of them are linearly independent. Therefore, by appropriately assigning row and column indices (i.e., $\begin{bmatrix} \boldsymbol{u} & \boldsymbol{r} \end{bmatrix}_i$ and $\boldsymbol{z}_{\zeta,j}^{(\mu)}$), we have

$$
\hat{P}_{Z_\zeta^{(\mu)}|UR} = \frac{1}{q^{m(\mu-k)}} \underbrace{\begin{bmatrix} I_{q^{mk} \times q^{mk}} & \cdots & I_{q^{mk} \times q^{mk}} \end{bmatrix}}_{q^{m(\mu-k)} \text{ unit matrices}}. \tag{41}
$$

For a submessage $\boldsymbol{u}_{\{\xi\}}$, we have

$$
\hat{P}_{Z_\zeta^{(\mu)}|U_{\{\xi\}}} = \frac{1}{q^{m(\mu-\xi)}} \underbrace{\begin{bmatrix} I_{q^{m\xi} \times q^{m\xi}} & \cdots & I_{q^{m\xi} \times q^{m\xi}} \end{bmatrix}}_{q^{m(\mu-\xi)} \text{ unit matrices}}. \tag{42}
$$

Note that each $\hat{P}_{Z_{\zeta'}^{(\mu)}|U_{\{\xi\}}}$, where $\zeta' \ne \zeta$, is obtained from $\hat{P}_{Z_\zeta^{(\mu)}|U_{\{\xi\}}}$ by some permutation of the rows and columns.

Equation (42) shows that Eve can uniquely determines $\boldsymbol{u}_{\{\xi\}}$ from $\boldsymbol{z}_\zeta^{(\mu)}$ and

$$
I_3(U_{\{\xi\}}; Z') = \xi p_L(n_0, k) \log_2 q^m. \tag{43}
$$

Finally, since the variable $Z'$ covers all possible wiretapped words that Eve can acquire from the compromised nodes, $I(U_{\{\xi\}}; Z')$ is the maximum information on $U_{\{\xi\}}$ available for Eve per single shot use of $P_{Z'|U_{\{\xi\}}}$. As shown above, this is attained by the coset-based random guess strategy. ∎

## REFERENCES

[1] R. Ahlswede, N. Cai, S.-Y. Li, and R. Yeung, "Network information flow," *IEEE Trans. Inf. Theory*, vol. 46, no. 4, pp. 1204–1216, July 2000.

[2] C. H. Bennett and G. Brassard, "Quantum cryptography: Public key distribution and coin tossing," in *Proc. Int. Conf. on Computers, Systems and Signal Processing*, Dec. 1984, pp. 175–179.

[3] A. K. Ekert, "Quantum cryptography based on Bell's theorem," *Phys. Rev. Lett.*, vol. 67, pp. 661–663, Aug. 1991.

[4] N. Gisin, G. Ribordy, W. Tittel, and H. Zbinden, "Quantum cryptography," *Rev. Mod. Phys.*, vol. 74, pp. 145–195, Mar. 2002.

[5] A. D. Wyner, "The Wire-Tap Channel," *Bell Syst. Tech. J.*, vol. 54, no. 8, pp. 1355–1387, Oct. 1975.

[6] I. Csiszár and J. Körner, "Broadcast channels with confidential messages," *IEEE Trans. Inf. Theory*, vol. 24, no. 3, pp. 339–348, May 1978.

[7] U. Maurer, "Secret key agreement by public discussion from common information," *IEEE Trans. Inf. Theory*, vol. 39, no. 3, pp. 733–742, May 1993.

[8] R. Ahlswede and I. Csiszár, "Common randomness in information theory and cryptography. I. Secret sharing," *IEEE Trans. Inf. Theory*, vol. 39, no. 4, pp. 1121–1132, July 1993.

[9] H. Endo, T. S. Han, T. Aoki, and M. Sasaki, "Numerical study on secrecy capacity and code length dependence of the performances in optical wiretap channels," *IEEE Photonics J.*, vol. 7, no. 5, pp. 7903418, Oct. 2015.

[10] H. Endo *et al.*, "Free-space optical channel estimation for physical layer security," *Opt. Express*, vol. 24, no. 8, pp. 8940–8955, Apr. 2016.

[11] T.-L. Wang and I. B. Djordjevic, "Physical-layer security of a binary data sequence transmitted with Bessel–Gaussian beams over an optical wiretap channel," *IEEE Photonics J.*, vol. 10, no. 6, p. 7908611, Dec. 2018.

[12] M. Fujiwara *et al.*, "Free-space optical wiretap channel and experimental secret key agreement in 7.8 km terrestrial link," *Opt. Express*, vol. 26, no. 15, pp. 19513–19523, July 2018.

[13] H. Endo *et al.*, "Free space optical secret key agreement," *Opt. Express*, vol. 26, no. 18, pp. 23305–23332, Sep. 2018.

[14] H. Endo *et al.*, "Group key agreement over free-space optical links," *OSA Contin.*, vol. 3, no. 9, pp. 2525–2543, Sep. 2020.

[15] Z. Pan *et al.*, "Secret-key distillation across a quantum wiretap channel under restricted eavesdropping," *Phys. Rev. Applied*, vol. 14, p. 024044, Aug. 2020.

[16] H. Endo, T. Sasaki, M. Takeoka, M. Fujiwara, M. Koashi, and M. Sasaki, "Line-of-sight quantum key distribution with differential phase shift keying," *New J. Phys.*, vol. 24, no. 2, p. 025008, Feb. 2022.

[17] M. Peev *et al.*, "The SECOQC quantum key distribution network in vienna," *New J. Phys.*, vol. 11, no. 7, p. 075001, July 2009.

[18] M. Sasaki *et al.*, "Field test of quantum key distribution in the Tokyo QKD Network," *Opt. Express*, vol. 19, no. 11, pp. 10387–10409, May 2011.

[19] S. Wang *et al.*, "Field and long-term demonstration of a wide area quantum key distribution network," *Opt. Express*, vol. 22, no. 18, pp. 21739–21756, Sep. 2014.

[20] Q. Zhang, F. Xu, Y.-A. Chen, C.-Z. Peng, and J.-W. Pan, "Large scale quantum key distribution: challenges and solutions [invited]," *Opt. Express*, vol. 26, no. 18, pp. 24260–24273, Aug. 2018.

[21] S.-K. Liao *et al.*, "Satellite-to-ground quantum key distribution," *Nature*, vol. 549, pp. 43–47, Aug. 2017.

[22] S.-K. Liao *et al.*, "Satellite-relayed intercontinental quantum network," *Phys. Rev. Lett.*, vol. 120, p. 030501, Jan. 2018.

[23] Y.-A. Chen *et al.*, "An integrated space-to-ground quantum communication network over 4,600 kilometres," *Nature*, vol. 589, no. 7841, pp. 214–219, Jan. 2021.

[24] Y. Liu, H.-H. Chen, and L. Wang, "Physical layer security for next generation wireless networks: Theories, technologies, and challenges," *IEEE Commun. Surv. .Tutor.*, vol. 19, no. 1, pp. 347–376, Jan. 2017.

[25] T. R. Beals and B. C. Sanders, "Distributed relay protocol for probabilistic information-theoretic security in a randomly-compromised network," in *Proc. Int. Conf. Inf. Theoretic Security*, ser. Lect. Notes Comput. Sci., R. Safavi-Naini, Ed., vol. 5155. Springer, 2008, pp. 29–39.

[26] D. Elkouss, J. Martinez-Mateo, A. Ciurana, and V. Martin, "Secure optical networks based on quantum key distribution and weakly trusted repeaters," *J. Opt. Commun. Netw.*, vol. 5, no. 4, pp. 316–328, Apr. 2013.

[27] M. Fujiwara, G. Kato, and M. Sasaki, "Information theoretically secure data relay using QKD network," *IEEE Access*, vol. 12, pp. 141167–141178, Sep. 2024.

[28] N. Cai and R. W. Yeung, "Secure network coding on a wiretap network," *IEEE Trans. Inf. Theory*, vol. 57, no. 1, pp. 424–435, Jan. 2011.

[29] J. Feldman, T. Malkin, C. Stein, and R. A. Servedio, "On the capacity of secure network coding," in *Proc. 42nd Annu. Allerton Conf. Commun., Contr., Comput.*, Sep. 2004, pp. 63–68.

[30] S. Y. El Rouayheb and E. Soljanin, "On wiretap networks II," in *Proc. IEEE Int. Symp. Inf. Theory*, June 2007, pp. 551–555.

[31] S. Jaggi *et al.*, "Resilient network coding in the presence of Byzantine adversaries," *IEEE Trans. Inf. Theory*, vol. 54, no. 6, pp. 2596–2603, June 2008.

[32] D. Silva and F. R. Kschischang, "Universal secure network coding via rank-metric codes," *IEEE Trans. Inf. Theory*, vol. 57, no. 2, pp. 1124–1135, Feb. 2011.

[33] H. Yao, D. Silva, S. Jaggi, and M. Langberg, "Network codes resilient to jamming and eavesdropping," *IEEE/ACM Trans. Netw.*, vol. 22, no. 6, pp. 1978–1987, Dec. 2014.

[34] S.-Y. Li, R. Yeung, and N. Cai, "Linear network coding," *IEEE Trans. Inf. Theory*, vol. 49, no. 2, pp. 371–381, Feb. 2003.

[35] T. Ho *et al.*, "A random linear network coding approach to multicast," *IEEE Trans. Inf. Theory*, vol. 52, no. 10, pp. 4413–4430, Oct. 2006.

[36] P. A. Chou, Y. Wu, and K. Jain, "Practical network coding," in *Proc. 41st Annu. Allerton Conf. Commun., Control, Comput.*, Oct. 2003, pp. 40–49, invited paper.

[37] P. Delsarte, "Bilinear forms over a finite field, with applications to coding theory," *J. Comb. Theory, Ser. A*, vol. 25, no. 3, pp. 226–241, Nov. 1978.

[38] E. M. Gabidulin, "Theory of codes with maximum rank distance," *Probl. Inf. Transm.*, vol. 21, no. 1, pp. 1–12, Mar. 1985.

[39] J. Kurihara, R. Matsumoto, and T. Uyematsu, "Relative generalized rank weight of linear codes and its applications to network coding," *IEEE Trans. Inf. Theory*, vol. 61, no. 7, pp. 3912–3936, July 2015.

[40] H. Yamamoto, "Secret sharing system using $(k, l, n)$ threshold scheme," *Electron. Commun. Jpn. I*, vol. 69, no. 9, pp. 46–54, Sep. 1986.

[41] M. Sasaki, M. Fujiwara, T. S. Han, K. Li, O. Hambrey, and A. Esumi, "Supplementary Material: Path-Controlled Secure Network Coding," 2025, [Online]. Available: https://siglead.com/papers/PUSNEC/SN_v1.0.pdf.

[42] T. Wang, C. Q. Wu, Y. Wang, A. Hou, and H. Cao, "Multi-path routing for maximum bandwidth with k edge-disjoint paths," in *Proc. 2018 14th Int. Wireless Commun. Mobile Comput. Conf.*, Aug. 2018, pp. 1178–1183.

[43] M. A. Ribeiro, I. A. Carvalho, and A. H. Pereira, "The widest k-set of disjoint paths problem," *RAIRO-Oper. Res.*, vol. 57, no. 1, pp. 87–97, Mar. 2023.

[44] M. Fujiwara, R. Nojima, T. Tsurumaru, S. Moriai, M. Takeoka, and M. Sasaki, "Long-term secure distributed storage using quantum key distribution network with third-party verification," *IEEE Trans. Quantum Eng.*, vol. 3, pp. 1–11, Feb. 2022.

[45] O. Ore, "On a special class of polynomials," *Trans. Amer. Math. Soc.*, vol. 35, no. 3, pp. 559–584, July 1933.

[46] M. Iwamoto and H. Yamamoto, "Strongly secure ramp secret sharing schemes for general access structures," *Inf. Process. Lett.*, vol. 97, no. 2, pp. 52–57, June 2006.

[47] M. Nishiara and K. Takizawa, "Strongly secure secret sharing scheme with ramp threshold based on Shamir's polynomial interpolation scheme," *IEICE Trans. Fundam. Electron. Commun. Comput. Sci.*, vol. J92, no. 12, pp. 1009–1013, Dec. 2009.

[48] R. Lidl and H. Niederreiter, *Finite Fields, ser. Encyclopedia of Mathematics and its Applications*. USA: Cambridge University Press, 1996.

[49] T. S. Han, H. Endo, and M. Sasaki, "Reliability and secrecy functions of the wiretap channel under cost constraint," *IEEE Trans. Inf. Theory*, vol. 60, no. 11, pp. 6819–6843, Nov. 2014.

[50] T. S. Kelso, "NORAD GP Element Sets/Current Data/Active Satellites (24 January 2024)," https://celestrak.org/NORAD/elements/.

[51] satellitemap.space, https://satellitemap.space/index.html.

[52] "3D View of Multicast Graphs in Space Networks," https://snv.rocketworks.co.jp/3dview/.

Supplementary Material

# Path-Controlled Secure Network Coding

Masahide Sasaki, Te Sun Han, Mikio Fujiwara, Kai Li, Oliver Hambrey, and Atsushi Esumi

## Supplementary Note 1: Related works on strong ramp SNC

Secure network coding can be seen as a generalization of secret sharing schemes (SSSs) [1], or the wiretap channel II [2] to network coding. In a $(k, n)$-threshold SSS [3], a secret message $u$ is encoded into $n$ shares in such a way that any $k-1$ or less shares do not leak out any information of the secret message $u$ (perfect secrecy) while $u$ can be decrypted by collecting $k$ out of $n$ shares. In any SSS with perfect secrecy, the size of each share must be greater than or equal to that of the secret message.

In order to improve the efficiency of SSSs, ramp SSSs are proposed [4]–[6]. In the $(k, L, n)$-ramp SSS, the secret that consists of multiple elements, $\begin{bmatrix} u_1 & \dots & u_L \end{bmatrix}$, is encoded into $n$ shares. The size of each share is the same as that of each message $u_i$, i.e., reduced to $1/L$ of the original message size. On the other hand, the scheme has a trade-off between coding efficiency and secrecy. In fact, while any $k$ shares determine the secret, any less than or equal to $k-L$ shares have no information about it, and any $k-L+j$ $(1 \leq j < L)$ shares have partial information about it.

It is then an important concern how the secret partially leaks out. It was pointed out in [7] that some of sub-secrets can be completely decrypted from $k-L+j$ shares. This is fatal especially when sub-secrets contain confidential information. Yamamoto had formulated in [6] rigorous characterization of secrecy of ramp SSSs which ensures that when $\mu$ shares are eavesdropped, any $k-\mu$ sub-secrets can be kept completely secret. Schemes which meet this criterion are referred to as *strongly secure* ramp SSSs. Nishiara and Takizawa proposed a method to construct strongly secure ramp SSSs based on Reed-Solomon codes [8].

In the context of SNC, a similar idea to ramp SSS was investigated in [9], which formulated SNC against Eve's guessing, referred to as *weakly* SNC. This scheme was extended to a *universal* weakly SNC, using MRD codes in [10]. A construction method of *strongly* (ramp) SNC was provided in [11] [1]. The term *weakly* secure in [9], [10] is actually equivalent to the term *strongly* secure in ramp SNC as indicated in Proposition 5 of [10].

All of these works focus on secrecy protection, but do not have error-control capability. Silva and Kschischang formulated universal SNC with MRD codes including error-control capability [12]. This scheme, however, suffers from the secrecy loophole pointed out by [7], and hence could not always guarantee strong ramp secrecy.

Kurihara et al. introduced the method proposed by Nishiara and Takizawa [8] to Silva-Kschischang scheme, and revised

---

[1] More precisely, *strongly $k$-secure* network coding was formulated in [11]. In terms of the parameters in our work, it corresponds to *strongly $\mu_0$-secure* network coding described in Supplementary Note 12.

it such that the strong secrecy can always be guaranteed in ramp SNC at the expense of a longer code length [13].

## Supplementary Note 2: Tutorial examples for secure and reliable multicast

### I. Secure and reliable 1-to-3 multicast by distributed relay

A minimal scheme of 1-to-3 multicast is illustrated in Supplementary Fig. 1a. Alice $A$ multicasts a message $u$ to Bobs $B_1, B_2$, and $B_3$. The message is relayed via Charlie 1, 2, then triplicated at Charlie 3, and finally distributed to Bobs. It is obvious that if just one of Charlies is compromised, or a packet of $u$ is lost in a link, then IT-secure multicast fails.

Secrecy can be improved by using distributed transmission with redundant paths. The simplest example is 2-path secure distributed scheme of Supplementary Fig. 1b. Let the message $u \in \mathbb{F}_{q^M}$ be an element of the $M$-degree extension field of the ground field $\mathbb{F}_q$. Alice first generates a random number sequence $r \in \mathbb{F}_{q^M}$ (masking key) locally with the same size as $u$. The XOR encrypted message $u \oplus r$ and the masking key $r$ are transmitted over the first and second paths, respectively. Each Bob combines $u \oplus r$ and $r$ by XOR operation to decrypt the original message. Under the threshold assumption that Eve can only successfully eavesdrop Charlie(s) in one of the paths, *perfect secrecy* of the message is ensured. This scheme, however, does not work if a packet is erased in one of the paths (the message cannot be decrypted).

Erasure tolerance can be improved by preparing three disjoint-multipath paths and employing erasure correction with, e.g., [3, 2] RS code (Supplementary Fig. 1c). Moreover, a ramp scheme has been introduced to save the number of OTP keys consumed in the links. The message is divided into two blocks of the same size $u = u_0 || u_1$ ($u_0, u_1 \in \mathbb{F}_{q^m}$) where $m = M/2$, assuming $M$ is even. It is then encoded into a codeword

$$\begin{bmatrix} x_0 & x_1 & x_2 \end{bmatrix} = \begin{bmatrix} u_0 & u_1 \end{bmatrix} G^{[3,2]}, \tag{S1}$$

with a generator matrix

$$G^{[3,2]} = \begin{bmatrix} 1 & 1 & 1 \\ 1 & \alpha & \alpha^2 \end{bmatrix},$$

where $\alpha$ is a root of $x^2 + x + 1 = 0$ which generates a finite field $\mathbb{F}_4 = \{0, 1, \alpha, \alpha^2\}$. The symbols $x_0, x_1$, and $x_2$ are then transmitted over the first, second, and third path. At each Bob, the message can be decoded if two symbols are collected. For example, if $x_2$ is lost, each Bob collects $x_0$ and $x_1$ to decode the message as

$$\begin{bmatrix} x_0 & x_1 \end{bmatrix} G_{(0,1)}^{[3,2]\,-1} = \begin{bmatrix} u_0 & u_1 \end{bmatrix}, \tag{S2}$$

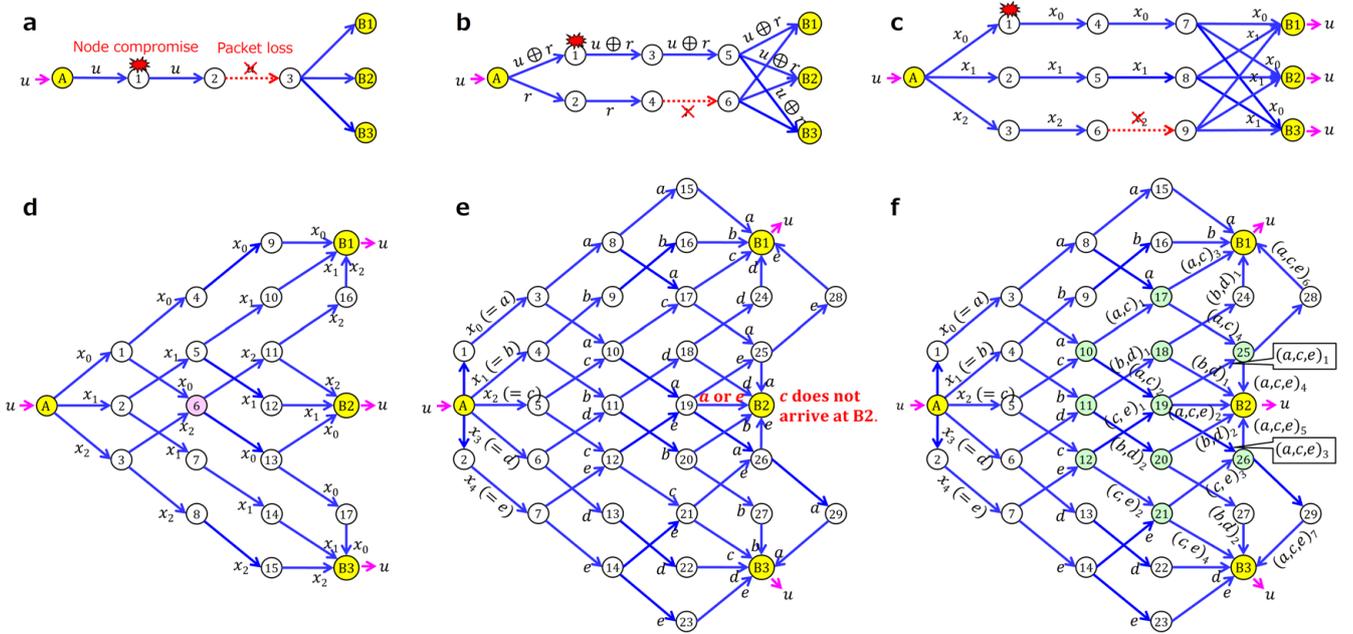

**Supplementary Fig. 1. Examples of 1-to-3 multicast graphs and distributed schemes. a** The minimal scheme with no redundant paths. **b** The 2-path secure distributed scheme with a masking key via two disjoint paths. **c** The 3-path RSDS via three disjoint paths. **d** The 3-degree RSDS for far-apart terminals without network coding at Charlies. **e** The 5-degree RSDS for far-apart terminals without network coding at Charlies. For ease of presentation, the five symbols $x_0, x_1, x_2, x_3, x_4$, are denoted as $a, b, c, d, e$. **f** The 5-degree RSDS for far-apart terminals with network coding at Charlies. Notations $(a, c)_1$, $(a, c)_2$, and so on, abbreviate randomly mixed packets $r_1 a + r_2 c$, $r_3 a + r_4 c$, and so on.

with

$$G^{[3,2]\,-1}_{(0,1)} = \begin{bmatrix} \alpha^2 & \alpha \\ \alpha & \alpha \end{bmatrix}.$$

Because the data size transmitting on each link becomes half, the number of OTP keys consumed in the network is not 3 times as large as that in the minimal scheme, but $3/2$ being saved in half.

If one of the symbols is eavesdropped from the compromised node, Eve is still confused about the value of each sub-message $u_0, u_1$ individually, provided that the message $u$ is uniformly distributed over $\mathbb{F}_{q^M}$. Mathematically the mutual information between a sub-message and a codeword symbol vanishes, i.e., $I(U_0; X_j) = I(U_1; X_j) = 0$ for $j = 0, 1, 2$, where $U_i$ and $X_j$ are random variables taking values $u_i \in \mathbb{F}_{q^m}$ and $x_j \in \mathbb{F}_{q^m}$, respectively. On the other hand, the mutual information between the whole message and a codeword symbol remains finite, i.e., $I(U_0, U_1; X_j) = (M/2) \log_2 q$ for $j = 0, 1, 2$. This does not, however, mean that the message is leaked to Eve, but reflects the fact that the possible range of $u$, over which Eve should search by brute force, reduces from $\mathbb{F}_{q^M}$ to $\mathbb{F}_{q^{M/2}}$. Note that no *meaningful* information on $u$ itself is leaked to Eve yet. Thus, this scheme ensures both information theoretic secrecy and reliability under the threshold assumptions that Eve can wiretap only one of $x_0$, $x_1$, and $x_2$ and that only one of $x_0$, $x_1$, and $x_2$ is erased. This scheme is called the *ramp* secure distributed scheme (RSDS). Details of calculation of the mutual information in this paragraph are given in Supplementary Note 2-Section III below.

In the examples of Supplementary Fig. 1a, b, and c, the essential multicast is made in the last-hop Charlies. However, this kind of multicast would not be a sensible way, when Bobs are far apart from each other. Actually, for B3 (B1) to acquire $x_0$ ($x_2$), last-hop link between Charlie 7 and B3 (Charlie 9 and B1) in Supplementary Fig. 1c should become very long, and hence would require further relay nodes. A more reasonable graph for far-apart Bobs may look like Supplementary Fig. 1d. The symbols $x_0$, $x_1$, and $x_2$ are duplicated at Charlies even from the first hop nodes (Charlie 1, 2, 3) and delivered to Bobs. In order to convey $x_0$ to B3 and $x_2$ to B1 with fewer hops, two paths crossing at Charlie 6 are prepared.

Unfortunately, however, while erasure tolerance increases, secrecy protection is insufficient yet. In fact, if Charlie 6, which is the joint node, were compromised, Eve could decrypt $u$ from $x_0$ and $x_2$. In order to realize both secrecy and reliability, the degree of distribution $n_0$ as well as the message length $k$ should be increased.

An immediate candidate is $[n_0, k] = [4, 3]$. This scheme can correct one erasure and keep secrecy tolerance at Charlie two input links. It cannot, however, correct an error when one of $x_0, ..., x_3$ is changed. On the other hand, the scheme of $[n_0, k] = [5, 3]$ can correct one error or two erasures. So, we consider 5-degree RSDS based on this scheme as depicted in Supplementary Fig. 1e. In Charlies, an input symbol is duplicated, or two input symbols are routed to two output links by conventional packet switching. This scheme, however, cannot deliver $x_2$ (denoted as $c$) to B2, because $c$ is duplicated only at Charlie 5, and distributed only to B1 and B3. Then $c$ is lost at B2 by default.

This problem can be resolved by introducing network coding to Charlies as in Supplementary Fig. 1f. Specifically, Charlie 10 converts the inputs $a$, $c$ into their random linear combinations $r_1 a + r_2 c$, $r_3 a + r_4 c$, abbreviated as $(a, c)_1$, $(a, c)_2$. Similarly, Charlie 12 converts the inputs $c$, $e$ into $(c, e)_1$, $(c, e)_2$. As a result, duplication and routing of input packets can be performed at once at each Charlie, which enables simpler configuration of multicast graph. In contrast, in the case of conventional packet switching of Supplementary Fig. 1e, further paths need to be added to realize multicast, e.g., paths from B1 to B2 to send symbol $c$ after B1 completes its decoding.

Thus, network coding is essential for flexible design and path control of multicast graphs to cope with both secrecy and reliability. Then $[n_0, k]$ RS codes or conventional Hamming distance-based codes do not work well for network coded packets but MRD codes need to be employed for controlling rank errors and erasures.

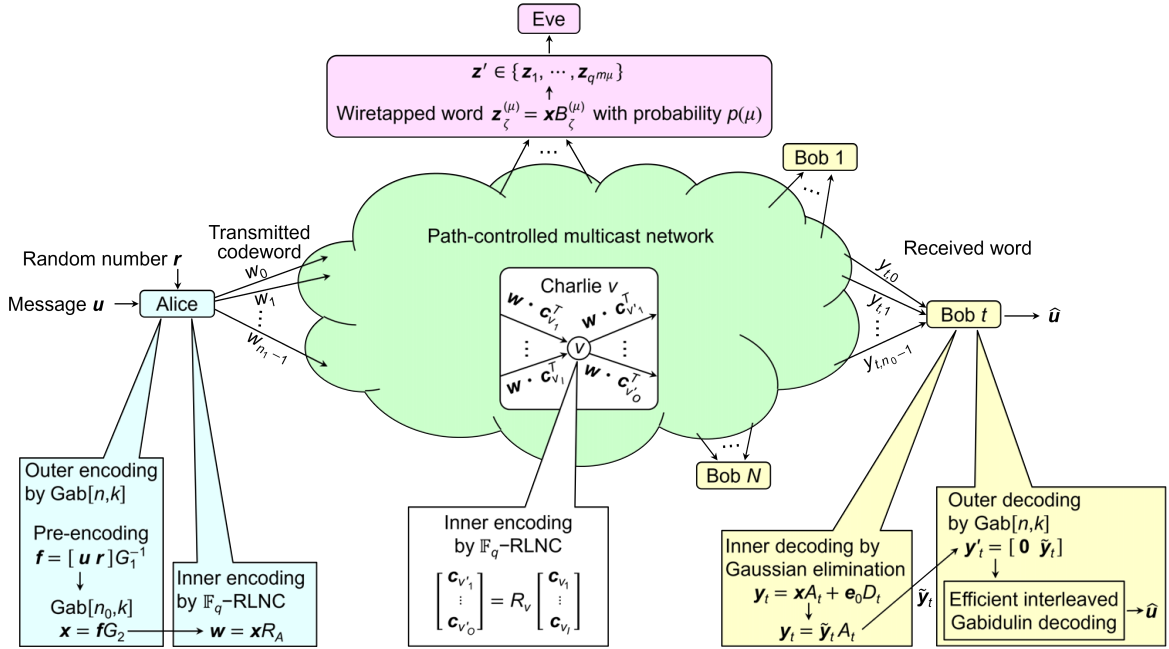

**Supplementary Fig. 2. A schematic diagram of the PUSNEC.** The scheme consists of usr-secrecy and error-control coding based on Gabidulin codes in Alice and Bobs as an outer code, and $\mathbb{F}_q$-RLNC in Alice, Charlies and Bobs as an inner code. Alice, Charlies, and Bobs are connected by a path-controlled multicast network. $\text{Gab}[n,k]$ and $\text{Gab}[n_0,k]$ denote $[n,k]$ and $[n_0,k]$ Gabidulin codes.

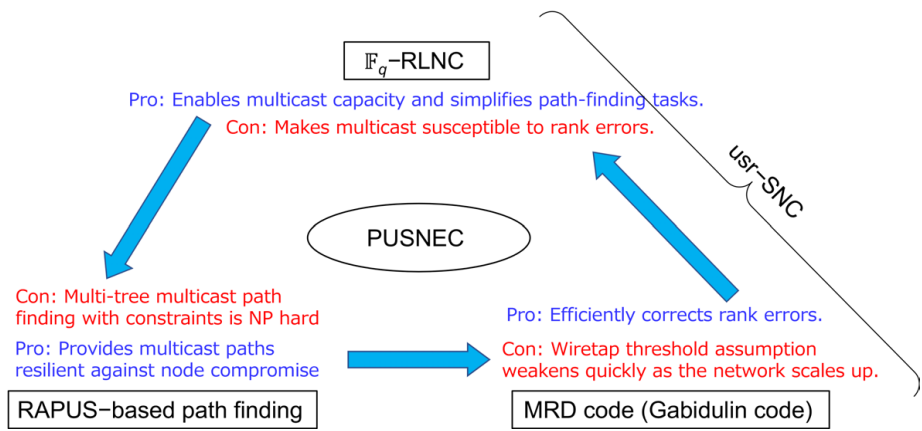

**Supplementary Fig. 3. Three techniques constituting the PUSNEC and their roles (Pros and Cons).**

## II. GENERALIZATION TO THE PUSNEC

Generalization to scalable multicast networks that achieve ITS and exhibit high resilience against errors and tampering can be made by the PUSNEC as shown in Supplementary Fig. 2.

The PUSNEC consists of three techniques: MRD code (especially Gabidulin code), $\mathbb{F}_q$-RLNC, and the RAPUS-based path control as depicted in Supplementary Fig. 3.

$\mathbb{F}_q$-RLNC is essential for realizing multicast capacity. On the other hand, its mixing and duplicating nature makes multicast susceptible to errors. In the worst case, an erroneous packet occurring only on a certain link may pollute all received packets at Bob(s), causing a burst error that cannot be corrected by standard error-control codes.

These errors usually have linear relationships, can be viewed as *rank errors*, and efficiently corrected by MRD codes. $\mathbb{F}_q$-RLNC and MRD code constitutes the usr-SNC. Unfortunately, the usr-SNC alone falls short of satisfying the wiretap threshold assumption as the network scales up.

The RAPUS-based path control enhances resilience against node-based Eve, widening usr-SNC's applicability. $\mathbb{F}_q$-RLNC simplifies the RAPUS-based path finding algorithms. The combination of these three techniques enables IT-secure, reliable, and resource-efficient multicast.

## III. WIRETAP CHANNEL MATRIX AND LEAKAGE INFORMATION OF 3-PATH RSDS

We describe how to quantify leakage information of the 3-path RSDS shown in Supplementary Fig. 1c. To be more specific, we consider the case that one of Charlies in the top path is compromised and the symbol $x_0$ is tapped by Eve. Our interest is the two kinds of mutual information $I(U_0; X_0)$

**Supplementary Table I**
**Wiretap channel matrix $[P(x_0|u_0)]$.**

|       |            | $x_0$ |     |          |            |
|-------|------------|-------|-----|----------|------------|
|       |            | 0     | 1   | $\alpha$ | $\alpha^2$ |
| $u_0$ | 0          | 1/4   | 1/4 | 1/4      | 1/4        |
|       | 1          | 1/4   | 1/4 | 1/4      | 1/4        |
|       | $\alpha$   | 1/4   | 1/4 | 1/4      | 1/4        |
|       | $\alpha^2$ | 1/4   | 1/4 | 1/4      | 1/4        |

**Supplementary Table II**
**Wiretap channel matrix $[P(x_0|u_0, u_1)]$. THE BLANKS IN THE MATRIX MEANS 0.**

|           |                    | $x_0$ |     |          |            |
|-----------|--------------------|-------|-----|----------|------------|
|           |                    | 0     | 1   | $\alpha$ | $\alpha^2$ |
| $u_0 u_1$ | 00                 | 1     |     |          |            |
|           | $\alpha\alpha$     | 1     |     |          |            |
|           | $\alpha^2\alpha^2$ | 1     |     |          |            |
|           | 11                 | 1     |     |          |            |
|           | $\alpha^2\alpha$   |       | 1   |          |            |
|           | 10                 |       | 1   |          |            |
|           | 01                 |       | 1   |          |            |
|           | $\alpha\alpha^2$   |       | 1   |          |            |
|           | $1\alpha^2$        |       |     | 1        |            |
|           | $\alpha^2\alpha$   |       |     | 1        |            |
|           | $\alpha 0$         |       |     | 1        |            |
|           | $0\alpha$          |       |     | 1        |            |
|           | $\alpha 1$         |       |     |          | 1          |
|           | $0\alpha^2$        |       |     |          | 1          |
|           | $1\alpha$          |       |     |          | 1          |
|           | $\alpha^2 0$       |       |     |          | 1          |

($= I(U_1; X_0) = 0$) and $I(U_0, U_1; X_0)$. The former is given by

$$I(U_0; X_0) = \sum_{u_0} P(u_0) \sum_{x_0} P(x_0|u_0) \log_2 \frac{P(x_0|u_0)}{\sum_{u_0'} P(x_0|u_0')},$$
(S3)

where $P(u_0)$ is a priori probability of $u_0$, which is assumed to be equal here, $P(u_0) = 1/q^m$, and $P(x_0|u_0)$ is the conditional probability that Eve's observation is $x_0$, given $u_0$.

Let us consider a concrete example of $q = 2$ and $m = 2$, and the message elements $u_0, u_1 \in \mathbb{F}_4$. Since $x_0 = u_0 + u_1$ by Eq. (S1), the wiretap channel matrix specified by conditional probability $P(x_0|u_0)$ is given as Supplementary Table I. This means that if Eve observes $x_0$ to be one of $\{0, 1, \alpha, \alpha^2\}$, $u_0 = 0, 1, \alpha, \alpha^2$ are equally likely, that is, no information is obtained. Actually, by substituting $P(x_0|u_0)$ of Supplementary Table I into Eq. (S3), we immediately see $I(U_0; X_0) = 0$.

Although Eve is uncertain about $u_0$ or $u_1$ itself, she can know some information on the whole message $\begin{bmatrix} u_0 & u_1 \end{bmatrix}$. To see this, we consider the wiretap channel matrix $[P(x_0|u_0, u_1)]$, which is given by Supplementary Table II. As seen, $\begin{bmatrix} u_0 & u_1 \end{bmatrix}$ can be classified into four groups, each of which is characterized by the sum $u_0 + u_1 = 0, 1, \alpha, \alpha^2$, and corresponds to each value of $x_0$. Thus observing $x_0$, Eve can tell which of the four groups the transmitted message belongs to. This is actually the information leaked to Eve, and its amount is measured by the mutual information to be $I(U_0, U_1; X_0) = \log_2 4$, as directly derived from $[P(x_0|u_0, u_1)]$ of Supplementary Table II. This is what should be paid for improving the transmission efficiency by the ramp scheme.

If Eve collects two symbols, she can decrypt the message as exemplified in Eq. (S2). This means that the wiretapped symbols $\begin{bmatrix} x_0 & x_1 \end{bmatrix}$ has 1-to-1 relation with the message $\begin{bmatrix} u_0 & u_1 \end{bmatrix}$ as shown by the wiretap channel matrix $[P(x_0, x_1|u_0, u_1)]$ given in Supplementary Table III.

**Supplementary Table III**
**Wiretap channel matrix** $[P(x_0, x_1|u_0, u_1)]$. THE BLANKS IN THE MATRIX MEANS 0.

| | | $x_0\,x_1$ | | | | | | | | | | | | | | | |
|---|---|---|---|---|---|---|---|---|---|---|---|---|---|---|---|---|---|
| | | 00 | 01 | $0\alpha$ | $0\alpha^2$ | 10 | 11 | $1\alpha$ | $1\alpha^2$ | $\alpha0$ | $\alpha1$ | $\alpha\alpha$ | $\alpha\alpha^2$ | $\alpha^20$ | $\alpha^21$ | $\alpha^2\alpha$ | $\alpha^2\alpha^2$ |
| $u_0\,u_1$ | 00 | 1 | | | | | | | | | | | | | | | |
| | $\alpha\alpha$ | | 1 | | | | | | | | | | | | | | |
| | $\alpha^2\alpha^2$ | | | 1 | | | | | | | | | | | | | |
| | 11 | | | | 1 | | | | | | | | | | | | |
| | $\alpha^2\alpha$ | | | | | 1 | | | | | | | | | | | |
| | 10 | | | | | | 1 | | | | | | | | | | |
| | 01 | | | | | | | 1 | | | | | | | | | |
| | $\alpha\alpha^2$ | | | | | | | | 1 | | | | | | | | |
| | $1\alpha^2$ | | | | | | | | | 1 | | | | | | | |
| | $\alpha^21$ | | | | | | | | | | 1 | | | | | | |
| | $\alpha0$ | | | | | | | | | | | 1 | | | | | |
| | $0\alpha$ | | | | | | | | | | | | 1 | | | | |
| | $\alpha1$ | | | | | | | | | | | | | 1 | | | |
| | $0\alpha^2$ | | | | | | | | | | | | | | 1 | | |
| | $1\alpha$ | | | | | | | | | | | | | | | 1 | |
| | $\alpha^20$ | | | | | | | | | | | | | | | | 1 |

# Supplementary Note 3: Efficient implementation of Gabidulin codes

Although there have been a fair amount of theoretical studies on efficient decoding algorithms of Gabidulin codes, only a few have been reported on software implementations. We used two kinds of techniques for efficient implementations, i.e., parallel processing over a smaller field (interleaving) and arithmetic operations with appropriate bases.

## I. INTERLEAVED GABIDULIN CODES

In our case, $m$ corresponds to the packet length, which is usually much larger than the code length $n$. We can then take $m = ln$ for some integer $l$, divide each packet into $l$ components of length $n$, and encode each component by Gab$[n, k]$ over the *smaller* field $\mathbb{F}_{q^n}$ in parallel. In other word, we form the vertically interleaved Gabidulin code [14], [15], which we denote as iGab$[n, k]$,

$$\mathrm{iGab}[n,k] \equiv \left\{ \boldsymbol{x}' \Big| \boldsymbol{x}' = \begin{bmatrix} \boldsymbol{x}'^{(0)} \\ \vdots \\ \boldsymbol{x}'^{(l-1)} \end{bmatrix}, \boldsymbol{x}'^{(i)} = \boldsymbol{f}^{(i)}\hat{G} \right\}, \tag{S4}$$

where each of $l$ components is constructed by evaluating the linearized polynomial $f^{(i)}(x)$ over $\mathbb{F}_{q^n}$ at the common points $\hat{g}_0, \hat{g}_1, ..., \hat{g}_{n-1} \in \mathbb{F}_{q^n}$, which are $\mathbb{F}_q$-linearly independent and constitute the generator matrix

$$\hat{G} = \begin{bmatrix} \hat{g}_0^{[0]} & \hat{g}_1^{[0]} & \cdots & \hat{g}_{n-1}^{[0]} \\ \hat{g}_0^{[1]} & \hat{g}_1^{[1]} & \cdots & \hat{g}_{n-1}^{[1]} \\ \vdots & \vdots & \ddots & \vdots \\ \hat{g}_0^{[k-1]} & \hat{g}_1^{[k-1]} & \cdots & \hat{g}_{n-1}^{[k-1]} \end{bmatrix} \in \mathbb{F}_{q^n}^{k \times n}. \tag{S5}$$

The iGab$[n, k]$ is known to significantly reduce computational complexity through optimized parallel processing over the smaller field, while maintaining the same error correction capability as the Gab$[n, k]$ over $\mathbb{F}_{q^m}$ [16]. Note that the iGab$[n, k]$ spans the Cartesian product space $\mathbb{F}_{q^n}^l$ (which is not necessarily the field), hence it is also referred to as the Cartesian product of Gabidulin codes.

Technical details of implementing vertically interleaved Gabidulin codes are explained in Supplementary Note 13.

The outer code of the PUSNEC is constructed based on iGab$[n, k]$. We divide $\boldsymbol{u}$, $\boldsymbol{r}$, and $\boldsymbol{f}$ vertically into $l$ components of the same row height $n$

$$\boldsymbol{u} = \begin{bmatrix} \boldsymbol{u}^{(0)} \\ \vdots \\ \boldsymbol{u}^{(l-1)} \end{bmatrix}, \boldsymbol{r} = \begin{bmatrix} \boldsymbol{r}^{(0)} \\ \vdots \\ \boldsymbol{r}^{(l-1)} \end{bmatrix}, \boldsymbol{f} = \begin{bmatrix} \boldsymbol{f}^{(0)} \\ \vdots \\ \boldsymbol{f}^{(l-1)} \end{bmatrix}, \tag{S6}$$

where $\boldsymbol{u}^{(i)} \in \mathbb{F}_{q^n}^{k_0}$, $\boldsymbol{r}^{(i)} \in \mathbb{F}_{q^n}^{\mu_0}$, and $\boldsymbol{f}^{(i)} \in \mathbb{F}_{q^n}^k$.

The pre-encoding is made by

$$\boldsymbol{f}^{(i)} = \begin{bmatrix} \boldsymbol{u}^{(i)} & \boldsymbol{r}^{(i)} \end{bmatrix} \hat{G}_1^{-1}, \quad i = 0, ..., l-1, \tag{S7}$$

where

$$\hat{G}_1 = \begin{bmatrix} \hat{g}_0^{[0]} & \hat{g}_1^{[0]} & \cdots & \hat{g}_{k-1}^{[0]} \\ \hat{g}_0^{[1]} & \hat{g}_1^{[1]} & \cdots & \hat{g}_{k-1}^{[1]} \\ \vdots & \vdots & \ddots & \vdots \\ \hat{g}_0^{[k-1]} & \hat{g}_1^{[k-1]} & \cdots & \hat{g}_{k-1}^{[k-1]} \end{bmatrix} \in \mathbb{F}_{q^n}^{k \times k}. \tag{S8}$$

Practical and efficient implementation of Eq. (S7) is explained in Supplementary Note 13-Subsection V-A.

Then the Gab$[n_0, k]$ encoding over the smaller field $\mathbb{F}_{q^n}$ is performed for $n_0$-degree distribution

$$\boldsymbol{x}^{(i)} = \boldsymbol{f}^{(i)}\hat{G}_2 \in \mathbb{F}_{q^n}^{n_0}, \quad i = 0, ..., l-1, \tag{S9}$$

where

$$\hat{G}_2 = \begin{bmatrix} \hat{g}_{k_1}^{[0]} & \cdots & \hat{g}_{n-1}^{[0]} \\ \hat{g}_{k_1}^{[1]} & \cdots & \hat{g}_{n-1}^{[1]} \\ \vdots & \ddots & \vdots \\ \hat{g}_{k_1}^{[k-1]} & \cdots & \hat{g}_{n-1}^{[k-1]} \end{bmatrix} \in \mathbb{F}_{q^n}^{k \times n_0}. \tag{S10}$$

The $l$ components are combined to an outer codeword

$$\boldsymbol{x} = \boldsymbol{f}\hat{G}_2 = \begin{bmatrix} \boldsymbol{x}^{(0)} \\ \vdots \\ \boldsymbol{x}^{(l-1)} \end{bmatrix} \in \mathbb{F}_{q^n}^{l \times n_0}, \tag{S11}$$

where $\mathbb{F}_{q^n}^{l \times n_0}$ represents a set of $l \times n_0$ matrices over $\mathbb{F}_{q^n}$. The $\boldsymbol{x}$ is further encoded into an inner codeword for transmission $\boldsymbol{w} = \boldsymbol{x}R_A \in \mathbb{F}_{q^n}^{l \times n_1}$ by $\mathbb{F}_q$-RLNC as Eq. (4). The $n_1$ symbols are sent out into $n_1$ links.

The received word at each Bob is a length-$n_0$ sequence $\boldsymbol{y} \in \mathbb{F}_{q^n}^{l \times n_0}$. This is first decoded by the $\mathbb{F}_q$-linear random network decoder, and then converted into the outer word $\tilde{\boldsymbol{y}} \in \mathbb{F}_{q^n}^{l \times n_0}$. The $\tilde{\boldsymbol{y}}$ is then decoded by the iGab$[n, k]$ decoder, which deals with $\tilde{\boldsymbol{y}}$ as a part of the length-$n$ word $\boldsymbol{y}' = \begin{bmatrix} \boldsymbol{0} & \tilde{\boldsymbol{y}} \end{bmatrix}$ with $k_1$ erasures expressed by $\boldsymbol{0}$. The iGab$[n, k]$ decoder first divides the input vertically into $l$ components

$$\boldsymbol{y}' = \begin{bmatrix} \boldsymbol{y}'^{(0)} \\ \vdots \\ \boldsymbol{y}'^{(l-1)} \end{bmatrix}, \tilde{\boldsymbol{y}}^{(i)} \in \mathbb{F}_{q^n}^{l \times n}. \tag{S12}$$

Then, this word is processed by our newly developed efficient decoder of iGab$[n, k]$, whose details are described in Supplementary Note 13-Subsection V-B.

In our implementation, we set $l = 3$ and constructed 3-fold interleaved Gabidulin codes iGab$[n, k]$ over $\mathbb{F}_{q^n}^l$. For the 5-degree PUSNEC ($n_0 = 5$), we used $q = 256$ and $n = 9, 11$. For the 6-degree PUSNEC ($n_0 = 6$), we used $q = 32$ and $n = 14$.

## II. Optimal self-dual normal basis

The elements of $\mathbb{F}_{q^n}$ are represented as $n$-dimensional vectors over $\mathbb{F}_q$ with regard to a *normal basis* of the form $\mathcal{B} = \{\beta^{[0]}, \beta^{[1]}, ..., \beta^{[n-1]}\}$. The generator matrix of Gabidulin codes can be given by the $q$-cyclic form of the normal basis [17]. All the Frobenius $q$-power evaluation are performed using cyclic shift (negligible complexity) [18].

The multiplication table $T \in \mathbb{F}_q^{n \times n}$ should have less non-zero entries. The number of non-zero entries $C_T$ of $T$ is called the complexity of the basis and is lower bounded by $C_T \geq 2n - 1$. In case $C_T = 2n - 1$, $\mathcal{B}$ is called optimal [19]. Although normal bases exist over any finite field, the optimal normal bases exist only for certain choices of $q$ and $n$. The optimal normal bases are special cases of the so-called *Gauss periods* [20], [21] and can be constructed by the methods given in [19], [22]. For any given normal basis $\mathcal{B}$, there exists a unique dual basis $\mathcal{B}^\perp$. If a basis is dual to itself, $\mathcal{B} = \mathcal{B}^\perp$, we call it a *self-dual* basis. Using a *self-dual* basis, both the generator matrix $\hat{G}$ and the parity-check matrix $\hat{H}$ can be formed with the same elements, and hence stored in a single memory.

For further ease of computation, we take $q$ to be a power of 2, i.e., $q = 2^w$ and searched appropriate combinations of $(n, w)$ for the optimal self-dual normal bases, such that all the entries in the multiplication table can be expressed as sequences of only 0 and 1. Such combinations are summarized in Supplementary Table IV.

## III. Selection guidelines of coding parameters of Gabidulin codes

We call Gabidulin codes with these combinations *good* Gabidulin codes. The code lengths of good Gabidulin codes are restricted to $n = 3, 5, 6, 9, 11, 14, 23, ....$ We employ the optimal self-dual normal bases thus obtained, and could make encoding and decoding much faster.

Supplementary Table V summarizes selection guidelines for good Gabidulin codes for given sets of $n_0$, $k_0$, and $\mu_0$. The first and second columns start from $n_0 = 5$ and $k = 3$. In fact, the case of $k = 2$ is outside the scope of this paper, because the indegree of each Charlie needs to be $in(v_i) = 1$ according to Eq. (6), which means that there is no room to employ network coding, making it much harder to configure multicast graphs. In the case of $k = 3$, the wiretap threshold is $\mu \leq 2$. In order to correct errors ($\tau > 0$), we need to take $n_0 \geq 5$ according to Eq. (9). The third column shows the maximum possible values of $2\tau + \rho$ as the error-control capability. The sixth column indicates the key consumption index. The seventh column presents sets of the code length and dimension $[n, k]$ of Gabidulin codes. The last column indicates whether the code length is minimal or redundant for ensuring strong ramp secrecy.

In the following, we mention typical features of Supplementary Table V. In the case of $(n_0, k) = (5, 3)$, the error/erasure-control capability is specified by $2\tau + \rho = 2$, that is, one rank error or two rank erasures can be corrected. We further distinguish three cases $(k_0, \mu_0) = (1, 2), (2, 1), (3, 0)$.

**Supplementary Table IV**
**Combination of $(n, w)$ for the optimal self-dual normal bases.**

| | $w$ | | | | | | | | | |
| | 1 | 2 | 3 | 4 | 5 | 6 | 7 | 8 | 9 | 10 |
|---|---|---|---|---|---|---|---|---|---|---|
| 1 | ○ | ○ | ○ | ○ | ○ | ○ | ○ | ○ | ○ | ○ |
| 2 | ○ | | ○ | | ○ | | ○ | | ○ | |
| 3 | ○ | ○ | | ○ | | ○ | | ○ | | ○ |
| 4 | | | | | | | | | | |
| 5 | ○ | ○ | | ○ | | ○ | | ○ | | ○ |
| 6 | ○ | | | | ○ | | ○ | | | |
| 7 | | | | | | | | | | |
| 8 | | | | | | | | | | |
| 9 | ○ | | | | | | | ○ | | |
| 10 | | | | | | | | | | |
| 11 | ○ | | | | | | | ○ | | |
| 12 | | | | | | | | | | |
| 13 | | | | | | | | | | |
| 14 | ○ | | ○ | | | | | ○ | | |
| 15 | | | | | | | | | | |
| 16 | | | | | | | | | | |
| 17 | | | | | | | | | | |
| 18 | ○ | | | | ○ | | | | | |
| 19 | | | | | | | | | | |
| 20 | | | | | | | | | | |
| 21 | | | | | | | | | | |
| 22 | | | | | | | | | | |
| 23 | ○ | ○ | | ○ | | ○ | | ○ | | ○ |
| 24 | | | | | | | | | | |
| 25 | | | | | | | | | | |

(left column header: $n$)

- For $(k_0, \mu_0) = (1, 2)$, we can use Gab$[6, 3]$ that has the minimal code length for strong secrecy ($n = k_0 + n_0$). Note that this case is not the ramp scheme. Using this code, we can realize perfect secrecy for $\mu \leq 2$.

- For $(k_0, \mu_0) = (2, 1)$, although the minimal code length is $k_0 + n_0 = 7$, good Gabidulin codes are not available. Therefore, we have to use Gab$[9, 3]$ even though the number of symbols is redundant. This code can ensure perfect secrecy for $\mu = 1$, and strong ramp secrecy for $\mu = 2$.

- For $(k_0, \mu_0) = (3, 0)$, we use Gab$[9, 3]$ again although the minimal code length is $k_0 + n_0 = 8$. In this case, perfect secrecy cannot be ensured, but strong ramp secrecy holds for $\mu \leq 2$.

From the viewpoint of reducing OTP-key consumption, it is better to use larger $k_0$, especially when the degree of distributed transmission $n_0$ becomes larger. To quickly capture to what extent the coding scheme can potentially save OTP-key consumption, we may use the key consumption index defined as

$$C_{key} \equiv \frac{n_0}{k_0}. \quad (S13)$$

Total OTP-key consumption in the whole network should be larger than $C_{key} N$, where $N$ is the number of Bobs. $C_{key}$ is minimized to be 5/3 for $(k_0, \mu_0) = (3, 0)$.

In the case of $(n_0, k) = (6, 3)$, we use Gab$[9, 3]$ for all of $(k_0, \mu_0) = (1, 2), (2, 1), (3, 0)$. Compared with the previous case of $(n_0, k) = (5, 3)$, the error/erasure-control capability is stronger, namely $2\tau + \rho = 3$, meaning that we can correct one rank error and one rank erasure, or three rank erasures. The secrecy performance is the same as that of $(n_0, k) = (5, 3)$ for each $(k_0, \mu_0) = (1, 2), (2, 1), (3, 0)$. $C_{key}$ is minimized to be 2 for $(k_0, \mu_0) = (3, 0)$ which is slightly larger than that of $(n_0, k) = (5, 3)$.

**Supplementary Table V**
**Selection guidelines of good Gabidulin codes for given sets of $n_0$, $k_0$, and $\mu_0$.**

| $n_0$ | $k$ | $2\tau+\rho$ | $k_0$ | $\mu_0$ | $C_{\text{key}}$ | code | code length |
|---|---|---|---|---|---|---|---|
| 5 | 3 | 2 | 1 | 2 | 5 | [6, 3] | minimal |
|  |  |  | 2 | 1 | 5/2 | [9, 3] | redundant |
|  |  |  | 3 | 0 | 5/3 | [9, 3] | redundant |
| 6 | 3 | 3 | 1 | 2 | 6 | [9, 3] | redundant |
|  |  |  | 2 | 1 | 3 | [9, 3] | redundant |
|  |  |  | 3 | 0 | 2 | [9, 3] | minimal |
|  | 4 | 2 | 1 | 3 | 6 | [9, 4] | redundant |
|  |  |  | 2 | 2 | 3 | [9, 4] | redundant |
|  |  |  | 3 | 1 | 2 | [9, 4] | minimal |
|  |  |  | 4 | 0 | 3/2 | [11, 4] | redundant |
| 7 | 3 | 4 | 1 | 2 | 7 | [9, 3] | redundant |
|  |  |  | 2 | 1 | 7/2 | [9, 3] | minimal |
|  |  |  | 3 | 0 | 7/3 | [11, 3] | redundant |
|  | 4 | 3 | 1 | 3 | 7 | [9, 4] | redundant |
|  |  |  | 2 | 2 | 7/2 | [9, 4] | minimal |
|  |  |  | 3 | 1 | 7/3 | [11, 4] | redundant |
|  |  |  | 4 | 0 | 7/4 | [11, 4] | minimal |
|  | 5 | 2 | 1 | 4 | 7 | [9, 5] | redundant |
|  |  |  | 2 | 3 | 7/2 | [9, 5] | minimal |
|  |  |  | 3 | 2 | 7/3 | [11, 5] | redundant |
|  |  |  | 4 | 1 | 7/4 | [11, 5] | minimal |
|  |  |  | 5 | 0 | 7/5 | [14, 5] | redundant |
| 8 | 3 | 5 | 1 | 2 | 8 | [9, 3] | minimal |
|  |  |  | 2 | 1 | 4 | [9, 3] | redundant |
|  |  |  | 3 | 0 | 8/3 | [11, 3] | minimal |
|  | 4 | 4 | 1 | 3 | 8 | [9, 4] | minimal |
|  |  |  | 2 | 2 | 4 | [11, 4] | redundant |
|  |  |  | 3 | 1 | 8/3 | [11, 4] | minimal |
|  |  |  | 4 | 0 | 2 | [14, 4] | redundant |
|  | 5 | 3 | 1 | 4 | 8 | [9, 5] | minimal |
|  |  |  | 2 | 3 | 4 | [11, 5] | redundant |
|  |  |  | 3 | 2 | 8/3 | [11, 5] | minimal |
|  |  |  | 4 | 1 | 2 | [14, 5] | redundant |
|  |  |  | 5 | 0 | 8/5 | [14, 5] | minimal |
|  | 6 | 2 | 1 | 5 | 8 | [9, 6] | minimal |
|  |  |  | 2 | 4 | 4 | [11, 6] | redundant |
|  |  |  | 3 | 3 | 8/3 | [11, 6] | minimal |
|  |  |  | 4 | 2 | 2 | [14, 6] | redundant |
|  |  |  | 5 | 1 | 8/5 | [14, 6] | redundant |
|  |  |  | 6 | 0 | 4/3 | [14, 6] | minimal |

In the case of $(n_0, k) = (6, 4)$, the wiretap threshold for secrecy can be larger such that $\mu \leq 3$. This allows 3-link amplification at Charlies, leading to richer topologies for multicast. On the other hand, if this is used in many Charlies, the tolerance for node-based Eve becomes weaker. So one needs to design graph topologies carefully in a balance between multicast efficiency and secrecy. For $(k_0, \mu_0) = (1, 3), (2, 2), (3, 1)$, we use Gab[9, 4], while for $(k_0, \mu_0) = (4, 0)$, we use Gab[11, 4]. The error-control capability is weaker than that of $(n_0, k) = (6, 3)$. $C_{\text{key}}$ is minimized to be 1.5 for $(k_0, \mu_0) = (4, 0)$ which is smaller than those of $(n_0, k) = (5, 3), (6, 3)$. If errors and erasures are expected to be relatively smaller, the scheme of $(n_0, k) = (6, 4)$ would be useful for supporting a larger number of Bobs.

As the degree of distribution $n_0$ increases, network resource overheads become larger. So we think that the schemes of $(n_0, k) = (5, 3), (6, 3), (6, 4)$ with the 2-link amplification would be sensible.

## IV. PROCESSING TIME OF THE DECODER

Gabidulin codes were programmed in c-language on a computer running Linux Ubuntu 22.04.3 with a CPU of AMD Ryzen 7 5700G and a 16GB memory of DDR4 with 3.2GHz clock.

Supplementary Table VI shows the decoding time of iGab[9, 3] over $\mathbb{F}_{256^9}$ per transmission at each Bob in several cases. It ranged from $307\,\mu s$ to $344\,\mu s$, which was short enough to conduct massive transmission simulation. Since the basic code unit size is $m \times n_0 = 135$ bytes per transmission, the decoder throughput is 3.1∼3.5 Mbits per second. When the link error rate $e_{L2}$ (the node erasure rate $\epsilon_N$) decreases from $10^{-2}$ to $10^{-3}$, the decoding time decreases slightly.

For comparison, the decoding time of RS[243, 81] over $\mathbb{F}_{2^8}$, which has the same coding rate and almost the same code length in bits as iGab[9, 3], is also shown for each case. One might feel odd that the decoding time of the RS code is longer than that of Gabidulin code. However, this indicates that correcting rank errors/erasures is a very demanding task for RS codes, requiring more operations to correct many symbol errors/erasures. It is worth mentioning that the decoding time of the RS code for $e_{L2} = 10^{-2}$ is shorter than that for $e_{L2} = 10^{-3}$. This is because, when the link error rate is larger, the RS decoder can more quickly decide that the errors are beyond the error-correction capability and stop the decoding process.

# Supplementary Note 4: Codeword structure of good Gabidulin codes

The output sequence $\boldsymbol{x}'$ of length $n$ ($= k_1 + n_0$) from the encoder of good Gab[n, k], Eq. (5), includes $\boldsymbol{u}$, a part of $\boldsymbol{r}$, and $\boldsymbol{x}$.

In the case of $k_0 \leq k_1 < k$, the first $k - k_1$ columns of $G_2$ are actually the last $k - k_1$ columns of $G_1$. For $k_1 = k_0$, the code length $n$ can be minimal for ensuring strong ramp secrecy, and the output sequence becomes

$$\boldsymbol{x}' = \begin{bmatrix} \boldsymbol{u} & \boldsymbol{x} \end{bmatrix}. \tag{S14}$$

For $k_0 < k_1 < k$,

$$\boldsymbol{x}' \equiv \begin{bmatrix} \boldsymbol{u} & \boldsymbol{r}_{(k_1 - k_0)} & \boldsymbol{x} \end{bmatrix}, \tag{S15}$$

where $\boldsymbol{r}_{(k_1 - k_0)} \equiv \begin{bmatrix} r_0 & r_1 & \dots & r_{k_1 - k_0 - 1} \end{bmatrix}$ is the ordered sequence of the first $k_1 - k_0$ symbols of $\boldsymbol{r}$. The remaining last

| Model | Condition | iGab[9,3] | | RS[243,81] | |
|---|---|---|---|---|---|
| | | Decoding time ($\mu s$) | $\sigma$ | Decoding time ($\mu s$) | $\sigma$ |
| Terrestrial (Bob 6) | $e_{L2} = 10^{-2}$ | 334.2 | 30.9 | 486.8 | 50.5 |
| | $e_{L2} = 10^{-3}$ | 313.1 | 20.8 | 526.9 | 38.7 |
| | $\epsilon_N = 10^{-2}$ | 307.8 | 14.9 | 534.7 | 33.6 |
| | $\epsilon_N = 10^{-3}$ | 307.3 | 15.6 | 534.1 | 34.9 |
| Space (Bob 13) | $e_{L2} = 10^{-2}$ | 343.9 | 29.5 | 477.4 | 55.9 |
| | $e_{L2} = 10^{-3}$ | 321.9 | 27.8 | 511.1 | 37.3 |
| | $\epsilon_N = 10^{-2}$ | 321.3 | 21.5 | 532.0 | 32.4 |
| | $\epsilon_N = 10^{-3}$ | 312.6 | 22.2 | 529.4 | 39.4 |

$k - k_1$ symbols $\boldsymbol{r}_{(k-k_1)} \equiv \begin{bmatrix} r_{k_1-k_0} & r_{k_1-k_0+1} & \cdots & r_{\mu_0-1} \end{bmatrix}$ constitutes the first $k - k_1$ symbols of $\boldsymbol{x}$

$$\begin{bmatrix} x_0 & x_1 & \cdots & x_{k-k_1-1} \end{bmatrix} = \boldsymbol{r}_{(k-k_1)}. \quad \text{(S16)}$$

For $k \leq k_1$, on the other hand,

$$\boldsymbol{x}' \equiv \begin{bmatrix} \boldsymbol{u} & \boldsymbol{r} & \boldsymbol{r}_R & \boldsymbol{x} \end{bmatrix}, \quad \text{(S17)}$$

where $\boldsymbol{r}_R \equiv \begin{bmatrix} f(g_k) & \cdots & f(g_{k_1-1}) \end{bmatrix}$ is a sequence of redundant symbols which appear because good Gabidulin codes were available only for the code length $n > k_0 + \mu_0 + n_0$. In this case, there are no common symbols between $\boldsymbol{x}$ and $\boldsymbol{r}$.

# Supplementary Note 5: Secrecy metrics and their relation

Let $P_U(\boldsymbol{u})$ and $P_{Z|U}(\boldsymbol{z}|\boldsymbol{u})$ be a probability distribution of $\boldsymbol{u}$ and a conditional probability distribution of $\boldsymbol{z}$, given $\boldsymbol{u}$, respectively. The mutual information is given by

$$I(U;Z) = \sum_{\boldsymbol{u}} P_U(\boldsymbol{u}) D\left(P_{Z|U}(\cdot|\boldsymbol{u})||P_Z(\cdot)\right), \quad \text{(S18)}$$

where

$$D\left(P_{Z|U}(\cdot|\boldsymbol{u})||P_Z(\cdot)\right) \equiv \sum_{\boldsymbol{z}} P_{Z|U}(\boldsymbol{z}|\boldsymbol{u}) \log \frac{P_{Z|U}(\boldsymbol{z}|\boldsymbol{u})}{P_Z(\boldsymbol{z})}, \quad \text{(S19)}$$

is the Kullback-Leibler distance, and

$$P_Z(\boldsymbol{z}) = \sum_{\boldsymbol{u}} P_U(\boldsymbol{u}) P_{Z|U}(\boldsymbol{z}|\boldsymbol{u}), \quad \text{(S20)}$$

is the probability distribution of $\boldsymbol{z}$ at Eve.

The divergence distance is defined and expressed by

$$\begin{aligned} \delta(U;Z) &= \sum_{\boldsymbol{u}} P_U(\boldsymbol{u}) D\left(P_{Z|U}(\cdot|\boldsymbol{u})||\pi_Z(\cdot)\right) \\ &= I(U;Z) + D\left(P_Z(\cdot)||\pi_Z(\cdot)\right), \end{aligned} \quad \text{(S21)}$$

which is referred to as Pythagorean theorem, where $\pi_Z$ represents a prescribed target probability distribution $\pi_Z(\boldsymbol{z})$ at Eve. The choice of $\pi_Z$ will depend on the application. By Pinsker inequality, we have

$$\partial(U;Z)^2 \leq 2\delta(U;Z). \quad \text{(S22)}$$

Since $\delta(U;Z) \to 0$ implies $I(U;Z) \to 0$ and $\partial(U;Z) \to 0$, the divergence distance is the strongest secrecy criterion [23]. Its operational meaning is discussed in [24] such that $I(U;Z)$ is interpreted as a measure of "non-confusion" and $D(P_Z||\pi_Z)$ as a measure of "non-stealth." The target distribution $\pi_Z$ is often chosen as the distribution that Eve expects to observe when Alice is not communicating useful messages, such as the background noise distribution on $\boldsymbol{z}$. Thus by making $\delta(U;Z) \to 0$, we can not only keep the message secret from Eve but also hide the presence of meaningful communication.

We apply the divergence distance criterion to our usr-SNC. Since we assume that Eve can acquire packets without errors and erasures, Eve sees no background noise. We may then set $\pi_Z(\boldsymbol{z})$ as the probability distribution at Eve when Alice inputs a random number sequence $\boldsymbol{r}' = \begin{bmatrix} r'_0 & r'_1 & \cdots & r'_{k-1} \end{bmatrix}$ to the Gabidulin encoder, instead of the message $\boldsymbol{u}$ and the masking key $\boldsymbol{r}$. Since both $\begin{bmatrix} \boldsymbol{u} & \boldsymbol{r} \end{bmatrix}$ and $\boldsymbol{r}'$ distribute uniformly over $\mathbb{F}_{q^m}^k$, there is no difference between $P_Z(\boldsymbol{z})$ and $\pi_Z(\boldsymbol{z})$, and hence $D(P_Z||\pi_Z) = 0$, i.e., one achieves stealth. In such a case, the divergence distance coincides with the mutual information.

# Supplementary Note 6: Analytical evaluation of LII and FER in multicast over disjoint multi-hop paths

In the case of multicast over disjoint paths, we can evaluate the LII and FER analytically. Consider 1-to-$N$ multicast over $n_0$-disjoint $\eta$-hop paths with a uniform erasure rate $\epsilon$ and node compromise rate $\gamma$ as shown in Supplementary Fig. 4a. We assume that the link error rate is zero, i.e., $e_L = 0$.

A sequence of the message and masking key $\begin{bmatrix} \boldsymbol{u} & \boldsymbol{r} \end{bmatrix}$ is encoded into a codeword $\begin{bmatrix} x_0 & x_1 & \cdots & x_{n_0-1} \end{bmatrix}$ by $[n_0, k]$ MRD code such as Gabidulin code or maximum distance separable (MDS) code such as RS code.

Because the paths from Alice to each Bob are node-disjoint, there is nothing to do with $\mathbb{F}_q$-RLNC, and the $[n_0, k]$ MDS codes based on the Hamming metric and MRD codes based on the rank metric work equivalently as the ramp secret sharing scheme [4]–[6]. The codeword symbols $x_0, ..., x_{n_0-1}$ correspond to shares. Eve can decrypt $\begin{bmatrix} \boldsymbol{u} & \boldsymbol{r} \end{bmatrix}$ by collecting $k$ shares while any $k - 1$ or less shares do not leak out any meaningful information about $\begin{bmatrix} \boldsymbol{u} & \boldsymbol{r} \end{bmatrix}$.

For the $j$th path connecting Alice and Charlies $1j$, ..., $\eta j$, we can consider two events;

- the path is secure, i.e., no Charlies are compromised, whose probability is $(1 - \gamma)^\eta$,
- the path is compromised, i.e., one or more Charlies are compromised, whose probability is $1 - (1 - \gamma)^\eta$.

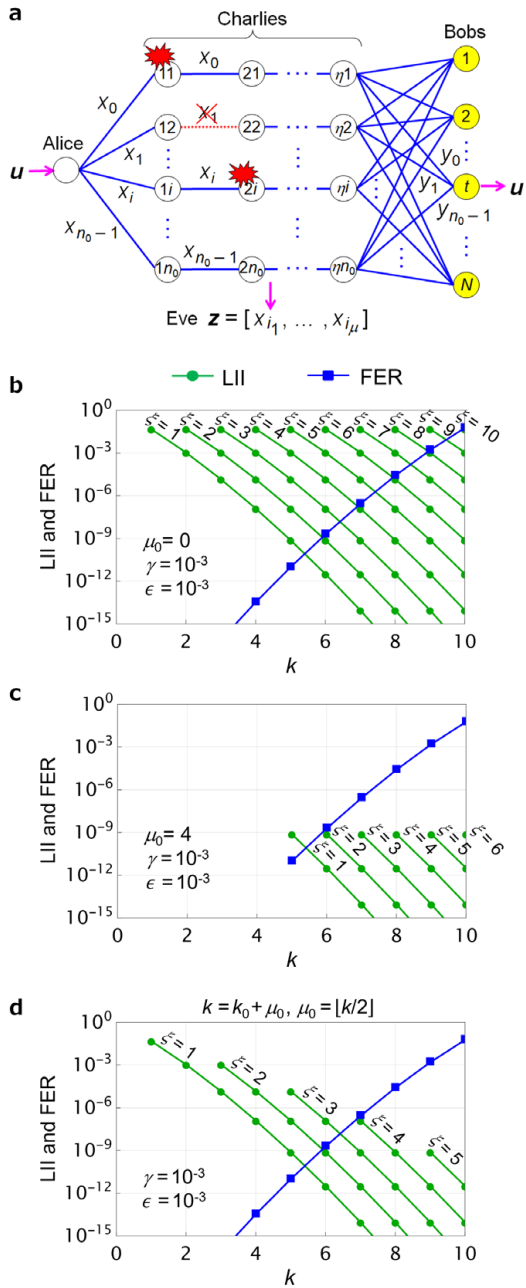

Eve $z = [x_{i_1}, \ldots, x_{i_{\mu}}]$

**Supplementary Fig. 4. Multicast over disjoint paths and the LII and FER as a function of $k$.** **a** A 1-to-$N$ multicast graph with $n_0$-disjoint $\eta$-hop paths. **b~d** The LII $I_L(\gamma, n_0, k, \xi)$ and FER $e_F(\epsilon, n_0, k)$ in the case of $\eta = 5$, $n_0 = 10$ and $k = 1 \sim 10$ with $\gamma = 10^{-3}$, $\epsilon = 10^{-3}$ ($e_L = e_N = 0$). **b** $\mu_0 = 0$ ($k = k_0$). **c** $\mu_0 = 4$ ($k = k_0 + 4$). **d** $\mu_0 = \lfloor k/2 \rfloor$, $k_0 = k - \mu_0$.

Then the $\mu$-wiretap probability is given by

$$p(\mu, \gamma, n_0, k) = {}_{n_0}C_{\mu}(1 - \gamma)^{\eta(n_0 - \mu)}[1 - (1 - \gamma)^{\eta}]^{\mu}. \quad (S23)$$

Using this quantity, the LII $I_L(\gamma, n_0, k, \xi)$ can be computed according to Theorem 1.

Next we evaluate the FER by using the fact that the $[n_0, k]$ MDS codes can correct erasures up to $n_0 - k$. We first consider the probability that erasures occur in $l$ paths among $n_0$ disjoint paths from Alice to the $\eta$-th hop Charlies $\eta 1$, ..., $\eta n_0$, which is given by

$$\varepsilon_1(\epsilon, n_0, l) = {}_{n_0}C_l(1 - \epsilon)^{\eta(n_0 - l)}[1 - (1 - \epsilon)^{\eta}]^{l}. \quad (S24)$$

We next consider the probability that $j$ erasures occur successively in the last hop links from Charlies $\eta 1$, ..., $\eta n_0$ to

Bob $t$, which is given by

$$\varepsilon_2(\epsilon, n_0 - l, j) = {}_{n_0 - l}C_j(1 - \epsilon)^{(n_0 - l - j)}\epsilon^{j}. \quad (S25)$$

Then, the probability that Bob $t$ can successfully correct erasures is given by

$$\sum_{l=0}^{n_0 - k} \varepsilon_1(\epsilon, n_0, l) \sum_{j=0}^{n_0 - k - l} \varepsilon_2(\epsilon, n_0 - l, j). \quad (S26)$$

Finally, the probability that all Bobs can successfully decode the message is given by

$$\sum_{l=0}^{n_0 - k} \varepsilon_1(\epsilon, n_0, l) \left[ \sum_{j=0}^{n_0 - k - l} \varepsilon_2(\epsilon, n_0 - l, j) \right]^{N}. \quad (S27)$$

Hence the FER is given by

$$e_F(\epsilon, n_0, k)$$
$$= 1 - \sum_{l=0}^{n_0 - k} \varepsilon_1(\epsilon, n_0, l) \left[ \sum_{j=0}^{n_0 - k - l} \varepsilon_2(\epsilon, n_0 - l, j) \right]^{N}. \quad (S28)$$

Their numerical behavior is shown in Supplementary Fig. 4b~d as a function of $k$ in the case of $n_0 = 10$. Strong ramp secrecy is characterized by the LII curves (green), such that, given $k$, the LII becomes larger for larger $\xi$. As $k$ increases, the LII becomes smaller, while the FER gets larger (blue curve). This illustrates the tradeoff between secrecy against Eve and reliability for Bobs quantitatively.

In the case of no masking key $\mu_0 = 0$ (Supplementary Fig. 4b), the message rate is maximized, while perfect secrecy of the *whole* message cannot be ensured but partial information on the message is gained by Eve in the sense that Eve's uncertainty regarding the message is reduced as exemplified in Supplementary Table II.

When the masking key is used (Supplementary Fig. 4c where $\mu_0 = 4$), the LII curves move downward compared with Supplementary Fig. 4b, namely, secrecy can be enhanced. For $\mu_0 = 4$, the LII and FER are defined for $k = 5 \sim 10$ because $k = k_0 + 4$ and $k_0 = 1 \sim 6$ ($\xi = 1 \sim 6$). The message rate is limited by $k_0/n_0 \leq 6/10$.

Supplementary Fig. 4d corresponds to the case with the balanced rate for the message and the masking key, i.e., $\mu_0 = \lfloor k/2 \rfloor$ and $k_0 = k - \mu_0$.

# Supplementary Note 7: 1-to-13 multicast paths on X-hop grid for LEO constellations

In the LEO satellite network model of Fig. 4, the satellite altitudes range from $500 \sim 600 \, \text{km}$ [26]. We assume that each LEO satellite for Charlie equips four antennas, which can support four links simultaneously.

The X-hop grid is generated on a plane at the average altitude of roughly $550 \, \text{km}$. The multicast path finding algorithm was run on this X-hop grid, and the virtual multicast graph $\tilde{\mathcal{G}}^{(13)}$ was derived as shown in Supplementary Fig. 5.

The virtual multicast graph $\tilde{\mathcal{G}}^{(13)}$ was then mapped back to a graph $\mathcal{G}^{(13)}$ in the LEO constellations, taking the three dimensional distribution of the satellites into account. The graph $\mathcal{G}^{(13)}$ thus mapped contained continuous segments of links without any branches or joints. Therefore the smoothing was applied to $\mathcal{G}^{(13)}$, and the smoother multicast graph $\bar{\mathcal{G}}^{(13)}$ with fewer hops was eventually derived as in Fig. 4.

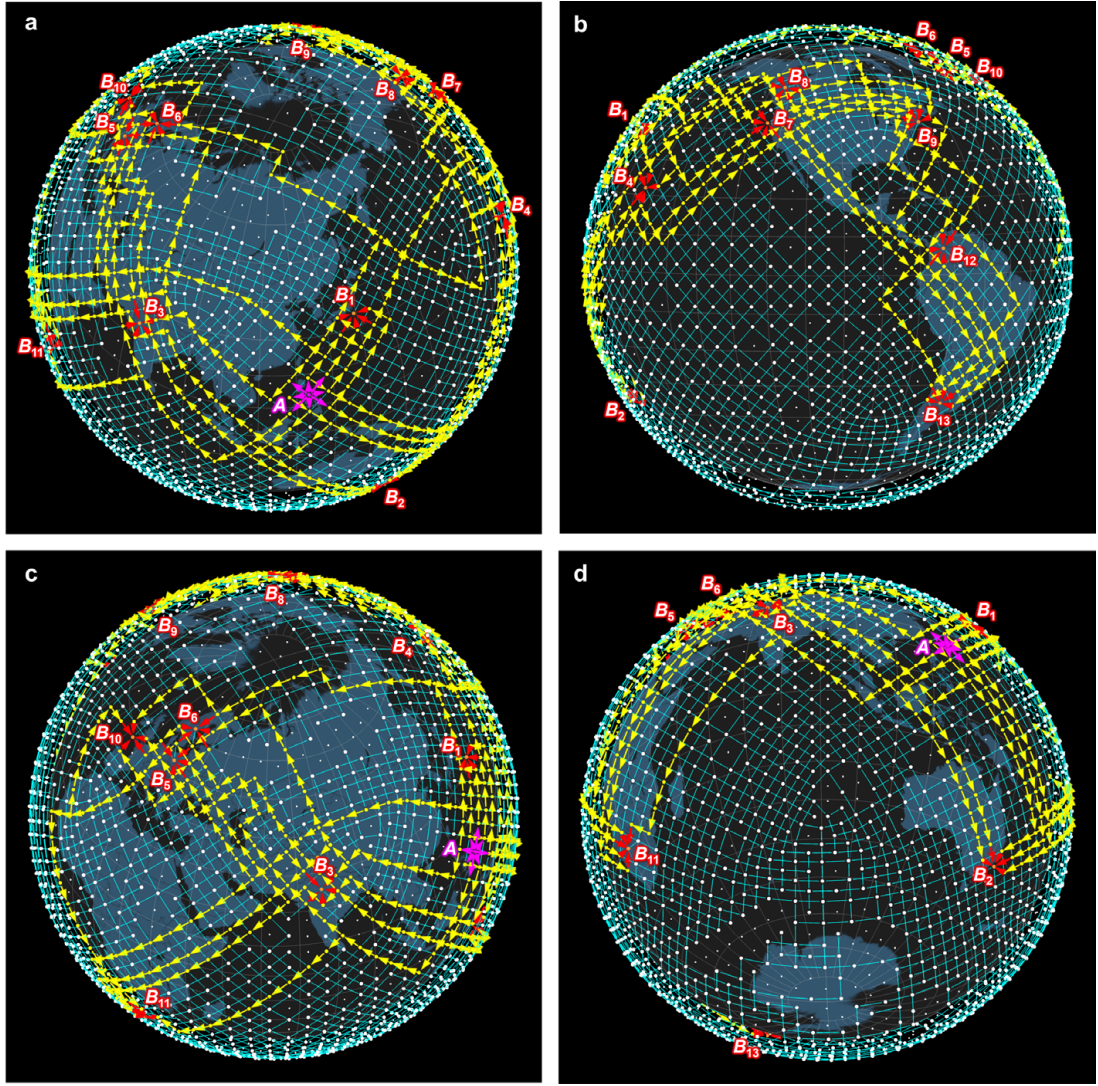

**Supplementary Fig. 5. The virtual 1-to-13 multicast graph on the X-hop grid corresponding to the multicast graph on the LEO constellations in Fig. 4.** The X-hop grid, viewed from four different directions—above Eastern Asia, the Eastern Pacific, Western Asia, and the Southern Indian Ocean—is represented by **a**, **b**, **c**, and **d**. A 3D interactive view is available in [25].

## Supplementary Note 8: Illustrative example of an error floor caused by finite-size effects

To illustrate the emergence of the error floor in FER-$B_t$ for the RS code, we consider a representative example that incorporates a transfer matrix and an error packet

$$A_t = \begin{bmatrix} 1 & 0 & 0 & 0 & 0 \\ a_1 & 1 & 0 & 0 & 0 \\ 0 & a_2 & 1 & 0 & 0 \\ 0 & 0 & a_3 & 1 & 0 \\ 0 & 0 & 0 & a_4 & a_5 \end{bmatrix}, \qquad (S29)$$

$$\boldsymbol{e_0}D_t = \begin{bmatrix} 0 & 0 & 0 & e & 0 \end{bmatrix}, \qquad (S30)$$

respectively in Eq. (7).

This means that an erasure occurs in the received symbol $y_{t,4}$ when $a_5 = 0$, and the received symbol $y_{t,3}$ suffers from the error $e$. The erasure can be expressed as $y_{t,4} = x_4 + F$ with

certain symbol $F$. The $\mathbb{F}_q$-random linear network decoder outputs the word

$$\tilde{\boldsymbol{y}}_t^T = \boldsymbol{x}^T + \begin{bmatrix} a_1 a_2 a_3 a_4 \\ -a_2 a_3 a_4 \\ a_3 a_4 \\ -a_4 \\ 1 \end{bmatrix} F \cdot \delta(a_5) + \begin{bmatrix} -a_1 a_2 a_3 \\ a_2 a_3 \\ -a_3 \\ 1 \\ 0 \end{bmatrix} e, \quad (S31)$$

where

$$\delta(a_5) = \begin{cases} 1, & a_5 = 0, \\ 0, & a_5 \neq 0. \end{cases} \qquad (S32)$$

Thus, the erasure $F$ pollutes all $x_i$, and the error does $x_0, x_1, x_2, x_3$. In terms of Hamming metric, they appear as 5 symbol erasures and 4 symbol errors, which exceed the correction capability of RS codes.

Increasing the code length does not resolve this issue. The probability that the transfer matrix $A_t$ becomes rank deficient (e.g., $a_5 = 0$ in Eq. (S29)) is roughly $1/q$ or slightly larger, which leads to the substantial error floor.

However, the second and third terms in Eq. (S31) actually correspond to a rank-1 erasure and error, both of which can be corrected by the Gabidulin code, thereby effectively suppressing the error floor (precisely, when $n_0 = 5$ and $k = 3$, two rank erasures or one rank error can be corrected).

# Supplementary Note 9: Performance of 5-degree PUSNEC for the 1-to-6 multicast in the terrestrial network

We present performance of the 5-degree PUSNEC in the 1-to-6 multicast in the terrestrial network shown in Fig. 1d. We assume that erasures and errors occur in all links at the rates $\epsilon$ and $e_{L2}$, respectively, and erasures occur in all Charies at the rate $\epsilon_N$.

Supplementary Figs. 6a and b show the PLP $p_L(\gamma, k)$ as a function of the node compromise rate $\gamma$, and the FER $e_F(\epsilon, k)$ as a function of the link erasure rate $\epsilon$ (with $e_{L2} = \epsilon_N = 0$), respectively, for the message dimension $k = 1 \sim 5$. The overall trends of them are the same as those of the 1-to-13 omni-directional multicast in the LEO satellite network shown in Figs. 5a and b. Actually $p_L(\gamma, k)$ and $e_F(\epsilon, k)$ consist of the flat region and monotonically decreasing region. However, upon closer inspection, we can notice some differences. As for the secrecy performance, as $\gamma$ decreases, $p_L(\gamma, k)$ for the terrestrial network starts to decrease sooner (Supplementary Fig. 6a) than it does for the LEO satellite network (Fig. 5a). This is simply because the total number of Charlies $N_C$ in the former is smaller than that in the latter. As for the reliability performance, on the other hand, as $\epsilon$ decreases, $e_F(\epsilon, k)$ for the terrestrial network starts to decrease later (Supplementary Fig. 6b) than it does for the LEO satellite network (Fig. 5b). This is simply because we assume that erasures occur in all links in the former while they occur only in the very last-hop links (the downlinks to the ground stations) in the latter. Notice that for $k = 5$ (no erasure correction capability), the FER floor for the terrestrial network is slightly smaller than that for the LEO satellite network. This is because $N_C$ in the former model is smaller than $N_C$ in the latter, and hence the size effect of the ground field in the $\mathbb{F}_q$-RLNC is weaker in the former.

The tradeoff relations between the PLP and the FER in terms of $k$ are shown in Supplementary Fig. 6c for four sets of $\gamma$ and $\epsilon$, those were taken such that the four curves are within the same range. The $k$-dependences of the LII $I_L(\gamma, k, \xi)$ for $\xi = 1 \sim 4$ and the FER $e_F(\epsilon, k)$ are explicitly shown in Supplementary Fig. 6d. Both Supplementary Figs. 6c and d look quite similar to Figs. 5d and f, respectively.

To summarize, the graph shapes of $p_L(\gamma, k)$, $e_F(\epsilon, k)$, and $I_L(\gamma, k, \xi)$ are almost identical for the terrestrial network and the LEO satellite network. Quantitatively, adjusting the values of $\gamma$ and $\epsilon$ for the terrestrial network makes $p_L(\gamma, k)$, $e_F(\epsilon, k)$, and $I_L(\gamma, k, \xi)$ nearly overlap with those for the LEO satellite network.

The reliability performances of the Gabidulin code and the RS code at the same coding rate with almost the same code length in bits ($\mathrm{Gab}[9, 3]$ over $\mathbb{F}_{256^{27}}$ versus $\mathrm{RS}[243, 81]$ over $\mathbb{F}_{2^8}$) are compared in Supplementary Figs. 7a, b, and c. The Gabidulin code greatly outperforms the RS code in error and erasure correction capabilities.

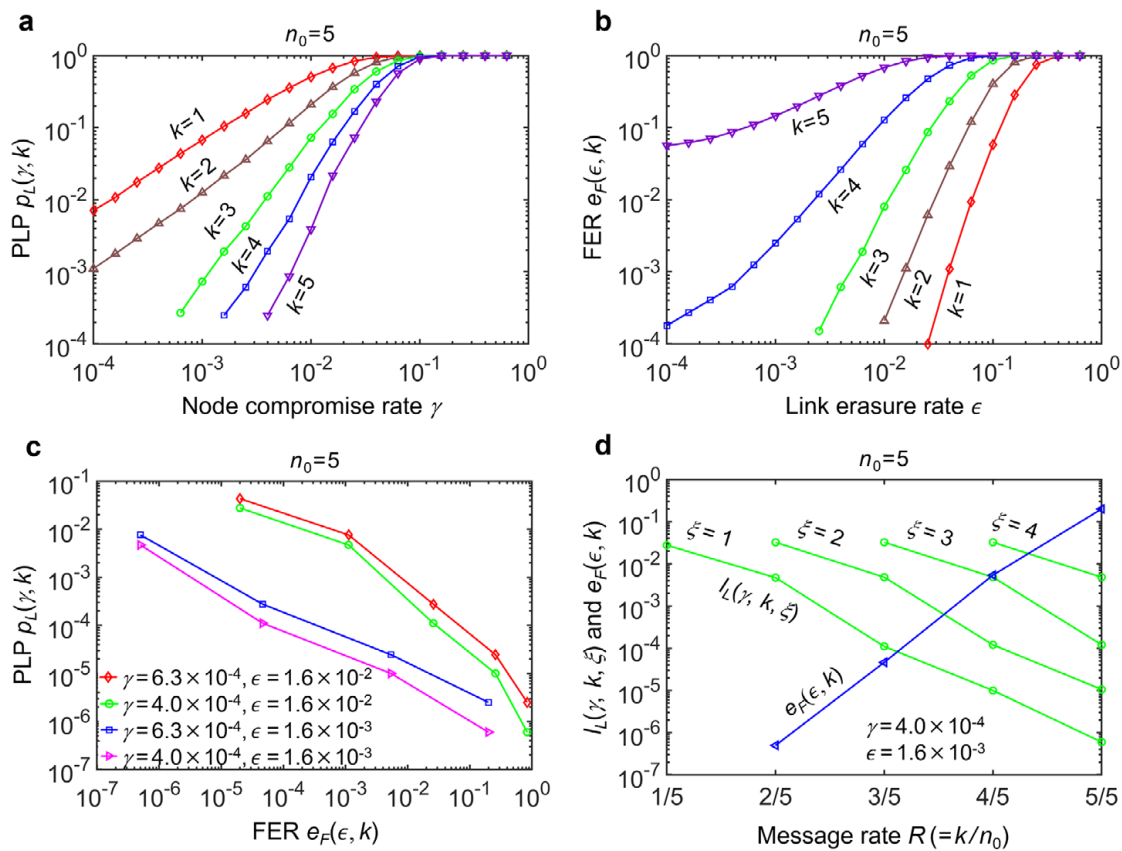

**Supplementary Fig. 6.** Secrecy and reliability performances of 1-to-6 multicast with the 5-degree PUSNEC in the terrestrial network of Fig. 1d. **a** The PLP $p_L(\gamma, k)$ as a function of the node compromise rate $\gamma$. **b** The FER $e_F(\epsilon, k)$ as a function of the link erasure rate $\epsilon$, setting $e_{L2} = \epsilon_N = 0$. **c** The tradeoff relations between the PLP and the FER for four sets of $(\gamma, \epsilon)$ ($e_{L2} = \epsilon_N = 0$). **d** The LII $I_L(\gamma, k, \xi)$ and FER $e_F(\epsilon, k)$ as a function of the message rate $R = k/n_0$ for a set of $(\gamma, \epsilon)$ ($e_{L2} = \epsilon_N = 0$).

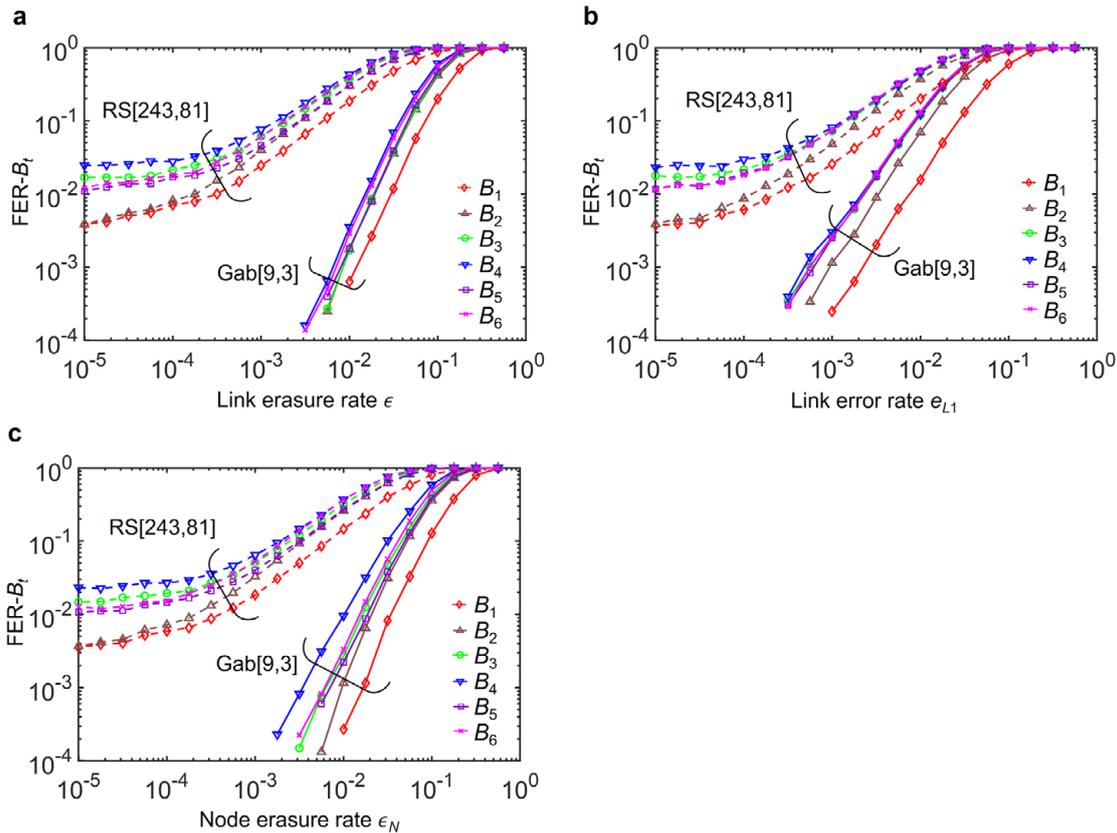

**Supplementary Fig. 7.** Comparison of FER-$B_t$ of Gabidulin code (solid lines) and RS code (dotted lines). The scheme is 1-to-6 multicast with the 5-degree PUSNEC in the terrestrial network of Fig. 1d. **a** FER-$B_t$ as a function of the link erasure rate $\epsilon$ at each Bob, setting $e_{L2} = \epsilon_N = 0$. **b** FER-$B_t$ as a function of the link error rate $e_{L2}$ at each Bob, setting $\epsilon = \epsilon_N = 0$. **c** FER-$B_t$ as a function of the node erasure rate $\epsilon_N$ at each Bob, setting $\epsilon = e_{L2} = 0$.

# Supplementary Note 10: Performances of 6-degree PUSNEC in LEO-MEO hybrid satellite network

A LEO-MEO hybrid satellite network with 6-degree multicast paths is shown in Supplementary Fig. 8. Alice is the LEO satellite above Philippines. Bobs are totally 13 ground stations in the northern hemisphere, i.e., in Guam, Shanghai, Tokyo, New Delhi, Haleakala, Big Pines, Vancouver, Matera, Nemea, Oberpfaffenhofen, London, Tenerife, and Washington.

For London, Oberpfaffenhofen, Matera, and Nemea, 5, 4, 3, and 4 link-disjoint paths could be found but no more. The missing paths can be supplemented with relays from certain Charlies $C_1$, $C_2$, $C_3$ to those cities via MEO satellites $M_1$, $M_2$, $M_3$. Because path configurations for MEO satellite relays are quite simple, the path search is specifically conducted for the LEO plane at the average altitude of roughly 550 km in the same way as in Supplementary Note 7.

The virtual 1-to-13 multicast graph derived in the X-hop grid was mapped back to a graph in the real network model, and converted to the final multicast graph after the smoothing was applied (Supplementary Fig. 8).

Errors and erasures are modeled in a similar manner to those in the LEO satellite network. We assume that there is no differences in the link budget and the noise characteristics between the LEO-ground and MEO-ground links (the received signal power is above the noise floor).

As in the 5-degree PUSNEC, we used the ramp scheme with $\mu_0 = 0$ (no masking key). Then a good Gabidulin code for the 6-degree PUSNEC can be found as $\mathrm{Gab}[n, k]$ over $\mathbb{F}_{q^n}$ with $n = 14$ and $q = 2, 2^3, 2^5, 2^9, \ldots$. The rationale is as follows: For ensuring the strong ramp secrecy, $n \geq k_0 + n_0$ should hold. We want to examine the cases with $k_0 = 1 \sim 6$, hence the code length must be $n = 12$ or longer. Unfortunately, good Gabidulin codes do not exist for $n = 12, 13$, but one is found for $n = 14$ as seen in Supplementary Table IV. So $n = 14$ is the minimum code length for our purpose. As for the field size $q$, we took $q = 2^5 = 32$. The reason is as follows. The larger the ground field size $q$ is, the more the FER due to the field size effect can be suppressed. However, the processing time also increases, requiring an enormous amount of time to conduct multicast simulation experiments with sufficient statistical data.

Supplementary Figs. 9a and b show the PLP $p_L(\gamma, k)$ as a function of the node compromise rate $\gamma$, and the FER $e_F(\epsilon, k)$ as a function of the link erasure rate $\epsilon$ (with $e_{L1} = \epsilon_N = e_{L2} = 0$), respectively, for the message dimension $k = 1 \sim 6$. The tradeoff relations between the PLP and the FER in terms of $k$ are shown in Supplementary Fig. 9c for four sets of $\gamma$ and $\epsilon$. The $k$-dependences of the LII $I_L(\gamma, k, \xi)$ for $\xi = 1 \sim 5$ and the FER $e_F(\epsilon, k)$ are explicitly shown in Supplementary Fig. 9d. The overall trends of them are the same as those of the LEO satellite network (Figs. 5a and b) and the terrestrial network (Supplementary Figs. 6a and b). However, upon closer inspection, the error floors due to the field size effect became more apparent in Supplementary Fig. 9b (especially for $k = 4, 5, 6$) than those in Fig. 5b (the LEO satellite network) and Supplementary Fig. 6b (the terrestrial network), because the field size is $q = 32$, which is smaller than $q = 256$ used for the LEO satellite network and the terrestrial network.

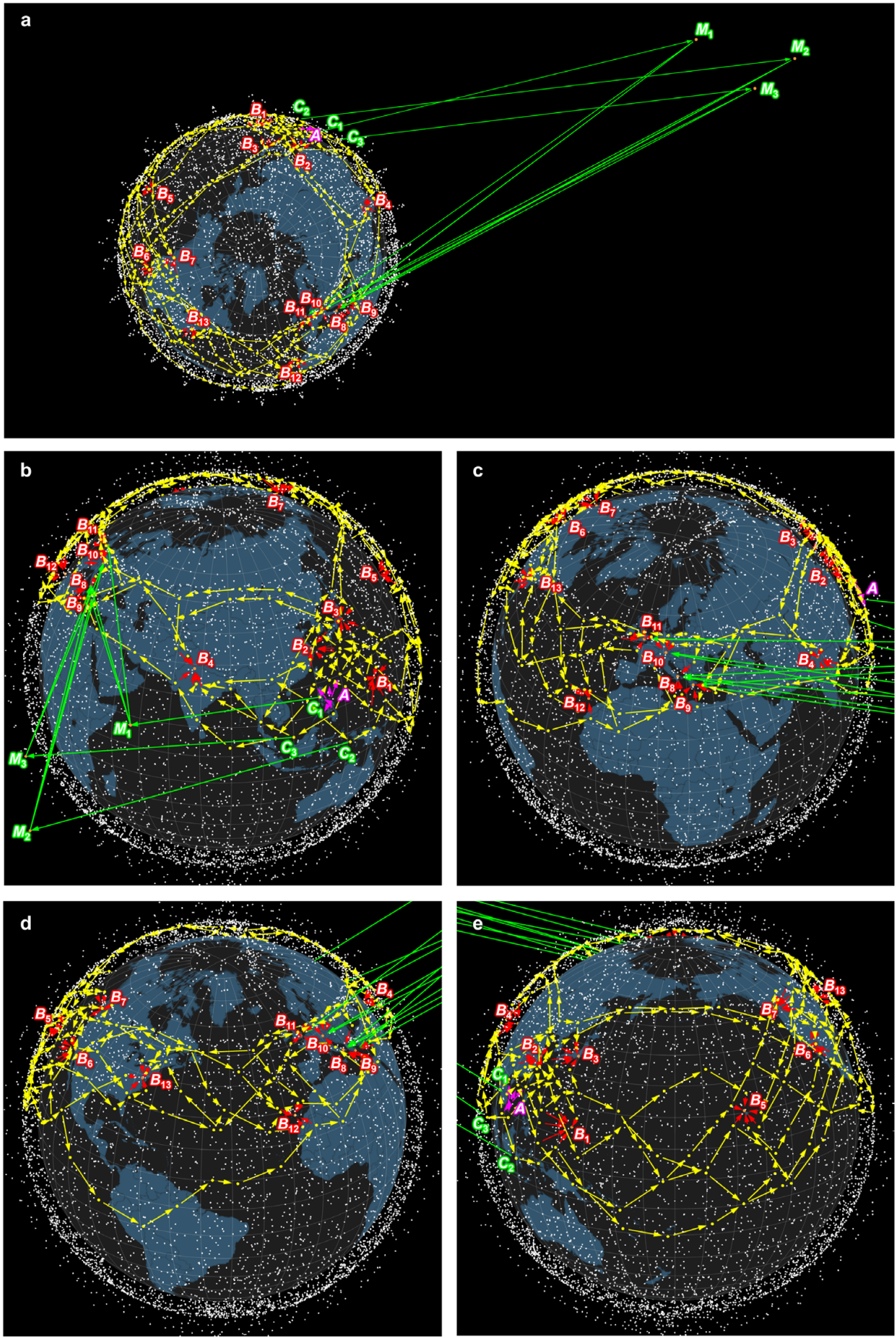

**Supplementary Fig. 8. The 1-to-13 multicast graph for the 6-degree PUSNEC over the northern hemisphere using LEO and MEO constellations.** **a**, **b**, **c**, **d**, and **e** represent views from five different directions above North Pole, India, Europe, Central Pacific and Central Atlantic. Alice is the LEO satellite above Philippines as in Fig. 4. Bobs are totally 13 ground stations in the northern hemisphere, i.e., in Guam ($B_1$), Shanghai ($B_2$), Tokyo ($B_3$), New Delhi ($B_4$), Haleakala ($B_5$), Big Pines ($B_6$), Vancouver ($B_7$), Matera ($B_8$), Nemea ($B_9$), Oberpfaffenhofen ($B_{10}$), London ($B_{11}$), Tenerife ($B_{12}$), and Washington ($B_{13}$). London, Oberpfaffenhofen, Matera, and Nemea are supported by the relay path(s) via MEO satellites $M_1$, $M_2$, $M_3$. LEO-to-MEO uplink points $C_1$, $C_2$, $C_3$ are selected from Charlies who have only 1-input and 1-output LEO-to-LEO links. A 3D interactive view is available in [25].

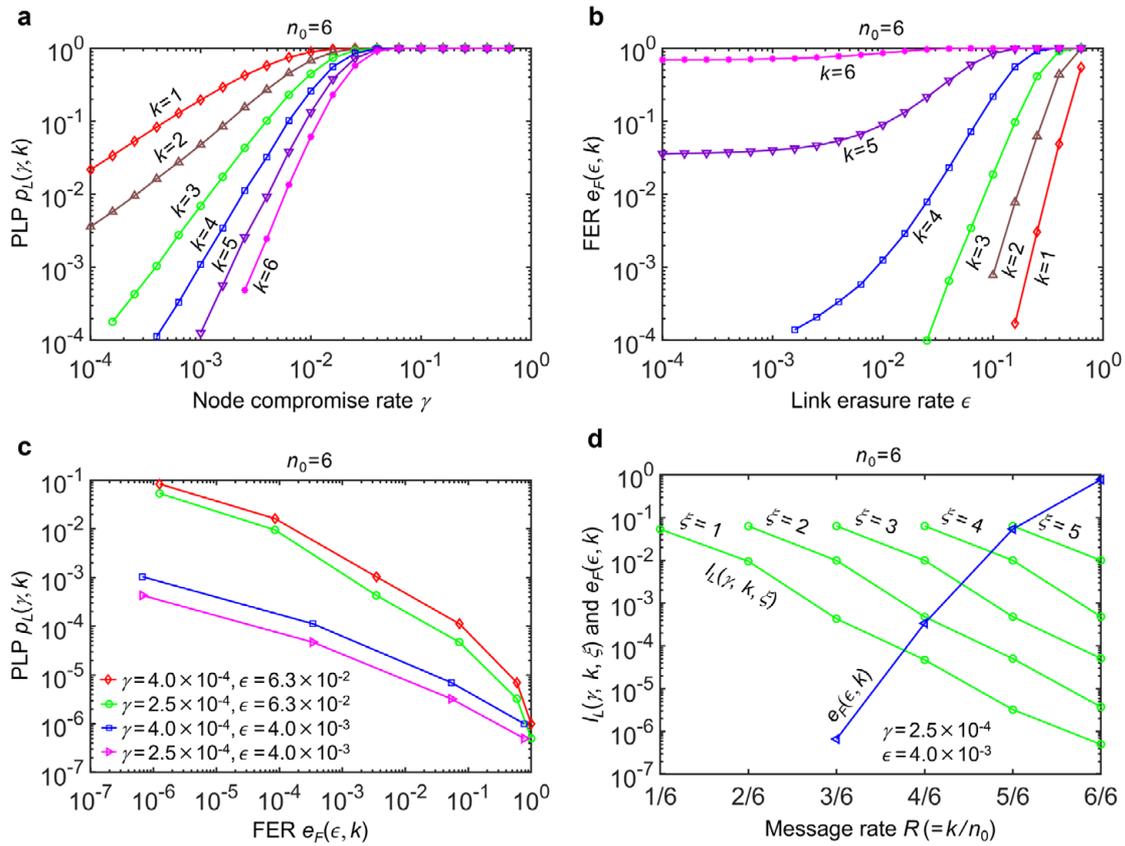

**Supplementary Fig. 9. Secrecy and reliability performances of 1-to-13 omni-directional multicast with the 6-degree PUSNEC in the LEO-MEO hybrid satellite network. a** The PLP $p_L(\gamma, k)$ as a function of the node compromise rate $\gamma$. **b** The FER $e_F(\epsilon, k)$ as a function of the link erasure rate $\epsilon$, setting $e_{L1} = \epsilon_N = e_{L2} = 0$. **c** The tradeoff relations between the PLP and the FER due to the link erasure, $e_F(\epsilon, k)$, for four sets of $(\gamma, \epsilon)$ ($e_{L1} = \epsilon_N = e_{L2} = 0$). **d** The LII $I_L(\gamma, k, \xi)$ and the FER due to the link erasure, $e_F(\epsilon, k)$, as a function of the message rate $R = k/n_0$ for a set of $(\gamma, \epsilon)$ ($e_{L1} = \epsilon_N = e_{L2} = 0$).

The reliability performances of the Gabidulin code and the RS code at the same coding rate with almost the same code length in bits (Gab[14, 3] over $\mathbb{F}_{32^{42}}$ versus RS[294, 63] over $\mathbb{F}_{2^{10}}$) are compared in Supplementary Figs. 10a, b, c, and d. The Gabidulin code greatly outperforms the RS code in error and erasure correction capabilities. Note that the error floors due to the size effect appear remarkably in the FER-$B_t$ for the RS code. The Gabidulin code succeeded in suppressing these error floors impressively.

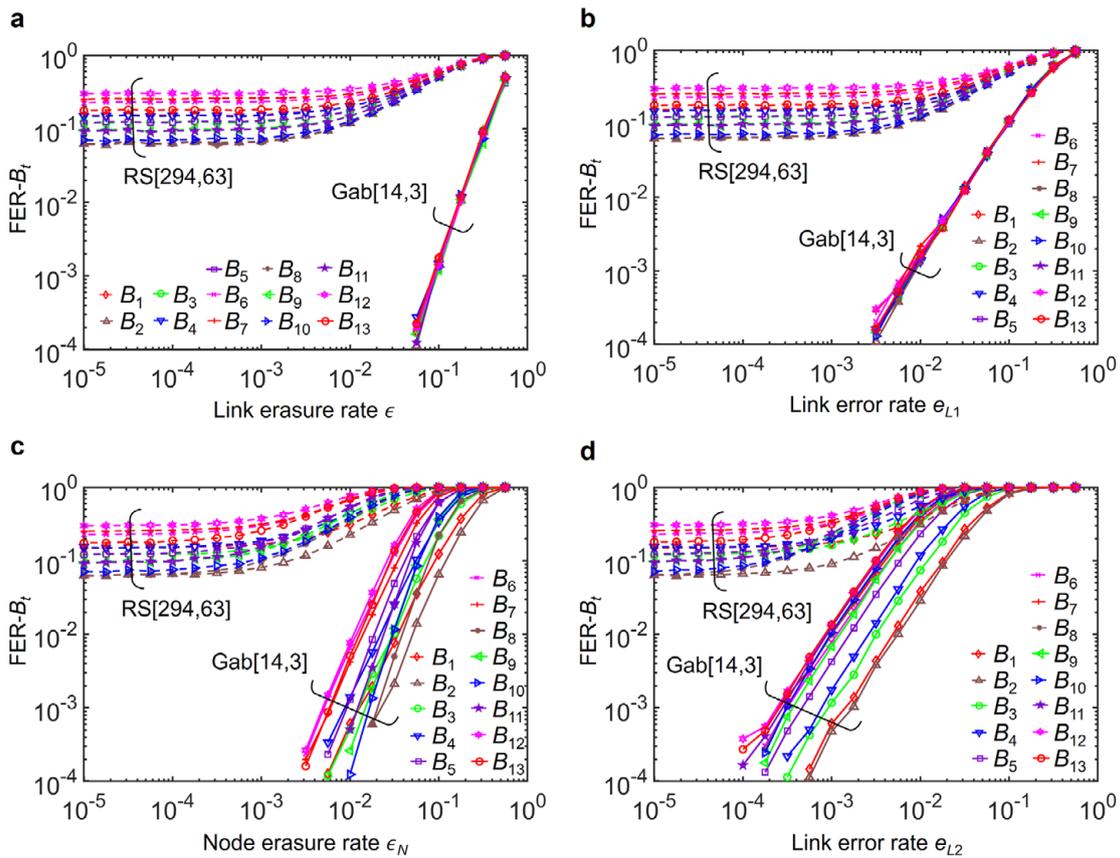

**Supplementary Fig. 10. Comparison of FER-$B_t$ for Gabidulin code (solid lines) and RS code (dotted lines).** The scheme is 1-to-13 omni-directional multicast with the 6-degree PUSNEC in the LEO-MEO hybrid satellite network. **a** FER-$B_t$ as a function of the link erasure rate $\epsilon$ at each Bob, setting $e_{L1} = \epsilon_N = e_{L2} = 0$. **b** FER-$B_t$ as a function of the link error rate $e_{L1}$ at each Bob, setting $\epsilon = \epsilon_N = e_{L2} = 0$. **c** FER-$B_t$ as a function of the node erasure rate $\epsilon_N$ at each Bob, setting $\epsilon = e_{L1} = e_{L2} = 0$. **d** FER-$B_t$ as a function of the link error rate $e_{L2}$ at each Bob, setting $\epsilon = e_{L1} = \epsilon_N = 0$.

**Supplementary Table VII**
**Performance comparison of various cryptographic scheme combinations with the PUSNEC.** RSA: Rivest-Shamir-Adleman public-key cryptography, ECDSA: Elliptic Curve Digital Signature Algorithm, AES: Advanced Encryption Standard, WC: Wegman-Carter message authentication code.

|  | Type I | Type II | Type III | Type IV |
|---|---|---|---|---|
| User authentication | RSA/ECDSA | PQC | PQC | WC |
| Link encryption | AES | AES | QKD/PLS | QKD/PLS |
| Multicast scheme | IP multicast | PUSNEC | PUSNEC | PUSNEC |
| Authenticity | CS | CS | CS | ITS |
| Data confidentiality | CS | CS | ITS | ITS |
| Multicast capacity | Unattainable | Attainable | Attainable | Attainable |
| Data availability | Low | High | High | High |
| Implementation cost | Low (software) | Low (software) | High (hardware+software) | High (hardware+software) |

# Supplementary Note 11: Benchmark of cryptographic scheme combinations with the PUSNEC

The PUSNEC can be effectively integrated with not only ITS schemes but also computational security (CS)-based cryptographic technologies. Supplementary Table VII presents a comparison of various cryptographic scheme combinations with the PUSNEC. Type I represents the current IP multicast scheme, which fails to achieve multicast capacity or long-term security. Type II, the PUSNEC-overlaid PQC scheme, addresses these limitations. Both Type I and Type II are primarily software-based solutions, resulting in relatively low implementation costs. Type III is a hybrid scheme incorporating user authentication via PQC with CS, link encryption via QKD/PLS with ITS, and multicast via the PUSNEC. Type IV is a fully ITS-based scheme, utilizing Wegman-Carter message authentication codes for user authentication, QKD/PLS for link encryption, and the PUSNEC for multicast communication. Type III and IV entail additional hardware costs due to QKD/PLS devices. Ultimately, the selection and implementation of cryptographic technologies should align with the specific needs and operational policies of users, ensuring that each approach is applied in the most effective manner.

# Supplementary Note 12: Strong ramp secrecy in threshold-based wiretap model

For reader's convenience, we review the characterization of strong ramp secrecy in the threshold-based wiretap model.

We can see that the matrix $\bar{G}$ is nonsingular according to the following two lemmas.

**Lemma 3.** *[17, Theorem 2]. Assume that $m \geq n$. A linear $[n, k]$ code over $\mathbb{F}_{q^m}$ with parity-check matrix $H \in \mathbb{F}_{q^m}^{(n-k) \times n}$ is an MRD code if and only if*

$$\operatorname{rank}_{q^m} HT = n - k \qquad (S33)$$

*for any full-rank matrix $T \in \mathbb{F}_q^{n \times (n-k)}$.*

**Lemma 4.** *[17, Theorem 3]. Let $G \in \mathbb{F}_{q^m}^{k \times n}$ be a generator matrix of the $[n, k]$ code $\mathcal{C}$. The code $\mathcal{C}$ is an MRD code if and only if*

$$\operatorname{rank}_{q^m} GT = k \qquad (S34)$$

*for any full-rank matrix $T \in \mathbb{F}_q^{n \times k}$.*

*Proof.* It is well known that if a code $\mathcal{C}$ is an $[n, n-k]$ MRD code, so is the dual code $\mathcal{C}^{\perp}$ and it has parameters $[n, k]$. The generator matrix of the dual code $\mathcal{C}^{\perp}$ coincides with the parity-check matrix $H \in \mathbb{F}_{q^m}^{k \times n}$ of the code $\mathcal{C}$, for which Lemma 3 can be rewritten

$$\operatorname{rank}_{q^m} HT = k, \qquad (S35)$$

for any full-rank matrix $T \in \mathbb{F}_q^{n \times k}$. □

In the threshold-based wiretap model, we assume a fixed number of wiretapped packets $\mu$. We may also assume that the wiretapped packet configuration $\zeta$ is also given and fixed.

Let $Z$ denote Eve's random variable, whose realization $\boldsymbol{z}$ is drawn from a set

$$\boldsymbol{z} \in \left\{ \boldsymbol{z}_{\zeta,1}^{(\mu)}, ..., \boldsymbol{z}_{\zeta, q^{m\mu}}^{(\mu)} \text{ for given } \mu, \zeta \right\}. \qquad (S36)$$

Strong ramp secrecy is measured by the mutual information

$$\begin{aligned} I^{(\mu)}&(U_{\{\xi\}}; Z) \\ &= \frac{1}{q^{m\xi}} \sum_{\boldsymbol{u}_{\{\xi\}}} \sum_{\boldsymbol{z}} P_{Z|U_{\{\xi\}}}(\boldsymbol{z}|\boldsymbol{u}_{\{\xi\}}) \\ &\quad \times \log \frac{P_{Z|U_{\{\xi\}}}(\boldsymbol{z}|\boldsymbol{u}_{\{\xi\}})}{\frac{1}{q^{m\xi}} \sum_{\boldsymbol{u}'_{\{\xi\}}} P_{Z|U_{\{\xi\}}}(\boldsymbol{z}|\boldsymbol{u}_{\{\xi\}})}, \end{aligned} \qquad (S37)$$

which is characterized in the following way (see the process of the proof in Appendix).

Case 1: $\mu \leq \mu_0$

The whole message $\boldsymbol{u}$ can be kept secret:

$$I^{(\mu)}(U_{\{\xi\}}; Z) = 0, \quad 1 \leq \xi \leq k_0. \qquad (S38)$$

Case 2: $\mu_0 < \mu \leq k - 1$

Any less than or equal to $k - \mu$ message symbols can be kept secret. However, $\boldsymbol{z}$ has finite correlation with any $k - \mu + 1$ message symbols and more:

$$I^{(\mu)}(U_{\{\xi\}}; Z) = \begin{cases} 0, & 1 \leq \xi \leq k - \mu, \\ (\xi - k + \mu) \log_2 q^m, & k - \mu < \xi \leq k_0. \end{cases} \qquad (S39)$$

The second line of Eq. (S39) means that, by observing $\boldsymbol{z}$, Eve can tell which of $q^{m\xi}/q^{m(k-\mu)}$ groups the sub-message $\boldsymbol{u}_{\{\xi\}}$ belongs to. Note that there still remain $q^{m(k-\mu)}$ equally likely sequences, and hence Eve is still uncertain about $\boldsymbol{u}_{\{\xi\}}$ itself, i.e., no meaningful information is obtained.

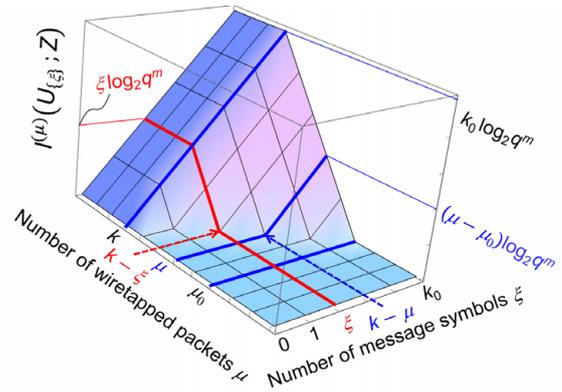

**Supplementary Fig. 11. The mutual information $I^{(\mu)}(U_{\{\xi\}}; Z)$ in terms of $\xi$ and $\mu$.**

Case 3: $k \leq \mu$

$$I^{(\mu)}(U_{\{\xi\}}; Z) = \xi \log_2 q^m. \qquad (S40)$$

Eve can decode the secret message $\boldsymbol{u}$ completely.

The above behavior is graphically depicted in Supplementary Fig. 11.

# Supplementary Note 13: Technical details of implementing vertically interleaved Gabidulin codes

## I. INTRODUCTION

A rank-metric code $\mathcal{C}$ of length $n$ is a set of vectors of length $n$ over the extension field $\mathbb{F}_{q^m}$ or equivalently a set of $m \times n$ matrices over the ground field $\mathbb{F}_q$. Let $\mathbb{F}_q^{m \times n}$ denote the set of all $m \times n$ matrices over $\mathbb{F}_q$. The rank weight of each codeword is the rank of its matrix representation computed over $\mathbb{F}_q$, and the rank distance between matrices $\boldsymbol{x}, \boldsymbol{y} \in \mathbb{F}_q^{m \times n}$ is the rank of their difference

$$d(\boldsymbol{x}, \boldsymbol{y}) = \operatorname{rank}_q(\boldsymbol{x} - \boldsymbol{y}), \tag{S41}$$

where the index $q$ emphasizes that the rank is computed over $\mathbb{F}_q$. Let $d$ be the minimum rank distance among all pairs of distinct codewords of $\mathcal{C}$. Then the cardinality of $\mathcal{C}$ is bounded by

$$|\mathcal{C}| \leq q^{\max\{n,m\}(\min\{n,m\}-d+1)}. \tag{S42}$$

MRD codes are ones that achieve this bound. They are known to exist for all choices of parameters $q, n, m$ and $d \leq \min\{n, m\}$ [17]. Let $k$ be the message dimension of $\mathcal{C}$. When $m \geq n$, an $[n, k]$ MRD code with distance $d$ satisfies the Singleton bound $d \leq n - k + 1$ with equality, and hence is also a maximum distance separable (MDS) code. The universality property of MRD codes comes from the fact that the MRD codes are nonlinear over $\mathbb{F}_q$ but linear over the extension field $\mathbb{F}_{q^m}$ [2].

Gabidulin codes are an important and well-studied subclass of MRD codes. Efficient decoding algorithms for Gabidulin codes have been developed by extending the well established methods for RS codes to the rank-metric version. Two representative methods are known: a "standard" method based on the Berlekamp-Massey algorithm (BMA) (or the extended Euclidean algorithm) [17], [27], and a method based on a Welch-Berlekamp-like algorithm (WBA) [28]. The BMA is more efficient for high rate codes, while the WBA is more suitable for low rate codes [18]. In contrast to that the WBA is more specific to RS codes, the BMA itself can be applied to various codes, requires less memory overheads, and hence is regarded as the standard method.

The standard method for Gabidulin codes defined over $\mathbb{F}_{q^m}$ were studied in [29], [30], showing that the complexity is reduced to $\mathcal{O}(m^3)$ over the ground field $\mathbb{F}_q$. Puchinger and Wachter-Zeh showed faster methods, reducing the complexity to $\mathcal{O}(m^{2.69})$ over $\mathbb{F}_q$, which is converted to a sub-quadratic order over $\mathbb{F}_{q^m}$ [31].

Loidreau and Overbeck introduced interleaved Gabidulin codes, which can be effective when errors occur in the same space [14]. This can be viewed in analogy with interleaved RS codes for correcting burst errors. Efficient decoding algorithms for interleaved Gabidulin codes were studied in [15], [32], [33]. The Cartesian product of many shorter Gabidulin codes with the same distance was considered in the context of reducing the decoding complexity significantly in [16]. This can also be viewed as an interleaved Gabidulin code.

A valuable example on software implementations of Gabidulin codes was given by Kunz et al. [34]. These authors implemented two decoders for the Gabidulin code of length $n = 113$ and dimension $k = 3$ over $\mathbb{F}_{q^m}$ with $q = 2$ and $m = 127$; one is the WBA decoder which is currently used in the cryptosystem called Rank Quasi-Cyclic (RQC) [35] and the other is the Transform Domain Decoder by Silva and Kschischang [29]. They analyzed CPU times for main functions in the decoding algorithms, and showed that knowing the number of operations is not enough in estimating practical decoding complexities.

In what follows, we first outline our techniques for interleaved Gabidulin codes and introduce some notations in Section II. We next provide basic guides to Gabidulin decoding in Sections III and IV. We then present our implementation techniques of iGab$[n, k]$ for the PUSNEC in Section V. The reader familiar with Gabidulin codes may skip Sections II, III, IV, and directly go to Section V. We also provide, for reader's convenience, basic notions and technical tools involved in the decoding of Gabidulin codes in Section VI.

## II. OUTLINE

### A. Our techniques for interleaved Gabidulin codes

The overall decoding complexity of an $[n, k]$ Gabidulin code over $\mathbb{F}_{q^m}$ is known to be $\mathcal{O}(dm)$ operations over $\mathbb{F}_{q^m}$, where $d \ (= n - k + 1)$ is the minimum distance of the code. When $m$ is sufficiently larger than $n$ such that one can take $m = ln$ for some integer $l$, the decoding complexity can be significantly reduced to $\mathcal{O}(dm)$ operations over the *smaller* field $\mathbb{F}_{q^n}$ by using the vertically interleaved Gabidulin code iGab$[n, k]$ over the Cartesian product space $\mathbb{F}_{q^n}^l$, while keeping the same error correction capability [16]. Decoding is made for each Gab$[n, k]$ code over $\mathbb{F}_{q^n}$ individually (the independent component-wise decoder). In [16], it was suggested that further reduction of decoding complexity may be possible since all components $\boldsymbol{y}'^{(0)}, ..., \boldsymbol{y}'^{(l-1)}$ in the received word at each Bob

$$\boldsymbol{y}' = \begin{bmatrix} \boldsymbol{y}'^{(0)} \\ \vdots \\ \boldsymbol{y}'^{(l-1)} \end{bmatrix}, \tag{S43}$$

will share the same set of error locations. Unfortunately, however, efficient strategies have not been reported nor demonstrated with any concrete software implementations yet to our best knowledge [2].

We have developed efficient implementation techniques of iGab$[n, k]$ over $\mathbb{F}_{q^n}^l$. Our techniques are based on several novel modifications of the standard method, and applied to both the encoder and the decoder of iGab$[n, k]$ designed for the PUSNEC.

In the encoder, the pre-encoding process is carried out for ensuring the universal strong ramp secrecy. This process includes the matrix inversion, whose complexity usually costs $\mathcal{O}(k^3)$ over $\mathbb{F}_{q^n}$. Instead of directly computing this matrix inversion, we apply one of heavily used technique in the Gabidulin decoder, Gabidulin's recursive algorithm (GRA), which allows us to reduce the complexity to $\mathcal{O}(k^2)$ operations over $\mathbb{F}_{q^n}$.

---

[2] In fact, decoding over the Cartesian product space $\mathbb{F}_{q^n}^l$ requires special attention, different from decoding over the extension field $\mathbb{F}_{q^m}$. For example, the existence of multiplicative inverses cannot be guaranteed in the Cartesian product space $\mathbb{F}_{q^n}^l$ unlike the extension field $\mathbb{F}_{q^m}$. As another aspect, although both codeword sets in $\mathbb{F}_{q^n}^l$ and $\mathbb{F}_{q^m}$ can have the same maximal rank distance, the set of codewords of iGab$[n, k]$ over $\mathbb{F}_{q^n}^l$ is not the same as that of the non-interleaved Gabidulin code over $\mathbb{F}_{q^m}$. So, decoding a received word encoded with the non-interleaved Gabidulin code using the iGab$[n, k]$ decoder (or vice versa) will not succeed in general.

In the decoder, the computational complexity is dominated by the division operation appearing in error-location dependent computations. Most error locations are common to all the components $\boldsymbol{y}'^{(0)}, ..., \boldsymbol{y}'^{(l-1)}$. Furthermore, the division operation does not appear in processes which require component-wise computations such as syndrome-dependent ones. Therefore, it must be possible to complete the error-location dependent computations (nearly) once, instead of performing them $l$ times independently for each component, thereby reducing the overall decoding complexity roughly by a factor of $1/l$. This kind of *common-error-location aware decoder* is embodied by three kinds of modifications from the independent component-wise decoder as follows.

The first modification concerns the step of finding the rank of errors and synthesizing the error location polynomial (ELP), namely the Linearized Berlekamp Massey Algorithm (LBMA). More specifically, instead of running the decoder of $\mathrm{Gab}[n, k]$ over $\mathbb{F}_{q^n}$ in parallel or $l$ times independently, we process the received components $\boldsymbol{y}'^{(0)}, ..., \boldsymbol{y}'^{(l-1)}$ recursively, using the results from the previous components for the current component. Actually, the error locations output from the LBMA for the 0th component $\boldsymbol{y}'^{(0)}$ cover almost all error locations of all the components. This allows one to eliminate redundant computations from the first component $\boldsymbol{y}'^{(1)}$ onward. The method is referred to as the LBMA for Cartesian product space (CPSLBMA).

The second modification concerns the step of finding rootspace of ELP, which determines the error locations. Unlike the independent component-wise decoder, this step can be common to all the components and hence can be carried out only once.

The third modification concerns the step of determining the error values, which uses GRA. We split the computations in GRA into two parts; the first part is the computations depending only on the error locations, and the second part is those depending on the syndromes, which are component dependent. The first part can be carried out only once, and the result can be fed into the second part of the component-wise GRA, thereby reducing the number of computations compared to the independent component-wise GRA.

### B. $q$-cyclic Gabidulin codes

Gabidulin codes are defined by evaluating degree-bounded linearized polynomials at $\mathbb{F}_q$-linearly independent evaluation points in the extension field. (Details of linearized polynomials are given in Subsection VI-E.) In this work, we adopt a specific class of $q$-cyclic Gabidulin codes $\mathrm{Gab}[n, k]$ over $\mathbb{F}_{q^n}$ and construct the vertically interleaved Gabidulin code $\mathrm{iGab}[n, k]$ over $\mathbb{F}_{q^n}^l$.

Let $\beta \in \mathbb{F}_{q^n}$ generate a normal basis of $\mathbb{F}_{q^n}$ over $\mathbb{F}_q$. Then $\beta^{[0]}, ..., \beta^{[n-1]}$ is a basis of $\mathbb{F}_{q^n}$ over $\mathbb{F}_q$, and as such the parity check matrix $\check{H} \in \mathbb{F}_{q^n}^{(n-k) \times n}$ given by

$$\check{H} = \begin{bmatrix} \beta^{[k]} & \beta^{[k+1]} & \dots & \beta^{[k-1]} \\ \beta^{[k+1]} & \beta^{[k+2]} & \dots & \beta^{[k]} \\ \vdots & \vdots & \ddots & \vdots \\ \beta^{[n-1]} & \beta^{[0]} & \dots & \beta^{[n-2]} \end{bmatrix}, \quad (S44)$$

generates a Gab$[n, k]$ code over $\mathbb{F}_{q^n}$. Gabidulin codes generated by a normal basis are typically referred to as $q$-cyclic

Gabidulin codes. Let $\beta^\perp \in \mathbb{F}_{q^n}$ be the dual element of $\beta$, then the generator matrix $\check{G} \in \mathbb{F}_{q^n}^{k \times n}$ is given by

$$\check{G} = \begin{bmatrix} \beta^{\perp [0]} & \beta^{\perp [1]} & \dots & \beta^{\perp [n-1]} \\ \beta^{\perp [1]} & \beta^{\perp [2]} & \dots & \beta^{\perp [0]} \\ \vdots & \vdots & \ddots & \vdots \\ \beta^{\perp [k-1]} & \beta^{\perp [k]} & \dots & \beta^{\perp [k-2]} \end{bmatrix}, \quad (S45)$$

Due to the duality of $\beta$ and $\beta^\perp$, we have

$$\check{G}\check{H}^T = 0, \quad (S46)$$

as required. In the later sections, we often use a general expression for the parity check matrix as

$$\hat{H} = \begin{bmatrix} \hat{h}_0^{[k]} & \hat{h}_1^{[k]} & \dots & \hat{h}_{n-1}^{[k]} \\ \hat{h}_0^{[k+1]} & \hat{h}_1^{[k+1]} & \dots & \hat{h}_{n-1}^{[k+1]} \\ \vdots & \vdots & \ddots & \vdots \\ \hat{h}_0^{[n-1]} & \hat{h}_1^{[n-1]} & \dots & \hat{h}_{n-1}^{[n-1]} \end{bmatrix}, \quad (S47)$$

with the basis $\hat{h}_0, ..., \hat{h}_{n-1} \in \mathbb{F}_{q^n}$ over $\mathbb{F}_q$ which is dual to the basis $\hat{g}_0, ..., \hat{g}_{n-1} \in \mathbb{F}_{q^n}$ over $\mathbb{F}_q$ used in the generator matrix of $\mathrm{Gab}[n, k]$

$$\hat{G} = \begin{bmatrix} \hat{g}_0^{[0]} & \hat{g}_1^{[0]} & \dots & \hat{g}_{n-1}^{[0]} \\ \hat{g}_0^{[1]} & \hat{g}_1^{[1]} & \dots & \hat{g}_{n-1}^{[1]} \\ \vdots & \vdots & \ddots & \vdots \\ \hat{g}_0^{[k-1]} & \hat{g}_1^{[k-1]} & \dots & \hat{g}_{n-1}^{[k-1]} \end{bmatrix}. \quad (S48)$$

### C. Notations

A codeword of $\mathrm{iGab}[n, k]$ over $\mathbb{F}_{q^n}^l$ at Alice is expressed as $\begin{bmatrix} \boldsymbol{x}'^{(0)} \\ \vdots \\ \boldsymbol{x}'^{(l-1)} \end{bmatrix} \in \mathbb{F}_{q^n}^l$ where $\boldsymbol{x}'^{(i)} \in \mathbb{F}_{q^n}$ is a codeword of $\mathrm{Gab}[n, k]$ over $\mathbb{F}_{q^n}$ such that

$$\boldsymbol{x}'^{(i)}\hat{H}^T = \boldsymbol{0}. \quad (S49)$$

A received word at each Bob is of length $n_0$ and represented as $\boldsymbol{y} \in \mathbb{F}_{q^n}^l$. This is first decoded by the decoder of $\mathbb{F}_q$-RLNC, and converted into the $\mathbb{F}_q$-RLNC output word $\tilde{\boldsymbol{y}} \in \mathbb{F}_{q^n}^l$. The $\tilde{\boldsymbol{y}}$ is then decoded by the $\mathrm{iGab}[n, k]$ decoder, which deals with $\tilde{\boldsymbol{y}}$ as a part of the length-$n$ word $\boldsymbol{y}' = \begin{bmatrix} \boldsymbol{0} & \tilde{\boldsymbol{y}} \end{bmatrix}$ with $k_1$ erasures expressed by $\boldsymbol{0}$. The $\mathrm{iGab}[n, k]$ decoder first divides the input vertically into $l$ components $\begin{bmatrix} \boldsymbol{y}'^{(0)} \\ \vdots \\ \boldsymbol{y}'^{(l-1)} \end{bmatrix} \in \mathbb{F}_{q^n}$, and then processes the above components row-wise sequentially based on the algorithms described in Section V.

Sections III and IV are devoted to tutorial guides for decoding of a *single* Gabidulin code. To save mathematical notation, we denote the single Gabidulin code as $\mathrm{Gab}[n, k]$ over $\mathbb{F}_{q^n}$, Alice's message as $\boldsymbol{u} \in \mathbb{F}_{q^n}^k$ (setting $k = k_0, \mu_0 = 0$), and a codeword and received word as $\boldsymbol{x}' \in \mathbb{F}_{q^n}^n$ and $\boldsymbol{y}' \in \mathbb{F}_{q^n}^n$, respectively. In Section V, they actually correspond to each message component $\boldsymbol{u}^{(i)}$, each codeword component $\boldsymbol{x}'^{(i)}$, and each received word component $\boldsymbol{y}'^{(i)}$.

As for the elements of the generator matrix $\hat{G}$ and the parity check matrix $\hat{H}$ over $\mathbb{F}_{q^n}$, to avoid complexity and improve readability, we will omit the hat notation and consider $h_0, ..., h_{n-1}$ and $g_0, ..., g_{n-1}$ as elements defined over $\mathbb{F}_{q^n}$ from here on.

## III. ERROR CORRECTION: Gab$[n, k]$ OVER $\mathbb{F}_{q^n}$

Suppose that Alice wants to transmit a message $\boldsymbol{u} \in \mathbb{F}_{q^n}^n$ reliably to Bob. The Gabidulin encoder generates a codeword by

$$\boldsymbol{x}' = \boldsymbol{u}\hat{G}. \tag{S50}$$

Bob receives an output $\boldsymbol{y}'$ from a channel. The received word is the sum of the codeword $\boldsymbol{x}'$ and an error sequence $\boldsymbol{e}$

$$\boldsymbol{y}' = \boldsymbol{x}' + \boldsymbol{e}. \tag{S51}$$

In general, the goal of the decoder is to find, from the received word $\boldsymbol{y}'$, the number of the errors, the locations of the errors, and the error values at those positions. In the Gabidulin decoder, this goal is represented in terms of matrix-based notions, i.e., the rank of $\boldsymbol{e}$ (some non-negative integer $\tau$), the error values

$$\boldsymbol{a} = \begin{bmatrix} a_0 & \dots & a_{\tau-1} \end{bmatrix} \in \mathbb{F}_{q^n}^\tau, \tag{S52}$$

and the error locations

$$B = \begin{bmatrix} \boldsymbol{B}_0 \\ \vdots \\ \boldsymbol{B}_{\tau-1} \end{bmatrix} \in \mathbb{F}_q^{\tau \times n}, \tag{S53}$$

those give a decomposition of the error sequence $\boldsymbol{e}$ as

$$\boldsymbol{e} = \boldsymbol{a}B. \tag{S54}$$

The error values $a_0, \dots, a_{\tau-1} \in \mathbb{F}_{q^n}$ are linearly independent over $\mathbb{F}_q$, and $B$ is full rank, i.e., the error locations $\boldsymbol{B}_0, \dots, \boldsymbol{B}_{\tau-1}$ are linearly independent over $\mathbb{F}_q$.

This decomposition is not unique, but any of them is good for decoding. The decomposition of $\boldsymbol{e}$ leads to two strategies to determine $\boldsymbol{e}$ [16]. One is to find $\boldsymbol{a}$ (a basis of the column space of the error as a possible decomposition) and then use $\boldsymbol{a}$ to determine $B$ (which fixes the row space). This strategy was originally proposed by Gabidulin [17], and referred to as the Error Span method. The other is to determine $B$ and then use $B$ to determine $\boldsymbol{a}$. This strategy is referred to as the Error Location method.

Both methods are typically referred to as syndrome decoders and share an almost identical flow. Although the Error Span method is generally preferred for error decoding over $\mathbb{F}_{q^n}$ because the Error Location method requires two additional steps involving $q$-power shifts, we shall see later that when decoding errors and erasures or when decoding over the Cartesian product space $\mathbb{F}_{q^n}^\ell$, the Error Location method becomes naturally preferable. A flow diagram of the two methods for Gabidulin decoding is presented in Supplementary Fig. 12. Both methods can be split into 6 steps.

1) Compute the syndromes using the received word (and in the case of the Error Location method further compute the reversed syndromes using the syndromes via additional $q$-power shifts).
2) Synthesize the error span polynomial (ESP) using the syndromes via the LBMA [27], [36] [3] or synthesize the ELP using the reversed syndromes via the LBMA.
3) Determine a basis of the rootspace of the ESP or ELP.
4) Determine the error locations using the syndromes and the basis of the ESP via GRA (see Subsubsection VI-G2 for details) or determine the error values using the reversed syndromes and the basis of the ELP via GRA and additional $q$-power shifts.
5) Resolve the error using the basis of the ESP and the error locations or the basis of the ELP and the error values.
6) Determine the message from Alice using the received word and the error.

Greater detail of each step in the flow diagrams are presented next along with justification of why they lead to successful decoding in the case when $\tau \le \lfloor \frac{n-k}{2} \rfloor$.

### A. Syndromes: step 1a)

For decoding via both Error Span and Error Location methods, the first step is to compute the syndromes $s_0, \dots, s_{n-k-1} \in \mathbb{F}_{q^n}$ from the received word symbols $y_0', \dots, y_{n-1}' \in \mathbb{F}_{q^n}$. The syndrome $s_i \in \mathbb{F}_{q^n}$ is defined by

$$s_i = \sum_{v=0}^{n-1} y_v' h_v^{[k+i]}, \quad i = 0, \dots, n-k-1. \tag{S55}$$

Terms of the form $h_v^{[k+i]}$ do not depend on the received word, so can be precomputed and stored in memory. As such, $(n-k)(n-1)$ $\mathbb{F}_{q^n}$-additions and $(n-k)n$ $\mathbb{F}_{q^n}$-multiplications are required to compute the syndromes. Notice that in the case of a $q$-cyclic Gabidulin code, the $\mathbb{F}_{q^n}$-multiplication can be replaced with $\mathbb{F}_{q^n}$-fast multiplication. In order to justify steps 2), 4) and 5), we observe an alternative expression for the syndromes. Splitting the received word into its codeword and error component, we see

$$s_i = \sum_{v=0}^{n-1} x_v' h_v^{[k+i]} + \sum_{v=0}^{n-1} e_v h_v^{[k+i]}. \tag{S56}$$

Notice the first term on the right-hand side is zero because $\boldsymbol{x}'$ is a codeword. Using the error decomposition, we can express the syndromes in terms of the error values $a_0, \dots, a_{\tau-1}$ and the transformed error locations $d_0, \dots, d_{\tau-1}$

$$\begin{aligned} s_i &= 0 + \sum_{v=0}^{n-1} e_v h_v^{[k+i]} \\ &= \sum_{v=0}^{n-1} \left( \sum_{j=0}^{\tau-1} a_j B_{v,j} \right) h_v^{[k+i]} \\ &= \sum_{j=0}^{\tau-1} a_j d_j^{[i]}, \quad i = 0, \dots, n-k-1, \end{aligned} \tag{S57}$$

where

$$d_j = \sum_{v=0}^{n-1} B_{v,j} h_v^{[k]}, \quad j = 0, \dots, \tau-1, \tag{S58}$$

are the transformed error locations [4].

Eq. (S57) is a system of $n-k$ equations with totally $2\tau$ unknown variables $a_0, \dots, a_{\tau-1}$ and $d_0, \dots, d_{\tau-1}$. Note that the rank $\tau$ of $\boldsymbol{e}$ is also unknown. The system Eq. (S57) has many solutions for specified $\tau$. For $\tau \le \lfloor \frac{n-k}{2} \rfloor$, however, all the solutions lead to the same error sequence $\boldsymbol{e}$. Therefore, it is sufficient to find one solution of the system. Thus, the decoding problem actually reduces to finding a solution of Eq. (S57) for the smallest possible value of $\tau$.

---

[3] Gabidulin originally suggested in [17] using the Linearized Extended Euclidean Algorithm (LEEA) in place of the LBMA.

[4] This can also be written as the transformation of $\boldsymbol{B}_0, \dots, \boldsymbol{B}_{\tau-1}$ by the first column of the parity check matrix $\hat{H}$ as $\boldsymbol{d} = \hat{\boldsymbol{H}}_0^T B$

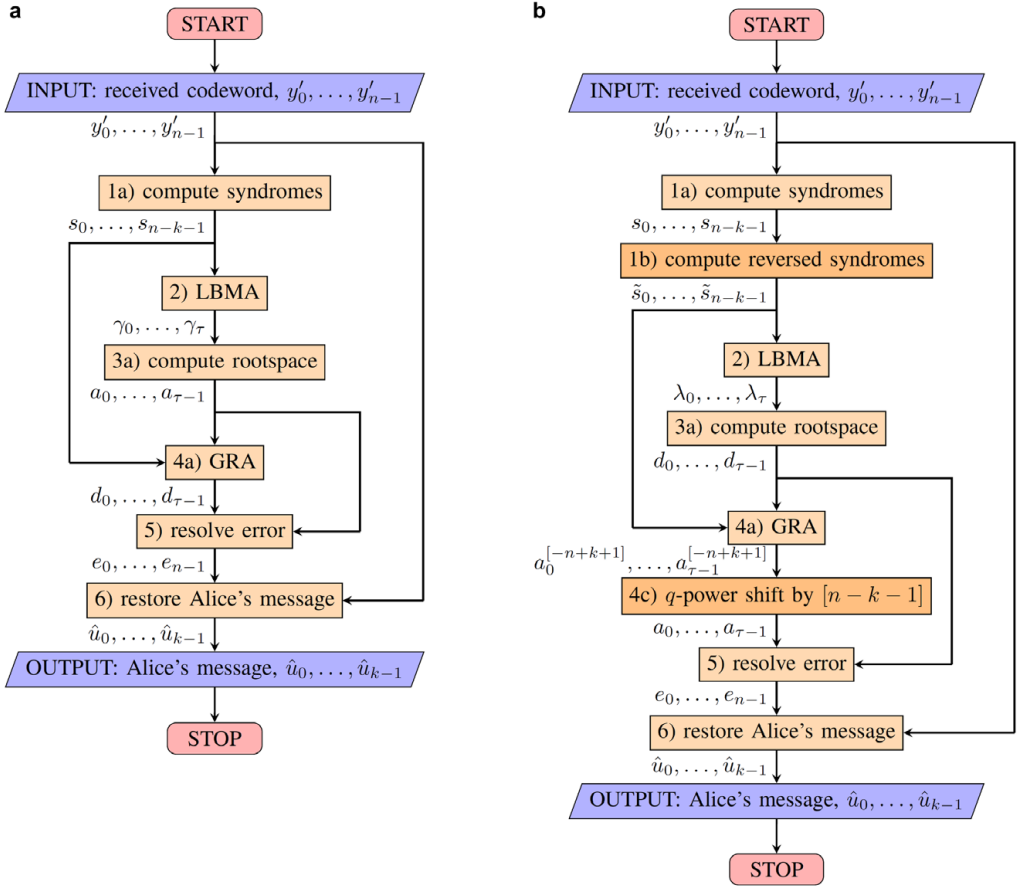

**Supplementary Fig. 12.** Flow diagrams of Gabidulin decoding. **a** Error Span method. **b** Error Location method.

## B. Reversed syndromes: step 1b)

For decoding via the Error Location method, we must use the syndromes $s_0, \ldots, s_{n-k-1}$ to compute the reversed syndromes $\tilde{s}_0, \ldots, \tilde{s}_{n-k-1} \in \mathbb{F}_{q^n}$. This step is not necessary for decoding via the Error Span method. For each $i = 0, \ldots, n-k-1$, the reversed syndrome $\tilde{s}_i \in \mathbb{F}_{q^n}$ is defined by

$$\tilde{s}_i = s_{n-k-1-i}^{[i-n+k+1]}. \quad (S59)$$

Computing the reversed syndromes from the syndromes requires $[-1]$ $q$-power operations of $\frac{1}{2}(n-k)(n-k-1)$. Similar to the syndromes, the reversed syndromes can be expressed in terms of the error values $a_0, \ldots, a_{\tau-1}$ and the transformed error locations $d_0, \ldots, d_{\tau-1}$ because

$$
\begin{aligned}
\tilde{s}_i &= \left( \sum_{j=0}^{\tau-1} a_j d_j^{[n-k-1-i]} \right)^{[i-n+k+1]} \\
&= \sum_{j=0}^{\tau-1} d_j a_j^{[i-n+k+1]}, \quad i = 0, \ldots, n-k-1. (S60)
\end{aligned}
$$

## C. Determining the ESP or ELP: step 2)

Each of the syndrome equations, Eq. (S57) or Eq. (S60) is nonlinear in $d_0, \ldots, d_{\tau-1}$ or $a_0, \ldots, a_{\tau-1}$, respectively, and does not have an obvious solution. So, the second step is to convert Eq. (S57) or Eq. (S60) to a linear system of equations with $\tau$ unknown variables, and determine the smallest possible value of $\tau$.

### 1) Determining the ESP from the syndromes using the LBMA

In order to eliminate the variables $d_0, \ldots, d_{\tau-1}$ from Eq. (S57), we compute $q$-power of the syndromes and make a matrix equation

$$
\begin{bmatrix}
s_\tau^{[0]} & \cdots & s_{n-k-1}^{[0]} \\
s_{\tau-1}^{[1]} & \cdots & s_{n-k-2}^{[1]} \\
\vdots & \ddots & \vdots \\
s_0^{[\tau]} & \cdots & s_{n-k-\tau-1}^{[\tau]}
\end{bmatrix}
=
\begin{bmatrix}
a_0^{[0]} & \cdots & a_{\tau-1}^{[0]} \\
a_0^{[1]} & \cdots & a_{\tau-1}^{[1]} \\
\vdots & \ddots & \vdots \\
a_0^{[\tau]} & \cdots & a_{\tau-1}^{[\tau]}
\end{bmatrix} D, \quad (S61)
$$

where

$$
D =
\begin{bmatrix}
d_0^{[\tau]} & \cdots & d_0^{[n-k-1]} \\
\vdots & \ddots & \vdots \\
d_{\tau-1}^{[\tau]} & \cdots & d_{\tau-1}^{[n-k-1]}
\end{bmatrix}. \quad (S62)
$$

Using the last $\tau$ rows of Eq. (S61) to solve for $D$, we have

$$
D =
\begin{bmatrix}
a_0^{[1]} & \cdots & a_{\tau-1}^{[1]} \\
\vdots & \ddots & \vdots \\
a_0^{[\tau]} & \cdots & a_{\tau-1}^{[\tau]}
\end{bmatrix}^{-1}
\begin{bmatrix}
s_{\tau-1}^{[1]} & \cdots & s_{n-k-2}^{[1]} \\
\vdots & \ddots & \vdots \\
s_0^{[\tau]} & \cdots & s_{n-k-\tau-1}^{[\tau]}
\end{bmatrix}. \quad (S63)
$$

Notice $\begin{bmatrix} a_0^{[1]} & \cdots & a_{\tau-1}^{[1]} \\ \vdots & \ddots & \vdots \\ a_0^{[\tau]} & \cdots & a_{\tau-1}^{[\tau]} \end{bmatrix}$ is invertible because it is the transpose of the linearized Vandermonde matrix generated by $a_0, \ldots, a_{\tau-1}$, which are linearly independent over $\mathbb{F}_q$, and hence it is full rank by Lemma 19 in Subsubsection VI-G1 (a square and full rank matrix is invertible). Substituting this expression for $D$ into the first line of Eq. (S61) we obtain

$$\begin{bmatrix} s_\tau^{[0]} & \cdots & s_{n-k-1}^{[0]} \end{bmatrix}$$

$$= \begin{bmatrix} a_0^{[0]} & \cdots & a_{\tau-1}^{[0]} \end{bmatrix} \begin{bmatrix} a_0^{[1]} & \cdots & a_{\tau-1}^{[1]} \\ \vdots & & \vdots \\ a_0^{[\tau]} & \cdots & a_{\tau-1}^{[\tau]} \end{bmatrix}^{-1}$$

$$\cdot \begin{bmatrix} s_{\tau-1}^{[1]} & \cdots & s_{n-k-2}^{[1]} \\ \vdots & \ddots & \vdots \\ s_0^{[\tau]} & \cdots & s_{n-k-\tau-1}^{[\tau]} \end{bmatrix}. \quad \text{(S64)}$$

We define

$$\begin{bmatrix} \gamma_1 & \cdots & \gamma_\tau \end{bmatrix}$$

$$\equiv -\begin{bmatrix} a_0^{[0]} & \cdots & a_{\tau-1}^{[0]} \end{bmatrix} \begin{bmatrix} a_0^{[1]} & \cdots & a_{\tau-1}^{[1]} \\ \vdots & \ddots & \vdots \\ a_0^{[\tau]} & \cdots & a_{\tau-1}^{[\tau]} \end{bmatrix}^{-1}, \quad \text{(S65)}$$

set $\gamma_0 = 1$, and rearrange the form as

$$\begin{bmatrix} \gamma_0 & \gamma_1 & \cdots & \gamma_\tau \end{bmatrix} \begin{bmatrix} a_0^{[0]} & \cdots & a_{\tau-1}^{[0]} \\ a_0^{[1]} & \cdots & a_{\tau-1}^{[1]} \\ \vdots & \ddots & \vdots \\ a_0^{[\tau]} & \cdots & a_{\tau-1}^{[\tau]} \end{bmatrix} = \mathbf{0}. \quad \text{(S66)}$$

Thus $a_0, \ldots, a_{\tau-1}$ can be regarded as roots of an $\mathbb{F}_{q^n}$-linearized polynomial

$$\Gamma(x) = \sum_{v=0}^{\tau} \gamma_v x^{[v]}, \quad \text{(S67)}$$

which is actually the ESP. Using the coefficients of the ESP, Eq. (S64) is rearranged to give

$$\begin{bmatrix} \gamma_0 & \gamma_1 & \cdots & \gamma_\tau \end{bmatrix} \begin{bmatrix} s_\tau^{[0]} & \cdots & s_{n-k-1}^{[0]} \\ s_{\tau-1}^{[1]} & \cdots & s_{n-k-2}^{[1]} \\ \vdots & \ddots & \vdots \\ s_0^{[\tau]} & \cdots & s_{n-k-\tau-1}^{[\tau]} \end{bmatrix} = \mathbf{0}, \quad \text{(S68)}$$

or concisely

$$\sum_{v=0}^{\tau} \gamma_v s_{i-v}^{[v]} = 0, \quad \forall i = \tau, \ldots, n-k-1. \quad \text{(S69)}$$

This is referred to as the key equation for the ESP, and is used to determine $\tau$ and $\gamma_1, \ldots, \gamma_\tau$.

When $2\tau \le n - k$, the key equation has the unique solution because $\begin{bmatrix} s_{\tau-1}^{[1]} & \cdots & s_{2\tau-2}^{[1]} \\ \vdots & \ddots & \vdots \\ s_0^{[\tau]} & \cdots & s_{\tau-1}^{[\tau]} \end{bmatrix}$ is nonsingular [27]. The key equation can be efficiently solved by using an iterative technique, i.e., the LBMA, which deals with an equivalent problem of finding the shortest linear feedback shift register that generates the known sequence of syndromes. This technique also overcomes the problem of not knowing $\tau$. Specifically, the LBMA returns the shortest $\mathbb{F}_{q^n}$-linearized feedback shift register ($\mathbb{F}_{q^n}$-LFSR) $\gamma_1, \ldots, \gamma_\tau \in \mathbb{F}_{q^n}$ for the given sequence of syndromes $s_0, \ldots, s_{n-k-1}$ [27], [37], [38]. (See Subsection VI-F for details.) Thus the ESP can be determined.

### 2) Determining the ELP from the reversed syndromes using the LBMA

In order to eliminate the variables $a_0, \ldots, a_{\tau-1}$ from Eq. (S60), we compute $q$-power of the reversed syndromes and make a matrix equation

$$\begin{bmatrix} \tilde{s}_\tau^{[0]} & \cdots & \tilde{s}_{n-k-1}^{[0]} \\ \tilde{s}_{\tau-1}^{[1]} & \cdots & \tilde{s}_{n-k-2}^{[1]} \\ \vdots & \ddots & \vdots \\ \tilde{s}_0^{[\tau]} & \cdots & \tilde{s}_{n-k-\tau-1}^{[\tau]} \end{bmatrix} = \begin{bmatrix} d_0^{[0]} & \cdots & d_{\tau-1}^{[0]} \\ d_0^{[1]} & \cdots & d_{\tau-1}^{[1]} \\ \vdots & \ddots & \vdots \\ d_0^{[\tau]} & \cdots & d_{\tau-1}^{[\tau]} \end{bmatrix} A, \quad \text{(S70)}$$

where

$$A = \begin{bmatrix} a_0^{[\tau-n+k+1]} & \cdots & a_0^{[0]} \\ \vdots & \ddots & \vdots \\ a_{\tau-1}^{[\tau-n+k+1]} & \cdots & a_{\tau-1}^{[\tau]} \end{bmatrix}. \quad \text{(S71)}$$

Using the last $\tau$ rows of Eq. (S70) to solve for $A$, we have

$$A = \begin{bmatrix} d_0^{[1]} & \cdots & d_{\tau-1}^{[1]} \\ \vdots & \ddots & \vdots \\ d_0^{[\tau]} & \cdots & d_{\tau-1}^{[\tau]} \end{bmatrix}^{-1} \begin{bmatrix} \tilde{s}_{\tau-1}^{[1]} & \cdots & \tilde{s}_{n-k-2}^{[1]} \\ \vdots & \ddots & \vdots \\ \tilde{s}_0^{[\tau]} & \cdots & \tilde{s}_{n-k-\tau-1}^{[\tau]} \end{bmatrix}. \quad \text{(S72)}$$

Notice $\begin{bmatrix} d_0^{[1]} & \cdots & d_{\tau-1}^{[1]} \\ \vdots & \ddots & \vdots \\ d_0^{[\tau]} & \cdots & d_{\tau-1}^{[\tau]} \end{bmatrix}$ is invertible because it is the transpose of the linearized Vandermonde matrix generated by $d_0^{[1]}, \ldots, d_{\tau-1}^{[1]}$ which are linearly independent over $\mathbb{F}_q$, and hence it is full rank by Lemma 19 in Subsubsection VI-G1 (a square and full rank matrix is invertible). Substituting this expression for $A$ into the first line of Eq. (S70) we obtain

$$\begin{bmatrix} \tilde{s}_\tau^{[0]} & \cdots & \tilde{s}_{n-k-1}^{[0]} \end{bmatrix}$$

$$= \begin{bmatrix} d_0^{[0]} & \cdots & d_{\tau-1}^{[0]} \end{bmatrix} \begin{bmatrix} d_0^{[1]} & \cdots & d_{\tau-1}^{[1]} \\ \vdots & \ddots & \vdots \\ d_0^{[\tau]} & \cdots & d_{\tau-1}^{[\tau]} \end{bmatrix}^{-1}$$

$$\cdot \begin{bmatrix} \tilde{s}_{\tau-1}^{[1]} & \cdots & \tilde{s}_{n-k-2}^{[1]} \\ \vdots & \ddots & \vdots \\ \tilde{s}_0^{[\tau]} & \cdots & \tilde{s}_{n-k-\tau-1}^{[\tau]} \end{bmatrix}. \quad \text{(S73)}$$

We define

$$\begin{bmatrix} \lambda_1 & \cdots & \lambda_\tau \end{bmatrix} =$$

$$-\begin{bmatrix} d_0^{[0]} & \cdots & d_{\tau-1}^{[0]} \end{bmatrix} \begin{bmatrix} d_0^{[1]} & \cdots & d_{\tau-1}^{[1]} \\ \vdots & \ddots & \vdots \\ d_0^{[\tau]} & \cdots & d_{\tau-1}^{[\tau]} \end{bmatrix}^{-1}, \quad \text{(S74)}$$

set $\lambda_0 = 1$, and rearrange the form as

$$\begin{bmatrix} \lambda_0 & \lambda_1 & \cdots & \lambda_\tau \end{bmatrix} \begin{bmatrix} d_0^{[0]} & \cdots & d_{\tau-1}^{[0]} \\ d_0^{[1]} & \cdots & d_{\tau-1}^{[1]} \\ \vdots & \ddots & \vdots \\ d_0^{[\tau]} & \cdots & d_{\tau-1}^{[\tau]} \end{bmatrix} = \mathbf{0}. \quad \text{(S75)}$$

Thus $d_0, \ldots, d_{\tau-1}$ can be regarded as roots of an $\mathbb{F}_{q^n}$-linearized polynomial

$$\Lambda(x) = \sum_{v=0}^{\tau} \lambda_v x^{[v]}. \quad \text{(S76)}$$

which is actually the ELP. Using the coefficients of the ELP, Eq. (S73) is rearranged to give

$$\begin{bmatrix} \lambda_0 & \lambda_1 & \dots & \lambda_\tau \end{bmatrix} \begin{bmatrix} \tilde{s}_\tau^{[0]} & \dots & \tilde{s}_{n-k-1}^{[0]} \\ \tilde{s}_{\tau-1}^{[1]} & \dots & \tilde{s}_{n-k-2}^{[1]} \\ \vdots & & \vdots \\ \tilde{s}_0^{[\tau]} & \dots & \tilde{s}_{n-k-\tau-1}^{[\tau]} \end{bmatrix} = \mathbf{0} \quad \text{(S77)}$$

or concisely

$$\sum_{v=0}^{\tau} \lambda_v \tilde{s}_{i-v}^{[v]} = 0, \quad \forall i = \tau, \dots, n-k-1. \quad \text{(S78)}$$

This is referred to as the key equation for the ELP, and is used to determine $\tau$ and $\lambda_1, \dots, \lambda_\tau$.

When $2\tau \leq n-k$, the key equation has the unique solution because $\begin{bmatrix} \tilde{s}_{\tau-1}^{[1]} & \dots & \tilde{s}_{2\tau-2}^{[1]} \\ \vdots & \ddots & \vdots \\ \tilde{s}_0^{[\tau]} & \dots & \tilde{s}_{\tau-1}^{[\tau]} \end{bmatrix}$ is nonsingular [27]. The key equation can be solved efficiently by using the LBMA, which returns the shortest $\mathbb{F}_{q^n}$-LFSR $\lambda_1, \dots, \lambda_\tau$ for the given sequence of syndromes $\tilde{s}_0, \dots, \tilde{s}_{n-k-1}$ [27], [37], [38]. (See Subsection VI-F for details.) Thus the ELP can be determined.

### D. Determining the rootspace of the ESP or ELP: step 3)

Step 3) of decoding is to determine the error values $a_0, \dots, a_{\tau-1}$ as a basis of the rootspace of the ESP

$$\text{rootspace}(\Gamma) = \text{span}\{a_0, \dots, a_{\tau-1}\}, \quad \text{(S79)}$$

in the case of the Error Span method, or the error locations $d_0, \dots, d_{\tau-1}$ as a basis of the rootspace of the ELP

$$\text{rootspace}(\Lambda) = \text{span}\{d_0, \dots, d_{\tau-1}\}, \quad \text{(S80)}$$

in the case of the Error Location method. This is done using Supplementary Algorithm 2 in Subsubsection VI-E3 for details.

### E. Determining the error locations via GRA: step 4a) for ESP

For decoding using the Error Span method, recall that the error values $a_0, \dots, a_{\tau-1}$, the error locations $d_0, \dots, d_{\tau-1}$ and the syndromes $s_0, \dots, s_{\tau-1}$ are related as follows:

$$s_i = \sum_{j=0}^{\tau-1} a_j d_j^{[i]}, \quad \text{(S81)}$$

for each $i = 0, \dots, \tau-1$ (instead of $i = 0, \dots, n-k-1$ in Eq. (S57)). Notice this system of equations is of the form solvable by GRA (see Subsubsection VI-G2), where the error values $a_0, \dots, a_{\tau-1}$ act as $z_0, \dots, z_{\tau-1}$ and are linearly independent over $\mathbb{F}_q$. The syndromes $s_0, \dots, s_{\tau-1}$ remain as they are, and the error locations $d_0, \dots, d_{\tau-1}$ act as $x_0, \dots, x_{\tau-1}$ and are the variables to solve. Because both the error values $a_0, \dots, a_{\tau-1}$ and the syndromes $s_0, \dots, s_{\tau-1}$ are already known, we can pass them as the respective inputs and GRA will return the error locations $d_0, \dots, d_{\tau-1}$ as the output.

### F. Determining the error values via GRA: step 4a), 4c) for ELP

For decoding using the Error Location method, recall that the error locations $d_0, \dots, d_{\tau-1}$, the error values $a_0, \dots, a_{\tau-1}$ and the reversed syndromes $\tilde{s}_0, \dots, \tilde{s}_{\tau-1}$ are related as follows:

$$\tilde{s}_i = \sum_{v=0}^{\tau-1} d_v a_v^{[i-n+k+1]}, \quad \text{(S82)}$$

for each $i = 0, \dots, \tau-1$ (instead of $i = 0, \dots, n-k-1$ in Eq. (S60)). Notice this system of equations is of the form solvable by GRA (see Subsubsection VI-G2), where the error locations $d_0, \dots, d_{\tau-1}$ act as $z_0, \dots, z_{\tau-1}$ and are linearly independent over $\mathbb{F}_q$, the reversed syndromes $\tilde{s}_0, \dots, \tilde{s}_{\tau-1}$ act as $s_0, \dots, s_{\tau-1}$, and the $[-n+k+1]$ $q$-power of the error values $a_0^{[-n+k+1]}, \dots, a_{\tau-1}^{[-n+k+1]}$ act as $x_0, \dots, x_{\tau-1}$ and are the variables to solve. Because both the error locations $d_0, \dots, d_{\tau-1}$ and the reversed syndromes $\tilde{s}_0, \dots, \tilde{s}_{\tau-1}$ are already known, we can pass them as the respective inputs and GRA will return $[-n+k+1]$ $q$-power error values $a_0^{[-n+k+1]}, \dots, a_{\tau-1}^{[-n+k+1]}$ as the output, completing step 4a). The error values $a_0, \dots, a_{\tau-1}$ can be determined by performing an additional $[n-k-1]$ $q$-power operation to each output of the GRA, completing step 4c). (Note that step 4b) will appear in Section IV.)

### G. Resolving the error: step 5)

Both methods have now determined the error values $a_0, \dots, a_{\tau-1}$ and the transformed error locations $d_0, \dots, d_{\tau-1}$. Therefore, both methods now conclude with exactly the same finale.

We can determine the error locations $\boldsymbol{B}_0, \dots, \boldsymbol{B}_{\tau-1}$ from $d_0, \dots, d_{\tau-1}$ by using Eq. (S58). To do this, we write Eq. (S58) in matrix form

$$\begin{bmatrix} h_0^{[k]} & \dots & h_{n-1}^{[k]} \\ \vdots & \ddots & \vdots \\ h_0^{[k-1]} & \dots & h_{n-1}^{[k-1]} \end{bmatrix} B = \begin{bmatrix} d_0^{[0]} & \dots & d_{\tau-1}^{[0]} \\ \vdots & \ddots & \vdots \\ d_0^{[n-1]} & \dots & d_{\tau-1}^{[n-1]} \end{bmatrix}, \quad \text{(S83)}$$

where the cyclic property of the $q$-power operator $h_j^{[i]} = h_j^{[i \bmod n]}$ over $\mathbb{F}_{q^n}$ is used (see Subsection VI-A). Multiplying both sides on the left by the matrix $\begin{bmatrix} g_0^{[k]} & \dots & g_0^{[k-1]} \\ \vdots & \ddots & \vdots \\ g_{n-1}^{[k]} & \dots & g_{n-1}^{[k-1]} \end{bmatrix}$, we can determine the error locations

$$B = \begin{bmatrix} g_0^{[k]} & \dots & g_0^{[k-1]} \\ \vdots & \ddots & \vdots \\ g_{n-1}^{[k]} & \dots & g_{n-1}^{[k-1]} \end{bmatrix} \begin{bmatrix} d_0^{[0]} & \dots & d_{\tau-1}^{[0]} \\ \vdots & \ddots & \vdots \\ d_0^{[n-1]} & \dots & d_{\tau-1}^{[n-1]} \end{bmatrix}, \quad \text{(S84)}$$

thanks to the fact that $g_0, \dots, g_{n-1}$ is the dual basis of $h_0, \dots, h_{n-1}$ meaning that

$$\sum_{v=0}^{n-1} h_i^{[v]} g_j^{[v]} = \delta_{i,j}, \quad \text{(S85)}$$

(see Subsection VI-D).

When Gabidulin codes are $q$-cyclic, however, the error locations can be determined more straightforwardly. In that case, $h_0, \ldots, h_{n-1}$ is a normal basis generated by $\beta$. Let the $\beta$-normal basis representation of $d_j$ be

$$d_j = \sum_{u=0}^{n-1} D_{j,u} \beta^{[u]}. \tag{S86}$$

Then Eq. (S58) can be rewritten as

$$\sum_{u=0}^{n-1} D_{j,u} \beta^{[u]} = \sum_{v=0}^{n-1} B_{j,v} \beta^{[k+v]}, \tag{S87}$$

for $j = 0, \ldots, \tau - 1$. Because $\beta^{[0]}, \ldots, \beta^{[n-1]}$ are linearly independent over $\mathbb{F}_q$, we equate $q$-powers of $\beta$ to get for each $v = 0, \ldots, n-1$

$$B_{j,v} = D_{j,(v+k) \bmod n}, \tag{S88}$$

namely $B_j$ is the $[-k]$ $q$-power of the $\beta$-normal basis representation of $d_j$.

Once $B$ is thus determined, the error $e$ can be resolved since for each $v = 0, \ldots, n-1$

$$e_v = \sum_{j=0}^{\tau-1} a_j B_{v,j}. \tag{S89}$$

### H. Restoring Alice's message: step 6)

The estimated codeword $\hat{x}'$ is given by

$$\hat{x}' = y' - e. \tag{S90}$$

According to Eq. (S50), the estimated Alice's message $\hat{u}$ can be given by multiplying $x'$ on the right by the $k \times n$ matrix

$$\hat{H}' = \begin{bmatrix} h_0^{[0]} & h_1^{[0]} & \cdots & h_{n-1}^{[0]} \\ h_0^{[1]} & h_1^{[1]} & \cdots & h_{n-1}^{[1]} \\ \vdots & \vdots & \ddots & \vdots \\ h_0^{[k]} & h_1^{[k]} & \cdots & h_{n-1}^{[k]} \end{bmatrix}, \tag{S91}$$

such that

$$\begin{aligned} \hat{x}' \, \hat{H}'^T &= \hat{u}\hat{G} \, \hat{H}'^T \\ &= \hat{u} I_{k \times k} \\ &= \hat{u}, \end{aligned} \tag{S92}$$

where the second equality is due to the duality

$$\sum_{v=0}^{n-1} g_v^{[i]} h_v^{[j]} = \delta_{i,j}, \tag{S93}$$

of the bases $g_0, \ldots, g_{n-1}$ and $h_0, \ldots, h_{n-1}$ (see Subsection VI-D).

## IV. Error and erasure correction: $\mathrm{Gab}[n,k]$ over $\mathbb{F}_{q^n}$

In certain cases, some (or all) of the error locations may be known before decoding takes place. Typical cases include situations where some symbols (packets) of the codeword are not received (and/or not transmitted for some purposes), or when the $\mathbb{F}_q$-RLNC fails to deliver a sufficient number of packets to Bob(s) for decodability. Such *known* locations are referred to as erasure locations, and erasure sequence, denoted $\hat{e} \in \mathbb{F}_{q^n}^n$, can be incorporated into the errata model as follows:

$$\hat{e} = \hat{a}\hat{B}, \tag{S94}$$

where for some *known* non-negative integer $\rho$

$$\hat{a} = [\hat{a}_0, \ldots, \hat{a}_{\rho-1}] \in \mathbb{F}_{q^n}^{\rho}, \tag{S95}$$

are the erasure values and

$$\hat{B} = \begin{bmatrix} \hat{B}_0 \\ \vdots \\ \hat{B}_{\rho-1} \end{bmatrix} \in \mathbb{F}_q^{\rho \times n}, \tag{S96}$$

are the known erasure locations such that $\begin{bmatrix} B \\ \hat{B} \end{bmatrix} \in \mathbb{F}_q^{(\tau+\rho) \times n}$ is full rank. Then the received word $y'$ is given by the addition of the error sequence $e$ and the erasure sequence $\hat{e}$ to the codeword $x'$

$$y' = x' + e + \hat{e}. \tag{S97}$$

When $2\tau + \rho \leq n - k$, the error and erasure sequences can be uniquely determined. An overview of error and erasure decoding via the ELP, split into seven steps, follows.

0) a) Compute the transformed erasure locations $\hat{d}_0, \ldots, \hat{d}_{\rho-1} \in \mathbb{F}_{q^n}$ from the erasure locations $\hat{B}_0, \ldots, \hat{B}_{\rho-1} \in \mathbb{F}_{q^n}$.

   b) Synthesize the erasUre Location Polynomial (ULP) $\hat{\Lambda}(x)$ from the transformed erasure locations $\hat{d}_0, \ldots, \hat{d}_{\rho-1}$ using Richter-Plass algorithm (RPA) [27] (Supplementary Algorithm 3).

1) a) Compute the syndromes $s_0, \ldots, s_{n-k-1} \in \mathbb{F}_{q^n}$ from the received word $y'_0, \ldots, y'_{n-1} \in \mathbb{F}_{q^n}$.

   b) Compute the reversed syndromes $\tilde{s}_0, \ldots, \tilde{s}_{n-k-1} \in \mathbb{F}_{q^n}$ from the syndromes $s_0, \ldots, s_{n-k-1}$.

   c) Compute the filtered reversed syndromes $\hat{\tilde{s}}_0, \ldots, \hat{\tilde{s}}_{n-k-\rho-1} \in \mathbb{F}_{q^n}$ from the reversed syndromes $\tilde{s}_0, \ldots, \tilde{s}_{n-k-1}$ and the ULP $\hat{\Lambda}(x)$.

2) Synthesize the ELP $\Lambda(x)$ from the filtered reversed syndromes $\hat{\tilde{s}}_0, \ldots, \hat{\tilde{s}}_{n-k-\rho-1}$ using the LBMA.

3) a) Determine a basis $l_0, \ldots, l_{\tau-1} \in \mathbb{F}_{q^n}$ of the rootspace of the ELP.

   b) Compute the transformed error locations $d_0, \ldots, d_{\tau-1} \in \mathbb{F}_{q^n}$ from the basis $l_0, \ldots, l_{\tau-1}$ and the ULP $\hat{\Lambda}(x)$.

4) a) Compute $q$-power shifted error values $a_0^{[\rho-n+k+1]}, \ldots, a_{\tau-1}^{[\rho-n+k+1]} \in \mathbb{F}_{q^n}$ from $l_0, \ldots, l_{\tau-1}$ and the filtered reversed syndromes $\hat{\tilde{s}}_0, \ldots, \hat{\tilde{s}}_{\tau-1}$ using GRA.

   b) Compute $q$-power shifted error values $a_0^{[-n+k+1]}, \ldots, a_{\tau-1}^{[-n+k+1]} \in \mathbb{F}_{q^n}$ from $q$-power shifted error values $a_0^{[\rho-n+k+1]}, \ldots, a_{\tau-1}^{[\rho-n+k+1]}$.

   c) Determine the error values $a_0, \ldots, a_{\tau-1} \in \mathbb{F}_{q^n}$ from the $q$-power shifted error values $a_0^{[-n+k+1]}, \ldots, a_{\tau-1}^{[-n+k+1]}$.

   d) Compute modified reversed syndromes $\bar{s}_0, \ldots, \bar{s}_{\rho-1} \in \mathbb{F}_{q^n}$ from the reversed syndromes $\tilde{s}_0, \ldots, \tilde{s}_{\rho-1}$, the transformed error locations $d_0, \ldots, d_{\tau-1}$ and the error values $a_0, \ldots, a_{\tau-1}$.

   e) Compute $q$-power shifted erasure values $\hat{a}_0^{[-n+k+1]}, \ldots, \hat{a}_{\rho-1}^{[-n+k+1]} \in \mathbb{F}_{q^n}$ from the erasure locations $\hat{d}_0, \ldots, \hat{d}_{\rho-1}$ and the modified reversed syndromes $\bar{s}_0, \ldots, \bar{s}_{\rho-1}$ using GRA.

   f) Determine the erasure values $\hat{a}_0, \ldots, \hat{a}_{\rho-1} \in \mathbb{F}_{q^n}$ from the $q$-power shifted erasure values $\hat{a}_0^{[-n+k+1]}, \ldots, \hat{a}_{\rho-1}^{[-n+k+1]}$.

5) a) Resolve the error $e \in \mathbb{F}_{q^n}^n$ from the transformed error locations $d_0, \ldots, d_{\tau-1}$ and the error values $a_0, \ldots, a_{\tau-1}$.

b) Resolve the erasures $\hat{e} \in \mathbb{F}_{q^n}^n$ from the transformed erasure locations $\hat{d}_0, \ldots, \hat{d}_{\rho-1}$ and the erasure values $\hat{a}_0, \ldots, \hat{a}_{\rho-1}$.

6) Restore Alice's message $u_0, \ldots, u_{k-1} \in \mathbb{F}_{q^n}$ from the received word $y_0', \ldots, y_{n-1}'$, the error $e$ and the erasure $\hat{e}$.

A flow diagram of the decoding method is given in Supplementary Fig. 13. Notice that steps 1a), 1b), 2), 3a), 4a), 4c), 5a), and 6) constitute the steps of the error decoder. The remaining steps 0a), 0b), 1c), 3b), 4b), 4d)-4f), and 5b) are the additional steps required to incorporate the decoding of erasures. The justification of the additional steps required for error and erasure decoding is given in the following sections.

### A. Determining the ULP: step 0a), 0b)

Note that the addition of the erasure term means that each reversed syndrome $\tilde{s}_i$ for $i = 0, \ldots, n-k-1$ can be expressed as

$$\tilde{s}_i = \sum_{j=0}^{\tau-1} d_j a_j^{[i-n+k+1]} + \sum_{j=0}^{\rho-1} \hat{d}_j \hat{a}_j^{[i-n+k+1]}, \qquad (\text{S98})$$

where for each $j = 0, \ldots, \rho-1$, $\hat{d}_j \in \mathbb{F}_{q^n}$ is the transformed erasure location

$$\hat{d}_j = \sum_{v=0}^{n-1} \hat{B}_{v,j} h_v^{[k]}. \qquad (\text{S99})$$

By following the same argument from Eq. (S60) and Eq. (S70) to Eq. (S76), $\hat{d}_0, \ldots, \hat{d}_{\rho-1}$ can be regarded as roots of the ULP

$$\hat{\Lambda}(x) = \sum_{v=0}^{\rho} \hat{\lambda}_v x^{[v]}, \qquad (\text{S100})$$

with coefficients $\hat{\lambda}_0, \ldots, \hat{\lambda}_\rho \in \mathbb{F}_{q^n}$ where $\hat{\lambda}_0 = 1$ and $\hat{\lambda}_\rho \neq 0$. Note that the transformed erasure locations $\hat{d}_0, \ldots, \hat{d}_{\rho-1}$ are known. We can then synthesize the ULP $\hat{\Lambda}(x)$ as the minimal $\mathbb{F}_{q^n}$-linearized polynomial of the given set $\left\{ \hat{d}_0, \ldots, \hat{d}_{\rho-1} \right\}$ (Subsubsections VI-E4 and VI-E5), by using RPA (Supplementary Algorithm 3). As such, for all $j = 0, \ldots, \rho-1$

$$\hat{\Lambda}\left(\hat{d}_j\right) = 0. \qquad (\text{S101})$$

### B. Filtering the reversed syndromes with the ULP: step 1c)

The ULP can be used to filter the erasure locations out of the reversed syndromes, resulting in the filtered reversed syndromes containing only terms associated with the error. For each $i = 0, \ldots, n-k-\rho-1$, we define the filtered reversed syndrome $\hat{\tilde{s}}_i \in \mathbb{F}_{q^n}$ based on the ULP $\hat{\Lambda}(x)$ as

$$\hat{\tilde{s}}_i = \sum_{v=0}^{\rho} \hat{\lambda}_v \tilde{s}_{\rho+i-v}^{[v]}. \qquad (\text{S102})$$

To see the filtered reversed syndromes contain terms only associated with the error and not the erasure, we use Eq. (S98) to show

$$
\begin{aligned}
\hat{\tilde{s}}_i &= \sum_{v=0}^{\rho} \hat{\lambda}_v \tilde{s}_{\rho+i-v}^{[v]} \\
&= \sum_{v=0}^{\rho} \hat{\lambda}_v \left( \sum_{j=0}^{\tau-1} d_j a_j^{[\rho+i-v-n+k+1]} \right. \\
&\qquad\qquad \left. + \sum_{j=0}^{\rho-1} \hat{d}_j \hat{a}_j^{[\rho+i-v-n+k+1]} \right)^{[v]} \\
&= \sum_{v=0}^{\rho} \hat{\lambda}_v \sum_{j=0}^{\tau-1} d_j^{[v]} a_j^{[\rho+i-n+k+1]} \\
&\qquad\qquad + \sum_{v=0}^{\rho} \hat{\lambda}_v \sum_{j=0}^{\rho-1} \hat{d}_j^{[v]} \hat{a}_j^{[\rho+i-n+k+1]} \\
&= \sum_{j=0}^{\tau-1} \left( \sum_{v=0}^{\rho} \hat{\lambda}_v d_j^{[v]} \right) a_j^{[\rho+i-n+k+1]} \\
&\qquad\qquad + \sum_{j=0}^{\rho-1} \left( \sum_{v=0}^{\rho} \hat{\lambda}_v \hat{d}_j^{[v]} \right) \hat{a}_j^{[\rho+i-n+k+1]} \\
&= \sum_{j=0}^{\tau-1} \hat{\Lambda}(d_j) a_j^{[\rho+i-n+k+1]} + \sum_{j=0}^{\rho-1} \hat{\Lambda}\left(\hat{d}_j\right) \hat{a}_j^{[\rho+i-n+k+1]} \\
&= \sum_{j=0}^{\tau-1} \hat{\Lambda}(d_j) a_j^{[\rho+i-n+k+1]} + \sum_{j=0}^{\rho-1} 0 \cdot \hat{a}_j^{[\rho+i-n+k+1]} \\
&= \sum_{j=0}^{\tau-1} \hat{\Lambda}(d_j) a_j^{[\rho+i-n+k+1]}.
\end{aligned}
$$
$$(\text{S103})$$

### C. Determining the ELP from the filtered reversed syndromes: step 2)

Comparing Eq. (S103) with Eq. (S60), we notice that $\hat{\Lambda}(d_0), \ldots, \hat{\Lambda}(d_{\tau-1})$ can be interpreted as the filtered transformed error locations. Therefore, by following the same reasoning as in Eq. (S76), Eq. (S78), and Eq. (S80), we see that if $\tau \leq \lfloor \frac{n-k-\rho}{2} \rfloor$, passing $\hat{\tilde{s}}_0, \ldots, \hat{\tilde{s}}_{n-k-\rho-1}$ as input into the LBMA will result in the output $\mathbb{F}_{q^n}$-LFSR $\lambda_0, \ldots, \lambda_\tau$ with $\lambda_0 = 1$ and $\lambda_\tau \neq 0$ such that

$$\text{rootspace}(\Lambda) = \text{Span}\left\{ \hat{\Lambda}(d_0), \ldots, \hat{\Lambda}(d_{\tau-1}) \right\}, \qquad (\text{S104})$$

where

$$\Lambda(x) = \sum_{v=0}^{\tau} \lambda_v x^{[v]}. \qquad (\text{S105})$$

This is because $\hat{\Lambda}(d_0), \ldots, \hat{\Lambda}(d_{\tau-1})$ are linearly independent over $\mathbb{F}_q$ as a consequence of $\boldsymbol{B}_0, \ldots, \boldsymbol{B}_{\tau-1}, \hat{\boldsymbol{B}}_0, \ldots, \hat{\boldsymbol{B}}_{\rho-1}$ being linearly independent over $\mathbb{F}_q$, and hence $d_0, \ldots, d_{\tau-1} \notin$ rootspace $\left(\hat{\Lambda}\right)$. The ELP $\Lambda(x)$ is thus synthesized.

### D. Determining the error locations from the ELP and ULP: step 3a), 3b)

The filtered transformed error locations $\hat{\Lambda}(d_0), \ldots, \hat{\Lambda}(d_{\tau-1})$ are determined by finding a basis of rootspace$(\Lambda)$ using Supplementary Algorithm 2 in the same fashion as step 3) of the error decoder.

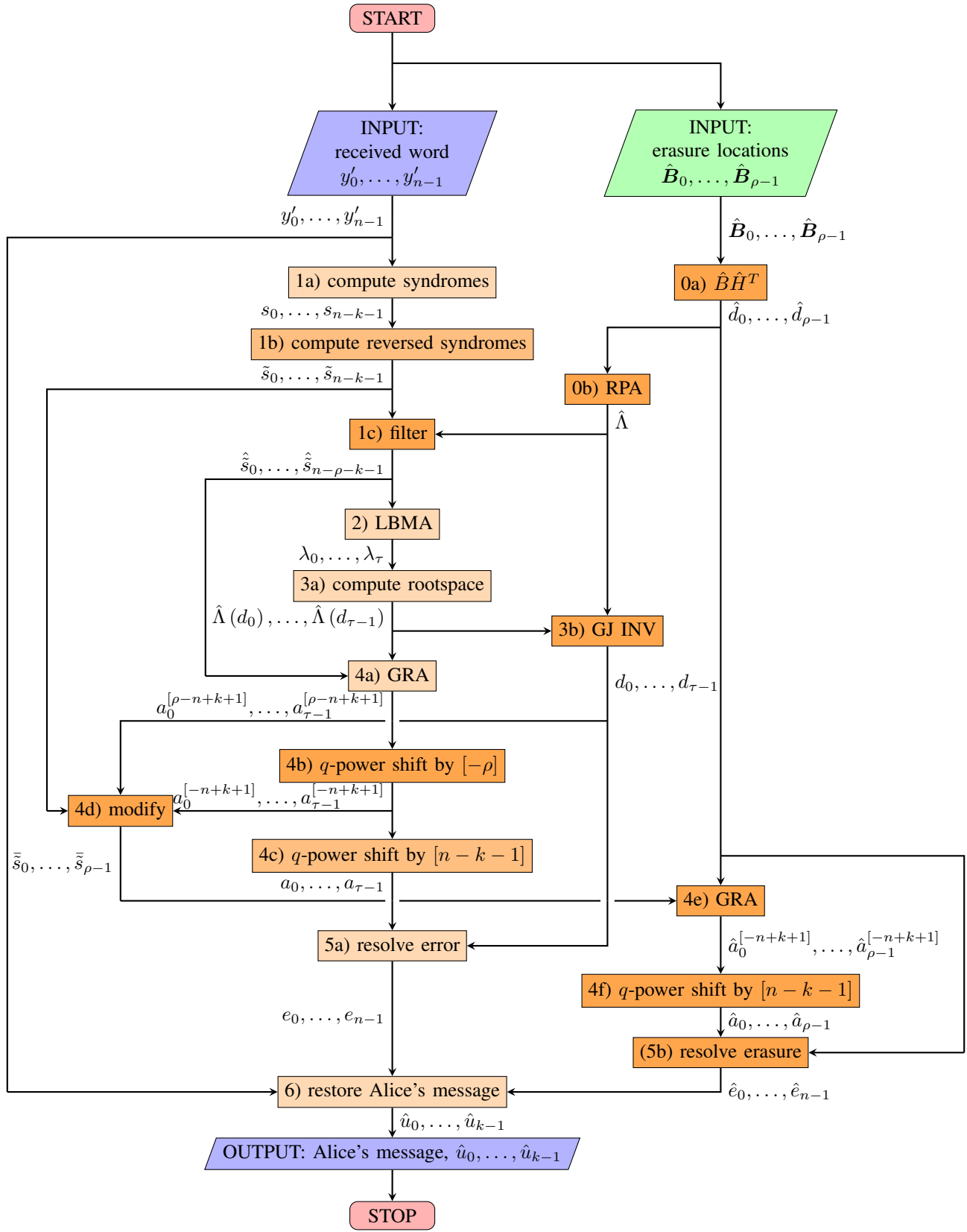

**Supplementary Fig. 13.** Flow diagram of error-erasure decoder via the Error Location method.

For each $j = 0, \ldots, \tau - 1$, let

$$l_j \equiv \hat{\Lambda}(d_i) = \begin{bmatrix} L_{0,j} & \ldots & L_{n-1,j} \end{bmatrix}^T \in \mathbb{F}_q^n, \quad (S106)$$

$$d_j = \begin{bmatrix} D_{0,j} & \ldots & D_{n-1,j} \end{bmatrix}^T \in \mathbb{F}_q^n, \quad (S107)$$

be the $\beta$-normal basis representations of of $l_j$ and $d_j$, respectively. We have

$$
\begin{aligned}
\sum_{i=0}^{n-1} L_{i,j} \beta^{[i]} &= \hat{\Lambda}(d_j) \\
&= \sum_{v=0}^{\rho} \hat{\lambda}_v d_j^{[v]} \\
&= \sum_{v=0}^{\rho} \hat{\lambda}_v \left( \sum_{u=0}^{n-1} D_{u,j} \beta^{[u]} \right)^{[v]} \\
&= \sum_{u=0}^{n-1} D_{u,j} \sum_{v=0}^{\rho} \hat{\lambda}_v \beta^{[v+u]} \\
&= \sum_{u=0}^{n-1} D_{u,j} \hat{\Lambda}\left( \beta^{[u]} \right) \\
&= \sum_{u=0}^{n-1} D_{u,j} \left( \sum_{i=0}^{n-1} \Psi_{i,u} \beta^{[i]} \right) \\
&= \sum_{i=0}^{n-1} \sum_{u=0}^{n-1} \Psi_{i,u} D_{u,j} \beta^{[i]}, \quad (S108)
\end{aligned}
$$

where for each $u = 0, \ldots, n-1$, $\boldsymbol{\Psi}_u = \begin{bmatrix} \Psi_{0,u} & \ldots & \Psi_{n-1,u} \end{bmatrix}^T \in \mathbb{F}_q^n$ is the $\beta$-normal basis representation of $\hat{\Lambda}\left( \beta^{[u]} \right)$. Since $\beta^{[0]}, \ldots, \beta^{[n-1]}$ are linearly independent over $\mathbb{F}_q$, we must have

$$L_{i,j} = \sum_{u=0}^{n-1} \Psi_{i,u} D_{u,j}, \quad (S109)$$

for each $i, j = 0, \ldots, n-1$, or in matrix form

$$L = \Psi D, \quad (S110)$$

where $L, D \in \mathbb{F}_q^{n \times \tau}$ are the matrices whose columns are $l_0, \ldots, l_{\tau-1}$ and $d_0, \ldots, d_{\tau-1}$, respectively, and $\Psi \in \mathbb{F}_q^{n \times n}$ is the matrix whose columns are $\boldsymbol{\Psi}_0, \ldots, \boldsymbol{\Psi}_{n-1}$. Therefore

$$D = \Psi^+ L + \hat{\Psi}, \quad (S111)$$

where $\Psi^+ \in \mathbb{F}_q^{n \times n}$ is the pseudoinverse of $\Psi$ and $\hat{\Psi} \in \mathbb{F}_q^{n \times \tau}$ such that $\Psi \hat{\Psi} = 0$. Note that the solution to Eq. (S111) is not unique. We are free to choose any $\hat{\Psi}$. In practice, we compute $D$ using Gauss-Jordan inversion over $\mathbb{F}_q$ on the system given in Eq. (S110) because $L$ is already computed as the $\beta$-normal basis representation of a basis of rootspace $(\Lambda)$, and $\Psi$ can be computed as the matrix of $\beta$-normal basis representations of the ULP evaluated at $\beta^{[u]}$ for each $u = 0, \ldots, n-1$. This means we are effectively choosing $\hat{\Psi} = 0$. Note that because $D$ is the matrix of $\beta$-normal basis representations of $d_0, \ldots, d_{\tau-1}$, the error locations are now determined.

### E. Determining the error values via GRA: step 4a)-4c)

Notice that from Eq. (S103) and because $\hat{\Lambda}(d_0), \ldots, \hat{\Lambda}(d_{\tau-1})$ and $\tilde{\hat{s}}_0, \ldots, \tilde{\hat{s}}_{\tau-1}$ are already computed, we can use GRA to solve for $q$-power shifted error values $a_0^{[\rho-n+k+1]}, \ldots, a_{\tau-1}^{[\rho-n+k+1]}$. A $q$-power shift by $[n - k - \rho - 1]$ of the output of GRA, $a_0^{[\rho-n+k+1]}, \ldots, a_{\tau-1}^{[\rho-n+k+1]}$, will recover the error values

$a_0, \ldots, a_{\tau-1}$. In practice, however, it is useful to firstly apply a $q$-power shift of $[-\rho]$ to recover $a_0^{[-n+k+1]}, \ldots, a_{\tau-1}^{[-n+k+1]}$ since we will see that these $q$-power shifted error values are used to recover the erasure values, and then apply the further $q$-power shift of $[n - k - 1]$ to $a_0^{[-n+k+1]}, \ldots, a_{\tau-1}^{[-n+k+1]}$ to recover $a_0, \ldots, a_{\tau-1}$.

### F. Determining the erasure values via GRA: step 4d)-4f)

Rearranging Eq. (S98), notice for each $i = 0, \ldots, n-k-1$

$$\tilde{s}_i - \sum_{j=0}^{\tau-1} d_j a_j^{[i-n+k+1]} = \sum_{j=0}^{\rho-1} \hat{d}_j \hat{a}_j^{[i-n+k+1]}. \quad (S112)$$

Using the first $\rho$ of these equations and because $\tilde{s}_0, \ldots, \tilde{s}_{\rho-1}$, $d_0, \ldots, d_{\tau-1}$ and $a_0^{[-n+k+1]}, \ldots, a_{\tau-1}^{[-n+k+1]}$ are already computed, we can compute the modified reversed syndromes $\bar{\tilde{s}}_i \in \mathbb{F}_{q^n}$ for each $i = 0, \ldots, \rho-1$ as

$$\bar{\tilde{s}}_i = \tilde{s}_i - \sum_{j=0}^{\tau-1} d_j a_j^{[i-n+k+1]}. \quad (S113)$$

This gives the system of equations for each $i = 0, \ldots, \rho-1$

$$\bar{\tilde{s}}_i = \sum_{j=0}^{\rho-1} \hat{d}_j \hat{a}_j^{[i-n+k+1]}. \quad (S114)$$

Notice this system of equations can be solved for $\hat{a}_0^{[-n+k+1]}, \ldots, \hat{a}_{\rho-1}^{[-n+k+1]}$ by GRA with already computed inputs $\hat{d}_0, \ldots, \hat{d}_{\rho-1}$ and $\bar{\tilde{s}}_0, \ldots, \bar{\tilde{s}}_{\rho-1}$. Applying a $q$-power shift of $[n - k - 1]$ to the output of GRA, $\hat{a}_0^{[-n+k+1]}, \ldots, \hat{a}_{\rho-1}^{[-n+k+1]}$, will recover the erasure values $\hat{a}_0, \ldots, \hat{a}_{\rho-1}$.

### G. Resolving the error and erasure: step 5b), 5c)

The error $e$ is resolved in exactly the same way as the error decoder, using the error values $a_0, \ldots, a_{\tau-1}$ and the error locations $d_0, \ldots, d_{\tau-1}$. The same method for resolving the error can also be used to resolve the erasure $\hat{e}$, using the erasure values $\hat{a}_0, \ldots, \hat{a}_{\rho-1}$ and the erasure locations $\hat{d}_0, \ldots, \hat{d}_{\rho-1}$. However, it should be noted that because $\hat{B}_0, \ldots, \hat{B}_{\rho-1}$ is already known, it is not necessary to compute $\bar{B}_0, \ldots, \bar{B}_{\rho-1}$ from $\hat{d}_0, \ldots, \hat{d}_{\rho-1}$.

### H. Restoring Alice's message: step 6)

In a similar fashion to the error decoder, the estimated codeword $\hat{x}'$ can be constructed from the received message $y'$, the error $e$ and the erasure $\hat{e}$ by

$$\hat{x}' = y' - e - \hat{e}. \quad (S115)$$

Now $\hat{x}'$ is known, and we use the same method as the error decoder to restore Alice's message $\hat{u}$.

## V. Efficient Encoding and Decoding of Vertically Interleaved Gabidulin Codes

Based on the methods presented in Section III and IV, especially the Error Location method, we describe our implementation techniques of vertically interleaved Gabidulin codes over $\mathbb{F}_{q^n}$, iGab$[n, k]$, for the PUSNEC.

### A. Efficient encoding of $\text{iGab}[n,k]$

Hereafter we denote the input to $\text{iGab}[n,k]$ as $\boldsymbol{u}' = \begin{bmatrix} \boldsymbol{u} & \boldsymbol{r} \end{bmatrix}$, which is divided into $l$ components

$$\boldsymbol{u}' = \begin{bmatrix} \boldsymbol{u}'^{(0)} \\ \vdots \\ \boldsymbol{u}'^{(l-1)} \end{bmatrix}, \quad \boldsymbol{u}'^{(i)} \in \mathbb{F}_{q^n}^k. \tag{S116}$$

The encoding can be executed for each component in parallel independently. The pre-encoding for each component

$$\boldsymbol{f}^{(i)} = \boldsymbol{u}'^{(i)} \hat{G}_1^{-1}, \tag{S117}$$

includes the matrix inversion whose complexity usually costs $\mathcal{O}(k^3)$ over $\mathbb{F}_{q^n}$. The complexity can be reduced to $\mathcal{O}(k^2)$ operations over $\mathbb{F}_{q^n}$ by applying GRA for eliminating variables to

$$\boldsymbol{f}^{(i)} \hat{G}_1 = \boldsymbol{u}'^{(i)}, \tag{S118}$$

instead of performing the matrix inversion directly. Actually, since $\text{Gab}[n,k]$ is $q$-cyclic, namely the generator matrix is given by Eq. (S45), the above equation reads the system of equations represented by a $q$-power Vandermonde matrix in the following form

$$\sum_{v=0}^{k-1} \beta^{[k-1+v]} \left( f_v^{(i)} \right)^{[j]} = \left( u'^{(i)}_{k-1-j} \right)^{[j]}, \tag{S119}$$

for each $j = 0, \ldots, k-1$. This system can be solved for $\boldsymbol{f}^{(i)}$ via direct matrix multiplication by GRA.

### B. Efficient decoding of $\text{iGab}[n,k]$

At each Bob, the received word of length $n_0$, $\boldsymbol{y} \in \mathbb{F}_{q^n}^{n_0}$, is first decoded by the $\mathbb{F}_q$-RLNC decoder, and then converted into the output word $\tilde{\boldsymbol{y}} \in \mathbb{F}_{q^n}^l$. Then 0's of length $k_1$ are attached to $\tilde{\boldsymbol{y}}$, and the resulting word of length $n$, $\boldsymbol{y}' = \begin{bmatrix} \boldsymbol{0} & \tilde{\boldsymbol{y}} \end{bmatrix}$, is input into the $\text{iGab}[n,k]$ decoder. The $\text{iGab}[n,k]$ decoder first divides the input vertically into $l$ components $\begin{bmatrix} \boldsymbol{y}'^{(0)} \\ \vdots \\ \boldsymbol{y}'^{(l-1)} \end{bmatrix} \in \mathbb{F}_{q^n}^l$. Each component is the sum of the codeword, the error and erasure sequences

$$\boldsymbol{y}'^{(i)} = \boldsymbol{x}'^{(i)} + \boldsymbol{e}^{(i)} + \hat{\boldsymbol{e}}^{(i)}. \tag{S120}$$

Each component of the error sequence is decomposed as

$$\boldsymbol{e}^{(i)} = \boldsymbol{a}^{(i)} B, \tag{S121}$$

with the error values

$$\boldsymbol{a}^{(i)} = \begin{bmatrix} a_0^{(i)} & \cdots & a_{\tau-1}^{(i)} \end{bmatrix} \in \mathbb{F}_{q^n}^\tau, \tag{S122}$$

and the error locations

$$B = \begin{bmatrix} \boldsymbol{B}_0 \\ \vdots \\ \boldsymbol{B}_{\tau-1} \end{bmatrix} \in \mathbb{F}_q^{\tau \times n}. \tag{S123}$$

Note that $B$ is taken to be common to all components of vertical interleaving. Actually, we will see later how the ELP common to all the components can be determined. Similarly, each component of the erasure sequence is decomposed as

$$\hat{\boldsymbol{e}}^{(i)} = \hat{\boldsymbol{a}}^{(i)} \hat{B}, \tag{S124}$$

with the erasure values

$$\hat{\boldsymbol{a}}^{(i)} = \begin{bmatrix} \hat{a}_0^{(i)} & \cdots & \hat{a}_{\rho-1}^{(i)} \end{bmatrix} \in \mathbb{F}_{q^n}^\rho, \tag{S125}$$

and the erasure locations

$$\hat{B} = \begin{bmatrix} \hat{\boldsymbol{B}}_0 \\ \vdots \\ \hat{\boldsymbol{B}}_{\rho-1} \end{bmatrix} \in \mathbb{F}_q^{\rho \times n}. \tag{S126}$$

A naive decoding of $\text{iGab}[n,k]$ is to run the decoder of $\text{Gab}[n,k]$ over $\mathbb{F}_{q^n}$ (Section III and IV) in parallel or $l$ times independently (the independent component-wise decoder as shown in Supplementary Fig. 14a). However, an efficient decoding can be realized based on the Error Location method, by taking advantage of the fact that the components $\boldsymbol{y}'^{(0)}, \ldots, \boldsymbol{y}'^{(l-1)}$ typically share all or almost all of error and erasure locations. Actually, $l$ times computations of error and erasure locations in the independent component-wise decoder are redundant and unnecessary. Instead, once computations for determining the error and erasure locations are carried out for the 0th component $\boldsymbol{y}'^{(0)}$, the result can effectively be used for the remaining components recursively, thereby reducing the number of computations from $\boldsymbol{y}'^{(1)}$ onwards. Furthermore, once the ELP is derived, the rootspace and information on the error and erasure locations can be shared between all the components, and then syndrome-dependent computations for error values are performed component-wise. In this way, the overall number of computations in step 2) $\sim$ 4) can be reduced. This is the common-error-location aware decoder (Supplementary Fig. 14b).

The six steps of it are explained below (see also a flow diagram in Supplementary Fig. 15), focusing on error correction with regard to $\boldsymbol{e}^{(i)}$'s. Extension to error and erasure correction is straightforward (Section IV).

#### 1) Syndromes and reversed syndromes: step 1a), 1b)

This step is component-wise. Each component of the syndromes is given by

$$s_j^{(i)} = \sum_{v=0}^{\tau-1} a_v^{(i)} d_v^{[j]}, \quad j = 0, \ldots, n-k-1, \tag{S127}$$

with the transformed error locations to be present in all the components

$$d_v = \sum_{u=0}^{n-1} B_{u,v} h_u^{[k]}, \quad v = 0, \ldots, \tau-1. \tag{S128}$$

Each component of the reversed syndromes is given by

$$\begin{aligned} \tilde{s}_j^{(i)} &= \left( s_{n-k-1-j}^{(i)} \right)^{[j-n+k+1]} \\ &= \sum_{v=0}^{\tau-1} d_v \left( a_v^{(i)} \right)^{[j-n+k+1]}, \end{aligned} \tag{S129}$$

for $j = 0, \ldots, n-k-1$, which is referred to as the reversed syndrome equation.

#### 2) Determining the ELP over $\mathbb{F}_{q^n}^l$ by CPSLBMA: step 2)

Given the reversed syndromes, we have the key equation for the ELP for each component $i$ $(= 0, \ldots, l-1)$

$$\sum_{v=0}^{\tau_i} \lambda_v^{(i)} \left( \tilde{s}_{j-v}^{(i)} \right)^{[v]} = 0, \quad j = \tau_i, \ldots, n-k-1. \tag{S130}$$

We set $\lambda_0^{(i)} = 1$. In order to determine the ELP common to all the components, we solve the above key equations recursively from the 0th to the $(l-1)$th component as described below (Supplementary Fig. 16). This method is actually CPSLBMA.

For the 0th component, the key equation is solved by the LBMA just as in step 2) in Subsubsection III-C2, given

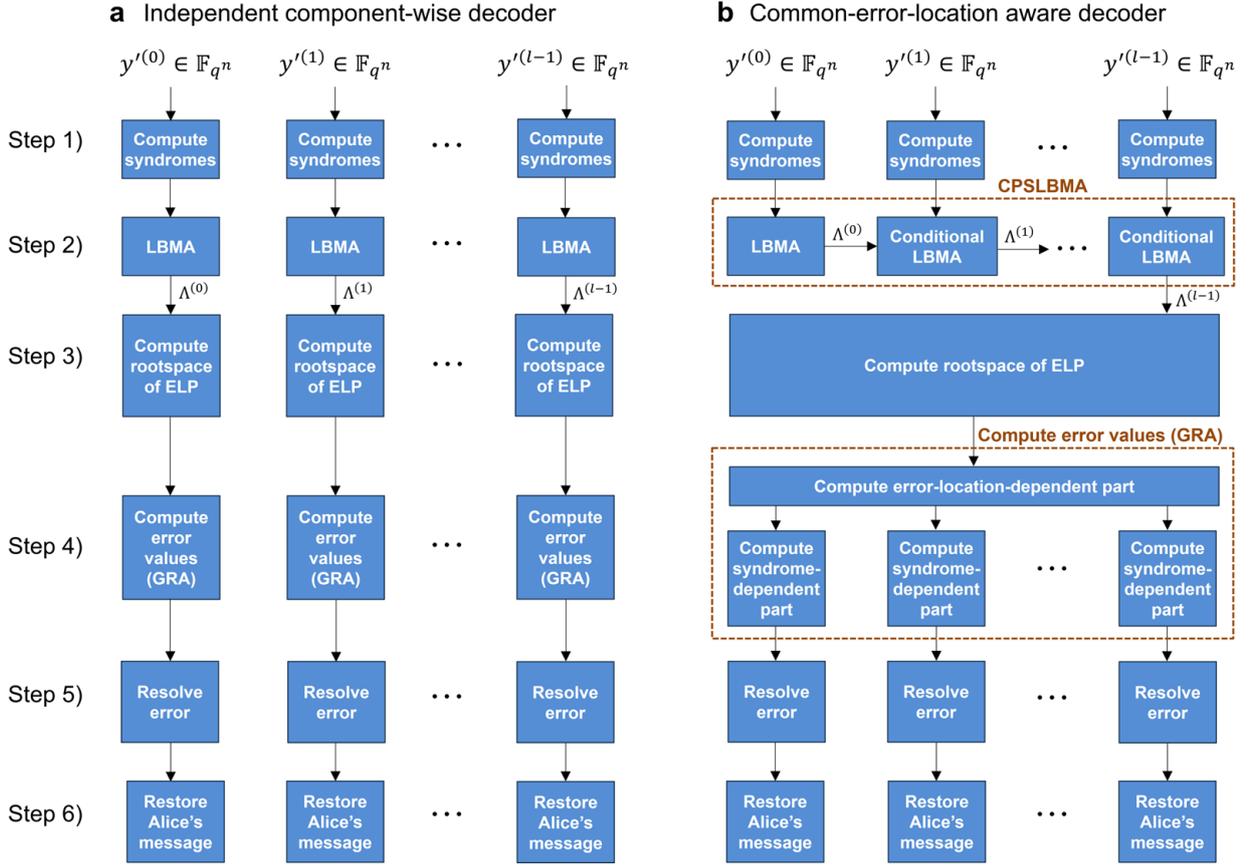

**a Independent component-wise decoder**

**b Common-error-location aware decoder**

**Supplementary Fig. 14. Flow diagrams of the interleaved Gabidulin decoder. a** The independent component-wise decoder. **b** The common-error-location aware decoder.

$\hat{s}_0^{(0)}, \ldots, \hat{s}_{n-k-1}^{(0)}$. The LBMA outputs the shortest $\mathbb{F}_{q^n}$-LFSR of length $\tau_0$, i.e., $\lambda_0^{(0)}, \ldots, \lambda_{\tau_0}^{(0)} \in \mathbb{F}_{q^n}$. The ELP thus obtained for the 0th component

$$\Lambda^{(0)}(x) = \sum_{v=0}^{\tau_0} \lambda_v^{(0)} x^{[v]} \tag{S131}$$

normally covers most of roots that we want to compute for the Cartesian product space $\mathbb{F}_{q^n}^l$.

For the 1st component, given $\hat{s}_0^{(1)}, \ldots, \hat{s}_{n-k-1}^{(1)}$, the modified syndromes are defined as

$$\hat{\hat{s}}_i^{(1)} = \sum_{j=0}^{\tau_0} \lambda_j^{(0)} \left( \hat{s}_{i+\tau_0-j}^{(1)} \right)^{[j]}, \tag{S132}$$

for $i = 0, 1, \ldots, n-k-1-\tau_0$. Notice that the number of syndromes reduces from $n-k$ to $n-k-\tau_0$. For the modified syndromes $\hat{\hat{s}}_0^{(1)}, \ldots, \hat{\hat{s}}_{n-k-1-\tau_0}^{(1)}$, the corresponding key equation becomes

$$\sum_{v=0}^{\tau_1'} \lambda_v'^{(1)} \left( \hat{\hat{s}}_{j-v}^{(1)} \right)^{[v]} = 0, \tag{S133}$$

for $j = \tau_1', \ldots, n-k-1-\tau_0$. This key equation is then solved by using the LBMA conditioned on the modified syndromes in the following way:

- If the modified syndromes are all-zeros (most of the case they are), then the results of the 0th component are output as they are, namely, $\Lambda^{(1)}(x) = \Lambda^{(0)}(x)$.
- If the modified syndromes are not all-zeros, then the LBMA is executed, and $\Lambda'^{(1)}(x)$ is retrieved. This polynomial is used to construct the ELP for the 1st

component by the operation of symbolic product with $\Lambda^{(0)}(x)$

$$\begin{aligned} \Lambda^{(1)}(x) &= \Lambda'^{(1)}(x) \otimes \Lambda^{(0)}(x) \\ &= \Lambda'^{(1)} \left( \Lambda^{(0)}(x) \right). \end{aligned} \tag{S134}$$

$\Lambda^{(1)}(x)$ has $\tau_1 = \tau_0 + \tau_1'$ roots in which $\tau_0$ roots are the roots of $\Lambda^{(0)}(x)$, namely the $\tau_0$ transformed error locations of the syndromes input to the 0th component. The remaining $\tau_1'$ roots are the $\tau_1'$ transformed error locations of the modified syndromes calculated from the syndromes input to the first component. See more detail about the description regarding the roots of polynomial synthesized by symbolic product in Subsection VI-H.

The result of length $\tau_1$ and the coefficients $\lambda_0^{(1)}, \ldots, \lambda_{\tau_1}^{(1)} \in \mathbb{F}_{q^n}$ of $\Lambda^{(1)}(x)$ are output.

The process is repeated recursively. Suppose that after $i$ recursions, $\Lambda^{(i-1)}(x)$ with coefficients $\lambda_0^{(i-1)}, \ldots, \lambda_{\tau_{i-1}}^{(i-1)}$ were output from the conditional LBMA for the $(i-1)$th component. Then the conditional LBMA is executed for the $i$th component with inputs $\hat{s}_0^{(i)}, \ldots, \hat{s}_{n-k-1}^{(i)}$, those are converted to the smaller number of syndromes $\hat{\hat{s}}_0^{(i)}, \ldots, \hat{\hat{s}}_{n-k-1-\tau_{i-1}}^{(i)}$ by using $\lambda_0^{(i-1)}, \ldots, \lambda_{\tau_{i-1}}^{(i-1)}$. The output from the conditional LBMA $\lambda_0^{(i)}, \ldots, \lambda_{\tau_i}^{(i)}$ is the shortest $\mathbb{F}_{q^n}$-LFSR for the reversed syndromes of all previous components from 0th to $i$th. This can be proven in a similar manner to the approach taken by Richter and Plass for efficient rank error and erasure correction by the LBMA with erasure location information (Section 4 of [27]).

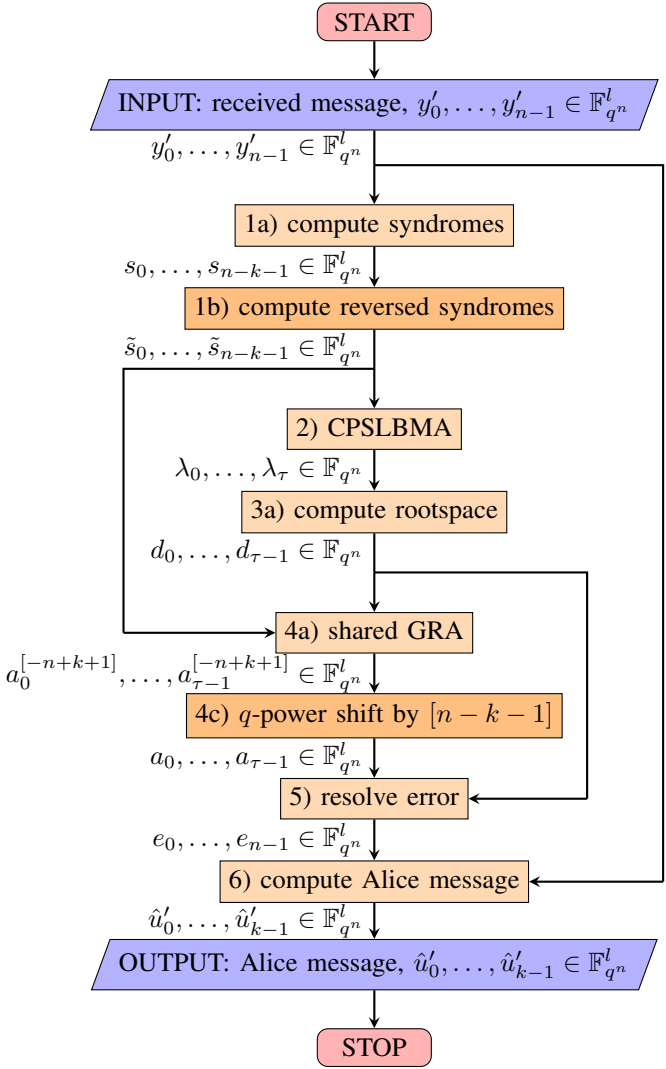

**Supplementary Fig. 15.** Flow diagram for efficient error decoding by the common-error-location aware decoder for iGab$[n, k]$.

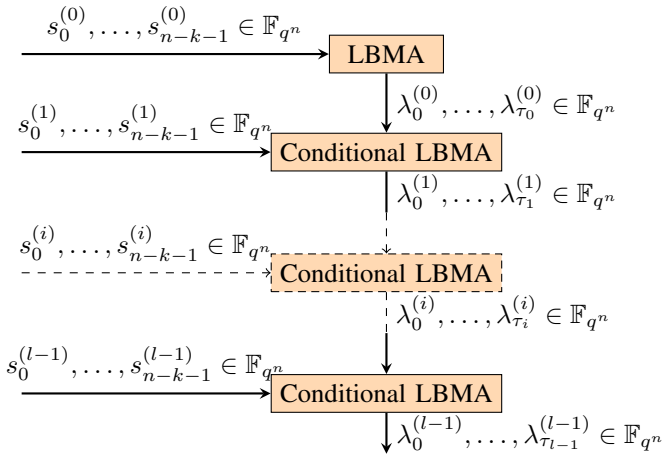

**Supplementary Fig. 16.** Flow diagram for the CPSLBMA.

After $l$ recursions, CPSLBMA will return the unique solution of ELP

$$\Lambda^{(l-1)}(x) = \sum_{v=0}^{\tau_{l-1}} \lambda_v^{(l-1)} x^{[v]} \tag{S135}$$

to the key equation Eq. (S130), whose rootspace is spanned by the error locations present in all of the $l$ components [5].

It should be noted that although this is guaranteed in the case $\tau \leq \lfloor \frac{(n-k)}{2} \rfloor$, there do exist certain error patterns where iGab$[n, k]$ is capable of correcting (probabilistically) errors up to $\lfloor \frac{l}{l+1}(n-k) \rfloor$ beyond the threshold [14], [32], and such $\lambda_0, \ldots, \lambda_\tau$ can still be determined using the algorithm above. This is not true for the corresponding decoder over the extension field $\mathbb{F}_{q^m}$.

*3) Determining a basis of rootspace of the ELP: step 3a)*

Since CPSLBMA has determined only one ELP, we are required to carry out the rootspace determination algorithm only once [6]. As seen in Subsection III-D, when $\tau \leq \lfloor \frac{(n-k)}{2} \rfloor$, the rootspace of the ELP is equal to the subspace spanned by the transformed error locations $d_0, \ldots, d_{\tau-1} \in \mathbb{F}_{q^n}$. In the same manner, we find a basis of the rootspace and set this basis equal to $d_0, \ldots, d_{\tau-1}$.

*4) Determining the error values by the shared GRA: step 4a)*

Once a common set of the transformed error locations $d_0, \ldots, d_{\tau-1}$ have been determined, their corresponding error values of each component $a_0^{(i)}, \ldots, a_{\tau-1}^{(i)} \in \mathbb{F}_{q^n}$ must be determined so as to satisfy the reversed syndrome equation, Eq. (S129). More specifically, they are related as follows:

$$\tilde{s}_j^{(i)} = \sum_{v=0}^{\tau-1} d_v \left( a_v^{(i)} \right)^{[j-n+k+1]}, \tag{S136}$$

for $j = 0, \ldots, \tau - 1$ (instead of $j = 0, \ldots, n-k-1$).

This can be done by using GRA (Subsection VI-G2) for each component separately. In each component's GRA, given $\tilde{s}_0^{(i)}, \ldots, \tilde{s}_{\tau-1}^{(i)}$ and $d_0, \ldots, d_{\tau-1}$ as input, we find $\left( a_0^{(i)} \right)^{[-n+k+1]}, \ldots, \left( a_{\tau-1}^{(i)} \right)^{[-n+k+1]}$ as the output, and then apply a $q$-power operation of order $[n-k-1]$ to each of them. The computational complexity is $\mathcal{O}(\tau^2)$ per each component.

However, notice that a significant proportion of the computation in each GRA depends only on the input $d_0, \ldots, d_{\tau-1}$ common to all the components. More precisely, we allow the error locations determined in step 3a) to be present in all the components, even if some of error locations may not be present in some of the components. Actually, if an error location is not present in a component, its error value for that component will be determined as zero. We may then split computations in GRA into two parts; the first part is the computations depending only on the error locations that are component-independent, and the second part is the computations depending on the syndromes that are component dependent. The first part can be carried out only once, and the result fed into the second part of the component-wise GRA. The method is referred to as the shared GRA.

Along with the notations in Subsection VI-G2, the shared GRA proceeds as follows.

In the first part, we set component-independent inputs as

$$A_j^{(0)} = d_j, \ j = 0, 1, \ldots, \tau - 1, \tag{S137}$$

and conduct the recursive computations. At the first iteration, we have

$$A_j^{(1)} = A_j^{(0)} - R^{(0)} A_j^{(0)[-1]}, \ j = 1, 2, \ldots, \tau - 1, \tag{S138}$$

---

[5] In contrast to CPSLBMA, the independent component-wise decoder determines $l$ ELPs (one ELP for each component), each ELP having rootspace spanned by the error locations present in its component only.

[6] However, the independent component-wise decoder will perform the rootspace determination algorithm $l$ times.

where

$$R^{(0)} = A_0^{(0)} \left( A_0^{(0)^{-1}} \right)^{[-1]}. \tag{S139}$$

Note that the number of elements reduces from $\tau$ to $\tau - 1$. At the $(v+1)$th iteration during the recursive computations, we have

$$A_j^{(v+1)} = A_j^{(v)} - R^{(v)} A_j^{(v)^{[-1]}},$$
$$j = v+1, v+2, \ldots, \tau - 1, \tag{S140}$$
$$R^{(v)} = A_v^{(v)} \left( A_v^{(v)^{-1}} \right)^{[-1]}. \tag{S141}$$

Finally, at the $(\tau - 1)$th iteration, we have

$$A_{\tau-1}^{(\tau-1)} = A_{\tau-1}^{(\tau-2)} - R^{(\tau-2)} A_{\tau-1}^{(\tau-2)^{[-1]}}, \tag{S142}$$
$$R^{(\tau-2)} = A_{\tau-2}^{(\tau-2)} \left( A_{\tau-2}^{(\tau-2)^{-1}} \right)^{[-1]}. \tag{S143}$$

The results $\{A_j^{(v+1)} | v = 0, 1, \ldots, \tau-2; j = v+1, v+2, \ldots, \tau-1\}$ are shared by all the components.

In the second part, we set component-dependent inputs for $i = 0, \ldots, l-1$ as

$$Q_j^{(i,0)} = \tilde{s}_j^{(i)}, \ j = 0, 1, \ldots, \tau - 1, \tag{S144}$$

and conduct the recursive computations. At the first iteration, we have

$$Q_j^{(i,1)} = Q_j^{(i,0)} - R^{(0)} \, Q_{j+1}^{(i,0)^{[-1]}}, \ j = 0, 1, \ldots, \tau - 2. \tag{S145}$$

Note that the number of elements reduces from $\tau$ to $\tau - 1$. At the $(v+1)$th iteration during the recursive computations, we have

$$Q_j^{(i,v+1)} = Q_j^{(i,v)} - R^{(v)} Q_{j+1}^{(i,v)^{[-1]}},$$
$$j = 0, 1, \ldots, \tau - 1 - (v+1), \tag{S146}$$

and arrive at the system of $\tau - v - 1$ equations

$$Q_j^{(i,v+1)} = \sum_{u=v+1}^{\tau-1} A_u^{(v+1)} x_u^{[j]},$$
$$j = 0, 1, \ldots, \tau - 1 - (v+1). \tag{S147}$$

Finally, at the $(\tau - 1)$th iteration, we have

$$Q_0^{(i,\tau-1)} = Q_0^{(i,\tau-2)} - R^{(\tau-2)} \, Q_1^{(i,\tau-2)^{[-1]}}, \tag{S148}$$

and arrive at an equation

$$Q_0^{(i,\tau-1)} = A_{\tau-1}^{(\tau-1)} x_{\tau-1}^{[0]}. \tag{S149}$$

The variable $x_{\tau-1}^{[0]}$ can then be solved, providing the error value

$$\left( a_{\tau-1}^{(i)} \right)^{[-n+k+1]} = A_{\tau-1}^{(\tau-1)^{-1}} Q_0^{(i,\tau-1)}. \tag{S150}$$

We then have recursively

$$\left( a_j^{(i)} \right)^{[-n+k+1]} = A_j^{(j)^{-1}} \left( Q_0^{(i,j)} - \sum_{v=j+1}^{\tau-1} A_v^{(j)} \left( a_v^{(i)} \right)^{[-n+k+1]} \right), \tag{S151}$$

for each $j = \tau - 2, \ldots, 0$. By applying a $q$-power shift by $[n-k-1]$, we get $a_0^{(i)}, \ldots, a_{\tau-1}^{(i)}$, and finally obtain the error values of all the components $a_0, \ldots, a_{\tau-1} \in \mathbb{F}_{q^n}^l$.

### 5) Resolving the error: step 5)

The error locations $\boldsymbol{B}_0, \ldots, \boldsymbol{B}_{\tau-1}$ can be determined from $d_0, \ldots, d_{\tau-1}$ in the same manner as in Subsection III-G. The error sequence $\boldsymbol{e}$ can be resolved by Eq. (S121).

### 6) Restoring Alice's message: step 6)

The estimated codeword is given as $\hat{\boldsymbol{x}}' = \boldsymbol{y}' - \boldsymbol{e}$. In the same way as in Subsection III-H, we have the estimated coefficient vector $\hat{\boldsymbol{f}} = \hat{\boldsymbol{x}}' \, \hat{H}'^T$. The estimated Alice's message is finally restored as $\hat{\boldsymbol{u}}' = \hat{\boldsymbol{f}} \hat{G}_1$.

## VI. BASIC NOTIONS AND TOOLS TO DEAL WITH GABIDULIN CODES

Let $\mathbb{Z}$ be the set of integers. $\mathbb{Z}_{\geq 0}$ denotes the set of non-negative integers

$$\mathbb{Z}_{\geq 0} \stackrel{\text{def}}{=} \{ z \in \mathbb{Z} : z \geq 0 \}. \tag{S152}$$

$\mathbb{N}$ denotes the set of natural numbers

$$\mathbb{N} \stackrel{\text{def}}{=} \{ z \in \mathbb{Z} : z > 0 \}. \tag{S153}$$

### A. $q$-power operator

For any $i \in \mathbb{Z}$, the $q$-power (or Frobenius power) operator, denoted by $^{[i]}$, is defined such that for any $b \in \mathbb{F}_{q^n}$

$$b^{[i]} = b^{q^i} \tag{S154}$$

In particular, because $b \in \mathbb{F}_{q^n}$, $b^{q^n-1} = 1$, so it is the case that

$$\begin{aligned} b^{[n]} &= b^{q^n} \\ &= b^{q^n - 1} b \\ &= 1 \cdot b \\ &= b. \end{aligned} \tag{S155}$$

Therefore,

$$b^{[i]} = b^{[i \bmod n]}. \tag{S156}$$

In what follows, we shall see that the cyclic property of the $q$-power operator is of significant importance for Gabidulin codes. The $q$-power operator is also linear. To see this, we first need to prove an important fact relating to number theory.

**Lemma 5.** Let $p \in \mathbb{N}$ be prime. Then for any $\omega \in \mathbb{N}$ and $i = 1, \ldots, p^\omega - 1$, $p$ divides $\begin{pmatrix} p^\omega \\ i \end{pmatrix}$.

*Proof.* We begin by observing

$$\begin{pmatrix} p^\omega \\ i \end{pmatrix} = \frac{p^\omega}{i} \begin{pmatrix} p^\omega - 1 \\ i - 1 \end{pmatrix}. \tag{S157}$$

So multiplying through by $i$, we have

$$i \begin{pmatrix} p^\omega \\ i \end{pmatrix} = p^\omega \begin{pmatrix} p^\omega - 1 \\ i - 1 \end{pmatrix}. \tag{S158}$$

By the fundamental theorem of arithmetic, $i$ has a unique prime power representation, namely

$$i = rp^\varsigma, \tag{S159}$$

where $r \in \mathbb{N}$ is the product of all primes in its prime power representation apart from $p$ and $\varsigma \in \mathbb{Z}_{\geq 0}$ is the order of $p$. Note that $0 \leq \varsigma < \omega$ because $i < p^\omega$ and $\text{hcf}\{r, p\} = 1$, because $p$ is not a prime factor of $r$. Substituting our expression for $i$ into Eq. (S158), we get

$$rq^\varsigma \begin{pmatrix} p^\omega \\ i \end{pmatrix} = p^\omega \begin{pmatrix} p^\omega - 1 \\ i - 1 \end{pmatrix}. \tag{S160}$$

Dividing through by $q^\varsigma$ we get

$$r \begin{pmatrix} p^\omega \\ i \end{pmatrix} = pp^{\omega-\varsigma-1} \begin{pmatrix} p^\omega - 1 \\ i - 1 \end{pmatrix}. \quad (S161)$$

Notice that because $\varsigma < \omega$, $\omega - \varsigma - 1 \geq 0$, so $p^{\omega-\varsigma-1} \in \mathbb{N}$. Eq. (S161) shows that either $p$ must divide $r$ or $p$ must divide $\begin{pmatrix} p^\omega \\ i \end{pmatrix}$, but since $\mathrm{hcf}\{r,p\} = 1$, $p$ does not divide $r$, so the only option is that $p$ divides $\begin{pmatrix} p^\omega \\ i \end{pmatrix}$ as required. $\qquad \square$

Establishing Lemma 5 enables us to prove the following lemma closely related to the linearity of the $q$-power operator.

**Lemma 6.** *Let* $b_0, b_1 \in \mathbb{F}_{q^n}$ *and* $A_0, A_1 \in \mathbb{F}_q$, *then*

$$(A_0 b_0 + A_1 b_1)^q = A_0 b_0^q + A_1 b_1^q. \quad (S162)$$

*Proof.* Firstly, notice that because $A_0 \in \mathbb{F}_q$, we have

$$A_0^{q-1} = 1. \quad (S163)$$

So

$$\begin{aligned} (A_0 b_0)^q &= A_0^q b_0^q \\ &= A_0^{q-1} A_0 b_0^q \\ &= 1 \cdot A_0 b_0^q \\ &= A_0 b_0^q. \end{aligned} \quad (S164)$$

Similarly, we also have

$$(A_1 b_1)^q = A_1 b_1^q. \quad (S165)$$

Let $b_0' = A_0 b_0$ and $b_1' = A_1 b_1$. Notice that $b_0', b_1' \in \mathbb{F}_{q^n}$, so if we can show for any $b_0', b_1' \in \mathbb{F}_{q^n}$

$$(b_0' + b_1')^q = b_0'^q + b_1'^q, \quad (S166)$$

then substituting in $b_0' = A_0 b_0$ and $b_1' = A_1 b_1$ and using Eqs. (S164) and (S165), we get the desired result. To show Eq. (S166), notice that

$$(b_0' + b_1')^q = \sum_{i=0}^{q} \begin{pmatrix} q \\ i \end{pmatrix} \cdot b_0'^i b_1'^{q-i}, \quad (S167)$$

where $\begin{pmatrix} q \\ i \end{pmatrix} \cdot b_0'^i b_1'^{q-i}$ means to add $b_0'^i b_1'^{q-i}$, $\begin{pmatrix} q \\ i \end{pmatrix}$ times. But notice that because $q = p^\omega$ for some $w \in \mathbb{N}$ and $p$ prime, $p$ divides $\begin{pmatrix} q \\ i \end{pmatrix}$ for any $i = 1, \ldots, q-1$ due to Lemma 5. Therefore, addition in $\mathbb{F}_{q^n}$ yields

$$\begin{pmatrix} q \\ i \end{pmatrix} \cdot b_0'^i b_1'^{q-i} = 0. \quad (S168)$$

We therefore have

$$\begin{aligned} (b_0' + b_1')^q &= \sum_{i=0}^{q} \begin{pmatrix} q \\ i \end{pmatrix} \cdot b_0'^i b_1'^{q-i} \\ &= 1 \cdot b_0'^0 b_1'^q + \sum_{i=1}^{q-1} 0 \cdot b_0'^i b_1'^{q-i} + 1 \cdot b_0'^q b_1'^0 \\ &= b_0'^q + b_1'^q. \end{aligned} \quad (S169)$$

$\qquad \square$

The linearity of the $q$-power operator now follows as a corollary of Lemma 6.

**Corollary 7.** *Let* $b_0, b_1 \in \mathbb{F}_{q^n}$ *and* $A_0, A_1 \in \mathbb{F}_q$, *then for any* $i \in \mathbb{Z}$,

$$(A_0 b_0 + A_1 b_1)^{[i]} = A_0 b_0^{[i]} + A_1 b_1^{[i]}. \quad (S170)$$

*Proof.* Apply Lemma 6 to $(A_0 b_0 + A_1 b_1)$ $i$-times to get the required result. $\qquad \square$

## B. Bases of $\mathbb{F}_{q^n}$ over $\mathbb{F}_q$

We shall observe that finding a basis of $\mathbb{F}_{q^n}$ over the ground field $\mathbb{F}_q$ allows us to represent elements of $\mathbb{F}_{q^n}$ as an $n$-dimensional vector over $\mathbb{F}_q$. Later we shall observe that a vector representation of the extension field over the subfield provides useful methods for arithmetic operations and gives rise to a notion of distance between vectors over $\mathbb{F}_{q^n}$, namely rank distance.

**Definition 8.** *For* $m \in \mathbb{N}$, $b_0, \ldots, b_{m-1} \in \mathbb{F}_{q^n}$ *are linearly independent over* $\mathbb{F}_q$ *if for any* $A_0, \ldots A_{m-1} \in \mathbb{F}_q$

$$\sum_{j=0}^{m-1} A_j b_j = 0 \Leftrightarrow A_j = 0 \text{ for all } j = 0, \ldots, m-1. \quad (S171)$$

If a collection of $n$ elements of $\mathbb{F}_{q^n}$ is linearly independent over $\mathbb{F}_{q^n}$, then that collection forms a basis of $\mathbb{F}_{q^n}$ over $\mathbb{F}_q$. As a result, if $h_0, \ldots, h_{n-1}$ is a basis of $\mathbb{F}_{q^n}$ over $\mathbb{F}_q$, we can express any element $b \in \mathbb{F}_{q^n}$ as a $\mathbb{F}_q$-linear combination of $h_0, \ldots, h_{n-1}$. That is to say, for any $b \in \mathbb{F}_{q^n}$, there exists $B_0, \ldots, B_{n-1} \in \mathbb{F}_q$ such that

$$b = \sum_{i=0}^{n-1} B_i h_i. \quad (S172)$$

$\boldsymbol{B} = \begin{bmatrix} B_0 & \ldots & B_{n-1} \end{bmatrix}^T \in \mathbb{F}_q^n$, an $n$-dimension vector over $\mathbb{F}_q$, is referred to as the $\{h_0, \ldots, h_{n-1}\}$-basis representation of $b$.

**Lemma 9.** *For any* $m \in \mathbb{N}$, *let* $b_0, \ldots, b_{m-1} \in \mathbb{F}_{q^n}$ *and* $\boldsymbol{B}_0, \ldots, \boldsymbol{B}_{m-1} \in \mathbb{F}_q^n$ *such that for each* $j = 0, \ldots, m-1$, $\boldsymbol{B}_j$ *is the* $\{h_0, \ldots, h_{n-1}\}$-*basis representation of* $b_j$. *Then* $b_0, \ldots, b_{m-1}$ *are linearly independent over* $\mathbb{F}_q$ *if and only if* $\boldsymbol{B}_0, \ldots, \boldsymbol{B}_{m-1}$ *are linearly independent over* $\mathbb{F}_q$.

*Proof.* To show $b_0, \ldots, b_{m-1}$ linearly independent over $\mathbb{F}_q$ $\Rightarrow \boldsymbol{B}_0, \ldots, \boldsymbol{B}_{m-1}$ linearly independent over $\mathbb{F}_q$, we need to show that for any $A_0, \ldots, A_{m-1} \in \mathbb{F}_q$

$$\sum_{j=0}^{m-1} A_j B_{i,j} = 0 \text{ for all } i = 0, \ldots, n-1$$
$$\Leftrightarrow A_j = 0 \text{ for all } j = 0, \ldots, m-1,$$

under the assumption that $b_0, \ldots, b_{m-1}$ are linearly independent over $\mathbb{F}_q$. $\Leftarrow$ clearly holds, so we need to show $\Rightarrow$. Suppose for all $i = 0, \ldots, n-1$, we have

$$\sum_{j=0}^{m-1} A_j B_{i,j} = 0. \quad (S173)$$

Then multiplying the $i$th equation by $h_i$ and summing over $i = 0, \ldots, n-1$, we have

$$\begin{aligned} 0 &= \sum_{i=0}^{n-1} \left( \sum_{j=0}^{m-1} A_j B_{i,j} \right) h_i \\ &= \sum_{j=0}^{m-1} A_j \left( \sum_{i=0}^{n-1} B_{i,j} h_i \right) \\ &= \sum_{j=0}^{m-1} A_j b_j. \end{aligned} \quad (S174)$$

Under the assumption that $b_0, \ldots, b_{m-1}$ are linearly independent over $\mathbb{F}_q$, this means $A_j = 0$ for all $j = 0, \ldots, m-1$. To show $\boldsymbol{B}_0, \ldots, \boldsymbol{B}_{m-1}$ linearly independent over $\mathbb{F}_q$ $\Rightarrow$ $b_0, \ldots, b_{m-1}$ linearly independent over $\mathbb{F}_q$, we need to show that for any $A_0, \ldots, A_{m-1} \in \mathbb{F}_q$

$$\sum_{j=0}^{m-1} A_j b_j = 0 \Leftrightarrow A_j = 0 \text{ for all } j = 0, \ldots, m-1, \quad \text{(S175)}$$

under the assumption $\boldsymbol{B}_0, \ldots, \boldsymbol{B}_{m-1}$ are linearly independent over $\mathbb{F}_q$. $\Leftarrow$ clearly holds, so we need to show $\Rightarrow$. Suppose we have

$$\sum_{j=0}^{m-1} A_j b_j = 0. \quad \text{(S176)}$$

Then expressing each $b_j$ in its $\{h_0, \ldots, h_{n-1}\}$-basis representation, we have

$$
\begin{aligned}
0 &= \sum_{j=0}^{m-1} A_j b_j \\
&= \sum_{j=0}^{m-1} A_j \left( \sum_{i=0}^{n-1} B_{i,j} h_i \right) \\
&= \sum_{i=0}^{n-1} \left( \sum_{j=0}^{m-1} A_j B_{i,j} \right) h_i. \quad \text{(S177)}
\end{aligned}
$$

But because $h_0, \ldots, h_{n-1}$ is a basis of $\mathbb{F}_{q^n}$ over $\mathbb{F}_q$, $h_0, \ldots, h_{n-1}$ are linearly independent over $\mathbb{F}_q$ and so the only way this equation can hold is if $\sum_{j=0}^{m-1} A_j B_{i,j} = 0$ for all $i = 0, \ldots, n-1$. But since $\boldsymbol{B}_0, \ldots, \boldsymbol{B}_{m-1}$ are linearly independent over $\mathbb{F}_q$, if $\sum_{j=0}^{m-1} A_j B_{i,j} = 0$ for all $i = 0, \ldots, n-1$, then $A_j = 0$ for all $j = 0, \ldots, m-1$. $\square$

Notice Lemma 9 is equivalent to the following. Let the matrix $B \in \mathbb{F}_q^{n \times m}$ be formed by the concatenation of $\boldsymbol{B}_0, \ldots, \boldsymbol{B}_{m-1}$, namely

$$B = \begin{bmatrix} \boldsymbol{B}_0 & \ldots & \boldsymbol{B}_{m-1} \end{bmatrix}, \quad \text{(S178)}$$

then $b_0, \ldots, b_{m-1}$ are linearly independent over $\mathbb{F}_q$ if and only if $\mathbf{B}$ is full rank.

### C. Normal Bases of $\mathbb{F}_{q^n}$ over $\mathbb{F}_q$

Let $\beta \in \mathbb{F}_{q^n}$ such that $\beta^{[0]}, \ldots, \beta^{[n-1]} \in \mathbb{F}_{q^n}$ are linearly independent over $\mathbb{F}_q$. We say $\beta$ generates a normal basis of $\mathbb{F}_{q^n}$ over $\mathbb{F}_q$ and we typically refer to $\beta^{[0]}, \ldots, \beta^{[n-1]}$ as the $\beta$-normal basis of $\mathbb{F}_{q^n}$ over $\mathbb{F}_q$. Given a normal basis generated by some $\beta \in \mathbb{F}_{q^n}$, since $\beta^{[0]}, \ldots, \beta^{[n-1]}$ form a basis of $\mathbb{F}_{q^n}$ over $\mathbb{F}_q$, any $b \in \mathbb{F}_{q^n}$ has a $\beta$-normal basis representation $\begin{bmatrix} B_0 & \ldots & B_{n-1} \end{bmatrix}^T \in \mathbb{F}_q^n$ such that

$$b = \sum_{i=0}^{n-1} B_i \beta^{[i]}. \quad \text{(S179)}$$

In addition to being able to represent the element $b \in \mathbb{F}_{q^n}$ of the extension field as an $n$-dimensional vector of the subfield $\mathbb{F}_q$, the $\beta$-normal basis representation allows for efficient arithmetic computation especially when $q^n$ is large. Details of the necessary normal basis arithmetic required for Gabidulin encoding and decoding are presented next.

#### 1) q-power in the normal basis
Notice that for any $i = 0, \ldots, n-1$,

$$
\begin{aligned}
b^{[i]} &= \left( \sum_{j=0}^{n-1} B_j \beta^{[j]} \right)^{[i]} \\
&= \sum_{j=0}^{n-1} B_j \beta^{[j+i]} \\
&= \sum_{j=0}^{n-1} B_{(j-i) \bmod n} \beta^{[j]}. \quad \text{(S180)}
\end{aligned}
$$

So the $\beta$-normal basis representation of $b^{[i]}$ is a cyclic shift forward by $i$ positions of the $\beta$-normal basis representation of $b$.

#### 2) Addition in the normal basis
Notice for any $b_0, b_1 \in \mathbb{F}_{q^n}$ with $\beta$-normal basis representations $\begin{bmatrix} B_{0,0} & \ldots & B_{n-1,0} \end{bmatrix}^T \in \mathbb{F}_q^n$ and $\begin{bmatrix} B_{0,1} & \ldots & B_{n-1,1} \end{bmatrix}^T \in \mathbb{F}_q^n$, respectively

$$
\begin{aligned}
b_0 + b_1 &= \sum_{i=0}^{n-1} B_{i,0} \beta^{[i]} + \sum_{i=0}^{n-1} B_{i,1} \beta^{[i]} \\
&= \sum_{i=0}^{n-1} \left( B_{i,0} + B_{i,1} \right) \beta^{[i]}. \quad \text{(S181)}
\end{aligned}
$$

So the $\beta$-normal basis representation of $b_0 + b_1$ is the component-wise subfield addition of the $\beta$-normal basis representations of $z_0$ and $z_1$, namely $\begin{bmatrix} (B_{0,0} + B_{0,1}) & \ldots & (B_{n-1,0} + B_{n-1,1}) \end{bmatrix}^T \in \mathbb{F}_q^n$ and requires $n$ $\mathbb{F}_q$-additions.

#### 3) Multiplication in the normal basis
Notice for any $b_0, b_1 \in \mathbb{F}_{q^n}$ with $\beta$-normal basis representations $\begin{bmatrix} B_{0,0} & \ldots & B_{n-1,0} \end{bmatrix}^T \in \mathbb{F}_q^n$ and $\begin{bmatrix} B_{0,1} & \ldots & B_{n-1,1} \end{bmatrix}^T \in \mathbb{F}_q^n$, respectively

$$
\begin{aligned}
b_0 b_1 &= \left( \sum_{i=0}^{n-1} B_{i,0} \beta^{[i]} \right) \left( \sum_{j=0}^{n-1} B_{j,1} \beta^{[j]} \right) \\
&= \sum_{i=0}^{n-1} \sum_{j=0}^{n-1} B_{i,0} B_{j,1} \beta^{[i]} \beta^{[j]}. \quad \text{(S182)}
\end{aligned}
$$

For each $i, j = 0, \ldots, n-1$, we define $t_{i,j} \in \mathbb{F}_q^n$ such that

$$\beta^{[i]} \beta^{[j]} = \sum_{v=0}^{n-1} t_{i,j}^{(v)} \beta^{[v]}, \quad \text{(S183)}$$

where $t_{i,j}^{(v)} \in \mathbb{F}_q$ is the $v$th component of $t_{i,j}$. Then

$$
\begin{aligned}
b_0 b_1 &= \sum_{i=0}^{n-1} \sum_{j=0}^{n-1} B_{i,0} B_{j,1} \beta^{[i]} \beta^{[j]} \\
&= \sum_{i=0}^{n-1} \sum_{j=0}^{n-1} B_{i,0} B_{j,1} \left( \sum_{v=0}^{n-1} t_{i,j}^{(v)} \beta^{[v]} \right) \\
&= \sum_{v=0}^{n-1} \left( \sum_{i=0}^{n-1} \sum_{j=0}^{n-1} B_{i,0} B_{j,1} t_{i,j}^{(v)} \right) \beta^{[v]}. \quad \text{(S184)}
\end{aligned}
$$

But notice the following property of $t_{i,j}$. For any $u \in \mathbb{Z}$, we have

$$
\begin{aligned}
\sum_{v=0}^{n-1} t_{i,j}^{(v)} \beta^{[v]} &= \beta^{[i]} \beta^{[j]} \\
&= \left( \beta^{[i-u]} \beta^{[j-u]} \right)^{[u]} \\
&= \left( \beta^{[(i-u) \bmod n]} \beta^{[(j-u) \bmod n]} \right)^{[u]} \\
&= \left( \sum_{v=0}^{n-1} t_{(i-u) \bmod n, (j-u) \bmod n}^{(v)} \beta^{[v]} \right)^{[u]} \\
&= \sum_{v=0}^{n-1} t_{(i-u) \bmod n, (j-u) \bmod n}^{(v)} \beta^{[v+u]} \\
&= \sum_{v=u}^{n-1+u} t_{(i-u) \bmod n, (j-u) \bmod n}^{(v-u)} \beta^{[v]} \\
&= \sum_{v=0}^{n-1} t_{(i-u) \bmod n, (j-u) \bmod n}^{((v-u) \bmod n)} \beta^{[v]}. \quad \text{(S185)}
\end{aligned}
$$

Equating coefficients of $\beta^{[v]}$, because $\beta^{[0]}, \ldots, \beta^{[n-1]}$ are linearly independent over $\mathbb{F}_q$, we must have that for all $v = 0, \ldots, n-1$

$$
t_{i,j}^{(v)} = t_{(i-u) \bmod n, (j-u) \bmod n}^{((v-u) \bmod n)}. \quad \text{(S186)}
$$

And so setting $u = v$, we get that for all $v = 0, \ldots, n-1$,

$$
t_{i,j}^{(v)} = t_{(i-v) \bmod n, (j-v) \bmod n}^{(0)}. \quad \text{(S187)}
$$

So substituting the expression for $t_{i,j}^{(v)}$ from Eq. (S187) into Eq. (S184), we arrive at

$$
b_0 b_1 = \sum_{v=0}^{n-1} \left( \sum_{i=0}^{n-1} \sum_{j=0}^{n-1} B_{i,0} B_{j,1} t_{(i-v) \bmod n, (j-v) \bmod n}^{(0)} \right) \beta^{[v]}. \quad \text{(S188)}
$$

So the $v$th component, $v = 0, \ldots, n-1$, of the $\beta$-normal basis representation of $b_0 b_1$ is given by

$$
\sum_{i=0}^{n-1} \sum_{j=0}^{n-1} B_{i,0} B_{j,1} t_{(i-v) \bmod n, (j-v) \bmod n}^{(0)} \quad \text{(S189)}
$$

and requires $n \left( C(T) - 1 \right)$ $\mathbb{F}_q$-additions and $2nC(T)$ $\mathbb{F}_q$-multiplications, where $C(T)$ is the number of non-zero entries in the matrix $T^{(0)} \in \mathbb{F}_q^{n \times n}$ such that the $(i,j)$th entry of $T^{(0)}$, denoted $T_{i,j}^{(0)}$ for $i,j = 0, \ldots, n-1$, is given by

$$
T_{i,j}^{(0)} = t_{i,j}^{(0)}. \quad \text{(S190)}
$$

### 4) Fast multiplication in the normal basis

Fast multiplication is the name given to the multiplication of $b \in \mathbb{F}_{q^n}$ by a normal basis element $\beta^{[j]}$. Many multiplications involved in the encoding and decoding of Gabidulin codes are of this form, so it is useful to describe an arithmetic of lower complexity than general normal basis multiplication. Let $\begin{bmatrix} B_0 & \ldots & B_{n-1} \end{bmatrix}^T \in \mathbb{F}_q^n$ be the $\beta$-normal basis representation of $b$, namely

$$
b = \sum_{i=0}^{n-1} B_i \beta^{[i]}. \quad \text{(S191)}
$$

Then

$$
\begin{aligned}
b \beta^{[j]} &= \left( \sum_{i=0}^{n-1} B_i \beta^{[i]} \right) \beta^{[j]} \\
&= \sum_{i=0}^{n-1} B_i \beta^{[i]} \beta^{[j]} \\
&= \sum_{v=0}^{n-1} \left( \sum_{i=0}^{n-1} B_i t_{i,j}^{(v)} \right) \beta^{[v]} \\
&= \sum_{v=0}^{n-1} \left( \sum_{i=0}^{n-1} B_i t_{(i-v) \bmod n, (j-v) \bmod n}^{(0)} \right) \beta^{[v]}. \quad \text{(S192)}
\end{aligned}
$$

So the $v$th component, $v = 0, \ldots, n-1$, of the $\beta$-normal basis representation of $b \beta^{[j]}$ is given by

$$
\sum_{i=0}^{n-1} B_i t_{(i-v) \bmod n, (j-v) \bmod n}^{(0)}, \quad \text{(S193)}
$$

and requires $\left( C(T) - n \right)$ $\mathbb{F}_{q^n}$-additions and $C(T)$ $\mathbb{F}_{q^n}$-multiplications. Note that in the case where $q = 2^w$ and the normal basis has a Gaussian period construction, $T^{(0)} \in \mathbb{F}_2^{n \times n}$, so $\mathbb{F}_{q^n}$-multiplication is not needed.

### 5) Inversion in the normal basis

Inversion of non-zero elements of $x \in \mathbb{F}_{q^n}$ is represented as follows:

$$
x^{-1} = x^{q^n - 2}. \quad \text{(S194)}
$$

However, directly computing the inverse of $x$ by multiplying $x$ $(q^n - 2)$ times is inefficient. This subsubsection presents Itoh's method for efficiently computing the inversion using $q$-power operations, which are merely cyclic shifts, and fewer $\mathbb{F}_{q^n}$-multiplications.

Notice

$$
q^n - 2 = \frac{(q^n - 1)}{q - 1} (q - 2) + \frac{(q^{n-1} - 1)}{q - 1} q. \quad \text{(S195)}
$$

So

$$
\begin{aligned}
x^{-1} &= x^{q^n - 2} \\
&= y^{q-2} z^q, \quad \text{(S196)}
\end{aligned}
$$

where

$$
y \equiv x^{\frac{q^n - 1}{q - 1}}, \quad \text{(S197)}
$$

and

$$
z \equiv x^{\frac{q^{n-1} - 1}{q - 1}}. \quad \text{(S198)}
$$

Notice

$$
y = x z^q, \quad \text{(S199)}
$$

so $y$ can be found by finding $z$ from $x$, applying a $q$-power shift to $z$ to find $z^q$ and then multiplying $z^q$ by $x$ to obtain $y$. Then notice

$$
\begin{aligned}
y &= x^{\frac{q^n - 1}{q - 1}} \\
&= \left( x^{q^n - 1} \right)^{\frac{1}{q - 1}} \\
&= 1^{\frac{1}{q - 1}}. \quad \text{(S200)}
\end{aligned}
$$

So $y$ is a $(q-1)$th root of unity. This means that $y$ must be in the subfield. Because $y \in \mathbb{F}_q$, we have $y^{q-1} = 1$, meaning

$$
y^{q-2} = y^{-1}. \quad \text{(S201)}
$$

Once we have found $y$, we can use $\mathbb{F}_q$-inversion (either a lookup table if $q$ is small, or otherwise) to find $y^{q-2}$

from $y$ and then $x^{-1}$ can be computed by multiplying $z^q$ by $y^{q-2}$. So, once $z$ has been computed from $x$, computing $x^{-1}$ requires a further $q$-power shift, 2 $\mathbb{F}_{q^n}$-multiplications and a $\mathbb{F}_q$-inversion.

To compute $z$ from $x$, notice that there exits unique $k_1, \ldots, k_t \in \mathbb{Z}$ with $t = \text{Ham}(n-1)$ [7] and $0 \le k_1 < \ldots < k_t = \lfloor \log_2(n-1) \rfloor$ such that

$$n - 1 = \sum_{i=1}^{t} 2^{k_i}. \tag{S202}$$

Therefore we have

$$
\begin{aligned}
\frac{q^{n-1} - 1}{q - 1} &= \frac{q^{\sum_{i=1}^{t} 2^{k_i}} - 1}{q - 1} \\
&= \sum_{i=1}^{t} \frac{\left(q^{2^{k_i}} - 1\right)}{q - 1} q^{\sum_{j=1}^{i-1} 2^{k_j}}, \tag{S203}
\end{aligned}
$$

meaning

$$
\begin{aligned}
z &= x^{\frac{q^{n-1} - 1}{q - 1}} \\
&= \prod_{i=1}^{t} \left( x^{\frac{q^{2^{k_i}} - 1}{q - 1}} \right)^{\left[ \sum_{j=1}^{i-1} 2^{k_j} \right]}. \tag{S204}
\end{aligned}
$$

For each $i = 1, \ldots, t$, let

$$z_i = \left( x^{\frac{q^{2^{k_i}} - 1}{q - 1}} \right)^{\left[ \sum_{j=1}^{i-1} 2^{k_j} \right]}. \tag{S205}$$

Now we have

$$z = \prod_{i=1}^{t} z_i, \tag{S206}$$

meaning that $(t-1)$ $\mathbb{F}_{q^n}$-multiplications are required to compute $z$ from $z_1, \ldots, z_t$.

So, the problem is how to compute the power of $x$ efficiently. Here, we explain an efficient iterative method to achieve this. Specifically, we compute $z_1, \ldots, z_t$ in an iterative manner. In each iteration, we determine $z_i$ from $z_{i-1}$ by introducing an additional iteration for the interval $[k_{i-1}, k_i)$ where $0 \le k_1 < \ldots < k_t = \lfloor \log_2(n-1) \rfloor$ to exploit efficient $q$-power operations (merely cyclic shifts). Note that in case of $i = 1$, the interval is $[0, k_1)$.

For $i = 1$, set $z_0 = x$ and

$$z_{1,0} = z_0. \tag{S207}$$

and for each $k = 0, \ldots, k_1 - 1$, compute $z_{1,k+1}$ from $z_{1,k}$ as follows:

$$z_{1,k+1} = z_{1,k}^{\left[ 2^k \right]} z_{1,k}, \tag{S208}$$

which is efficient because only $q$-power and a multiplication are involved. Then we will have $z_{1,k_1} = z_1$.

The rationale is as follows. Using the fact that for any $k \in \mathbb{Z}_{\ge 0}$,

$$\frac{q^{2^{k+1}} - 1}{q - 1} = \left( \frac{q^{2^k} - 1}{q - 1} \right) \left( q^{2^k} + 1 \right), \tag{S209}$$

[7] $\text{Ham}(n-1)$ means the Hamming weight of $n-1$, i.e., the number of 1s in the binary representation of $n-1$.

we can show by induction that

$$z_{1,k} = x^{\frac{q^{2^k} - 1}{q - 1}}, \tag{S210}$$

as follows: Notice it is true for $k = 0$ because

$$
\begin{aligned}
z_{1,0} &= z_0 \\
&= x \\
&= x^{\frac{q^{2^0} - 1}{q - 1}}. \tag{S211}
\end{aligned}
$$

If it is true for $k$, then it is true for $k+1$ because

$$
\begin{aligned}
z_{1,k+1} &= z_{1,k}^{\left[ 2^k \right]} z_{1,k} \\
&= z_{1,k}^{\left( q^{2^k} + 1 \right)} \\
&= \left( x^{\frac{q^{2^k} - 1}{q - 1}} \right)^{\left( q^{2^k} + 1 \right)} \\
&= x^{\frac{q^{2^{k+1}} - 1}{q - 1}}. \tag{S212}
\end{aligned}
$$

So when we get to $k = k_1 - 1$, we have

$$
\begin{aligned}
z_{1,k_1} &= x^{\frac{q^{2^{k_1}} - 1}{q - 1}} \\
&= z_1. \tag{S213}
\end{aligned}
$$

For $i = 2$, set

$$z_{2,k_1} = z_1, \tag{S214}$$

and for each $k = k_1, \ldots, k_2 - 1$, compute $z_{2,k+1}$ from $z_{2,k}$ as follows:

$$z_{2,k+1} = z_{2,k}^{\left[ 2^k \right]} z_{2,k}. \tag{S215}$$

When we get to $k = k_2 - 1$, we can obtain $z_2$ by raising $z_{2,k_2}$ to the $[2^{k_1}]$ $q$-power as

$$z_2 = z_{2,k_2}^{\left[ 2^{k_1} \right]}, \tag{S216}$$

whose rationale is given for general $i$ later.

For $i = 3$, set

$$z_{3,k_2} = z_2, \tag{S217}$$

and for each $k = k_2, \ldots, k_3 - 1$, compute $z_{3,k+1}$ from $z_{3,k}$ as follows:

$$z_{3,k+1} = z_{3,k}^{\left[ 2^k \right]} z_{3,k}. \tag{S218}$$

When we get to $k = k_3 - 1$, we can obtain $z_3$ by raising $z_{3,k_3}$ to the $[2^{k_2}]$ $q$-power as

$$z_3 = z_{3,k_3}^{\left[ 2^{k_2} \right]}, \tag{S219}$$

whose rationale is given for general $i$ below.

For $i = 4, \ldots, t$, set

$$z_{i,k_{i-1}} = z_{i-1}, \tag{S220}$$

and for each $k = k_{i-1}, \ldots, k_i - 1$, compute $z_{i,k+1}$ from $z_{i,k}$ as follows:

$$z_{i,k+1} = z_{i,k}^{\left[ 2^k \right]} z_{i,k}. \tag{S221}$$

When we get to $k = k_i - 1$, we can obtain $z_i$ by raising $z_{i,k_i}$ to the $[2^{k_{i-1}}]$ $q$-power as

$$z_i = z_{i,k_i}^{\left[ 2^{k_{i-1}} \right]}. \tag{S222}$$

The rationale is as follows. Using Eq. (S209), we can show by induction that

$$z_{i,k} = \left( x^{\frac{q^{2^k} - 1}{q - 1}} \right)^{\left[ \sum_{j=1}^{i-2} 2^{k_j} \right]}. \tag{S223}$$

Notice it is true for $k = k_{i-1}$ because

$$
\begin{aligned}
z_{i,k_{i-1}} &= z_{i-1} \\
&= \left(x^{\frac{q^{2^{k_{i-1}}}-1}{q-1}}\right)^{\left[\sum_{j=1}^{i-2} 2^{k_j}\right]}.
\end{aligned}
\tag{S224}
$$

If it is true for $k$, then it is true for $k+1$ because

$$
\begin{aligned}
z_{i,k+1} &= z_{i,k}^{[2^k]} z_{i,k} \\
&= z_{i,k}^{\left(q^{2^k}+1\right)} \\
&= \left(\left(x^{\frac{q^{2^k}-1}{q-1}}\right)^{\left[\sum_{j=1}^{i-2} 2^{k_j}\right]}\right)^{\left(q^{2^k}+1\right)} \\
&= \left(x^{\frac{q^{2^{k+1}}-1}{q-1}}\right)^{\left[\sum_{j=1}^{i-2} 2^{k_j}\right]}.
\end{aligned}
\tag{S225}
$$

So when we get to $k = k_i - 1$, we have

$$
z_{i,k_i} = \left(x^{\frac{q^{2^{k_i}}-1}{q-1}}\right)^{\left[\sum_{j=1}^{i-2} 2^{k_j}\right]}.
\tag{S226}
$$

This means

$$
\begin{aligned}
z_i &= \left(x^{\frac{q^{2^{k_i}}-1}{q-1}}\right)^{\left[\sum_{j=1}^{i-1} 2^{k_j}\right]} \\
&= z_{i,k_i}^{[2^{k_i-1}]}.
\end{aligned}
\tag{S227}
$$

Actually Eq. (S213) can also be aligned to this expression as

$$
z_1 = z_{1,k_1}^{[2^{k_0}]},
\tag{S228}
$$

by defining $k_0 = -\infty$, and $2^{-\infty} = 0$. Itoh's algorithm is summarized in Supplementary Algorithm 1. The algorithm requires $\left(2^{\lfloor \log_2(n-1)\rfloor} + n - 1\right)$ $q$-power operations, $\left(\lfloor \log_2(n-1)\rfloor + \mathrm{Ham}(n-1)\right)$ $\mathbb{F}_{q^n} \times \mathbb{F}_{q^n}$-multiplications, one $\mathbb{F}_q \times \mathbb{F}_{q^n}$-multiplication and one $\mathbb{F}_q$-inversion.

### D. Duality

Let $h_0, \ldots, h_{n-1} \in \mathbb{F}_{q^n}$ be a basis of $\mathbb{F}_{q^n}$ over $\mathbb{F}_q$. Then there exists unique basis $h_0^\perp, \ldots, h_{n-1}^\perp \in \mathbb{F}_{q^n}$ of $\mathbb{F}_{q^n}$ over $\mathbb{F}_q$ such that for all $i, j = 0, \ldots, n-1$,

$$
\sum_{v=0}^{n-1} \left(h_i h_j^\perp\right)^{[v]} = \delta_{i,j},
\tag{S229}
$$

where $\delta_{i,j}$ is the Kronecker delta operator such that

$$
\delta_{i,j} = \begin{cases} 1 & \text{if } i = j, \\ 0 & \text{otherwise}. \end{cases}
\tag{S230}
$$

$h_0^\perp, \ldots, h_{n-1}^\perp$ is called the dual basis of $h_0, \ldots, h_{n-1}$.

If $h_0^\perp, \ldots, h_{n-1}^\perp$ is the dual basis of $h_0, \ldots, h_{n-1}$, we must also have for all $i, j = 0, \ldots, n-1$,

$$
\sum_{v=0}^{n-1} h_v^{\perp [i]} h_v^{[j]} = \delta_{i,j}
\tag{S231}
$$

---

**Supplementary Algorithm 1** Itoh's $\mathbb{F}_{q^n}$-inversion algorithm

**input:** $x \in \mathbb{F}_{q^n} \setminus \{0\}$
**begin:**
  1) $z \leftarrow 1$, $z' \leftarrow x$, $n' \leftarrow n-1$, $i \leftarrow -\infty$, $k \leftarrow 0$
  2) **while** $n' > 1$ **do**
    a) **if** $n' \bmod 2 = 1$ **then**
      i) $z' \leftarrow z'^{[2^i]}$
      ii) $z \leftarrow z'z$
      iii) $i \leftarrow k$
      iv) $n' \leftarrow n'-1$
    b) $z' \leftarrow z'^{[2^k]} z'$
    c) $n' \leftarrow \frac{n'}{2}$
    d) $k \leftarrow k+1$
  3) $z \leftarrow z'^{[2^i]} z$
  4) $z \leftarrow z^{[1]}$
  5) $y \leftarrow xz$
  6) $y \leftarrow y^{-1}$
  7) $z \leftarrow yz$
  **end**
**output:** $z$

**note:**
  1) $2^{-\infty} = 0$.

---

To see this, notice from the duality of $h_0, \ldots, h_{n-1}$ and $h_0^\perp, \ldots, h_{n-1}^\perp$, we have the matrix equation

$$
\hat{H}\hat{H}^\perp = I_n
\tag{S232}
$$

where $\hat{H} \in \mathbb{F}_{q^n}^{n \times n}$ is the matrix

$$
\hat{H} = \begin{bmatrix} h_0^{[0]} & \cdots & h_0^{[n-1]} \\ \vdots & \ddots & \vdots \\ h_{n-1}^{[0]} & \cdots & h_{n-1}^{[n-1]} \end{bmatrix},
\tag{S233}
$$

$\hat{H}^\perp \in \mathbb{F}_{q^n}^{n \times n}$ is the matrix

$$
\hat{H}^\perp = \begin{bmatrix} h_0^{\perp \, [0]} & \cdots & h_{n-1}^{\perp \, [0]} \\ \vdots & \ddots & \vdots \\ h_0^{\perp \, [n-1]} & \cdots & h_{n-1}^{\perp \, [n-1]} \end{bmatrix},
\tag{S234}
$$

and $I_n \in \mathbb{F}_{q^n}^{n \times n}$ is the $n \times n$ identity matrix. This means that $\hat{H}$ is the left inverse of $\hat{H}^\perp$, so $\hat{H}$ must also be the right inverse of $\hat{H}^\perp$. Therefore the matrix equation

$$
\hat{H}^\perp \hat{H} = I_n,
\tag{S235}
$$

also holds, which yield the desired result. In the case where $h_0, \ldots, h_{n-1}$ is a normal basis generated by some $\beta \in \mathbb{F}_{q^n}$, it is the case that its unique dual basis $h_0^\perp, \ldots, h_{n-1}^\perp$ is also a normal basis generated by some $\beta^\perp \in \mathbb{F}_{q^n}$. In such a case, we refer to $\beta^\perp$ as the dual element of $\beta$.

### E. Linearized polynomials

Let $\mathbb{F}_{q^n}[x]$ be the linearized polynomial ring in $\mathbb{F}_{q^n}$ with indeterminate $x$.

A $\mathbb{F}_{q^n}$-linearized polynomial $f \in \mathbb{F}_{q^n}[x]$ of degree $L \in \mathbb{Z}_{\geq 0}$ is a polynomial of the form

$$
f(x) = \sum_{i=0}^{L} f_i x^{[i]},
\tag{S236}
$$

where $f_0, \ldots, f_L \in \mathbb{F}_{q^n}$ and $f_L \neq 0$.

*1) Linearized polynomials as a linear operator*

A $\mathbb{F}_{q^n}$-linearized polynomial is a linear transformation. This is summarized in the following lemma.

**Lemma 10.** *Let $f \in \mathbb{F}_{q^n}[x]$ be an $\mathbb{F}_{q^n}$-linearized polynomial of degree $L \in \mathbb{Z}_{\geq 0}$. Then for any $b_0, b_1 \in \mathbb{F}_{q^n}$ and $A_0, A_1 \in \mathbb{F}_q$,*

$$f\left(A_0 b_0 + A_1 b_1\right) = A_0 f\left(b_0\right) + A_1 f\left(b_1\right). \tag{S237}$$

*Proof.* Notice

$$
\begin{aligned}
f\left(A_0 b_0 + A_1 b_1\right) &= \sum_{i=0}^{L} f_i \left(A_0 b_0 + A_1 b_1\right)^{[i]} \\
&= \sum_{i=0}^{L} f_i \left(A_0 b_0^{[i]} + A_1 b_1^{[i]}\right) \\
&= A_0 \sum_{i=0}^{L} f_i b_0^{[i]} + \sum_{i=0}^{L} f_i b_1^{[i]} \\
&= A_0 f\left(b_0\right) + A_1 f\left(b_1\right), \tag{S238}
\end{aligned}
$$

as required. $\square$

*2) Rootspace of a linearized polynomial*

The rootspace of a $\mathbb{F}_{q^n}$-linearized polynomial $f \in \mathbb{F}_{q^n}[x]$ is the set of all $x \in \mathbb{F}_{q^n}$ such that $f(x) = 0$. In other words

$$\text{rootspace}\left(f\right) = \{x \in \mathbb{F}_{q^n} : f(x) = 0\}. \tag{S239}$$

**Lemma 11.** *Let $f \in \mathbb{F}_{q^n}[x]$ be a $\mathbb{F}_{q^n}$-linearized polynomial of degree $L \in \mathbb{Z}_{\geq 0}$. rootspace $(f)$ is a vector space over $\mathbb{F}_q$.*

*Proof.* To show rootspace $(f)$ is a vector space over $\mathbb{F}_q$, we need to show that for any $b_0, b_1 \in$ rootspace $(f)$ and $A_0, A_1 \in \mathbb{F}_q$, $A_0 b_0 + A_1 b_1 \in$ rootspace $(f)$. To do this notice

$$
\begin{aligned}
f\left(A_0 b_0 + A_1 b_1\right) &= A_0 f\left(b_0\right) + A_1 f\left(b_1\right) \\
&= A_0 \cdot 0 + A_1 \cdot 0 \\
&= 0. \tag{S240}
\end{aligned}
$$

$\square$

Now we have established the rootspace of a $\mathbb{F}_{q^n}$-linearized polynomial as a vector space, the rootspace has a basis and therefore a dimension. The following lemma relates the degree of the $\mathbb{F}_{q^n}$-linearized polynomial and the dimension of its rootspace.

**Lemma 12.** *(without proof) Let $f \in \mathbb{F}_{q^n}[x]$ be a $\mathbb{F}_{q^n}$-linearized polynomial of degree $L \in \mathbb{Z}_{\geq 0}$, then*

$$\dim\left(\text{rootspace}\left(f\right)\right) \leq L. \tag{S241}$$

*3) Computing the rootspace of a linearized polynomial*

In the error correction of Gabidulin codes, once the ESP or ELP is synthesized as a $\mathbb{F}_{q^n}$-linearized polynomial, a basis of its rootspace needs to be computed for determining the error values or error locations. We need to prove the following lemma to do this.

**Lemma 13.** *Let $f \in \mathbb{F}_{q^n}[x]$ be a $\mathbb{F}_{q^n}$-linearized polynomial of degree $L \in \mathbb{Z}_{\geq 0}$. Let $\beta \in \mathbb{F}_{q^n}$ generate a normal basis of $\mathbb{F}_{q^n}$ over $\mathbb{F}_q$. For each $j = 0, \ldots, n-1$, let $\boldsymbol{\Phi}_j = \begin{bmatrix} \Phi_{0,j} & \ldots & \Phi_{n-1,j} \end{bmatrix}^T \in \mathbb{F}_q^n$ be the $\beta$-normal basis representation of $f\left(\beta^{[j]}\right)$. Let $\Phi \in \mathbb{F}_q^{n \times n}$ be the $n \times n$ matrix formed from the concatenation of $\boldsymbol{\Phi}_0, \ldots, \boldsymbol{\Phi}_{n-1}$, namely*

$$\Phi = \begin{bmatrix} \boldsymbol{\Phi}_0 & \ldots & \boldsymbol{\Phi}_{n-1} \end{bmatrix}. \tag{S242}$$

*Let $\hat{\boldsymbol{\Phi}}_0, \ldots, \hat{\boldsymbol{\Phi}}_{\hat{L}-1} \in \mathbb{F}_q^n$ be a basis of ker $(F)$ over $\mathbb{F}_q$, where $\hat{L} = \text{null}\left(\Phi\right)$. Then $\hat{\phi}_0, \ldots, \hat{\phi}_{\hat{L}-1} \in \mathbb{F}_{q^n}$ such that for each $j = 0, \ldots, \hat{L}-1$*

$$\hat{\phi}_j = \sum_{i=0}^{n-1} \hat{\Phi}_{i,j} \beta^{[i]}, \tag{S243}$$

*is a basis of rootspace $(f)$.*

In other words, the $\beta$-normal basis representation of a basis of rootspace $(f)$ can be determined by finding a basis of ker $(\Phi)$, where $\Phi$ is the $n \times n$ matrix formed by the concatenation of the $\beta$-normal basis representation of the evaluation of $f$ at each basis vector $\beta^{[0]}, \ldots, \beta^{[n-1]}$.

*Proof.* Let $f_0, \ldots, f_L \in \mathbb{F}_{q^n}$ such that

$$f\left(x\right) = \sum_{v=0}^{L} f_v x^{[v]}. \tag{S244}$$

For each $v = 0, \ldots, L$, let $\boldsymbol{F}_v = \begin{bmatrix} F_{0,v} & \ldots & F_{n-1,v} \end{bmatrix}^T \in \mathbb{F}_q^n$ be the $\beta$-normal basis representation of $f_v$, namely

$$f_v = \sum_{w=0}^{n-1} F_{w,v} \beta^{[w]}. \tag{S245}$$

Firstly, let us establish the form of the $\beta$-normal basis representation of $f\left(\beta^{[j]}\right)$. For each $j = 0, \ldots, n-1$,

$$
\begin{aligned}
f\left(\beta^{[j]}\right) &= \sum_{v=0}^{L} f_v \beta^{[j+v]} \\
&= \sum_{v=0}^{L} \left(\sum_{w=0}^{n-1} F_{w,v} \beta^{[w]}\right) \beta^{[j+v]} \\
&= \sum_{v=0}^{L} \sum_{w=0}^{n-1} F_{w,v} \beta^{[w]} \beta^{[j+v]} \\
&= \sum_{i=0}^{n-1} \left(\sum_{v=0}^{L} \sum_{w=0}^{n-1} F_{w,v} t_{w,(j+v) \bmod n}^{(i)}\right) \beta^{[i]} \\
&= \sum_{i=0}^{n-1} \Phi_{i,j} \beta^{[i]}, \tag{S246}
\end{aligned}
$$

where for each $i = 0, \ldots, n-1$

$$\Phi_{i,j} = \sum_{v=0}^{L} \sum_{w=0}^{n-1} F_{w,v} t_{w,(j+v) \bmod n}^{(i)}. \tag{S247}$$

Notice this means $\boldsymbol{\Phi}_j = \begin{bmatrix} \Phi_{0,j} & \ldots & \Phi_{n-1,j} \end{bmatrix}^T \in \mathbb{F}_q^n$ is the $\beta$-normal basis representation of $f\left(\beta^{[j]}\right)$.

Secondly, let us established the form of the $\beta$-normal basis representation of an element in the rootspace of $f$. Suppose $b \in$ rootspace $(f)$. Let $\boldsymbol{B} = \begin{bmatrix} B_0 & \ldots & B_{n-1} \end{bmatrix}^T \in \mathbb{F}_q^n$ be the $\beta$-normal basis representation of $b$, namely

$$b = \sum_{j=0}^{n-1} B_j \beta^{[j]}. \tag{S248}$$

Then we have

$$
\begin{aligned}
0 &= f(b) \\
&= \sum_{v=0}^{L} f_v b^{[v]} \\
&= \sum_{v=0}^{L} \left( \sum_{w=0}^{n-1} F_{w,v} \beta^{[w]} \right) \left( \sum_{j=0}^{n-1} B_j \beta^{[j+v]} \right) \\
&= \sum_{j=0}^{n-1} B_j \sum_{v=0}^{L} \sum_{w=0}^{n-1} F_{w,v} \beta^{[w]} \beta^{[j+v]} \\
&= \sum_{i=0}^{n-1} \left( \sum_{j=0}^{n-1} B_j \sum_{v=0}^{L} \sum_{w=0}^{n-1} F_{w,v} t^{(i)}_{w,(j+v) \bmod n} \right) \beta^{[i]} \\
&= \sum_{i=0}^{n-1} \left( \sum_{j=0}^{n-1} B_j \Phi_{i,j} \right) \beta^{[i]}. \tag{S249}
\end{aligned}
$$

However, because $\beta^{[0]}, \dots, \beta^{[n-1]}$ are linearly independent over $\mathbb{F}_q$, this holds if and only if for each $i = 0, \dots, n-1$

$$
\sum_{j=0}^{n-1} B_j \Phi_{i,j} = 0, \tag{S250}
$$

or in matrix form if and only if

$$
\Phi \boldsymbol{B} = \boldsymbol{0}. \tag{S251}
$$

Hence we establish $b \in \text{rootspace}(f) \Leftrightarrow \boldsymbol{B} \in \ker(\Phi)$.

Thirdly, we need to show that if $\hat{\boldsymbol{\Phi}}_0, \dots, \hat{\boldsymbol{\Phi}}_{\hat{L}-1} \in \mathbb{F}_q^n$ is a basis of $\ker(\Phi)$ over $\mathbb{F}_q$, then $\hat{\phi}_0, \dots, \hat{\phi}_{\hat{L}-1} \in \mathbb{F}_{q^n}$, whereby for each $j = 0, \dots, \hat{L}-1$

$$
\hat{\phi}_j = \sum_{i=0}^{n-1} \hat{\Phi}_{i,j} \beta^{[i]}, \tag{S252}
$$

is a basis of $\text{rootspace}(f)$. Let $\hat{\boldsymbol{\Phi}}_0, \dots, \hat{\boldsymbol{\Phi}}_{\hat{L}-1} \in \mathbb{F}_q^n$ be a basis of $\ker(\Phi)$ over $\mathbb{F}_q$ such that for each $j = 0, \dots, \hat{L}-1$, $\hat{\boldsymbol{\Phi}}_j = \begin{bmatrix} \hat{\Phi}_{0,j} & \dots & \hat{\Phi}_{n-1,j} \end{bmatrix}^T$. From the previous part of the proof, we have that $\hat{\boldsymbol{\Phi}}_j \in \ker(\Phi) \Leftrightarrow \hat{\phi}_j \in \text{rootspace}(f)$. This means that $\hat{\phi}_0, \dots, \hat{\phi}_{\hat{L}-1} \in \text{rootspace}(f)$ and

$$
\dim(\text{rootspace}(f)) = \text{null}(\Phi), \tag{S253}
$$

for if the dimensions of the vector spaces were not the same, there would be an element in $\text{rootspace}(f)$ whose $\beta$-normal basis representation is not in $\ker(\Phi)$ or vice versa, which is not possible. So, all that remains to show is that $\hat{\phi}_0, \dots, \hat{\phi}_{\hat{L}-1}$ are linearly independent over $\mathbb{F}_q$. To show this we need to establish that for $A_0, \dots, A_{\hat{L}-1} \in \mathbb{F}_q$

$$
\sum_{j=0}^{\hat{L}-1} A_j \hat{\phi}_j = 0 \Leftrightarrow A_j = 0 \text{ for all } j = 0, \dots, \hat{L}-1. \tag{S254}
$$

$\Leftarrow$ clearly holds. To show $\Rightarrow$, notice that

$$
\begin{aligned}
0 &= \sum_{j=0}^{\hat{L}-1} A_j \hat{\phi}_j \\
&= \sum_{j=0}^{\hat{L}-1} A_j \left( \sum_{i=0}^{n-1} \hat{\Phi}_{i,j} \beta^{[i]} \right) \\
&= \sum_{i=0}^{n-1} \left( \sum_{j=0}^{\hat{L}-1} A_j \hat{\Phi}_{i,j} \right) \beta^{[i]}. \tag{S255}
\end{aligned}
$$

---

**Supplementary Algorithm 2** Algorithm to compute the rootspace of a $\mathbb{F}_{q^n}$-linearized polynomial

**input:** $\mathbb{F}_{q^n}$-linearized polynomial coefficients $f_0, \dots, f_L \in \mathbb{F}_{q^n}$ such that $f_L \neq 0$.

**begin:**

1) **for** each $j$ from $0$ to $n-1$ **do**

   a) $\phi_j \leftarrow \sum_{v=0}^{L} f_v \beta^{[j+v]}$

   b) determine $\boldsymbol{\Phi}_j = \begin{bmatrix} \Phi_{0,j} & \dots & \Phi_{n-1,j} \end{bmatrix}^T \in \mathbb{F}_q^n$ such that $\phi_j = \sum_{i=0}^{n} \Phi_{i,j} \beta^{[i]}$

2) $\Phi \leftarrow \begin{bmatrix} \boldsymbol{\Phi}_0 & \dots & \boldsymbol{\Phi}_{n-1} \end{bmatrix}$

3) $\hat{\boldsymbol{\Phi}}_0, \dots, \hat{\boldsymbol{\Phi}}_{\hat{L}-1} \leftarrow$ kernel-Gauss-Jordan $(\Phi)$

  **end**

**output:** $\hat{\phi}_0, \dots, \hat{\phi}_{\hat{L}-1} \in \mathbb{F}_{q^n}$ such that for each $j = 0, \dots, \hat{L}-1$, $\hat{\phi}_j = \sum_{i=0}^{n-1} \hat{\Phi}_{i,j} \beta^{[i]}$.

**notes:**

1) In the case that $f_0, \dots, f_L$ are input in $\beta$-normal basis representation, step 1(a) requires only $\mathbb{F}_{q^n}$-addition and $\mathbb{F}_{q^n}$-fast multiplication and returns $\phi_0, \dots, \phi_{n-1}$ in $\beta$-normal basis representation meaning that step 1(b) is already done.

2) In the case that the output is required in $\beta$-normal basis representation, the algorithm can output $\hat{\boldsymbol{\Phi}}_0, \dots, \hat{\boldsymbol{\Phi}}_{\hat{L}-1}$.

---

But $\beta^{[0]}, \dots, \beta^{[n-1]}$ are linearly independent over $\mathbb{F}_q$, so this holds if and only if for each $i = 0, \dots, n-1$

$$
\sum_{j=0}^{\hat{L}-1} A_j \hat{\Phi}_{i,j} = 0. \tag{S256}
$$

But because $\hat{\boldsymbol{\Phi}}_0, \dots, \hat{\boldsymbol{\Phi}}_{\hat{L}-1}$ are linearly independent over $\mathbb{F}_q$, we must have $A_j = 0$ for all $j = 0, \dots, \hat{L}-1$ as required. $\square$

The algorithm for determining the rootspace of a $\mathbb{F}_{q^n}$-linearized polynomial is given in Supplementary Algorithm 2. Given $\Phi \in \mathbb{F}_q^{n \times n}$, a basis of $\ker(\Phi)$ can be found by using Gauss-Jordan elimination. If the coefficients of $f$ are given in their $\beta$-normal basis form and $f_0 = 1$, $\Phi$ can be established using $Ln$ $\mathbb{F}_{q^n}$-additions and $Ln$ $\mathbb{F}_{q^n}$-fast multiplications.

The arithmetic complexity is given in Supplementary Table VIII and assumes that coefficients of $f$ are input in their $\beta$-normal basis representation and $f_0 = 1$.

**Supplementary Table VIII**
**Arithmetic operations required for computing the rootspace of a degree $L \in \mathbb{Z}_{\geq 0}$ $\mathbb{F}_{q^n}$-linearized polynomial with $x^{[0]} = 1$.**

| Type | $\mathbb{F}_{q^n}$ operations | $\mathbb{F}_q$ operations |
|---|---|---|
| $+$ | $Ln$ | $\frac{1}{2}(n-1)(n-L)(n+L-1)$ |
| $\times$ | $Ln$ (fast) | $\frac{1}{2}n(n-L)(n+L-1)$ |
| $-1$ | $0$ | $n-L$ |

*4) Minimal linearized polynomial*

In the erasure correction of Gabidulin codes, once the transformed erasure locations $\hat{d}_0, \dots, \hat{d}_{\rho-1}$ are computed from the erasure locations, the ULP needs to be synthesized by RPA. To do this, we need to find a $\mathbb{F}_{q^n}$-linearized polynomial of lowest degree whose rootspace is spanned by a given set $\left\{ \hat{d}_0, \dots, \hat{d}_{L-1} \right\} \subseteq \mathbb{F}_{q^n}$ for $0 \leq L \leq \rho$. This gives rise to the notion of the minimal linearized polynomial.

**Definition 14.** *Given a set* $\left\{ \hat{d}_0, \dots, \hat{d}_{L-1} \right\} \subseteq \mathbb{F}_{q^n}$, $f \in \mathbb{F}_{q^n}[x]$ *such that* $x^{[0]}$ *coefficient of* $f$ *is set to one is the minimal* $\mathbb{F}_{q^n}$-*linearized polynomial of* $\left\{ \hat{d}_0, \dots, \hat{d}_{L-1} \right\}$, *if* $f$ *is a* $\mathbb{F}_{q^n}$-*linearized polynomial of lowest degree such that*

$$rootspace\,(f) = span\left( \left\{ \hat{d}_0, \dots, \hat{d}_{L-1} \right\} \right). \quad (S257)$$

**Lemma 15.** *(without proof) Given a set* $\left\{ \hat{d}_0, \dots, \hat{d}_{L-1} \right\} \subseteq \mathbb{F}_{q^n}$, *the minimal* $\mathbb{F}_{q^n}$-*polynomial of* $\left\{ \hat{d}_0, \dots, \hat{d}_{L-1} \right\}$ *is unique.*

*5) Computing the minimal linearized polynomial*

Repeated application of the following lemma gives rise to RPA and computes the minimal $\mathbb{F}_{q^n}$-linearized polynomial of a set $\left\{ \hat{d}_0, \dots, \hat{d}_{L-1} \right\} \subseteq \mathbb{F}_{q^n}$ in $\mathcal{O}\left( L^2 \right)$.

**Lemma 16.** *(without proof) Let* $f' \in \mathbb{F}_{q^n}[x]$ *be the minimal* $\mathbb{F}_{q^n}$-*linearized polynomial of* $\left\{ \hat{d}_0, \dots, \hat{d}_{L-1} \right\} \subseteq \mathbb{F}_{q^n}$ *and let* $\hat{d}_L \in \mathbb{F}_{q^n}$. *Then*

$$f(x) = \begin{cases} f'(x) & \text{if } f'\left( \hat{d}_L \right) = 0, \\ f'(x) - f'\left( \hat{d}_L \right)^{1-q} f'(x)^q & \text{otherwise.} \end{cases} \quad (S258)$$

*is the minimal* $\mathbb{F}_{q^n}$-*linearized polynomial of* $\left\{ \hat{d}_0, \dots, \hat{d}_L \right\}$.

RPA is as follows: Set $f^{(0)} \in \mathbb{F}_{q^n}[x]$ such that

$$f^{(0)}(x) = x. \quad (S259)$$

Notice that

$$rootspace\left( f^{(0)} \right) = \{0\}. \quad (S260)$$

So $f^{(0)}$ is the minimal $\mathbb{F}_{q^n}$-linearized polynomial of $\emptyset$. Then proceeding iteratively through $i = 0, \dots, L-1$, set

$$f^{(i+1)}(x)$$
$$= \begin{cases} f^{(i)}(x) & \text{if } f^{(i)}\left( \hat{d}_i \right) = 0 \\ f^{(i)}(x) - f^{(i)}\left( \hat{d}_i \right)^{1-q} f^{(i)}(x)^q & \text{otherwise.} \end{cases} \quad (S261)$$

By Lemma 16, $f^{(i+1)}$ is the minimal $\mathbb{F}_{q^n}$-linearized polynomial of $\left\{ \hat{d}_0, \dots, \hat{d}_i \right\}$, meaning that $f^{(L)}$ is the minimal $\mathbb{F}_{q^n}$-linearized polynomial of $\left\{ \hat{d}_0, \dots, \hat{d}_{L-1} \right\}$. Output $f^{(L)}$.

RPA is summarized in Supplementary Algorithm 3 and its $\mathbb{F}_{q^n}$-arithmetic complexity in Supplementary Table IX. Notice that RPA also allows us to sift $\left\{ \hat{d}_0, \dots, \hat{d}_{L-1} \right\}$ such that we discard $\hat{d}_i$ if $f^{(i)}\left( \hat{d}_i \right) = 0$. This way we are left with a largest linearly independent subset of $\left\{ \hat{d}_0, \dots, \hat{d}_{L-1} \right\}$ because element $\hat{d}_i$ is discarded at iteration $i$ if and only if it is in the rootspace of $f^{(i)}$ and therefore in the space spanned by the previous elements $\hat{d}_0, \dots, \hat{d}_{i-1}$.

*F. Linearized feedback shift registers*

In the error correction of Gabidulin codes, once the key equation is set up by computing the syndromes/reversed syndromes, the synthesis of $\mathbb{F}_{q^n}$-LFSR is made to determine

**Supplementary Algorithm 3** Richter-Plass algorithm (RPA) for minimal $\mathbb{F}_{q^n}$-linearized polynomial synthesis

---

**input:** $\hat{d}_0, \dots, \hat{d}_{\rho-1} \in \mathbb{F}_{q^n}$

**begin:**

1) $f_0 \leftarrow 1, L \leftarrow 0$

2) **for** each $i$ from 0 to $\rho - 1$ **do**

    a) $r \leftarrow \sum_{j=0}^{L} f_j \hat{d}_i^{[j]}$

    b) **if** $r \neq 0$ **then**

        i) $r' \leftarrow -r \left( r^{-1} \right)^{[1]}$

        ii) $f_{L+1} \leftarrow r' f_L^{[1]}$

        iii) **for** each $j$ from $L$ to 1 step $-1$ **do**

            A) $f_j \leftarrow f_j + r' f_{j-1}^{[1]}$

        iv) $L \leftarrow L + 1$

**end**

**output:** $f(x) = \sum_{j=0}^{\rho} f_j x^{[j]}$

---

**Supplementary Table IX**
$\mathbb{F}_{q^n}$-**arithmetic operations required for the RPA.**

| Type | $\mathbb{F}_{q^n}$ operations |
|------|------|
| $+$ | $\rho(\rho - 1)$ |
| $\times$ | $\rho^2$ |
| $[1]$ | $\rho^2$ |
| $-1$ | $\rho$ |
| $-1$ | $\rho$ |

the coefficients of the ESP/ELP. In this subsection, we present an $\mathcal{O}\left( \tau^2 \right)$ method for synthesizing a shortest $\mathbb{F}_{q^n}$-LFSR in the style of the Berlekamp-Massey algorithm modified by Richter and Plass [27].

**Definition 17.** $\mathbb{F}_{q^n}$-*LFSR of length* $\tau$ *for a sequence* $s_0, \dots, s_{n-k-1} \in \mathbb{F}_{q^n}$ *with the feedback coefficients* $f_0, \dots, f_\tau \in \mathbb{F}_{q^n}$ *is a circuit specified by*

$$\sum_{j=0}^{\tau} f_j s_{i-j}^{[j]} = 0, \quad i = \tau, \dots, n-k-1. \quad (S262)$$

**Theorem 18.** *If a* $\mathbb{F}_{q^n}$-*LFSR of* $s_0, \dots, s_{n-k-1}$ *exists for length* $\tau \leq \left\lfloor \frac{n-k}{2} \right\rfloor$, *then the LBMA due to Richter and Plass [27] will return the unique shortest* $\mathbb{F}_{q^n}$-*LFSR* $f'_0, \dots, f'_{\tau'}$ *of* $s_0, \dots, s_{n-k-1}$, *such that* $f'_0 = 1$, $\tau' \leq \tau$ *and for all* $i = \tau', \dots, n-k-1$,

$$\sum_{j=0}^{\tau'} f'_j s_{i-j}^{[j]} = 0. \quad (S263)$$

*Proof.* (rough sketch)
Eq. (S262) (the key equation) can be written in matrix form

$$\begin{bmatrix} s_{\tau-1}^{[1]} & \cdots & s_0^{[\tau]} \\ \vdots & \ddots & \vdots \\ s_{n-k-2}^{[1]} & \cdots & s_{n-k-\tau-1}^{[\tau]} \end{bmatrix} \begin{bmatrix} f_1 \\ \vdots \\ f_\tau \end{bmatrix} = -f_0 \begin{bmatrix} s_0^{[0]} \\ \vdots \\ s_{n-k-1}^{[0]} \end{bmatrix}. \quad (S264)$$

Note that the matrix $\begin{bmatrix} s_{\tau-1}^{[1]} & \cdots & s_0^{[\tau]} \\ \vdots & \ddots & \vdots \\ s_{2\tau-2}^{[1]} & \cdots & s_{\tau-1}^{[\tau]} \end{bmatrix}$ is nonsingular, which can be seen from the $q$-power version of the syndrome equations, Eq. (S61) or Eq. (S70), with not only that $a_0, a_1, ..., a_{\tau-1}$ are linearly independent over $\mathbb{F}_q$ but also that $d_0, d_1, ..., d_{\tau-1}$ are also linearly independent over $\mathbb{F}_q$. Therefore, when $n - k - \tau \geq \tau$, Eq. (S264) has the unique solution up to a choice of $f_0$, leading to the condition of error correction capability of $2\tau \leq n - k$.

The solution can be efficiently found by the Berlekamp-Massey algorithm modified (linearized) for rank codes (LBMA) by Richter and Plass [27]. Along a similar line to [37], [38], Richter and Plass proved that the LBMA returns the shortest $\mathbb{F}_{q^n}$-LFSR $f'_0, ..., f'_{\tau'}$, such that $f'_0 = 1$ and $\tau' \leq \tau$ (Section 3 in [27]). □

The LBMA and its complexity are summarized in Supplementary Algorithm 4 and Supplementary Table X. The LBMA proceeds as follows:

1) $\mathbb{F}_{q^n}$-LFSR is initialized to 1. $\tau'$ is the current number of assumed errors and initialized to 0. Some auxiliary variables are introduced for iterations, including dummy coefficients $f_0, f_1, ...$ starting with $f_0 = 1$, a length $\tau$ initialized to 0, a copy of the last discrepancy $r$ initialized to 1, and a parameter $h$ to control the number of iterations initialized to 1.

2) $n - k$ is the total number of syndromes. The index of the syndromes $i$ is the main iterator, running from 0 to $n - k - 1$.

   a) Each iteration of the algorithm calculates a discrepancy $r'$.

   b) If $r' = 0$, the algorithm assumes that $\tau'$ and $f'_0, ..., f'_{\tau'}$ are correct for the moment, increments $h$, and continues.

   c) If $r' \neq 0$, the algorithm adjusts $f'_j$'s so that $r'$ would become 0. $\bar{f}_0, ..., \bar{f}_{\tau'}$ and $\bar{\tau}$ are variables for temporarily restoring the current $f'_0, ..., f'_{\tau'}$ and $\tau'$.

In this way, the LBMA outputs the shortest $\mathbb{F}_{q^n}$-LFSR $f'_0, ..., f'_{\tau'}$.

---

**Supplementary Algorithm 4** The LBMA for $\mathbb{F}_{q^n}$-LFSR synthesis

---

**input:** $s_0, ..., s_{n-k-1} \in \mathbb{F}_{q^n}$
**begin:**
 1) $f'_0 \leftarrow 1$, $\tau' \leftarrow 0$, $f_0 \leftarrow 1$, $\tau \leftarrow 0$, $r \leftarrow 1$, $h \leftarrow 1$
 2) **for** each $i$ from 0 to $n - k - 1$ **do**
   a) $r' \leftarrow \sum_{j=0}^{\tau'} f'_j s_{i-j}^{[j]}$
   b) **if** $r' = 0$ **then**
     i) $h \leftarrow h + 1$
   c) **else**
     i) **if** $\tau' < \tau + h$ **then**
       A) $(\bar{f}_0, ..., \bar{f}_{\tau'}) \leftarrow (f'_0, ..., f'_{\tau'})$
       B) $\bar{\tau} \leftarrow \tau'$
       C) $(f'_h, ..., f'_{\tau'}) \leftarrow (f'_0, ..., f'_{\tau'})$
          $- r' (r^{-1})^{[h]} (f_0^{[h]}, ..., f_{\tau'-h}^{[h]})$
       D) $(f'_{\tau'+1}, ..., f'_{\tau+h}) \leftarrow$
          $- r' (r^{-1})^{[h]} (f_{\tau'-h+1}^{[h]}, ..., f_\tau^{[h]})$
       E) $\tau' \leftarrow \tau + h$
       F) $(f_0, ..., f_{\bar{\tau}}) \leftarrow (\bar{f}_0, ..., \bar{f}_{\bar{\tau}})$
       G) $\tau \leftarrow \bar{\tau}$
       H) $r \leftarrow r'$
       I) $h \leftarrow 1$
     ii) **else**
       A) $(f'_h, ..., f'_{\tau+h}) \leftarrow (f'_h, ..., f'_{\tau+h})$
          $- r' (r^{-1})^{[h]} (f_0^{[h]}, ..., f_\tau^{[h]})$
       B) $h \leftarrow h + 1$
 **end**
**output:** $f'_0, ..., f'_{\tau'}$

---

### G. Linearized Vandermonde matrices and GRA

A $q$-power analogy of Vandermonde matrices exists and is described in this section along with a useful property of them. We also derive Gabidulin's recursive algorithm (GRA), which reduces computational complexity of solving a system of equations of size $\tau$ from $\mathcal{O}(\tau^3)$ to $\mathcal{O}(\tau^2)$, by exploiting the $q$-Vandermonde structure [17]. This is an algorithm used heavily for decoding and the pre-encoding of Gabidulin codes.

#### 1) Rank of a linearized Vandermonde matrix

For $\tau \in \mathbb{N}$, such that $\tau \leq n$, let $z_0, ..., z_{\tau-1} \in \mathbb{F}_{q^n}$. For any $v \in \mathbb{N}$, such that $v \leq n$, the matrix $Z \in \mathbb{F}_{q^n}^{\tau \times v}$ of the form

$$Z = \begin{bmatrix} z_0^{[0]} & \cdots & z_0^{[v-1]} \\ \vdots & \ddots & \vdots \\ z_{\tau-1}^{[0]} & \cdots & z_{\tau-1}^{[v-1]} \end{bmatrix} \qquad (S265)$$

is the $\tau \times v$ linearized Vandermonde matrix generated by $z_0, ..., z_{\tau-1}$.

**Lemma 19.** *Let $z_0, ..., z_{\tau-1} \in \mathbb{F}_{q^n}$ be linearly independent over $\mathbb{F}_q$. Then for any $v \in \mathbb{N}$, such that $v \leq n$, the $\tau \times v$ linearized Vandermonde matrix generated by $z_0, ..., z_{\tau-1}$ has rank $\min\{\tau, v\}$, i.e., is full rank.*

*Proof.* See [39, Lemma 3.15]. □

**Supplementary Table X**

**Maximum number of $\mathbb{F}_{q^n}$ operations required by the LBMA to determine the minimal $\mathbb{F}_{q^n}$-LFSR of length $\tau$ $\left(0 \le \tau \le \lfloor \frac{n-k}{2} \rfloor\right)$ for a sequence of length $n - k$.**

| Type | Number of operations | | |
|---|---|---|---|
| | $\tau = 0$ | $0 < 2\tau < n-k$ | $2\tau = n-k$ |
| $+$ | $0$ | $(n-k)\tau - \tau + 1$ | $2\tau^2 - \tau + 1$ |
| $\times$ | $0$ | $(n-k)\tau$ | $2\tau^2$ |
| [1] | $n-k-1$ | $2(n-k)\tau - 2\tau^2 - \tau + 1$ | $2\tau^2$ |
| $-1$ | $0$ | $\tau$ | $\tau$ |
| $-1$ | $0$ | $2\tau$ | $2\tau$ |

*2) GRA*

In decoding of Gabidulin codes, we often need to solve a system of equations for $x_0, \ldots, x_{\tau-1} \in \mathbb{F}_{q^n}$ of the form

$$s_i = \sum_{j=0}^{\tau-1} z_j x_j^{[i]}, \quad i = 0, \ldots, \tau-1 \qquad \text{(S266)}$$

where $s_0, \ldots, s_{\tau-1} \in \mathbb{F}_{q^n}$ and $z_0, \ldots, z_{\tau-1} \in \mathbb{F}_{q^n}$ are linearly independent over $\mathbb{F}_q$. In matrix form, this reads

$$\begin{bmatrix} s_0 & \cdots & s_{\tau-1} \end{bmatrix}$$
$$= \begin{bmatrix} z_0 & \cdots & z_{\tau-1} \end{bmatrix} \begin{bmatrix} x_0^{[0]} & \cdots & x_0^{[\tau-1]} \\ \vdots & \ddots & \vdots \\ x_{\tau-1}^{[0]} & \cdots & x_{\tau-1}^{[\tau-1]} \end{bmatrix}. \qquad \text{(S267)}$$

Let us first see a straightforward method to solve it. We apply the $[i - \tau + 1]$ $q$-power operation to the $(\tau-1-i)$th equation of the system to get the reversed system

$$s_{\tau-1-i}^{[i-\tau+1]} = \sum_{j=0}^{\tau-1} x_j z_j^{[i-\tau+1]}, \quad i = 0, \ldots, \tau-1. \qquad \text{(S268)}$$

In matrix form,

$$\begin{bmatrix} s_{\tau-1}^{[-\tau+1]} & \cdots & s_0^{[0]} \end{bmatrix}$$
$$= \begin{bmatrix} x_0 & \cdots & x_{\tau-1} \end{bmatrix} \begin{bmatrix} z_0^{[-\tau+1]} & \cdots & z_0^{[0]} \\ \vdots & \ddots & \vdots \\ z_{\tau-1}^{[-\tau+1]} & \cdots & z_{\tau-1}^{[0]} \end{bmatrix}. \qquad \text{(S269)}$$

This matrix form also appears in the pre-encoding of the PUSNEC. Notice that $\begin{bmatrix} z_0^{[-\tau+1]} & \cdots & z_0^{[0]} \\ \vdots & \ddots & \vdots \\ z_{\tau-1}^{[-\tau+1]} & \cdots & z_{\tau-1}^{[0]} \end{bmatrix}$ is the $\tau \times \tau$ linearized Vandermonde matrix generated by $z_0^{[-\tau+1]}, \ldots, z_{\tau-1}^{[-\tau+1]}$. Because $z_0, \ldots, z_{\tau-1} \in \mathbb{F}_{q^n}$ are linearly independent over $\mathbb{F}_q$, so are $z_0^{[-\tau+1]}, \ldots, z_{\tau-1}^{[-\tau+1]}$. Therefore by Lemma 19, the $\tau \times \tau$ matrix is full rank, and hence invertible. Thus we can finally determine a unique solution for $x_0, \ldots, x_{\tau-1}$, simply by inverting the $\tau \times \tau$ matrix in Eq. (S269), or by using Gaussian elimination. This standard approach generally requires computational complexity of $\mathcal{O}\left(\tau^3\right)$ over $\mathbb{F}_{q^n}$.

Let us next see that if GRA is used, the complexity can be reduced to $\mathcal{O}\left(\tau^2\right)$ over $\mathbb{F}_{q^n}$, by exploiting the $q$-Vandermonde structure of the involved matrix. Specifically we utilize the two celebrated properties of the $q$-power operations (Subsection VI-A):

- $q$-power can be easily computed by cyclic shifts as Eq. (S156).

- the linearity of $q$-power (Corollary 7).

GRA takes the sequences $z_0, \ldots, z_{\tau-1}$ and $s_0, \ldots, s_{\tau-1}$ as input, and returns the sequence $x_0, \ldots, x_{\tau-1}$ as output. Setting

$$\begin{aligned} Q_i^{(0)} &= s_i, \quad i = 0, \ldots, \tau-1, \\ A_j^{(0)} &= z_j, \quad j = 0, \ldots, \tau-1, \end{aligned}$$

we solve each equation $i = 1, \ldots, \tau-1$

$$Q_i^{(0)} = \sum_{j=0}^{\tau-1} A_j^{(0)} x_j^{[i]}, \qquad \text{(S270)}$$

for $x_0^{[i]}$ to get

$$x_0^{[i]} = A_0^{(0)^{-1}} \left( Q_i^{(0)} - \sum_{j=1}^{\tau-1} A_j^{(0)} x_j^{[i]} \right). \qquad \text{(S271)}$$

We plug the expression for $x_0^{[i]}$ into the $(i-1)$th equation to get

$$\begin{aligned} Q_{i-1}^{(0)} &= A_0^{(0)} \left( A_0^{(0)^{-1}} \left( Q_i^{(0)} - \sum_{j=1}^{\tau-1} A_j^{(0)} x_j^{[i]} \right) \right)^{[-1]} \\ &\quad + \sum_{j=1}^{\tau-1} A_j^{(0)} x_j^{[i-1]}. \end{aligned} \qquad \text{(S272)}$$

This is rearranged to

$$\begin{aligned} Q_{i-1}^{(0)} &- R^{(0)} Q_i^{(0)^{[-1]}} \\ &= \sum_{j=1}^{\tau-1} \left( A_j^{(0)} - R^{(0)} A_j^{(0)^{[-1]}} \right) x_j^{[i-1]}, \end{aligned} \qquad \text{(S273)}$$

where

$$R^{(0)} \equiv A_0^{(0)} \left( A_0^{(0)^{-1}} \right)^{[-1]}. \qquad \text{(S274)}$$

So setting

$$Q_i^{(1)} \equiv Q_i^{(0)} - R^{(0)} Q_{i+1}^{(0)^{[-1]}}, \ i = 0, \ldots, \tau-2 \qquad \text{(S275)}$$

$$A_j^{(1)} \equiv A_j^{(0)} - R^{(0)} A_j^{(0)^{[-1]}}, \ j = 1, \ldots, \tau-1. \qquad \text{(S276)}$$

we arrive at the system of equations

$$Q_i^{(1)} = \sum_{j=1}^{\tau-1} A_j^{(1)} x_j^{[i]}, \quad i = 0, \ldots, \tau-2. \qquad \text{(S277)}$$

Notice that this reduces the problem from a system of $\tau$ equations to a system of the same form, but with only $\tau - 1$ equations, and also the variable $x_0^{[i]}$ eliminated.

We repeat this procedure iteratively to further reduce the number of equations (we call it the forward iteration pass). Suppose that we have the system of $\tau - v$ equations

$$Q_i^{(v)} = \sum_{j=v}^{\tau-1} A_j^{(v)} x_j^{[i]}, \quad i = 0, 1, \ldots, \tau - v - 1. \quad \text{(S278)}$$

We take equations $i = 1, \ldots, \tau - v - 1$ from the above set and solve for $x_v^{[i]}$ to get

$$x_v^{[i]} = A_v^{(v)\,-1} \left( Q_i^{(v)} - \sum_{j=v+1}^{\tau-1} A_j^{(v)} x_j^{[i]} \right). \quad \text{(S279)}$$

We plug the expression for $x_v^{[i]}$ into the $(i-1)$th equation to get

$$
\begin{aligned}
Q_{i-1}^{(v)} &= A_v^{(v)} \left( A_v^{(v)\,-1} \left( Q_i^{(v)} - \sum_{j=v+1}^{\tau-1} A_j^{(v)} x_j^{[i]} \right) \right)^{[-1]} \\
&\quad + \sum_{j=v+1}^{\tau-1} A_j^{(v)} x_j^{[i-1]}.
\end{aligned}
\quad \text{(S280)}
$$

This is rearranged to

$$
\begin{aligned}
Q_{i-1}^{(v)} &- R^{(v)} Q_i^{(v)\,[-1]} \\
&= \sum_{j=v+1}^{\tau-1} \left( A_j^{(v)} - R^{(v)} A_j^{(v)\,[-1]} \right) x_j^{[i-1]}, \quad \text{(S281)}
\end{aligned}
$$

where

$$R^{(v)} \equiv A_v^{(v)} \left( A_v^{(v)\,-1} \right)^{[-1]}. \quad \text{(S282)}$$

So setting

$$
\begin{aligned}
Q_i^{(v+1)} &\equiv Q_i^{(v)} - R^{(v)} Q_{i+1}^{(v)\,[-1]}, \\
&\quad i = 0, 1, \ldots, \tau - 1 - (v+1), \quad \text{(S283)}
\end{aligned}
$$

$$
\begin{aligned}
A_j^{(v+1)} &\equiv A_j^{(v)} - R^{(v)} A_j^{(v)\,[-1]}, \\
&\quad j = v+1, v+2, \ldots, \tau - 1, \quad \text{(S284)}
\end{aligned}
$$

we arrive at

$$
\begin{aligned}
Q_i^{(v+1)} &= \sum_{j=v+1}^{\tau-1} A_j^{(v+1)} x_j^{[i]}, \\
&\quad i = 0, 1, \ldots, \tau - 1 - (v+1). \quad \text{(S285)}
\end{aligned}
$$

After iteration $v = \tau - 2$, we are left with the single equation

$$Q_0^{(\tau-1)} = A_{\tau-1}^{(\tau-1)} x_{\tau-1}^{[0]} \quad \text{(S286)}$$

allowing us to solve for $x_{\tau-1}$ by setting

$$x_{\tau-1} = A_{\tau-1}^{(\tau-1)\,-1} Q_0^{(\tau-1)}. \quad \text{(S287)}$$

We can then use our expressions for $x_v^{[0]}$ to proceed backward iteratively for $v = \tau - 2, \ldots, 0$ to get

$$x_v = A_v^{(v)\,-1} \left( Q_0^{(v)} - \sum_{j=v+1}^{\tau-1} A_j^{(v)} x_j \right). \quad \text{(S288)}$$

GRA and its complexity are summarized in Supplementary Algorithm 5 and Supplementary Table XI. The complexity of GRA is as follows: For each forward iteration $v = 0, \ldots, \tau - 2$, it is necessary to compute $A_v^{(v)\,-1}$ from $A_v^{(v)}$ requiring an $\mathbb{F}_{q^n}$-inversion. Computing

---

**Supplementary Algorithm 5** Gabidulin's recursive algorithm (GRA)

**input:**
$z_0, \ldots, z_{\tau-1} \in \mathbb{F}_{q^n}$ linearly independent over $\mathbb{F}_q$
$s_0, \ldots, s_{\tau-1} \in \mathbb{F}_{q^n}$

**begin:**
1) **for** each $i$ from 0 to $\tau - 1$ **do**
$\quad A_i^{(0)} \leftarrow z_i$
2) $S^{(0)} \leftarrow \left( A_0^{(0)} \right)^{-1}$
3) $R^{(0)} \leftarrow A_0^{(0)} S^{(0)\,[-1]}$
4) **for** each $i$ from 0 to $\tau - 1$ **do**
$\quad Q_i^{(0)} \leftarrow s_i$
5) **for** each $j$ from 1 to $\tau - 1$ **do**
$\quad$ a) **for** each $i$ from $j$ to $\tau - 2$ **do**
$\qquad A_i^{(j)} \leftarrow A_i^{(j-1)} - R^{(j-1)} \left( A_i^{(j-1)} \right)^{[-1]}$
$\quad$ b) $S^{(j)} \leftarrow \left( A_j^{(j)} \right)^{-1}$
$\quad$ c) $R^{(j)} \leftarrow A_j^{(j)} S^{(j)\,[-1]}$
$\quad$ d) **for** each $i$ from 0 to $\tau - 1 - j$ **do**
$\qquad Q_i^{(j)} \leftarrow Q_i^{(j-1)} - R^{(j-1)} \left( Q_{i+1}^{(j-1)} \right)^{[-1]}$
6) $A_{\tau-1}^{(\tau-1)} \leftarrow A_{\tau-1}^{(\tau-2)} - R^{(\tau-2)} \left( A_{\tau-1}^{(\tau-2)} \right)^{[-1]}$
7) $S^{(\tau-1)} \leftarrow \left( A_{\tau-1}^{(\tau-1)} \right)^{-1}$
8) $Q_0^{(\tau-1)} \leftarrow Q_0^{(\tau-2)} - R^{(\tau-2)} \left( Q_1^{(\tau-2)} \right)^{[-1]}$
9) $x_{\tau-1} \leftarrow S^{(\tau-1)} Q_0^{(\tau-1)}$
10) **for** each $j$ from 1 to $\tau - 1$ **do**
$\quad x_{\tau-1-j} \leftarrow S^{(\tau-1-j)} \left( Q_0^{(\tau-1-j)} \right.$
$\qquad \left. - \sum_{v=0}^{j-1} A_{\tau-1-v}^{(\tau-1-j)} x_{\tau-1-v} \right)$

**end**
**output:** $x_0, \ldots, x_{\tau-1}$

---

$R^{(v)}$ from $A_v^{(v)}$ and $A_v^{(v)\,-1}$ requires a $\mathbb{F}_{q^n}$-multiplication and a $[-1]$ shift. Computing $A_{v+1}^{(v+1)}, \ldots, A_{\tau-1}^{(v+1)}$ from $R^{(v)}$ and $A_{v+1}^{(v)}, \ldots, A_{\tau-1}^{(v)}$ requires $(\tau - v - 1)$ $\mathbb{F}_{q^n}$-multiplications, $(\tau - v - 1)$ $\mathbb{F}_{q^n}$-subtractions and $(\tau - v - 1)$ $[-1]$ shifts. Computing $Q_0^{(v+1)}, \ldots, Q_{\tau-v-2}^{(v+1)}$ from $R^{(v)}$ and $Q_0^{(v)}, \ldots, Q_{\tau-v-1}^{(v)}$ requires $(\tau - v - 1)$ $\mathbb{F}_{q^n}$-multiplications, $(\tau - v - 1)$ $\mathbb{F}_{q^n}$-subtractions and $(\tau - v - 1)$ $[-1]$ shifts. Computing $A_{\tau-1}^{(\tau-1)\,-1}$ from $A_{\tau-1}^{(\tau-1)}$ requires a $\mathbb{F}_{q^n}$-inversion. Computing $d_{\tau-1}$ from $Q_0^{(\tau-1)}$ and $A_{\tau-1}^{(\tau-1)\,-1}$ requires a $\mathbb{F}_{q^n}$-multiplication. For each backward iteration $v = \tau - 2, \ldots, 0$, computing $x_v$ from $Q_0^{(v)}$, $A_v^{(v)\,-1}$, $A_{v+1}^{(v)}, \ldots, A_{\tau-1}^{(v)}$ and $x_{v+1}, \ldots, x_{\tau-1}$ requires $(\tau - v)$ $\mathbb{F}_{q^n}$-multiplications and $(\tau - v - 1)$ $\mathbb{F}_{q^n}$-subtractions. In total, the GRA requires $\left( \frac{3}{2}\tau^2 + \frac{1}{2}\tau - 1 \right)$ $\mathbb{F}_{q^n}$-multiplications, $\left( \frac{3}{2}\tau^2 - \frac{3}{2}\tau \right)$ $\mathbb{F}_{q^n}$-subtractions, $(\tau^2 - 1)$ $[-1]$ shifts and $\tau$ $\mathbb{F}_{q^n}$-inversions.

**Supplementary Table XI**
**Arithmetic operations required for GRA.**

| Type | $\mathbb{F}_{q^n}$ operations |
|------|------------------------------|
| $-$ | $\frac{3}{2}\tau^2 - \frac{3}{2}\tau$ |
| $\times$ | $\frac{3}{2}\tau^2 + \frac{1}{2}\tau - 1$ |
| $[-1]$ | $\tau^2 - 1$ |
| $-1$ | $\tau$ |

### H. Roots of polynomial synthesized by symbolic product

In the common-error-location aware decoder for iGab$[n, k]$, the ELP common to all the components of interleaving is determined by CPSLBMA (Subsubsection V-B2). In CPSLBMA, ELPs are synthesized recursively, component to component, by the symbolic product of linearized polynomials (e.g., Eq. (S134)).

In this subsection, we provide the rationale for this process. More specifically, we explain why the modified syndromes and the updating of linearized polynomial synthesized by symbolic product can contain the expected roots. Namely, the roots of the updated polynomial contain all roots that can be obtained from the reversed syndromes in each component used up to that iterative step. We describe the step in the 0th component and the 1st component in detail and generalize the process.

We quickly review the 0th component. Given the reversed syndromes for each $i = 0, \ldots, n - k - 1$,

$$\tilde{s}_i^{(0)} = \sum_{v=0}^{\tau-1} d_v \left(a_v^{(0)}\right)^{[i-n+k+1]}, \quad (S289)$$

the LBMA will solve the key equation and return the shortest $\mathbb{F}_{q^n}$-LFSR of length $\tau_0$ ($0 \leq \tau_0 \leq \tau$), i.e., $\lambda_0^{(0)}, \ldots, \lambda_{\tau_0}^{(0)} \in \mathbb{F}_{q^n}$, determining the ELP for the 0th component

$$\Lambda^{(0)}(x) = \sum_{v=0}^{\tau_0} \lambda_v^{(0)} x^{[v]}. \quad (S290)$$

Without loss of generality, let a basis of rootspace of $\Lambda^{(0)}(x)$ be $d_{\tau-\tau_0}, \ldots, d_{\tau-1}$, which will be the last $\tau_0$ out of $\tau$ transformed error locations (if necessary, we can just re-order them), meaning that for $v = \tau - \tau_0, \ldots, \tau - 1$,

$$\Lambda^{(0)}(d_v) = 0. \quad (S291)$$

Now we come to the 1st component. Again, we have $n - k$ reversed syndromes $\hat{s}_0^{(1)}, \ldots, \hat{s}_{n-k-1}^{(1)}$. At here we don't directly use them in LMBA, instead, we firstly apply $\Lambda^{(0)}(x)$ obtained from the 0th component to the reversed syndromes from the 1st component to define the modified syndromes as

$$\hat{s}_i^{(1)} = \sum_{j=0}^{\tau_0} \lambda_j^{(0)} \left(\tilde{s}_{i+\tau_0-j}^{(1)}\right)^{[j]}, \quad (S292)$$

for $i = 0, \ldots, n - k - 1 - \tau_0$. Notice by this definition, we obtain the modified syndromes $\hat{s}_0^{(1)}, \ldots, \hat{s}_{n-k-1-\tau_0}^{(1)}$, and the number reduces from $n - k$ to $n - k - \tau_0$.

Here we explain why we do this. Notice that

$$
\begin{aligned}
\hat{s}_i^{(1)} &= \sum_{j=0}^{\tau_0} \lambda_j^{(0)} \left(\tilde{s}_{i+\tau_0-j}^{(1)}\right)^{[j]} \\
&= \sum_{j=0}^{\tau_0} \lambda_j^{(0)} \left(\sum_{v=0}^{\tau-1} d_v \left(a_v^{(1)}\right)^{[i+\tau_0-j-n+k+1]}\right)^{[j]} \\
&= \sum_{j=0}^{\tau_0} \lambda_j^{(0)} \sum_{v=0}^{\tau-1} d_v^{[j]} \left(a_v^{(1)}\right)^{[i+\tau_0-j-n+k+1+j]} \\
&= \sum_{v=0}^{\tau-1} \sum_{j=0}^{\tau_0} \lambda_j^{(0)} d_v^{[j]} \left(a_v^{(1)}\right)^{[i+\tau_0-n+k+1]} \\
&= \sum_{v=0}^{\tau-1} \Lambda^{(0)}(d_v) \left(a_v^{(1)}\right)^{[i+\tau_0-n+k+1]}. \quad (S293)
\end{aligned}
$$

Since we know for $v = \tau - \tau_0, \ldots, \tau - 1$,

$$\Lambda^{(0)}(d_v) = 0, \quad (S294)$$

which implies

$$\hat{s}_i^{(1)} = \sum_{v=0}^{\tau-\tau_0-1} \Lambda^{(0)}(d_v) \left(a_v^{(1)}\right)^{[i+\tau_0-n+k+1]}. \quad (S295)$$

Compare Eq. (S289) and Eq. (S295). The former has factor $d_v$ and the summation is done from 0 to $\tau - 1$, while the later has factor $\Lambda^{(0)}(d_v)$ and the summation is done from 0 to $\tau - \tau_0 - 1$. We should then know, if we input the modified syndromes $\hat{s}_0^{(1)}, \ldots, \hat{s}_{n-k-1-\tau_0}^{(1)}$ into the LBMA, it will output the minimal polynomial

$$\Lambda'^{(1)}(x) = \sum_{j=0}^{\tau_1'} \lambda_j'^{(1)} x^{[j]} \quad (S296)$$

with polynomial coefficients $\lambda_0'^{(1)}, \ldots, \lambda_{\tau_1'}'^{(1)} \in \mathbb{F}_{q^n}$ for some $0 \leq \tau_1' \leq \tau - \tau_0$. In particular, $\Lambda^{(0)}(d_{\tau-\tau_0-\tau_1'}), \ldots, \Lambda^{(0)}(d_{\tau-\tau_0-1})$ (note that it is not $d_{\tau-\tau_0-\tau_1'}, \ldots, d_{\tau-\tau_0-1}$) will be roots of $\Lambda'^{(1)}(x)$, meaning that for $v = \tau - \tau_0 - \tau_1', \ldots, \tau - \tau_0 - 1$,

$$\Lambda'^{(1)}\left(\Lambda^{(0)}(d_v)\right) = 0. \quad (S297)$$

On the other hand, from the result of the 0th component, we know for $v = \tau - \tau_0, \ldots, \tau - 1$,

$$
\begin{aligned}
\Lambda'^{(1)}\left(\Lambda^{(0)}(d_v)\right) &= \Lambda'^{(1)}(0) \\
&= 0. \quad (S298)
\end{aligned}
$$

It can be observed that both of left side of Eq. (S297) and Eq. (S298) are essentially exact symbolic product, where the former corresponds to the LBMA results of the 1st component, and the later corresponds to the LBMA results of the 0th component. Hence if we update the minimal polynomial based on symbolic product operation, namely,

$$\Lambda^{(1)}(x) = \Lambda'^{(1)}\left(\Lambda^{(0)}(x)\right), \quad (S299)$$

then the degree of the updated polynomial $\Lambda^{(1)}(x)$ is $\deg(\Lambda^{(1)}(x)) = \tau_1 = \tau_0 + \tau_1'$, and $d_{\tau-\tau_0-\tau_1'}, \ldots, d_{\tau-1}$ will be roots of $\Lambda^{(1)}(x)$. It means that for $v = \tau - \tau_1, \ldots, \tau - 1$,

$$\Lambda^{(1)}(d_v) = 0. \quad (S300)$$

Now, we can see the roots of the updated polynomial $\Lambda^{(1)}(x)$ contains all roots that can be obtained from the reversed syndromes in the 0th component and the 1st component.

In general, the modified syndromes for the $w$th component are given by

$$\hat{\tilde{s}}_i^{(w)} = \sum_{j=0}^{\tau_{w-1}} \lambda_j^{(w-1)} \left(\tilde{s}_{i+\tau_{w-1}-j}^{(w)}\right)^{[j]}, \qquad (S301)$$

for each $i = 0, \ldots, n-k-1-\tau_{w-1}$. Once again, we notice that

$$
\begin{aligned}
\hat{\tilde{s}}_i^{(w)} &= \sum_{j=0}^{\tau_{w-1}} \lambda_j^{(w-1)} \left(\tilde{s}_{i+\tau_{w-1}-j}^{(w)}\right)^{[j]} \\
&= \sum_{j=0}^{\tau_{w-1}} \lambda_j^{(w-1)} \left(\sum_{v=0}^{\tau-1} d_v \left(a_v^{(w)}\right)^{[i+\tau_{w-1}-j-n+k+1]}\right)^{[j]} \\
&= \sum_{j=0}^{\tau_{w-1}} \lambda_j^{(w-1)} \sum_{v=0}^{\tau-1} d_v^{[j]} \left(a_v^{(w)}\right)^{[i+\tau_{w-1}-j-n+k+1+j]} \\
&= \sum_{v=0}^{\tau-1} \sum_{j=0}^{\tau_{w-1}} \lambda_j^{(w-1)} d_v^{[j]} \left(a_v^{(w)}\right)^{[i+\tau_{w-1}-n+k+1]} \\
&= \sum_{v=0}^{\tau-1} \Lambda^{(w-1)}(d_v) \left(a_v^{(w)}\right)^{[i+\tau_{w-1}-n+k+1]}. \qquad (S302)
\end{aligned}
$$

Since we already know for each $v = \tau - \tau_{w-1}, \ldots, \tau-1$,

$$\Lambda^{(w-1)}(d_v) = 0, \qquad (S303)$$

meaning that

$$\hat{\tilde{s}}_i^{(w)} = \sum_{v=0}^{\tau-\tau_{w-1}-1} \Lambda^{(w-1)}(d_v) \left(a_v^{(w)}\right)^{[i+\tau_{w-1}-n+k+1]}. \qquad (S304)$$

So, if we input the modified syndromes $\hat{\tilde{s}}_0^{(w)}, \ldots, \hat{\tilde{s}}_{n-k-1-\tau_{w-1}}^{(w)}$ into the LBMA, it will output the minimal polynomial

$$\Lambda'^{(w)}(x) = \sum_{j=0}^{\tau_w'} \lambda'_j^{(w)} x^{[j]} \qquad (S305)$$

with polynomial coefficients $\lambda'_0^{(w)}, \ldots, \lambda'_{\tau_w'}^{(w)} \in \mathbb{F}_{q^n}$ for some $0 \leq \tau_w' \leq \tau - \tau_{w-1}$. In particular, $\Lambda^{(w-1)}(d_{\tau-\tau_{w-1}-\tau_w'}), \ldots, \Lambda^{(w-1)}(d_{\tau-\tau_{w-1}-1})$ (note that it is not $d_{\tau-\tau_{w-1}-\tau_w'}, \ldots, d_{\tau-\tau_{w-1}-1}$) will be roots of $\Lambda'^{(w)}$, meaning that for $v = \tau - \tau_{w-1} - \tau_w', \ldots, \tau-\tau_{w-1}-1$,

$$\Lambda'^{(w)}\left(\Lambda^{(w-1)}(d_v)\right) = 0. \qquad (S306)$$

Notice also that for $v = \tau - \tau_{w-1}, \ldots, \tau-1$,

$$
\begin{aligned}
\Lambda'^{(w)}\left(\Lambda^{(w-1)}(d_v)\right) &= \Lambda'^{(w)}(0) \\
&= 0. \qquad (S307)
\end{aligned}
$$

Hence if we update the minimal polynomial to

$$\Lambda^{(w)}(x) = \Lambda'^{(w)}\left(\Lambda^{(w-1)}(x)\right), \qquad (S308)$$

then $\deg(\Lambda^{(w)}(x)) = \tau_w = \tau_{w-1} + \tau_w' = \tau_0 + \tau_1' + \ldots + \tau_w'$, and $d_{\tau-\tau_0-\tau_1'-\ldots-\tau_w'}, \ldots, d_{\tau-1}$ will be roots of $\Lambda^{(w)}(x)$. It means that for $v = \tau - \tau_w, \ldots, \tau-1$,

$$\Lambda^{(w)}(d_v) = 0. \qquad (S309)$$

Therefore, similar to the 1st component, the roots of the updated polynomial contain all roots that can be obtained from the syndromes in each component used up to $w$th component.

In Subsubsection V-B2, in particular, with regard to Eq. (S134), we have mentioned that $\Lambda^{(1)}(x)$ has $\tau_1 = \tau_0 + \tau_1'$ roots in which $\tau_0$ roots are the roots of $\Lambda^{(0)}(x)$, namely the $\tau_0$ transformed error locations of the syndrome which was input to the 0th component. The remaining $\tau_1'$ roots are the $\tau_1'$ transformed error locations of the modified syndrome calculated from the syndrome input to the 1st component. This subsection further detailed the roots of polynomial synthesized by symbolic product, and the evidence of which was given here.

## SUPPLEMENTARY REFERENCES

[1] N. Cai and R. W. Yeung, "Secure network coding on a wiretap network," *IEEE Trans. Inf. Theory*, vol. 57, no. 1, pp. 424–435, Jan. 2011.

[2] S. El Rouayheb, E. Soljanin, and A. Sprintson, "Secure network coding for wiretap networks of type II," *IEEE Trans. Inf. Theory*, vol. 58, no. 3, pp. 1361–1371, Mar. 2012.

[3] A. Shamir, "How to share a secret," *Commun. ACM*, vol. 22, no. 11, pp. 612–613, Nov. 1979.

[4] G. R. Blakley and C. Meadows, "Security of ramp schemes," in *Proc. CRYPTO '84*. Springer-Verlag, Aug. 1985, pp. 242–268.

[5] R. J. McEliece and D. V. Sarwate, "On sharing secrets and Reed-Solomon codes," *Commun. ACM*, vol. 24, no. 9, pp. 583–584, Sep. 1981.

[6] H. Yamamoto, "Secret sharing system using $(k, l, n)$ threshold scheme," *Electron. Commun. Jpn. 1*, vol. 69, no. 9, pp. 46–54, Sep. 1986.

[7] M. Iwamoto and H. Yamamoto, "Strongly secure ramp secret sharing schemes for general access structures," *Inf. Process. Lett.*, vol. 97, no. 2, pp. 52–57, June 2006.

[8] M. Nishiara and K. Takizawa, "Strongly secure secret sharing scheme with ramp threshold based on Shamir's polynomial interpolation scheme," *IEICE Trans. Fundam. Electron. Commun. Comput. Sci.*, vol. J92, no. 12, pp. 1009–1013, Dec. 2009.

[9] K. Bhattad and K. R. Narayanan, "Weakly secure network coding," *Proc. NetCod*, vol. 104, pp. 8–20, Apr. 2005.

[10] D. Silva and F. R. Kschischang, "Universal weakly secure network coding," in *Proc. IEEE Inf. Theory Workshop Netw. Inf. Theory*, Volos, Greece, June 2009, pp. 281–285.

[11] K. Harada and H. Yamamoto, "Strongly secure linear network coding," *IEICE Trans. Fundam. Electron., Commun. Comput. Sci.*, vol. E91-A, no. 10, pp. 2720–2728, Oct. 2008.

[12] D. Silva and F. R. Kschischang, "Universal secure network coding via rank-metric codes," *IEEE Trans. Inf. Theory*, vol. 57, no. 2, pp. 1124–1135, Feb. 2011.

[13] J. Kurihara, R. Matsumoto, and T. Uyematsu, "Relative generalized rank weight of linear codes and its applications to network coding," *IEEE Trans. Inf. Theory*, vol. 61, no. 7, pp. 3912–3936, July 2015.

[14] P. Loidreau and R. Overbeck, "Decoding rank errors beyond the error correcting capability," in *Proc. 10th Int. Workshop Algebraic Combin. Coding Theory (ACCT)*, Zvenigorod, Russia, Sep. 2006, pp. 186–190.

[15] A. Wachter-Zeh and A. Zeh, "List and unique error-erasure decoding of interleaved Gabidulin codes with interpolation techniques," *Des. Codes Cryptogr.*, vol. 73, no. 2, pp. 547–570, Nov. 2014.

[16] D. Silva, F. R. Kschischang, and R. Koetter, "A rank-metric approach to error control in random network coding," *IEEE Trans. Inf. Theory*, vol. 54, no. 9, pp. 3951–3967, Sep. 2008.

[17] E. M. Gabidulin, "Theory of codes with maximum rank distance," *Probl. Inf. Transm.*, vol. 21, no. 1, pp. 1–12, Mar. 1985.

[18] M. Gadouleau and Z. Yan, "Complexity of decoding Gabidulin codes," in *Proc. 2008 42nd Annu. Conf. Inf. Sci. Syst.*, Princeton, NJ, USA, Mar. 2008, pp. 1081–1085.

[19] R. Mullin, I. Onyszchuk, S. Vanstone, and R. Wilson, "Optimal normal bases in $GF(p^n)$," *Discrete Appl. Math.*, vol. 22, no. 2, pp. 149–161, Dec. 1988.

[20] D. W. Ash, I. F. Blake, and S. A. Vanstone, "Low complexity normal bases," *Discrete Appl. Math.*, vol. 25, pp. 191–210, Nov. 1989.

[21] M. Christopoulou, T. Garefalakis, D. Panario, and D. Thomson, "Gauss periods as constructions of low complexity normal bases," *Des. Codes Cryptogr.*, vol. 62, no. 1, pp. 43–62, Jan. 2012.

[22] S. Gao, "Normal bases over finite fields," Ph.D. dissertation, University of Waterloo, Waterloo, Canada, 1993.

[23] T. S. Han, H. Endo, and M. Sasaki, "Reliability and secrecy functions of the wiretap channel under cost constraint," *IEEE Trans. Inf. Theory*, vol. 60, no. 11, pp. 6819–6843, Nov. 2014.

[24] J. Hou and G. Kramer, "Effective secrecy: Reliability, confusion and stealth," in *Proc. IEEE Int. Symp. Inf. Theory*, Honolulu, Hawaii, USA, Jun./Jul. 2014, pp. 601–605.

[25] "3D View of Multicast Graphs in Space Networks," https://snv.rocketworks.co.jp/3dview/.

[26] satellitemap.space, https://satellitemap.space/index.html.

[27] G. Richter and S. Plass, "Error and erasure decoding of rank-codes with a modified Berlekamp-Massey algorithm," in *Proc. Int. ITG Conf. Source Channel Coding (SCC)*, Erlangen, Germany, Jan. 2004, pp. 249–256.

[28] P. Loidreau, "A Welch–Berlekamp like algorithm for decoding Gabidulin codes," in *Coding and Cryptography*, Ø. Ytrehus, Ed., no. 3969. Berlin, Heidelberg: Springer, June 2006, pp. 36–45.

[29] D. Silva and F. R. Kschischang, "Fast encoding and decoding of Gabidulin codes," in *Proc. IEEE Int. Symp. Inf. Theory (ISIT)*, Seoul, Korea, June 2009, pp. 2858–2862.

[30] A. Wachter-Zeh, V. B. Afanassiev, and V. R. Sidorenko, "Fast decoding of Gabidulin codes," *Des. Codes Cryptogr.*, vol. 66, no. 1-3, pp. 57–73, Jan. 2013.

[31] S. Puchinger and A. Wachter-Zeh, "Sub-quadratic decoding of Gabidulin codes," in *Proc. IEEE Int. Symp. Inf. Theory (ISIT)*, Barcelona, Spain, July 2016, pp. 2554–2558.

[32] V. Sidorenko and M. Bossert, "Decoding interleaved Gabidulin codes and multisequence linearized shift-register synthesis," in *Proc. IEEE Int. Symp. Inf. Theory (ISIT)*, Austin, TX, USA, June 2010, pp. 1148–1152.

[33] H. Bartz, T. Jerkovits, S. Puchinger, and J. S. Rosenkilde, "Fast root finding for interpolation-based decoding of interleaved Gabidulin codes," in *Proc. 2019 IEEE Information Theory Workshop (ITW)*, Visby, Sweden, Aug. 2019, pp. 1–5.

[34] J. Kunz, J. Renner, G. Maringer, T. Schamberger, and A. Wachter-Zeh, "On software implementation of Gabidulin decoders," in *Proc. Int. Workshop Algebraic Combin. Coding Theory (ACCT)*, Albena, Bulgaria, Oct. 2020, pp. 95–101.

[35] D. Augot, P. Loidreau, and G. Robert, "Generalized Gabidulin codes over fields of any characteristic," *Des. Codes Cryptogr.*, vol. 86, no. 8, pp. 1807–1848, Aug. 2018.

[36] V. Sidorenko, G. Richter, and M. Bossert, "Linearized shift-register synthesis," *IEEE Trans. Inf. Theory*, vol. 57, no. 9, pp. 6025–6032, Sep. 2011.

[37] J. Massey, "Shift-register synthesis and BCH decoding," *IEEE Trans. Inf. Theory*, vol. 15, no. 1, pp. 122–127, Jan. 1969.

[38] K. Imamura and W. Yoshida, "A simple derivation of the Berlekamp-Massey algorithm and some applications (corresp.)," *IEEE Trans. Inf. Theory*, vol. 33, no. 1, pp. 146–150, Jan. 1987.

[39] R. Lidl and H. Niederreiter, *Finite Fields, ser. Encyclopedia of Mathematics and its Applications*. USA: Cambridge University Press, 1996.